\documentclass[12pt,titlepage]{article}
\renewcommand{\theequation}{\thesection.\arabic{equation}}
\renewcommand{\to}{\rightarrow}
\newcommand{\al}{\alpha}
\newcommand{\beq}{\begin{equation}}
\newcommand{\eeq}{\end{equation}}
\newcommand{\bea}{\begin{eqnarray}}
\newcommand{\eea}{\end{eqnarray}}
\newcommand{\eq}{\begin{equation}}
\newcommand{\en}{\end{equation}}
\newcommand{\eqa}{\begin{eqnarray}}
\newcommand{\ena}{\end{eqnarray}}
\newcommand{\rw}{\rightarrow}

\newcommand{\ti}{\tilde}
\def\e{{\rm e}}
\def\ln{{\rm ln}}
\def\tr{{\rm tr}}

\newcommand{\PR}[1]{Phys.\ Rev.\ {\bf #1}}
\newcommand{\PRL}[1]{Phys.\ Rev.\ Lett.\ {\bf #1}}

\newcommand{\TMP}[1]{Teor.\ Mat.\ Fiz.\ {\bf #1}}
\begin{document}
\thispagestyle{empty}
\begin{titlepage}
\addtolength{\baselineskip}{.7mm}
\thispagestyle{empty}
\begin{flushright}
DFTT 10/03\\
\end{flushright}
\vspace{10mm}
\begin{center}
{\large 
{\bf Random matrix theory and symmetric spaces
}}\\[15mm]
{
\bf 
M.~Caselle$^1$ and U.~Magnea$^2$ 
} \\
\vspace{5mm}
{\it $^1$Department of Theoretical Physics, University of Torino \\
and INFN, Sez. di Torino\\
Via P. Giuria 1, I-10125 Torino, Italy \\ 
$^2$Department of Mathematics, University of Torino \\ 
Via Carlo Alberto 10, I-10123 Torino, Italy}\\
caselle@to.infn.it \\
magnea@dm.unito.it
\\[6mm]
\vspace{13mm}
{\bf Abstract}\\[5mm]
\end{center}

In this review we discuss the relationship between random matrix
theories and symmetric spaces. We show that the integration manifolds
of random matrix theories, the eigenvalue distribution, and the Dyson and
boundary indices characterizing the ensembles are in strict
correspondence with symmetric spaces and the intrinsic characteristics
of their restricted root lattices. Several important results can be obtained
from this identification. In particular the Cartan classification of triplets of 
symmetric spaces with
positive, zero and negative curvature gives rise to a new
classification of random matrix ensembles.
The review is organized into two main parts. 
In Part~I the theory of symmetric spaces is
reviewed with particular emphasis on the ideas relevant for
appreciating the correspondence with random matrix theories.  
In Part~II we discuss various applications of symmetric spaces to random matrix
theories and in particular the new
classification of disordered systems derived from the classification
of symmetric spaces. We also review how the mapping from integrable
Calogero--Sutherland models to symmetric spaces can be used in the
theory of random matrices, with particular consequences for quantum
transport problems. We conclude indicating some interesting new directions of
research based on these identifications.

\end{titlepage}

\newpage
\setcounter{footnote}{0}

\tableofcontents
\section{Introduction}
\label{sec-Intro}
\setcounter{equation}{0}

The study of symmetric spaces has recently attracted interest in
various branches of physics, ranging from condensed matter physics to
lattice QCD.  This is mainly due to the gradual understanding during
the past few years of the deep connection between random matrix
theories and symmetric spaces. Indeed, this connection is a rather old
intuition, which traces back to Dyson \cite{DysonSS} and has
subsequently been pursued by several authors, notably by
H\"uffmann~\cite{Huff}.  Recently it has led to several interesting
results, like for instance a tentative classification of the
universality classes of disordered systems. The latter topic is the
main subject of this review.

The connection between random matrix theories and symmetric spaces is
obtained simply through the coset spaces defining the symmetry classes
of the random matrix ensembles.  Although Dyson was the first to
recognize that these coset spaces are symmetric spaces, the subsequent
emergence of new random matrix symmetry classes and their
classification in terms of Cartan's symmetric spaces is relatively
recent \cite{AltZ,Zirn,MCclass,TitBrou,Ivanov}.  Since symmetric
spaces are rather well understood mathematical objects, the main
outcome of such an identification is that several non--trivial results
concerning the behavior of the random matrix models, as well as the
physical systems that these models are expected to describe, can be
obtained.

In this context an important tool, that will be discussed in the
following, is a class of integrable models named Calogero--Sutherland
models~\cite{cs}. In the early eighties, Olshanetsky and Perelomov
showed that also these models are in one--to--one correspondence with
symmetric spaces through the reduced root systems of the latter
\cite{OlshPere}.  Thanks to this chain of identifications (random
matrix ensemble -- symmetric space -- Calogero--Sutherland model)
several of the results obtained in the last twenty years within the
framework of Calogero--Sutherland models can also be applied to random
matrix theories.

The aim of this review is to allow the reader to follow this chain of
correspondences.  To this end we will devote the first half of the
paper (sections \ref{sec-Lie} through \ref{sec-intmod}) to the
necessary mathematical background and the second part (sections
\ref{sec-RMT} through \ref{sec-beyond}) to the applications in 
random matrix theory.  In particular, in the last section we discuss
some open directions of research. The reader who is not
interested in the mathematical background could skip the first part
and go directly to the later sections where we list and discuss the
main results.

This review is organized as follows:

The first five sections of Part I (sections 
\ref{sec-Lie}--\ref{sec-Operators}) are devoted to an elementary
introduction to symmetric spaces.  As mentioned in the Abstract, these
sections consist of the material presented in \cite{SSmath}, which is
a self--contained introductory review of symmetric spaces from a
mathematical point of view. The material on symmetric spaces
should be accessible to physicists with only elementary background in
the theory of Lie groups. We have included quite a few examples to
illustrate all aspects of the material.  In the last section of
Part~I, section \ref{sec-intmod}, we briefly introduce the
Calogero--Sutherland models with particular emphasis on their
connection with symmetric spaces.
  
After this introductory material we then move on in Part~II to random
matrix theories and their connection with symmetric spaces (section
\ref{sec-RMT}).  Let us stress that this paper is not intended as an
introduction to random matrix theory, for which very good and thorough
references already exist \cite{Mehta,Beenakker,revVerb,GMW,JPA}. In this
review we will assume that the reader is already acquainted with the
topic, and we will only recall some basic information (definitions of
the various ensembles, main properties, and main physical
applications). The main goal of this section is instead to discuss the
identifications that give rise to the close relationship between
random matrix ensembles and symmetric spaces.
  
Section \ref{sec-appl} is devoted to a discussion of some of the
consequences of the above mentioned identifications. In particular we
will deduce, starting from the Cartan classification of symmetric
spaces, the analogous classification of random matrix ensembles. We
discuss the symmetries of the ensembles in terms of the underlying
restricted root system, and see how the orthogonal polynomials
belonging to a certain ensemble are determined by the root
multiplicities.  In this section we also give some examples of how the
connection between random matrix ensembles on the one hand, and
symmetric spaces and Calogero--Sutherland models on the other hand,
can be used to obtain new results in the theoretical description of
physical systems, more precisely in the theory of quantum
transport. 

The last section of the paper is devoted to some new results that
show that the mathematical tools discussed in this paper (or suitable
generalizations of these) can be useful for going beyond the symmetric
space paradigm, and to explore some new connections between random
matrix theory, group theory, and differential geometry. Here we discuss 
clustered solutions of the Dorokhov--Mello--Pereyra--Kumar equation, and
then we go on to discuss the most general Calogero--Sutherland
potential, given by the Weierstrass ${\cal P}$--function, and show
that it covers the three cases of symmetric spaces of positive, zero
and negative curvature.  
Finally, in the appendix we discuss some intriguing
exact results for the so called zonal spherical functions, which not only
play an
important role in our discussion, but are also of great relevance in several
other branches of physics.

There are some important and interesting topics that we will not
review because of lack of space and competence.  For these we refer
the reader to the existing literature. In particular we shall not
discuss:

\begin{itemize}
  
\item the supersymmetric approach to random matrix theories and in
  particular their classification in terms of supersymmetric spaces.
  Here we refer the reader to the original paper by M. Zirnbauer
  \cite{Zirn}, while a good introduction to the use of supersymmetry
  in random matrix theory and a complete set of the relevant references
  can be found in \cite{useSusy};

\item the very interesting topic of phase transitions. For this we refer to the
 recent and thorough review by G. Cicuta~\cite{Cicuta:2000};

\item the extension to two--dimensional models of the classification 
 of symmetric spaces, and more generally
 the methods of symmetric space analysis~\cite{2Dmodels};

\item the generalization of the classification of symmetric spaces 
 to non--hermitean random matrices~\cite{bl2001} 
 (see however a discussion in the concluding section \ref{sec-concl});

\item the so called q--ensembles~\cite{MCI};

\item the two--matrix models \cite{pm83} and multi--matrix 
 models \cite{ey98} and their continuum limit generalization.

\end{itemize}


The last item in the list given above is a very interesting topic,
which has several physical applications and would indeed deserve a
separate review. The common feature of these two-- and multi--matrix
models which is of relevance for the present review, is that they all
can be mapped onto suitably chosen Calogero--Sutherland systems. These
models represent a natural link to two classes of matrix theories
which are of great importance in high energy physics: on the one hand,
the matrix models describing two--dimensional quantum gravity
(possibly coupled to matter)~\cite{AmQG}, and on the other hand, the matrix models
pertaining to large $N$ QCD, which trace back to the original seminal
works of 't Hooft \cite{h74}.  In particular, a direct and explicit
connection exists between multi--matrix models (the so called
Kazakov--Migdal models) for large $N$ QCD \cite{km93} and the exactly
solvable models of two--dimensional QCD on the
lattice~\cite{rus90}. 

The mapping of these models to Calogero--Sutherland systems of the
type discussed in this review can be found for instance
in~\cite{camp93}.  The relevance of these models, and in particular of
their Calogero--Sutherland mappings, for the condensed matter systems
like those discussed in the second part of this review, was first
discussed in~\cite{sla94}. A recent review on this aspect, and more
generally on the use of Calogero--Sutherland models for
low-dimensional models, can be found in~\cite{pol98}.

We will necessarily be rather sketchy in discussing the many important
physical applications of the random matrix ensembles to be described
in section \ref{sec-RMT}.  We refer the reader to some excellent
reviews that have appeared in the literature during the last few
years: the review by Beenakker \cite{Beenakker} for the solid state
physics applications, the review by Verbaarschot \cite{revVerb} for
QCD--related applications, and \cite{GMW,JPA} for extensive reviews
including a historical outline.  

\newpage
\begin{center}
\Large{\bf Part I}
\end{center}
The theory of symmetric spaces has a long history in mathematics.  In
this first part of the paper we will introduce the reader to some of the most
fundamental concepts in the theory of symmetric spaces. We have tried
to keep the discussion as simple as possible without assuming any
previous familiarity of the reader with symmetric spaces. The review
should be particularly accessible to physicists. In the hope of
addressing a wider audience, we have almost completely avoided using
concepts from differential geometry, and we have presented the subject
mostly from an algebraic point of view.  In addition we have inserted
a large number of simple examples in the text, that will hopefully
help the reader visualize the ideas.  

Since our aim in Part~II will be
to introduce the reader to the application of symmetric spaces in
physical integrable systems and random matrix models, we have chosen
the background material presented here with this in mind. Therefore we
have put emphasis not only on fundamental issues but on subjects that
will be relevant in these applications as well.  Our treatment will be
somewhat rigorous; however, we skip proofs that can be found in the
mathematical literature and concentrate on simple examples that
illustrate the concepts presented.  The reader is referred to
Helgason's book \cite{Helgason} for a rigorous treatment; however,
this book may not be immediately accessible to physicists.  For the
reader with little background in differential geometry we recommend
the book by Gilmore \cite{Gilmore} (especially Chapter~9) for an
introduction to symmetric spaces of exceptional clarity.

In section~\ref{sec-Lie}, after reviewing the basics about Lie groups,
we will present some of the most important properties of root systems.
In section~\ref{sec-strSS} we define symmetric spaces and discuss
their main characteristics, defining involutive automorphisms,
spherical decomposition of the group elements, and the metric on the
Lie algebra.  We also discuss the algebraic structure of the coset
space.

In section~\ref{sec-realforms} we show how to obtain all the real
forms of a complex semisimple Lie algebra. The same techniques will
then be used to classify the real forms of symmetric spaces in
section~\ref{sec-claSS}. In this section we also define the curvature
of a symmetric space, and discuss triplets of symmetric spaces with
positive, zero and negative curvature, all corresponding to the same
symmetric subgroup.  We will see why curved symmetric spaces arise
from semisimple groups, whereas the flat spaces are associated to
non--semisimple groups.  In addition, in section~\ref{sec-claSS} we
will define restricted root systems.  The restricted root systems are
associated to symmetric spaces, just like ordinary root systems are
associated to groups. As we will discuss in detail in Part~II of this
paper, they are key objects when considering the integrability of
Calogero--Sutherland models.

In section \ref{sec-Operators} we discuss Casimir and Laplace
operators on symmetric spaces and mention some known properties of the
eigenfunctions of the latter, so called zonal spherical functions.
These functions play a prominent role in many physical applications.

The introduction to symmetric spaces we
present contains the basis for understanding the developments to be
discussed in more detail in Part~II. The reader already familiar with
symmetric spaces is invited to start reading in the last section of
Part~I, section \ref{sec-intmod}, where we give a brief introduction
to Calogero--Sutherland models.

\section{Lie groups and root spaces}
\label{sec-Lie}
\setcounter{equation}{0}

In this introductory section we define the basic concepts relating to
Lie groups. We will build on the material presented here when we
discuss symmetric spaces in the next section. The reader with a solid
background in group theory may want to skip most or all of this
section.

\subsection{Lie groups and manifolds}

A manifold can be thought of as the generalization of a surface, but
we do not in general consider it as embedded in a higher--dimensional
euclidean space.  A short introduction to differentiable manifolds can
be found in ref.~\cite{FosNigh}, and a more elaborate one in
refs.~\cite{Boothby} and \cite{3w} (Ch.~III). The points of an
$N$--dimensional manifold can be labelled by real coordinates
$(x^1,...,x^N)$. Suppose that we take an open set $U_\alpha $ of this
manifold, and we introduce local real coordinates on it. Let
$\psi_\alpha $ be the function that attaches $N$ real coordinates to
each point in the open set $U_\alpha $.  Suppose now that the manifold
is covered by overlapping open sets, with local coordinates attached
to each of them. If for each pair of open sets $U_\alpha $, $U_\beta
$, the fuction $\psi_\alpha \circ \psi_\beta^{-1}$ is differentiable
in the overlap region $U_\alpha \cap U_\beta $, it means that we can
go smoothly from one coordinate system to another in this region. Then
the manifold is differentiable.

Consider a group $G$ acting on a space $V$. We can think of $G$ as
being represented by matrices, and of $V$ as a space of vectors on
which these matrices act.  A group element $g\in G$ transforms the
vector $v\in V$ into $gv=v'$.  

If $G$ is a Lie group, it is also a differentiable manifold.  The fact
that a Lie group is a differentiable manifold means that for two group
elements $g$, $g'\in G$, the product $(g,g') \in G\times G \to gg'\in
G$ and the inverse $g\to g^{-1}$ are smooth ($C^\infty$) mappings,
that is, these mappings have continuous derivatives of all orders.

{\bf Example:} The space ${\bf R^n}$ is a smooth manifold and at the
same time an abelian group. The ``product'' of elements is addition
$(x,x')\to x+x'$ and the inverse of $x$ is $-x$. These operations are
smooth.  

{\bf Example:} 
The set $GL(n,R)$ of nonsingular real $n\times n$ matrices
$M$, ${\rm det}M\neq 0$, with matrix multiplication $(M,N)\to MN$ and
multiplicative matrix inverse $M\to M^{-1}$ is a non--abelian group
manifold. Any such matrix can be represented as $M=\e^{\sum_i t^iX_i}$
where $X_i$ are generators of the ${\bf GL(n,R)}$ algebra and $t^i$
are real parameters.

\subsection{The tangent space}

In each point of a differentiable manifold, we can define the tangent
space.  If a curve through a point $P$ in the manifold is parametrized
by $t\in {\bf R}$

\beq
x^a(t)=x^a(0)+\lambda^at \ \ \ \ \ \ \ \ a=1,...,N
\eeq

where $P=(x^1(0),...,x^N(0))$, then ${\bf
  \lambda}=(\lambda^1,...,\lambda^N)= (\dot{x}^1(0),...,\dot{x}^N(0))$
is a tangent vector at $P$. Here $\dot{x}^a(0) =\frac{\rm d}{{\rm
    d}t}x^a(t)|_{t=0}$.  The space spanned by all tangent vectors at
$P$ is the tangent space. In particular, the tangent vectors to the
coordinate curves (the curves obtained by keeping all the coordinates
fixed except one) through $P$ are called the natural basis for the
tangent space.  

{\bf Example:} In euclidean 3--space the natural basis is $\{ \hat
e_x,\hat e_y, \hat e_z\}$. On a patch of the unit 2--sphere parametrized
by polar coordinates it is $\{ \hat e_\theta,\hat e_\phi \}$.

For a Lie group, the tangent space at the origin is spanned by the
generators, that play the role of (contravariant) vector fields (also
called derivations), expressed in local coordinates on the group
manifold as $X=X^a(x)\partial_a$ (for an introduction to differential
geometry see ref.~\cite{SattW}, Ch. 5, or \cite{3w}). Here the partial
derivatives $\partial_a=\frac{\partial }{\partial x^a}$ form a basis
for the vector field.  That the generators span the tangent space at
the origin can easily be seen from the exponential map. Suppose $X$ is
a generator of a Lie group. The exponential map then maps $X$ onto
$\e^{tX}$, where $t$ is a parameter. This mapping is a one--parameter
subgroup, and it defines a curve $x(t)$ in the group manifold. The
tangent vector of this curve at the origin is then

\beq
\frac{\rm d}{{\rm d}t} \e^{tX}|_{t=0} =X
\eeq

All the generators together span the tangent space at the origin
(also called the identity element).

\subsection{Coset spaces}
\label{sec-cosets}

The isotropy subgroup $G_{v_0}$ of a group $G$ at the point $v_0\in V$
is the subset of group elements that leave $v_0$ fixed. The set of
points that can be reached by applying elements $g\in G$ to $v_0$ is
the orbit of $G$ at $v_0$, denoted $Gv_0$. If $Gv_0=V$ for one point
$v_0$, then this is true for every $v\in V$.  We then say that $G$
{\it acts transitively} on $V$.

In general, a symmetric space can be represented as a coset space.
Suppose $H$ is a subgroup of a Lie group $G$. The coset space $G/H$ is
the set of subsets of $G$ of the form $gH$, for $g\in G$. $G$ acts on
this coset space: $g_1(gH)$ is the coset $(g_1g)H$. We will refer to
the elements of the coset space by $g$ instead of by $gH$, when the
subgroup $H$ is understood from the context, because of the natural
mapping described in the next paragraph.  If $g\notin H$, $gH$
corresponds to a point on the manifold $G/H$ away from the origin,
whereas $hH=H$ ($h\in H$) is the identity element identified with the
origin of the symmetric space. This point is the north pole in the
example below.

If $G$ acts transitively on $V$, then $V=Gv$ for any $v\in V$. Since
the isotropy subgroup $G_{v_0}$ leaves a fixed point $v_0$ invariant,
$gG_{v_0}v_0=gv_0=v\in V$, we see that the action of the group $G$ on
$V$ defines a bijective action of elements of $G/G_{v_0}$ on $V$.
Therefore the space $V$ on which $G$ acts transitively, can be
identified with $G/G_{v_0}$, since there is one--to--one
correspondence between the elements of $V$ and the elements of
$G/G_{v_0}$. There is a natural mapping from the group element $g$
onto the point $gv_0$ on the manifold.

{\bf Example:} The $SO(2)$ subgroup of $SO(3)$ is the isotropy
subgroup at the north pole of a unit 2--sphere imbedded in
3--dimensional space, since it keeps the north pole fixed. On the
other hand, the north pole is mapped onto any point on the surface of
the sphere by elements of the coset $SO(3)/SO(2)$.  This can be seen
from the explicit form of the coset representatives.  As we will see
in eq.~(\ref{eq:cosetreps}) in subsection \ref{sec-algstr}, the
general form of the elements of the coset is

\beq
\label{eq:cosetrep}
M={\rm exp}\left(\begin{array}{cc} 0 & C \\ 
                          -C^T & 0 \end{array}\right)
=\left(\begin{array}{cc}\sqrt{I_2-XX^T} & X   \\ 
                                   -X^T &  \sqrt{1-X^TX} \end{array}\right)
\eeq

where $C$ is the matrix

\beq
C=\left(\begin{array}{c}t^2 \\ t^1\end{array}\right)
\eeq

and $t^1$, $t^2$ are real coordinates. $I_2$ in
eq.~(\ref{eq:cosetrep}) is the $2\times 2$ unit matrix.  For the coset
space $SO(3)/SO(2)$, $M$ is equal to

\beq
\label{eq:L1L2}
M={\rm exp}\left(\sum_{i=1}^2 t^iL_i\right), \ \ \ \ 
L_1= \frac{1}{2}\left(\begin{array}{ccc} 0&0&0\\ 0&0&1\\ 0&-1&0\end{array}\right), \ \ \ \ 
L_2=\frac{1}{2}\left(\begin{array}{ccc} 0&0&1\\ 0&0&0\\ -1&0&0\end{array}\right)
\eeq

The third $SO(3)$ generator

\beq
\label{eq:L3}
L_3=\frac{1}{2}\left(\begin{array}{ccc} 0&1&0\\ -1&0&0\\ 0&0&0\end{array}\right)
\eeq

spans the algebra of the stability subgroup $SO(2)$, that keeps the north pole 
fixed: 

\beq
\label{eq:fixedNP}
{\rm exp}(t^3L_3)\left(\begin{array}{c} 0 \\ 0 \\ 1 \end{array}\right)=
\left(\begin{array}{c} 0 \\ 0 \\ 1 \end{array}\right)
\eeq

The generators $L_i$ ($i=1,2,3$)
satisfy the $SO(3)$ commutation relations $[L_i,L_j]=\frac{1}{2}\epsilon_{ijk}L_k$. 
Note that since the $L_i$ and the $t^i$ are real, $C^\dagger =C^T$.

In (\ref{eq:cosetrep}), $M$ is a general representative of the coset
$SO(3)/SO(2)$.  By expanding the exponential we see that the explicit
form of $M$ is

\beq
M=\left(
  \begin{array}{ccc} 1+(t^2)^2\frac{({\rm cos}\sqrt{(t^1)^2+(t^2)^2}-1)}{(t^1)^2+(t^2)^2} &
                     t^1t^2\frac{({\rm cos}\sqrt{(t^1)^2+(t^2)^2}-1)}{(t^1)^2+(t^2)^2} &
                     t^2\frac{{\rm sin}\sqrt{(t^1)^2+(t^2)^2}}{\sqrt{(t^1)^2+(t^2)^2}}\\

                     t^1t^2\frac{({\rm cos}\sqrt{(t^1)^2+(t^2)^2}-1)}{(t^1)^2+(t^2)^2} &
                     1+(t^1)^2\frac{({\rm cos}\sqrt{(t^1)^2+(t^2)^2}-1)}{(t^1)^2+(t^2)^2} &
                     t^1\frac{{\rm sin}\sqrt{(t^1)^2+(t^2)^2}}{\sqrt{(t^1)^2+(t^2)^2}}\\

                     -t^2\frac{{\rm sin}\sqrt{(t^1)^2+(t^2)^2}}{\sqrt{(t^1)^2+(t^2)^2}} &
                     -t^1\frac{{\rm sin}\sqrt{(t^1)^2+(t^2)^2}}{\sqrt{(t^1)^2+(t^2)^2}} &
                     {\rm cos}\sqrt{(t^1)^2+(t^2)^2}\end{array}
\right)
\eeq

Thus the matrix $X=\left(\begin{array}{c}x\\ y\end{array}\right)$ is
given in terms of the components of $C$ by (cf.
eq.~(\ref{eq:functions})):

\beq 
X=\left(\begin{array}{c}x\\ y\end{array}\right)=\left(\begin{array}{c}
t^2\frac{{\rm sin}\sqrt{(t^1)^2+(t^2)^2}}{\sqrt{(t^1)^2+(t^2)^2}}\\
t^1\frac{{\rm sin}\sqrt{(t^1)^2+(t^2)^2}}{\sqrt{(t^1)^2+(t^2)^2}}\end{array}\right)
\eeq

Defining now $z={\rm cos}\sqrt{(t^1)^2+(t^2)^2}$, we see that 
the variables $x$, $y$, $z$ satisfy the equation of the 2--sphere:

\beq
x^2+y^2+z^2=1
\eeq

When the coset space representative $M$ acts on the north pole it is 
easily seen that the orbit is all of the 2--sphere:

\beq
M\left(\begin{array}{c} 0\\ 0\\ 1\end{array}\right) =
\left(\begin{array}{ccc}  .   &  .   & x \\
                          .   &  .   & y \\
                          .   &  .   & z \end{array}\right)
\left(\begin{array}{c} 0\\ 0\\ 1\end{array}\right) =
\left(\begin{array}{c} x\\ y\\ z\end{array}\right) 
\eeq

This shows that there is one--to--one correspondence between the
elements of the coset and the points of the 2--sphere. The coset
$SO(3)/SO(2)$ can therefore be identified with a unit 2--sphere
imbedded in 3--dimensional space.

\subsection{The Lie algebra and the adjoint representation}
\label{sec-Lie,adj}

A Lie algebra ${\bf G}$ is a vector space over a field $F$. 
Multiplication in the Lie algebra is given by the bracket
$[X,Y]$. It has the following properties:

\noindent [1] If $X$, $Y\in {\bf G}$, then $[X,Y]\in {\bf G}$,\\
\noindent [2] $[X,\alpha Y+\beta Z]=\alpha [X,Y]+\beta[X,Z]$ for $\alpha $,
$\beta \in F$, \\
\noindent [3] $[X,Y]=-[Y,X]$, \\
\noindent [4] $[X,[Y,Z]]+[Y,[Z,X]]+[Z,[X,Y]]=0$ (the Jacobi identity).

The algebra ${\bf G}$ generates a group through the exponential mapping.
A general group element is 

\beq
M={\rm exp}\left( \sum_it^iX_i\right);\ \ \ \ t^i\in F,\ X_i\in {\bf G}
\eeq
  
We define a mapping ${\rm ad} X$ from the Lie algebra to itself by
${\rm ad} X:Y\to [X,Y]$.  The mapping $X\to {\rm ad} X$ is a
representation of the Lie algebra called the adjoint representation.
It is easy to check that it is an automorphism: it follows from the
Jacobi identity that $[{\rm ad}X_i,{\rm ad}X_j]={\rm ad}[X_i,X_j]$.
Suppose we choose a basis $\{ X_i\}$ for ${\bf G}$. Then

\beq
\label{eq:adjr}
{\rm ad} X_i(X_j)=[X_i,X_j]=C^k_{ij}X_k
\eeq

where we sum over $k$. The $C^k_{ij}$ are called structure constants.
Under a change of basis, they transform as mixed tensor components.
They define the matrix $(M_i)_{jk}=C^j_{ik}$ associated with the
adjoint representation of $X_i$.  One can show that there exists a
basis for any complex semisimple algebra in which the structure
constants are real. This means the adjoint representation is real.
Note that the dimension of the adjoint representation is equal to the
dimension of the group.
 
{\bf Example:} Let's construct the adjoint representation of $SU(2)$.
The generators in the defining representation are

\beq
J_3=\frac{1}{2}\left( \begin{array}{cc} 1 & 0  \\ 
                                   0 & -1 \end{array} \right), \ \ \ \ \  
J_\pm =\frac{1}{2}\left( \left( \begin{array}{cc} 0 & 1  \\ 
                                   1 & 0  \end{array} \right)\pm 
i\left( \begin{array}{cc} 0 & -i \\ 
                                   i & 0  \end{array} \right)\right)
\eeq

and the commutation relations are

\beq
[J_3,J_\pm]=\pm J_\pm, \ \ \ \ \ \ \ [J_+,J_-]=2J_3
\eeq

The structure constants are therefore 
$C^+_{3+}=-C^+_{+3}=-C^-_{3-}=C^-_{-3}=1$, $C^3_{+-}=-C^3_{-+}=2$ and the
adjoint representation is given by $(M_3)_{++}=1$, $(M_3)_{--}=-1$, 
$(M_+)_{+3}=-1$, $(M_+)_{3-}=2$, $(M_-)_{-3}=1$, $(M_-)_{3+}=-2$, and all
other matrix elements equal to 0:

\beq
\label{eq:SU2adjoint}
M_3=\left(\begin{array}{ccc}0 & 0 & 0 \\0 & 1 & 0 \\ 0 & 0 & -1 \end{array}\right),
\ \ \ \ \ 
M_+=\left(\begin{array}{ccc}0 & 0 & 2 \\ -1 & 0 & 0\\ 0 & 0 & 0 \end{array}\right),
\ \ \ \ \ 
M_-=\left(\begin{array}{ccc} 0 & -2 & 0 \\ 0 & 0 & 0 \\ 1 & 0 & 0\end{array}\right),
\ \ \ \ \ 
\eeq

These representation matrices are real, have the same dimension as the
group, and satisfy the $SU(2)$ commutation relations $[M_3,M_\pm]=\pm
M_\pm $, $[M_+,M_-]=2M_3$.

\subsection{Semisimple algebras and root spaces}
\label{sec-rootsp}

In this paragraph we will briefly recall the basic facts about root
spaces and the classification of complex simple Lie algebras, to set
the stage for our discussion of real forms of Lie algebras and finally
symmetric spaces. 

An {\it ideal}, or {\it invariant subalgebra} ${\bf I}$ is a
subalgebra such that $[{\bf G},{\bf I}] \subset {\bf I}$.  An abelian
ideal also satisfies $[{\bf I},{\bf I}]=0$.  A {\it simple} Lie
algebra has no proper ideal. A {\it semisimple} Lie algebra is the
direct sum of simple algebras, and has no proper abelian ideal (by
proper we mean different from $\{ 0\}$).
  
A Lie algebra is a linear vector space over a field $F$, with an
antisymmetric product defined by the Lie bracket (cf. subsection
\ref{sec-Lie,adj}). If $F$ is the field of real, complex or quaternion
numbers, the Lie algebra is called a real, complex or quaternion
algebra.  A complexification of a real Lie algebra is obtained by
taking linear combinations of its elements with complex coefficients.
A real Lie algebra ${\bf H}$ is a real form of the complex algebra
${\bf G}$ if ${\bf G}$ is the complexification of ${\bf H}$.

In any simple algebra there are two kinds of generators: there is a
maximal abelian subalgebra, called the {\it Cartan subalgebra} ${\bf
  H_0}= \{ H_1,...,H_r\} $; $[H_i,H_j]=0$ for any two elements of the
Cartan subalgebra.  There are also raising and lowering operators denoted
$E_\alpha$. $\alpha $ is an $r$--dimensional vector $\alpha
=(\alpha_1,...,\alpha_r)$ and $r$ is the {\it rank} of the algebra.
\footnote{The rank of an algebra is defined through the secular
  equation (see subsection \ref{sec-Casi}).  For a non--semisimple
  algebra, the maximal number of mutually commuting generators can be
  greater than the rank of the algebra.}  The latter are
eigenoperators of the $H_i$ in the adjoint representation belonging to
eigenvalue $\alpha_i$: $[H_i,E_\alpha]=\alpha_iE_\alpha$.  For each
eigenvalue, or {\it root} $\alpha_i$, there is another eigenvalue
$-\alpha_i$ and a corresponding eigenoperator $E_{-\alpha}$ under the
action of $H_i$.

Suppose we represent each element of the Lie algebra
by an $n\times n$ matrix. Then $[H_i,H_j]=0$ means the matrices $H_i$
can all be diagonalized simultaneously. Their eigenvalues $\mu_i$ are
given by $H_i|\mu \rangle =\mu_i|\mu \rangle $, where the eigenvectors
are labelled by the {\it weight vectors} $\mu =(\mu_1,...,\mu_r)$
\cite{Georgi}.

A weight whose first non--zero component is positive is called a
positive weight.  Also, a weight $\mu $ is greater than another weight
$\mu' $ if $\mu - \mu' $ is positive. Thus we can define the highest
weight as the one which is greater than all the others. The highest
weight is unique in any representation.

The roots $\alpha_i \equiv \alpha(H_i)$ of the algebra ${\bf G}$
are the weights of the adjoint representation. Recall that in the
adjoint representation, the states on which the generators act are
defined by the generators themselves, and the action is defined by

\beq
X_a|X_b\rangle \equiv {\rm ad}X_a(X_b) \equiv [X_a,X_b]
\eeq

The roots are functionals on the Cartan subalgebra satisfying

\beq
{\rm ad}H_i(E_\alpha )=[H_i,E_\alpha]=\alpha (H_i)E_\alpha
\eeq

where $H_i$ is in the Cartan subalgebra.  The eigenvectors $E_\alpha$
are called the root vectors.  These are exactly the raising and
lowering operators $E_{\pm \alpha}$ for the weight vectors $\mu $.
There are canonical commutation relations defining the system of roots
belonging to each simple rank $r$--algebra. These are summarized
below: \footnote{For the reader who wants to understand more about the
  origin of the structure of Lie algebras, we recommend Chapter~7 of
  Gilmore \cite{Gilmore}.}

\beq
[H_i,H_j]=0,\ \ \ \ \ [H_i,E_\alpha]=\alpha_iE_\alpha, \ \ \ \ \ 
[E_\alpha,E_{-\alpha}]=\alpha_iH_i
\eeq

One can prove the fundamental relation \cite{SattW,Georgi}

\beq
\label{eq:fund}
\frac{2\alpha \cdot \mu}{\alpha^2}=-(p-q)
\eeq

where $\alpha $ is a root, $\mu $ is a weight, and $p$, $q$ are
positive integers such that $E_\alpha |\mu +p\alpha \rangle =0$,
$E_{-\alpha} |\mu -q\alpha \rangle =0$ \footnote{Here the scalar 
product $\cdot $ can be defined in terms of the metric on the Lie algebra.
For the adjoint representation, $\mu $ is a root $\beta $ and 

\beq
\frac{2\alpha \cdot \beta }{\alpha^2}=\frac{2K(H_\alpha, H_\beta )}
{K(H_\alpha, H_\alpha )}\equiv \frac{2 \beta (H_\alpha )}{\alpha  (H_\alpha )}
\eeq

\noindent where $K$ denotes the Killing form (see paragraph
\ref{sec-metric}).  There is always a unique element $H_\alpha $ in the
algebra such that $K(H,H_\alpha )=\alpha (H)$ for each $H\in {\bf H_0}$
(see for example \cite{SattW}, Ch.~10).  In general for a linear form
$\mu $ on the Lie algebra,

\beq
\frac{2\alpha \cdot \mu}{\alpha^2}=\frac{2\mu (H_\alpha)}
{\alpha (H_\alpha )}
\eeq

\noindent Then $\mu $ is a highest weight for some representation if and
only if this expression is an integer for each positive root $\alpha
$.}.  This relation gives rise to the strict properties of root
lattices, and permits the complete classification of all the complex
(semi)simple algebras.

Eq.~(\ref{eq:fund}) is true for any representation, but has
particularly strong implications for the adjoint representation. In
this case $\mu $ is a root.  As a consequence of eq.~(\ref{eq:fund}),
the possible angle between two root vectors of a simple Lie algebra is
limited to a few values: these turn out to be multiples of $\frac{\pi
  }{6}$ and $\frac{\pi }{4}$ (see e.g. \cite{Georgi}, Ch.~VI).  The
root lattice is invariant under reflections in the hyperplanes
orthogonal to the roots (the Weyl group).  As we will shortly see,
this is true not only for the root lattice, but for the weight lattice
of any representation.

Note that the roots $\alpha $ are real--valued linear functionals on
the Cartan subalgebra. Therefore they are in the space dual to ${\bf
  H_0}$.  A subset of the positive roots span the root lattice. These
are called simple roots. Obviously, since the roots are in the space
dual to ${\bf H_0}$, the number of simple roots is equal to the rank
of the algebra.

The same relation (\ref{eq:fund}) determines the highest weights of
all irreducible representations. Setting $p=0$, choosing a positive
integer $q$, and letting $\alpha $ run through the simple roots,
$\alpha=\alpha^i$ ($i=1,...,r$), we find the highest weights $\mu^i $
of all the irreducible representations corresponding to the given
value of $q$ \cite{Georgi}. For example, for $q=1$ we get the highest
weights of the $r$ fundamental representations of the group, each
corresponding to a simple root $\alpha^i$. For higher values of $q$ we
get the highest weights of higher--dimensional representations of the
same group.

The set of all possible simple root systems are classified by means of
Dynkin diagrams, each of which correspond to an equivalence class of
isomorphic Lie algebras.  The classical Lie algebras ${\bf
  SU(n+1,C)}$, ${\bf SO(2n+1,C)}$, ${\bf Sp(2n,C)}$ and ${\bf
  SO(2n,C)}$ correspond to root systems $A_n$, $B_n$, $C_n$, and
$D_n$, respectively.  In addition there are five exceptional algebras
corresponding to root systems $E_6$, $E_7$, $E_8$, $F_4$ and $G_2$.
Each of these complex algebras in general has several real forms
associated with it (see section \ref{sec-realforms}).  These real
forms correspond to the same Dynkin diagram and root system as the
complex algebra.  Since we will not make reference to Dynkin diagrams
in the following, we will not discuss them here.  The interested
reader can find sufficient material for example in the book by Georgi
\cite{Georgi}.

The (semi)simple complex algebra ${\bf G}$ decomposes into a direct
sum of root spaces \cite{SattW}:

\beq
\label{eq:diroot}
{\bf G}={\bf H_0}\oplus \sum_\alpha {\bf G_\alpha}
\eeq

where ${\bf G_\alpha}$ is generated by $\{ E_{\pm \alpha }\}$. This
will be evident in the example given below.

{\bf Example:} The root system $A_{n-1}$ corresponds to the complex
Lie algebra ${\bf SL(n,C)}$ and all its real forms. In a later section
we will see how to construct all the real forms associated with a
given complex Lie algebra. Let's see here explicitly how to construct
the root lattice of ${\bf SU(3,C)}$, which is one of the real forms of
${\bf SL(3,C)}$.

The generators are determined by the commutation relations. In physics
it is common to write the commutation relations in the form

\beq
[T_i,T_j]=if_{ijk}T_k
\eeq

(an alternative form is to define the generators as $X_i=iT_i$ and
write the commutation relations as $[X_i,X_j]=-f_{ijk}X_k$) where
$f_{ijk}$ are structure constants for the algebra ${\bf SU(3,C)}$.

Using the notation $g=\e^{it^aT_a}$ for the group elements (with $t^a$
real and a sum over $a$ implied), the generators $T_a$ in the
fundamental representation of this group are hermitean\footnote{Note
  that we have written an explicit factor of $i$ in front of the
  generators in the expression for the group elements. This is often
  done for compact groups; since the Killing form (subsection
  \ref{sec-metric}) has to be negative definite, the coordinates of
  the algebra spanned by the generators must be purely imaginary. Here
  we use this notation because it is conventional. If we absorb the
  factor of $i$ into the generators, we get {\it antihermitean}
  matrices $X_a=iT_a$; we will do this in the example in subsection
  \ref{sec-Inv} to comply with eq.~(\ref{eq:KP}). Of course, the
  matrices in the {\it algebra} are always antihermitean.}:
 
\beq
\label{eq:Gell-Mann}
\begin{array}{l}
T_1=\frac{1}{2}\left(\begin{array}{ccc} 0 & 1 & 0 \\
                                   1 & 0 & 0 \\
                                   0 & 0 & 0 \end{array}\right), \ \ \ \
T_2=\frac{1}{2}\left(\begin{array}{ccc} 0 & -i & 0 \\
                                   i & 0 & 0 \\
                                   0 & 0 & 0 \end{array}\right), \ \ \ \
T_3=\frac{1}{2}\left(\begin{array}{ccc} 1 & 0 & 0 \\
                                   0 & -1 & 0 \\
                                   0 & 0 & 0 \end{array}\right), \nonumber \\
\\
T_4=\frac{1}{2}\left(\begin{array}{ccc} 0 & 0 & 1 \\
                                   0 & 0 & 0 \\
                                   1 & 0 & 0 \end{array}\right), \ \ \ \  
T_5=\frac{1}{2}\left(\begin{array}{ccc} 0 & 0 & -i \\
                                   0 & 0 & 0 \\
                                   i & 0 & 0 \end{array}\right), \ \ \ \
T_6=\frac{1}{2}\left(\begin{array}{ccc} 0 & 0 & 0 \\
                                   0 & 0 & 1 \\
                                   0 & 1 & 0 \end{array}\right), \nonumber \\ 
\\ 
T_7=\frac{1}{2}\left(\begin{array}{ccc} 0 & 0 & 0 \\
                                   0 & 0 & -i \\
                                   0 & i & 0 \end{array}\right), \ \ \ \
T_8=\frac{1}{2\sqrt{3}}\left(\begin{array}{ccc} 1 & 0 & 0 \\
                                                0 & 1 & 0 \\
                                                0 & 0 & -2 
                                   \end{array}\right) \end{array}
\\ \nonumber
\eeq

In high energy physics the matrices $2T_a$ are known as Gell--Mann
matrices.  The generators are normalized in such a way that ${\rm
  tr}(T_aT_b)=\frac{1}{2} \delta_{ab}$. Note that $T_1$, $T_2$, $T_3$
form an ${\bf SU(2,C)}$ subalgebra.  We take the Cartan subalgebra to
be ${\bf H_0}=\{T_3,T_8\}$. The rank of this group is $r=2$.

Let's first find the weight vectors of the fundamental representation.
To this end we look for the eigenvalues $\mu_i $ of the operators in
the abelian subalgebra ${\bf H_0}$:

\beq 
T_3\left(\begin{array}{c} 1 \\ 0 \\ 0 \end{array}\right) = 
\frac{1}{2}\left(\begin{array}{c} 1 \\ 0 \\ 0 \end{array}\right),\ \ \ \ \ \ \
T_8\left(\begin{array}{c} 1 \\ 0 \\ 0 \end{array}\right) = 
\frac{1}{2\sqrt{3}}\left(\begin{array}{c} 1 \\ 0 \\ 0 \end{array}\right),
\eeq

therefore the eigenvector $(1\, 0\, 0)^T$ corresponds to the state
$|\mu \rangle $ where

\beq 
\mu \equiv (\mu_1,\mu_2)=\left(\frac{1}{2},\frac{1}{2\sqrt{3}}\right)
\eeq 

is distinguished by its eigenvalues under the operators $H_i$ of the
Cartan subalgebra. In the same way we find that $(0\, 1\, 0)^T$ and
$(0\, 0\, 1)^T$ correspond to the states labelled by weight vectors

\beq 
\mu' = \left(-\frac{1}{2},\frac{1}{2\sqrt{3}}\right) ,\ \ \ \ \ \ \ 
\mu''= \left(0,-\frac{1}{\sqrt{3}}\right)
\eeq

respectively. $\mu$, $\mu'$, and $\mu''$ are the weights of the
fundamental representation $\rho =D$ and they form an equilateral
triangle in the plane. The highest weight of the representation $D$ is
$\mu =\left(\frac{1}{2},\frac{1}{2\sqrt{3}}\right)$.

There is also another fundamental representation $\bar{D}$ of the
algebra ${\bf SU(3,C)}$, since it generates a group of rank 2.
Indeed, from eq.~(\ref{eq:fund}), for $p=0$, $q=1$, there is one
highest weight $\mu^i$, and one fundamental representation, for each
simple root $\alpha^i$.  The highest weight $\bar{\mu }$ of the
representation $\bar{D}$ is

\beq
\bar{\mu }= \left(\frac{1}{2},-\frac{1}{2\sqrt{3}}\right) 
\eeq

The highest weights of the representations corresponding to any
positive integer $q$ can be obtained as soon as we know the simple
roots.  Then, by operating with lowering operators on this weight, we
obtain other weights, on which we can further operate with lowering
operators until we have obtained all the weights in the
representation. For an example of this procedure see \cite{Georgi},
Ch.~IX.

Let's see now how to obtain the roots of ${\bf SU(3,C)}$. Each root
vector $E_\alpha $ corresponds to either a raising or a lowering
operator: $E_\alpha $ is the eigenvector belonging to the root
$\alpha_i \equiv \alpha (H_i)$ under the adjoint representation of
$H_i$, like in eq.~(\ref{eq:eig}).  Each raising or lowering operator
is a linear combination of generators $T_i$ that takes one state of
the fundamental representation to another state of the same
representation: $E_{\pm \alpha}|\mu \rangle = N_{\pm \alpha,\mu }|\mu
\pm \alpha \rangle $.  Therefore the root vectors $\alpha $ will be
differences of weight vectors in the fundamental representation.  We
find the raising and lowering operators $E_{\pm \alpha}$ to be

\beq
\label{eq:raislowSU3}
\begin{array}{l}
E_{\pm(1,0)}=\frac{1}{\sqrt{2}}(T_1\pm iT_2) \\ \\
E_{\pm(\frac{1}{2},\frac{\sqrt{3}}{2})}=\frac{1}{\sqrt{2}}(T_4\pm iT_5) \\ \\
E_{\pm(-\frac{1}{2},\frac{\sqrt{3}}{2})}=\frac{1}{\sqrt{2}}(T_6\pm iT_7) 
\end{array}
\\ \nonumber 
\eeq

These generate the subspaces ${\bf G_\alpha}$ in eq.~(\ref{eq:diroot}).
In the fundamental representation, we find using the Gell--Mann
matrices that these are matrices with only one non--zero element. For
example, the raising operator $E_\alpha $ that corresponds to the root
$\alpha = (1,0)$ is

\beq
\label{eq:explE}
E_{+(1,0)}=\frac{1}{\sqrt{2}}\left(\begin{array}{ccc} 0 & 1 & 0 \\
                                                      0 & 0 & 0 \\
                                                      0 & 0 & 0 \end{array}\right)
\eeq

This operator takes us from the state $|\mu' \rangle =|-\frac{1}{2},
\frac{1}{2\sqrt{3}}\rangle$ to the state $|\mu \rangle =
|\frac{1}{2},\frac{1}{2\sqrt{3}}\rangle $.  The components of the root
vectors of ${\bf SU(3,C)}$ are the eigenvalues $\alpha_i$ of these
under the adjoint representation of the Cartan subalgebra. That is,

\beq
\label{eq:eig}
H_i|E_\alpha \rangle \equiv {\rm ad}H_i (E_\alpha )\equiv 
[H_i,E_\alpha] =\alpha_i |E_\alpha \rangle 
\eeq

This way we easily find the roots: we can either explicitly use the
structure constants of $SU(3)$ in
$[T_a,T_b]=if_{abc}T_c=-iC^c_{ab}T_c$ (note the explicit factor of $i$
due to our conventions regarding the generators) or we can use an
explicit representation for $H_i$, $E_\alpha $ like in
eqs.~(\ref{eq:Gell-Mann}), (\ref{eq:raislowSU3}), (\ref{eq:explE}), to
calculate the commutators:

\beq
\begin{array}{l}
{\rm ad}H_1(E_{\pm(1,0)})= [H_1,E_{\pm(1,0)}] = 
[T_3,\frac{1}{\sqrt{2}}(T_1\pm iT_2)] = \frac{1}{\sqrt{2}}(iT_2 \pm T_1)=\pm
E_{\pm(1,0)} \equiv \alpha^\pm_1 E_{\pm(1,0)} \\ \\
{\rm ad}H_2(E_{\pm(1,0)})= [H_2,E_{\pm(1,0)}] = 
[T_8,\frac{1}{\sqrt{2}}(T_1\pm iT_2)] = 0  \equiv \alpha^\pm_2 E_{\pm(1,0)}
\end{array} 
\eeq

The root vector corresponding to the raising operator $E_{+(1,0)}$ is
thus $\alpha =(\alpha^+_1,\alpha^+_2) =(1,0)$ and the root vector
corresponding to the lowering operator $E_{-(1,0)}$ is $-\alpha
=(\alpha^-_1,\alpha^-_2)=(-1,0)$. These root vectors are indeed the
differences between the weight vectors $\mu =
\left(\frac{1}{2},\frac{1}{2\sqrt{3}}\right)$ and $\mu' =
\left(-\frac{1}{2},\frac{1}{2\sqrt{3}}\right)$ of the fundamental
representation.

In the same way we find the other root vectors $\left(\pm\frac{1}{2},
  \pm\frac{\sqrt{3}}{2}\right)$,
$\left(\mp\frac{1}{2},\pm\frac{\sqrt{3}}{2}\right)$, and $(0,0)$ (with
multiplicity 2), by operating with $H_1$ and $H_2$ on the remaining
$E_{\pm \alpha }$'s and on the $H_i$'s.  The last root with
multiplicity 2 has as its components the eigenvalues under $H_1$,
$H_2$ of the states $|H_1\rangle $ and $|H_2\rangle $: $H_i|H_j\rangle
= [H_i,H_j]=0$; $i$, $j\in \{1,2\}$.  The root vectors form a regular
hexagon in the plane. The positive roots are $(1,0)$,
$\alpha^1=\left(\frac{1}{2},\frac{\sqrt{3}}{2}\right)$ and
$\alpha^2=\left(\frac{1}{2},-\frac{\sqrt{3}}{2}\right)$.  The latter
two are simple roots.  $(1,0)$ is not simple because it is the sum of
the other positive roots. There are two simple roots, since the rank
of $SU(3)$ is 2 and the root lattice is two--dimensional.

The root lattice of $SU(3)$ is invariant under reflections in the
hyperplanes orthogonal to the root vectors. This is true of any weight
or root lattice; the symmetry group of reflections in hyperplanes
orthogonal to the roots is called the {\it Weyl group}.  It is
obtained from eq.~(\ref{eq:fund}): since for any root $\alpha $ and
any weight $\mu $, $2(\alpha \cdot \mu )/\alpha^2$ is the integer
$q-p$,

\beq
\label{eq:Weyl}
\mu'=\mu - \frac{2(\alpha \cdot \mu )}{\alpha^2}\alpha
\eeq

is also a weight. Eq.~(\ref{eq:Weyl}) is exactly the above mentioned
reflection, as can easily be seen.

\subsection{The Weyl chambers}
\label{sec-chambers}

The roots are linear functionals on the Cartan subalgebra. We may
denote the Cartan subalgebra by ${\bf H_0}$ and its dual space by
${\bf H_0^*}$. A Weyl reflection like the one in (\ref{eq:Weyl}) can be
defined not only for the weights or roots $\mu $ in the space ${\bf
  H_0^*}$, but for an arbitrary vector $q\in {\bf H_0^*}$ or, in all
generality, for a vector $q$ in an arbitrary finite--dimensional vector space:

\beq
\label{eq:s_alpha}
s_\alpha (q)=q-\alpha^*(q)\alpha
\eeq

Note that $q\in{\bf H_0^*}$ is in the space dual to ${\bf H_0}$ and
may denote a root.  In (\ref{eq:s_alpha}) the function $\alpha^*(q)$
is a linear functional on ${\bf H_0^*}$ such that $\alpha^*(\alpha
)=2$. We will be concerned only with the crystallographic case when
$\alpha^*(q)$ is integer. We denote the hyperplanes in ${\bf H_0^*}$
where the function $\alpha^*(q)$ vanishes by $H^{(\alpha )}$:

\beq
\label{eq:hyperplane}
H^{(\alpha )}=\{ q\in {\bf H_0^*}:\alpha^*(q)=0 \}
\eeq

$H^{(\alpha )}$ is orthogonal to the root $\alpha $, and 
$s_\alpha (q)$ is a reflection in this hyperplane. 

By identifying the dual spaces ${\bf H_0}$ and ${\bf H_0^*}$ (this is
possible since they have the same dimension), we can consider
hyperplanes like the ones in (\ref{eq:hyperplane}) in the space ${\bf H_0}$.
The role of the linear functional $\alpha^*(q)$ is then played by

\beq
\alpha^*(q)=\frac{2q\cdot \alpha}{\alpha^2}=
\frac{2q(H_\alpha)}{\alpha (H_\alpha)}
\eeq

where $\alpha (H_\alpha)=K(H_\alpha,H_\alpha)$.  Here $K$ is the
Killing form (a metric on the algebra to be defined in paragraph
\ref{sec-metric}) and $H_\alpha $ is the unique element in ${\bf H_0}$
such that $K(H,H_\alpha)=\alpha (H)$.  The open subsets of ${\bf H_0}$
where roots are nonzero are called {\it Weyl chambers}.  Consequently,
the walls of the Weyl chambers are the hyperplanes in ${\bf H_0}$
where the roots $q(H)$ are zero.

\subsection{The simple root systems}
\label{sec-srs}

We have just shown by an example, in subsection \ref{sec-rootsp}, 
how to obtain a root system of type
$A_n$.  In general, for any simple algebra the commutation relations
determine the Cartan subalgebra and raising and lowering operators,
that in turn determine a unique root system, and correspond to a given
Dynkin diagram. In this way we can classify all the simple algebras
according to the type of root system it possesses. The root systems
for the four infinite series of classical non--exceptional Lie groups
can be characterized as follows \cite{Georgi} (denote the
$r$--dimensional space spanned by the roots by ${\cal V}$ and let
$\{e_1,...e_n\}$ be a canonical basis in ${\bf R}^n$):

$A_{n-1}$: Let ${\cal V}$ be the hyperplane in ${\bf R}^n$ that passes
through the points $(1,0,0,...0)$, $(0,1,0,...,0)$, ...,
$(0,0,...,0,1)$ (the endpoints of the $e_i$, $i=1,...,n$). Then the
root lattice contains the vectors $\{ e_i-e_j, i\neq j \}$.

$B_n$:  Let ${\cal V}$ be ${\bf R}^n$; then the roots are $\{ \pm e_i,
\pm e_i \pm e_j, i\neq j \}$.

$C_n$: Let ${\cal V}$ be ${\bf R}^n$; then the roots are $\{ \pm 2e_i,
\pm e_i \pm e_j, i\neq j \}$.

$D_n$: Let ${\cal V}$ be ${\bf R}^n$; then the roots are $\{
\pm e_i \pm e_j, i\neq j \}$.

The root lattice $BC_n$, that we will discuss in conjunction with restricted
root systems, is the union of $B_n$ and $C_n$. It is characterized as follows:

$BC_n$:  Let ${\cal V}$ be ${\bf R}^n$; then the roots are $\{ \pm e_i, 
\pm 2e_i, \pm e_i \pm e_j, i\neq j \}$. 

Because this system contains both $e_i$ and $2e_i$, it is called
non--reduced (normally the only root collinear with $\alpha$ is
$-\alpha$).  However, it is irreducible in the usual sense, which
means it is not the direct sum of two disjoint root systems $B_n$ and
$C_n$.  This can be seen from the root multiplicities (cf. Table~\ref{tab1}).

The semisimple algebras are direct sums of simple ones. That means the
simple constituent algebras commute with each other, and the root
systems are direct sums of the corresponding simple root systems.
Therefore, knowing the properties of the simple Lie algebras, we also
know the semisimple ones.

\section{Symmetric spaces}
\label{sec-strSS}
\setcounter{equation}{0}

In the previous section, we have reminded ourselves of some elementary
facts concerning root spaces and the classification of the complex semisimple
algebras. In this section we will define and discuss symmetric spaces.

A symmetric space is associated to an involutive automorphism of a
given Lie algebra. As we will see, several different involutive
automorphisms can act on the same algebra. Therefore we normally have
several different symmetric spaces deriving from the same Lie algebra.
The involutive automorphism defines a symmetric subalgebra and a
remaining complementary subspace of the algebra.  Under general
conditions, the complementary subspace is mapped onto a symmetric
space through the exponential map. In the following subsections we
make these statements more precise. We discuss how the elements of the
Lie group can act as transformations on the elements of the symmetric
space. This naturally leads to the definition of two coordinate
systems on symmetric spaces: the spherical and the horospheric
coordinate systems. The radial coordinates associated to each element
of a symmetric space through its spherical or horospheric
decomposition will be of relevance when we discuss the radial parts of
differential operators on symmetric spaces in section
\ref{sec-Operators}. In the same section we explain why these
operators are important in applications to physical problems, and in
Part~II we will discuss some of their uses.


In all of this paper we will distinguish between compact and
non--compact symmetric spaces. In order to give a precise notion of
compactness, we will define the metric tensor on a Lie algebra in
terms of the Killing form in subsection \ref{sec-metric}. The latter
is defined as a symmetric bilinear trace form on the adjoint
representation, and is therefore expressible in terms of the structure
constants. We will give several examples of Killing forms later, as we
discuss the various real forms of a Lie algebra. The metric tensor
serves to define the curvature tensor on a symmetric space (subsection
\ref{sec-curv}). It is also needed in computing the Jacobian of the
transformation to radial coordinates. This Jacobian is relevant in
calculating the radial part of the Laplace--Beltrami operator (see
paragraph \ref{sec-Laplaceop}).

We will close this section with a discussion of the general algebraic
form of coset representatives in subsection \ref{sec-algstr}.  

\subsection{Involutive automorphisms}
\label{sec-Inv}

An automorphism of a Lie algebra ${\bf G}$ is a mapping from ${\bf G}$
onto itself such that it preserves the algebraic operations on the Lie
algebra. For example, if $\sigma $ is an automorphism, it preserves
multiplication: $[\sigma (X), \sigma (Y)]=\sigma ([X,Y])$, for $X$,
$Y\in {\bf G}$.

Suppose that the linear automorphism $\sigma:{\bf G}\to{\bf G}$ is
such that $\sigma^2=1$, but $\sigma $ is not the identity. That means
that $\sigma $ has eigenvalues $\pm 1$, and it splits the algebra
${\bf G}$ into orthogonal eigensubspaces corresponding to these
eigenvalues.  Such a mapping is called an {\it involutive
  automorphism}.

Suppose now that ${\bf G}$ is a compact simple Lie algebra, $\sigma $
is an involutive automorphism of ${\bf G}$, and ${\bf G}={\bf K}
\oplus {\bf P}$ where

\beq
\label{eq:KP}
\sigma (X)=X\ \  {\rm for}\ \  X\in {\bf K},\ \ \sigma (X)=-X\ \  
{\rm for}\ \  X\in {\bf P}
\eeq

From the properties of automorphisms mentioned above, it is easy to
see that ${\bf K}$ is a subalgebra, but ${\bf P}$ is not.  In fact,
the commutation relations

\beq
\label{eq:commrel}
[{\bf K},{\bf K}]\subset {\bf K},\ \  [{\bf K},{\bf P}]\subset {\bf P},\ \  
[{\bf P},{\bf P}]\subset {\bf K} 
\eeq

hold. A subalgebra ${\bf K}$ satisfying (\ref{eq:commrel}) is called a
{\it symmetric subalgebra}.  If we now multiply the elements in ${\bf
  P}$ by $i$ (the ``Weyl unitary trick''), we construct a new
noncompact algebra ${\bf G^*}={\bf K} \oplus i{\bf P}$. This is called
a {\it Cartan decomposition}, and ${\bf K}$ is a maximal compact
subalgebra of ${\bf G^*}$.  The coset spaces $G/K$ and $G^*/K$ are
{\it symmetric spaces}.

{\bf Example}: Suppose $G=SU(n,C)$, the group of unitary complex
matrices with determinant $+1$. The algebra of this group then
consists of complex antihermitean\footnote{See the footnote in
  subsection \ref{sec-rootsp}.}  matrices of zero trace (this follows
by differentiating the identities $UU^\dagger=1$ and ${\rm det}U=1$
with respect to $t$ where $U(t)$ is a curve passing through the
identity at $t=0$); a group element is written as $g=\e^{t^aX_a}$ with
$t^a$ real.  Therefore any matrix $X$ in the Lie algebra of this group
can be written $X=A+iB$, where $A$ is real, skew--symmetric, and
traceless and $B$ is real, symmetric and traceless.  This means the
algebra can be decomposed as ${\bf G}={\bf K}\oplus {\bf P}$, where
${\bf K}$ is the compact connected subalgebra ${\bf SO(n,R)}$
consisting of real, skew--symmetric and traceless matrices, and ${\bf
  P}$ is the subspace of matrices of the form $iB$, where $B$ is real,
symmetric, and traceless. ${\bf P}$ is not a subalgebra.

Referring to the example for ${\bf SU(3,C)}$ in subsection
\ref{sec-rootsp} we see, setting $X_a = iT_a$, that the $\{ X_a\} $
split into two sets under the involutive automorphism $\sigma $
defined by complex conjugation $\sigma =K$. This splits the compact
algebra ${\bf G}$ into ${\bf K}\oplus {\bf P}$, since ${\bf P}$
consists of imaginary matrices:

\beq
\label{eq:SU3}
\begin{array}{l}
{\bf K}=\{ X_2,X_5,X_7\}
=\left\{
\frac{1}{2}\left(\begin{array}{ccc} 0&1&0\\ -1&0&0\\ 0&0&0\end{array}\right),
\frac{1}{2}\left(\begin{array}{ccc} 0&0&1\\ 0&0&0\\ -1&0&0\end{array}\right),
\frac{1}{2}\left(\begin{array}{ccc} 0&0&0\\ 0&0&1\\ 0&-1&0\end{array}\right)
\right\}\\
\\
{\bf P}=\{ X_1,X_3,X_4,X_6,X_8\}\\ \\
=\left\{
\frac{i}{2}\left(\begin{array}{ccc} 0&1&0\\1&0&0\\0&0&0\end{array}\right),
\frac{i}{2}\left(\begin{array}{ccc} 1&0&0\\0&-1&0\\0&0&0\end{array}\right),
\frac{i}{2}\left(\begin{array}{ccc} 0&0&1\\0&0&0\\1&0&0\end{array}\right),
\frac{i}{2}\left(\begin{array}{ccc} 0&0&0\\0&0&1\\0&1&0\end{array}\right),
\frac{i}{2\sqrt{3}}\left(\begin{array}{ccc} 1&0&0\\0&1&0\\0&0&-2\end{array}\right)
\right\} \end{array} \nonumber \\ 
\nonumber \\ 
\eeq

${\bf K}$ spans the real subalgebra ${\bf SO(3,R)}$.  Setting $X_2
\equiv L_3$, $X_5 \equiv L_2$, $X_7 \equiv L_1$, the commutation
relations for the subalgebra are
$[L_i,L_j]=\frac{1}{2}\epsilon_{ijk}L_k$.  The Cartan subalgebra
$i{\bf H_0}=\{ X_3,X_8\} $ is here entirely in the subspace ${\bf P}$.

Going back to the general case of ${\bf G}={\bf SU(n,C)}$, we obtain
from ${\bf G}$ by the Weyl unitary trick the non--compact algebra
${\bf G^*}={\bf K}\oplus i{\bf P}$. $i{\bf P}$ is now the subspace of
real, symmetric, and traceless matrices $B$. The Lie algebra ${\bf
  G^*}={\bf SL(n,R)}$ is then the set of $n\times n$ real matrices of
zero trace, and generates the linear group of transformations
represented by real $n\times n$ matrices of unit determinant.

The involutive automorphism that split the algebra ${\bf G}$ above was
defined to be complex conjugation $\sigma =K$.  The involutive
automorphism that splits ${\bf G^*}$ is defined by $\tilde{\sigma
  }(g)=(g^T)^{-1}$ for $g\in G^*$, as we will now see.  On the level
of the algebra, $\tilde{\sigma }(g)=(g^T)^{-1}$ means $\tilde{\sigma
  }(X)=-X^T$. Suppose now $g=\e^{tX}\in G^*$ with $X$ real and
traceless and $t$ a real parameter.  If now $X$ is an element of the
subalgebra ${\bf K}$, we then have $\tilde{\sigma }(X)=+X$, i.e.
$-X^T=X$ and $X$ is skew--symmetric.  If instead $X\in i{\bf P}$, we
have $\tilde{\sigma }(X)=-X^T=-X$, i.e. $X$ is symmetric. The
decomposition ${\bf G^*} = {\bf K} \oplus i{\bf P}$ is the usual
decomposition of a ${\bf SL(n,R)}$ matrix in symmetric and
skew--symmetric parts.

$G/K=SU(n,C)/SO(n,R)$ is a symmetric space of compact type, and the
related symmetric space of non--compact type is
$G^*/K=SL(n,R)/SO(n,R)$.

\subsection{The action of the group on the symmetric space}
\label{sec-action}

Let $G$ be a semisimple Lie group and $K$ a compact symmetric
subgroup.  As we saw in the preceding paragraph, the coset spaces
$G/K$ and $G^*/K$ represent symmetric spaces. Just as we have defined
a Cartan subalgebra and the rank of a Lie algebra, we can define, in
an exactly analogous way, a Cartan subalgebra and the rank of a
symmetric space.  A Cartan subalgebra of a symmetric space is a
maximal abelian subalgebra of the subspace ${\bf P}$ (see paragraph
\ref{sec-restricted}), and the rank of a symmetric space is the number
of generators in this subalgebra.

If $G$ is connected and ${\bf G} = {\bf K} \oplus {\bf P}$ where ${\bf
  K}$ is a compact symmetric subalgebra, then each group element can
be decomposed as $g=kp$ (right coset decomposition) or $g=pk$ (left
coset decomposition), with $k\in K={\rm e}^{\bf K}$, $p\in P={\rm
  e}^{\bf P}$.  $P$ is not a subgroup, unless it is abelian and
coincides with its Cartan subalgebra. However, if the involutive
automorphism that splits the algebra is denoted $\sigma $, one can
show (\cite{Hermann}, Ch. 6) that $gp\sigma (g^{-1})\in P$.  This
defines $G$ as a transformation group on $P$.  Since $\sigma
(k^{-1})=k^{-1}$ for $k\in K$, this means

\beq
p'=kpk^{-1}\in P 
\eeq

if $k\in K$, $p\in P$. Now suppose there are no other elements in $G$
that satisfy $\sigma (g)=g$ than those in $K$.  This will happen if
the set of elements satisfying $\sigma (g)=g$ is connected. Then $P$
is isomorphic to $G/K$. Also, $G$ acts transitively on $P$ in the
manner defined above (cf. subsection \ref{sec-cosets}).  The tangent
space of $G/K$ at the origin (identity element) is spanned by the
subspace ${\bf P}$ of the algebra.

\subsection{Radial coordinates}
\label{sec-radial}

In this paragraph we define two coordinate systems frequently used on
symmetric spaces.  Let ${\bf G}={\bf K}\oplus {\bf P}$ be a Cartan
decomposition of a semisimple algebra and let ${\bf H_0}\subset {\bf
  P}$ be a maximal abelian subalgebra in the subspace ${\bf P}$.
Define $M$ to be the subgroup of elements in $K$ such that

\beq
M=\{ k\in K:kHk^{-1}=H,\ H\in {\bf H_0}\}
\eeq

This set is called the centralizer of ${\bf H_0}$ in $K$. Under
conjugation by $k\in K$, each element $H$ of the Cartan subalgebra is
preserved. Further, denote

\beq
M'=\{ k\in K:kHk^{-1}=H',\ H,\, H'\in {\bf H_0}\}
\eeq

This is a larger subgroup than $M$ that preserves the Cartan
subalgebra as a whole, but not necessarily each element separately,
and is called the normalizer of ${\bf H_0}$ in $K$.  If $K$ is a
compact symmetric subgroup of $G$, one can show (\cite{Hermann},
Ch.~6) that every element $p$ of $P\simeq G/K$ is conjugated with some
element $h={\rm e}^H$ for some $H\in {\bf H_0}$ by means of the
adjoint representation\footnote{Note that \beq {\rm e}^{K}H{\rm
    e}^{-K}={\rm e}^{{\rm ad}K}H\equiv \sum_{n=0}^\infty \frac{({\rm
      ad}K)^n}{n!}H \eeq } of the stationary subgroup $K$:

\beq
\label{eq:KAK}
p=khk^{-1} = kh\sigma (k^{-1})
\eeq

where $k\in K/M$ and $H$ is defined up to the elements in the factor
group $M'/M$. This factor group coincides with the Weyl group that was
defined in eq.~(\ref{eq:Weyl}): since the space ${\bf H_0}$ can be
identified with its dual space ${\bf H_0^*}$, we can identify $M'/M$
with the Weyl group of the restricted root system (see paragraph 
\ref{sec-restricted}).  The effect of the
Weyl group is to transform the algebra ${\bf H_0}\subset {\bf P}$ into
another Cartan subalgebra ${\bf H_0'}\subset {\bf P}$ conjugate with
the original one. This amounts to a permutation of the roots of the
restricted root lattice corresponding to a Weyl reflection.
Equation~(\ref{eq:KAK}) means that every element $g\in G$ can be
decomposed as $g=pk=k'hk'^{-1}k=k'hk''$, and this is very much like the
Euler angle decomposition of $SO(n)$.

Thus, if $x_0$ is the fixed point of the subgroup $K$, an arbitrary
point $x\in P$ can be written

\beq
x=khk^{-1}x_0=khx_0
\eeq

The coordinates $(k(x),h(x))$ are called spherical coordinates. $k(x)$
is the angular coordinate and $h(x)$ is the {\it spherical radial
  coordinate} of the point $x$.  Eq.~(\ref{eq:KAK}) defines the so
called spherical decomposition of the elements in the coset space.  Of
course, a similar reasoning is true for the space $P^*\simeq G^*/K$.

This means every matrix $p$ in the coset space $G/K$ can be {\it
  diagonalized} by a similarity transformation by the subgroup $K$,
and the radial coordinates are exactly the set of eigenvalues of the
matrix $p$.  These ``eigenvalues'' are not necessarily real numbers.
This is easily seen in the example in eq.~(\ref{eq:SU3}). It can also
be seen in the adjoint representation. Suppose the algebra ${\bf
  G}={\bf K}\oplus {\bf P}$ is compact.  From eq.~(\ref{eq:adjr}), in
the adjoint representation $H_i\in {\bf H_0}$ has the form

\beq
\label{eq:adjHi}
H_i=\left(\begin{array}{cccccccc} 0 & ...    &   & & & & & \\
                                . & \ddots &   & & & & & \\
                                . &        & 0 & & & & & \\
                                  &        &   & \alpha_i & & & & \\
                                  &        &   &          & -\alpha_i & & \\
                                  &        &   &          &           & \ddots & \\
                                  &        &   &          &           &        &  \eta_i & \\
                                  & & & & & &                                            & -\eta_i 
\end{array}\right)
\eeq

where the matrix is determined by the structure constants
($[H_i,H_j]=0$, $[H_i,E_{\pm \alpha }]=\pm \alpha_i E_{\pm \alpha}$
... and $\pm \alpha_i,...,\pm \eta_i$ are the roots corresponding to
$H_i$). Since the Killing form must be negative (see subsection
\ref{sec-metric}) for a compact algebra, the coordinates of the Cartan
subalgebra must be purely imaginary and the group elements
corresponding to ${\bf H_0}$ must have the form

\beq
\label{eq:adjrad}
\e^{i{\bf t\cdot H}}=\left(\begin{array}{cccccc} 1 & ...    &   & & & \\               
                   . & \ddots &   & & & \\
                   . &        & 1 & & & \\
                     &        &   & \e^{i{\bf t\cdot \alpha }} & & \\
                     &        &   &          & \ddots & \\
                     &        &   &          &        & 
\e^{-i{\bf t\cdot \eta }} \end{array}\right)
\eeq

with ${\bf t}=(t^1,t^2,...t^r)$ and $t^i$ real parameters.  In
particular, if the eigenvalues are real for $p \in P^*$, they are
complex numbers for $p\in P$.

{\bf Example}: In the example we gave in the preceding subsection, the
coset space $G^*/K$ $=SL(n,R)/SO(n)$ $\simeq P^*=\e^{i{\bf P}}$
consists of real positive--definite symmetric matrices.  Note that
${\bf G} = {\bf K} \oplus {\bf P}$ implies that $G$ can be decomposed
as $G=PK$ and $G^*$ as $G^*=P^*K$.  The decomposition $G^*=P^*K$ in
this case is the decomposition of a $SL(n,R)$ matrix in a
positive--definite symmetric matrix and an orthogonal one. Each
positive--definite symmetric matrix can be further decomposed: it can
be diagonalized by an $SO(n)$ similarity transformation. This is the
content of eq.~(\ref{eq:KAK}) for this case, and we know it to be true
from linear algebra.  Similarly, according to eq.~(\ref{eq:KAK}) the
complex symmetric matrices in $G/K$ $=SU(n,C)/SO(n)$ $\simeq P=\e^{\bf
  P}$ can be diagonalized by the group $K=SO(n)$ to a form where the
eigenvalues are similar to those in eq.~(\ref{eq:adjrad}).

In terms of the subspace ${\bf P}$ of the algebra, eq.~(\ref{eq:KAK})
amounts to saying that any two Cartan subalgebras ${\bf H_0}$, ${\bf
  H'_0}$ of the symmetric space are conjugate under a similarity
transformation by $K$, and we can choose the Cartan subalgebra in any
way we please.  However, the number of elements that we can
diagonalize simultaneously will always be equal to the rank of the
symmetric space.

There is also another coordinate system valid only for spaces of the
type $P^*\sim G^*/K$. This coordinate system is called {\it
  horospheric} and is based on the so called {\it Iwasawa
  decomposition} \cite{Hermann} of the algebra:

\beq
{\bf G}={\bf N^+}\oplus {\bf H_0} \oplus {\bf K}
\eeq

Here ${\bf K},\ {\bf H_0},\ {\bf N^+}$ are three subalgebras of ${\bf
  G}$. ${\bf K}$ is a maximal compact subalgebra, ${\bf H_0}$ is a
Cartan subalgebra, and

\beq
{\bf N^+}=\sum_{\alpha \in R^+}{\bf G'_\alpha}
\eeq

is an algebra of raising operators corresponding to the positive roots
$\alpha (H) >0$ with respect to ${\bf H_0}$ (${\bf G'_\alpha}$ is the
space generated by $E_\alpha $). As a consequence, the group
elements can be decomposed $g=nhk$, in an obvious notation. This means
that if $x_0$ is the fixed point of $K$, any point $x\in G^*/K$ can be
written
 
\beq
x=nhkx_0=nhx_0
\eeq

The coordinates $(n(x),h(x))$ are called horospheric
coordinates and the element $h=h(x)$ is called the
{\it horospheric projection} of the point $x$ or the {\it horospheric
  radial coordinate}.

\subsection{The metric on a Lie algebra} 
\label{sec-metric}
 
A metric tensor can be defined on a Lie algebra
\cite{Helgason,Gilmore,SattW,Hermann}. For our purposes, it will
eventually serve to define the curvature of a symmetric space and be
useful in computing the Jacobian of the transformation to radial
coordinates. In sections \ref{sec-Operators} and \ref{sec-RMT} we will
see the importance of this Jacobian in physical applications in
connection with the radial part of the Laplace--Beltrami operator.

If $\{ X_i\}$ form a basis for the Lie algebra ${\bf G}$, the metric
tensor is defined by

\beq
\label{eq:metric}
g_{ij}=K(X_i,X_j)\equiv \tr ({\rm ad}X_i {\rm ad}X_j)=C^r_{is}C^s_{jr}
\eeq

The symmetric bilinear form $K(X_i,X_j)$ is called the {\it Killing form}.  
It is intrinsically associated with the Lie algebra, and
since the Lie bracket is invariant under automorphisms of the algebra, 
so is the Killing form.

{\bf Example:} The generators $X_7 \equiv L_1$, $X_5 \equiv L_2$, $X_2
\equiv L_3$ of $SO(3)$ given in eq.~(\ref{eq:SU3}) obey the
commutation relations
$[L_i,L_j]=C_{ij}^kL_k=\frac{1}{2}\epsilon_{ijk}L_k$. From
eq.~(\ref{eq:metric}), the metric for this algebra is
$g_{ij}=-\frac{1}{2}\delta_{ij}$. The generators and the structure
constants can be normalized so that the metric takes the canonical
form $g_{ij}=-\delta_{ij}$.

Just like we defined the Killing form $K(X_i,X_j)$ for the algebra
${\bf G}$ in eq.~(\ref{eq:metric}) using the adjoint representation,
we can define a similar trace form $K_\rho$ and a metric tensor
$g_\rho $ for {\it any} representation $\rho $ by

\beq
\label{eq:g_rho}
g_{\rho,ij}=K_\rho (X_i,X_j) ={\rm tr}(\rho (X_i)\rho (X_j))
\eeq

where $\rho (X)$ is the matrix representative of the Lie algebra
element $X$.  If $\rho $ is an automorphism of ${\bf G}$, $K_\rho
(X_i,X_j)=K(X_i,X_j)$.

Suppose the Lie algebra is semisimple (this is true for all the
classical Lie algebras except the Lie algebras $GL(n,C)$, $U(n,C)$).
According to Cartan's criterion, {\it the Killing form is
  non--degenerate for a semisimple algebra}.  This means that ${\rm
  det}g_{ij}\neq 0$, so that the inverse of $g_{ij}$, denoted by
$g^{ij}$, exists. Since it is also real and symmetric, it can be
reduced to canonical form $g_{ij}={\rm diag}(-1,...,-1,1,...,1)$
with $p$ $-1$'s and $(n-p)$ $+1$'s, where $n$ is the dimension of the algebra. 

$p$ is an invariant of the quadratic form. In fact, for any real form
of a complex algebra, the trace of the metric, called the {\it
  character} of the particular real form (see below and in
\cite{Gilmore}) distinguishes the real forms from each other (though
it can be degenerate for the classical Lie algebras \cite{Gilmore}).
The character ranges from $-n$, where $n$ is the dimension of the
algebra, to $+r$, where $r$ is its rank. All the real forms of the
algebra have a character that lies in between these values. In subsection
\ref{sec-realforms1} we will see several explicit examples of Killing
forms.

A famous theorem by Weyl states that {\it a simple Lie group $G$ is
  compact, if and only if the Killing form on ${\bf G}$ is negative
  definite}. Otherwise it is non--compact. This is actually quite
intuitive and natural (see \cite{Gilmore}, Ch.~9, paragraph~I.2). On a
compact algebra, the metric can be chosen to be minus the Killing
form, if it is required to be positive--definite.

The metric on the Lie algebra can be extended to the whole coset space
$P\simeq G/K$, $P^*\simeq G^*/K$ as follows. At the origin of $G/K$
and $G^*/K$, the identity element $I$, the metric is identified with
the metric in the algebra, restricted to the respective tangent spaces
${\bf P}$, $i{\bf P}$.  Since the group acts transitively on the coset
space (cf. paragraph \ref{sec-cosets}), and the orbit of the origin is
the entire space, we can use a group transformation to map the metric
at the origin to any point $M$ in the space.  The metric tensor at $M$
will depend on the coset representative $M$.  It is given by

\beq
\label{eq:g(M)}
g_{rs}(M)=
g_{ij}(I)\frac{\partial x^i(I)}{\partial x^r(M)}\frac{\partial x^j(I)}{\partial x^s(M)}
\eeq

where $g_{ij}(I)$ is the metric at the origin (identity element) of
the coset space. (\ref{eq:g(M)}) follows from the invariance of the
line element $ds^2=g_{ij}dx^idx^j$ under translations.  If $\{ X_i\} $
is a basis in the tangent space, and $dM={\rm exp}(dx^iX_i)$ is a
coset representative infinitesimally close to the identity, we need to
know how $dx^i$ transforms under translations by the coset
representative $M$. We will not discuss that here, but some
generalities can be found for example in Ch.~9, paragraph V.4. of
ref.~\cite{Gilmore}. In general, it is not an easy problem unless the
coset has rank $1$.

{\bf Example:} The line element $ds^2$ on the radius--1 2--sphere
$SO(3)/SO(2)$ in polar coordinates is $ds^2= d\theta^2+{\rm
  sin}^2\theta \, d\phi^2$. The metric at the point $(\theta,\phi)$ is

\beq
\label{eq:metric_on_sphere}
g_{ij}=\left(\begin{array}{cc} 1 & 0 \\
                       0 & {\rm sin}^2\theta \end{array}\right),\ \ \ \ \ \ \  
g^{ij}=\left(\begin{array}{cc} 1 & 0 \\
                       0 & {\rm sin}^{-2}\theta \end{array}\right) 
\eeq

where the rows and columns are labelled in the order $\theta $, $\phi $.

The distance between points on the symmetric space is defined as
follows.  The length of a vector $X=\sum_it^iX_i$ in the tangent space
${\bf P}$ (this object is well--defined because ${\bf P}$ is endowed
with a definite metric) is identified with the length of the geodesic
connecting the identity element in the coset space with the element
$M={\rm exp}(X)$ \cite{Gilmore}.

\subsection{The algebraic structure of symmetric spaces}
\label{sec-algstr}

Except for the two algebras ${\bf SL(n,R)}$ and ${\bf SU^*(2n)}$ (and
their dual spaces related by the Weyl trick), for which the subspace
representatives of ${\bf K}$, ${\bf P}$ and $i{\bf P}$ consist of
square, irreducible matrices (for ${\bf SL(n,R)}$, we saw this in the
example in subsection \ref{sec-Inv} and for ${\bf SU(n,C)}$ explicitly
in eq.~(\ref{eq:SU3})), the matrix representatives of the subalgebra
${\bf K}$ and of the subspaces ${\bf P}$ and $i{\bf P}$ in the
fundamental representation consist of block--diagonal matrices $X\in
{\bf K}$, $Y\in {\bf P}$, $Y'\in i{\bf P}$ of the form \cite{Gilmore}

\beq
\label{eq:ssrep}
X=\left(\begin{array}{cc} A & 0 \\
                          0 & B \end{array}\right),\ \ \ \ \ \ 
Y=\left(\begin{array}{cc} 0 & C \\
                          -C^\dagger & 0 \end{array}\right),\ \ \ \ \ \
Y'=\left(\begin{array}{cc} 0 & \tilde{C} \\
                          \tilde{C}^\dagger & 0 \end{array}\right), 
\eeq
 
in the Cartan decomposition. Here $A^\dagger =-A$, $B^\dagger =-B$ and
$\tilde{C}=iC$.  In fact, for {\it any} finite--dimensional
representation, the matrix representatives of ${\bf K}$ and ${\bf P}$
are antihermitean (thus they become antisymmetric if the
representation of ${\bf P}$ is real) and as a consequence, those of
$i{\bf P}$ are hermitean (symmetric in case the representation of
$i{\bf P}$ is real) \cite{Gilmore}.  This is true irrespective of
whether the matrix representatives are block--diagonal or square.

The exponential maps of the subspaces ${\bf P}$ and $i{\bf P}$ are
isomorphic to coset spaces $G/K$ and $G^*/K$, respectively (see for
example \cite{Helgason,Hermann}).  The exponential map of the algebra
maps the subspaces ${\bf P}$ and $i{\bf P}$ into unitary and hermitean
matrices, respectively. In the fundamental representation, these
spaces are mapped onto \cite{Gilmore}

\bea
\label{eq:cosetreps}
{\rm exp}({\bf P})={\rm exp}{\left(\begin{array}{cc} 0 & C \\
                                       -C^\dagger & 0 \end{array}\right)}=
\left(\begin{array}{cc} \sqrt{I-XX^\dagger} & X \\
                         -X^\dagger & \sqrt{I-XX^\dagger}\end{array}\right) 
\nonumber \\
{\rm exp}(i{\bf P})={\rm exp}{\left(\begin{array}{cc} 0 & \tilde{C} \\
                                  \tilde{C}^\dagger & 0 \end{array}\right)}=
\left(\begin{array}{cc} \sqrt{I+\tilde{X}\tilde{X}^\dagger} & \tilde{X}\\
    \tilde{X}^\dagger & \sqrt{I+\tilde{X}\tilde{X}^\dagger}\end{array}\right)
\\ \nonumber
\eea

where $X$ is a spherical and $\tilde{X}$ a hyperbolic function of the 
submatrix $C$ ($\tilde{C}$):

\beq
\label{eq:functions}
X=C\frac{{\rm sin}\sqrt{C^\dagger C}}{\sqrt{C^\dagger C}},\ \ \ 
\tilde{X}=\tilde{C}\frac{{\rm sinh}\sqrt{\tilde{C}^\dagger \tilde{C}}}
{\sqrt{\tilde{C}^\dagger \tilde{C}}}
\eeq

This shows explicitly that the range of parameters parametrizing 
the two cosets is bounded for the compact coset and unbounded for the
non--compact coset, respectively. We already saw an explicit example 
of these formulas in subsection \ref{sec-cosets}.

\section{Real forms of semisimple algebras}
\label{sec-realforms}
\setcounter{equation}{0}

In this section we will introduce the tools needed to find all the
real forms of any (semi)simple algebra. The same tools will then be used
in the next section to find the real forms of a symmetric space.
When thinking of a real form, it is convenient to visualize it in
terms of its metric. As we saw in paragraph \ref{sec-metric} the trace
of the metric is called the character of the real form and it
distinguishes the real forms from each other. In the following
subsection we discuss various real forms of an algebra and we see how
to go from one form to another. In each case, we compute the metric
and the character explicitly. We also give the simplest possible example 
of this procedure, the rank--1 algebra. In subsection \ref{sec-realforms2}  
we enumerate the involutive automorphisms needed to classify all real forms
of semisimple algebras and again, we illustrate it with two examples.
 
\subsection{The real forms of a complex algebra}
\label{sec-realforms1}

In general a semisimple complex algebra has several distinct real
forms.  Recall from subsection \ref{sec-rootsp} that a real form of an
algebra is obtained by taking linear combinations of its elements with
real coefficients.  The real forms of the complex Lie algebra ${\bf
  G}$

\beq
\label{eq:complexalg}
\sum_i c^iH_i + \sum_\alpha c^\alpha E_\alpha \ \ \ \ \ \ \ (c^i,\ c^\alpha 
\ {\rm complex}),
\eeq

where ${\bf H_0}=\{ H_i\} $ is the Cartan subalgebra and $\{E_{\pm
\alpha }\}$ are the sets of raising and lowering operators, can be
classified according to all the involutive automorphisms of ${\bf G}$
obeying $\sigma^2=1$. Two distinctive real forms are the normal real
form and the compact real form.

The {\it normal real form} of the algebra (\ref{eq:complexalg}), which
is also the least compact real form, consists of the subspace in which
the coefficients $c^i$, $c^\alpha$ are real. The metric in this case
with respect to the bases $\{H_i,E_{\pm \alpha}\}$ is (with
appropriate normalization of the elements of the Lie algebra to make
the entries of the metric equal to $\pm 1$)

\beq
\label{eq:raislow_metric}
g_{ij}=\left( \begin{array}{cccccccc}  1 &        &    &   &    &        &   &    \\
                                         & \ddots &    &   &    &        &   &    \\
                                         &        &  1 &   &    &        &   &    \\
                                         &        &    & 0 & 1  &        &   &    \\
                                         &        &    & 1 & 0  &        &   &    \\
                                         &        &    &   &    & \ddots &   &    \\
                                         &        &    &   &    &        & 0 & 1  \\
                                         &        &    &   &    &        & 1 & 0  \end{array} \right)
\eeq

where the $r$ 1's on the diagonal correspond to the elements of the
Cartan subalgebra ($r$ is obviously the rank of the algebra), and the
$2\times 2$ matrices on the diagonal correspond to the pairs $E_{\pm
  \alpha}$ of raising and lowering operators. This structure reflects
the decomposition of the algebra ${\bf G}$ into a direct sum of the
root spaces: ${\bf G}={\bf H_0}\oplus \sum_\alpha {\bf G_\alpha}$.
This metric tensor can be transformed to diagonal form, if we choose
the generators to be

\beq
\label{eq:NRFd}
{\bf K}=\left\{\frac{(E_\alpha - E_{-\alpha})}{\sqrt{2}}\right\},\ \ \ \ \ \ \ 
i{\bf P}=\left\{H_i,\frac{(E_\alpha + E_{-\alpha})}{\sqrt{2}}\right\} 
\eeq

{\bf Example:} In our example with ${\bf SU(3,C)}$, ${\bf K}$ and
$i{\bf P}$ are exactly the subspaces spanned by $\{ X_2,X_5,X_7\} $
and $\{ iX_1,iX_3,iX_4,iX_6,iX_8\}$ (cf. eq.~(\ref{eq:SU3})), and
$(E_\alpha - E_{-\alpha})$ and $-i(E_\alpha + E_{-\alpha})$ are
exactly the Gell--Mann matrices (cf. eq.~(\ref{eq:raislowSU3})).

Then $g_{ij}$ takes the form

\beq
\label{eq:NRFd_metric}
g_{ij}=\left( \begin{array}{cccccccc}  1 &        &    &   &    &        &   &    \\
                                         & \ddots &    &   &    &        &   &    \\
                                         &        &  1 &   &    &        &   &    \\
                                         &        &    & 1 & 0  &        &   &    \\
                                         &        &    & 0 & -1 &        &   &    \\
                                         &        &    &   &    & \ddots &   &    \\
                                         &        &    &   &    &        & 1 & 0  \\
                                         &        &    &   &    &        & 0 & -1  \end{array} \right)
\eeq

where the entries with a minus sign correspond to the generators of
the compact subalgebra ${\bf K}$, the first $r$ entries equal to $+1$ 
correspond to the Cartan subalgebra, and the remaining ones to the operators
in $i{\bf P}$ {\it not} in the Cartan subalgebra.  
This is the diagonal metric tensor corresponding to 
the normal real form. The character of the normal real form is plus the rank 
of the algebra. 

The {\it compact real form} of ${\bf G}$ is obtained 
from the normal real form by the Weyl unitary trick:

\beq
\label{eq:compactrealform}
{\bf K}=\left\{\frac{(E_\alpha - E_{-\alpha})}{\sqrt{2}}\right\},\ \ \ \ \ \ \ 
{\bf P}=\left\{iH_i,\frac{i(E_\alpha + E_{-\alpha})}{\sqrt{2}}\right\} 
\eeq

The character of the compact real form is minus the dimension of the
algebra, and the metric tensor is $g_{ij}={\rm diag}(-1,...,-1)$.  

{\bf Example:} We will use as an example the well--known ${\bf
  SU(2,C)}$ algebra with Cartan subalgebra ${\bf H_0}=\{ J_3\}$ and
raising and lowering operators $\{ J_\pm \} $.  We have chosen the
normalization such that the non--zero entries of $g_{ij}$ are all
equal to $1$:

\beq
\begin{array}{c}
J_3=\frac{1}{2\sqrt{2}}\tau_3,\ \ \ \ \ \ \ 
J_\pm =\frac{1}{4}(\tau_1\pm i\tau_2)\end{array}
\eeq

where in the defining representation of ${\bf SU(2,C)}$

\beq
\tau_3= \left( \begin{array}{cc} 1 & 0  \\ 
                                   0 & -1 \end{array} \right),
\ \ \ \ 
\tau_1= \left( \begin{array}{cc} 0 & 1  \\ 
                                   1 & 0  \end{array} \right),
\ \ \ \ 
\tau_2= \left( \begin{array}{cc} 0 & -i \\ 
                                   i & 0  \end{array} \right)
\eeq

The normalization is such that 

\beq
\begin{array}{c}
[J_3,J_\pm]=\pm \frac{1}{\sqrt{2}}J_\pm,\ \ \ \ \ \ \ 
[J_+,J_-]=\frac{1}{\sqrt{2}}J_3\end{array}
\eeq

In equation (\ref{eq:SU2adjoint}) we constructed the adjoint
representation of this algebra, albeit with a different normalization.
Using the present normalization to set the entries of the metric equal
to 1, we see that the non--zero structure constants are
$C^+_{3+}=-C^+_{+3}=-C^-_{3-}=C^-_{-3}=C^3_{+-}=-C^3_{-+}=\frac{1}{\sqrt{2}}$.
The entries of the metric are given by eq.~(\ref{eq:metric}),
$g_{ij}=K(J_i,J_j)=C^r_{is}C^s_{jr}$ with summation over repeated
indices, so we see that the metric of the normal real form ${\bf
  SU(2,R)}$ in this basis is

\beq
g_{ij}=\left( \begin{array}{ccc} 1 & 0 & 0 \\
                                 0 & 0 & 1 \\
                                 0 & 1 & 0 \end{array} \right)
\eeq

where the rows and columns are labelled by $3,+,-$ respectively. This 
corresponds to eq.~(\ref{eq:raislow_metric}).

To pass now to a diagonal metric, we just have to set

\beq
\begin{array}{c}
\Sigma_3=J_3 \nonumber \\ \nonumber \\
\Sigma_1=\frac{J_++J_-}{\sqrt{2}}=\frac{1}{2\sqrt 2}\tau_1 \nonumber \\
\nonumber \\
\Sigma_2=\frac{J_+-J_-}{\sqrt{2}}=\frac{i}{2\sqrt 2}\tau_2 \end{array} 
\nonumber 
\eeq

like in eq.~(\ref{eq:NRFd}). The commutation relations then become

\beq
\label{eq:Sigmarels}
\begin{array}{c} 
[\Sigma_1,\Sigma_2]=-\frac{1}{\sqrt 2}\Sigma_3,\ \ \ \ \ \ \ 
[\Sigma_2,\Sigma_3]=-\frac{1}{\sqrt 2}\Sigma_1,\ \ \ \ \ \ \ 
[\Sigma_3,\Sigma_1]=\frac{1}{\sqrt 2}\Sigma_2\end{array}
\eeq

These commutation relations characterize the algebra ${\bf
  SO(2,1;R)}$.  From here we find the structure constants
$C^3_{12}=-C^3_{21}=C^1_{23}=-C^1_{32}=-C^2_{31}=C^2_{13}=
-\frac{1}{\sqrt 2}$ and the diagonal metric of the normal real form
with rows and columns labelled $3,1,2$ (in order to comply with the
notation in eq.~(\ref{eq:NRFd_metric})) is

\beq
\label{eq:SO21}
g_{ij}=\left( \begin{array}{ccc} 1 & 0 & 0 \\
                                 0 & 1 & 0 \\
                                 0 & 0 & -1 \end{array} \right)
\eeq

which is to be compared with eq.~(\ref{eq:NRFd_metric}).  According to
eq.~(\ref{eq:NRFd}), the Cartan decomposition of ${\bf G^*}$ is ${\bf
  G^*}={\bf K}\oplus i{\bf P}$ where ${\bf K}=\{\Sigma_2\}$ and $i{\bf
  P}=\{\Sigma_3,\Sigma_1\}$.  The Cartan subalgebra consists of $\Sigma_3$.

Finally, we arrive at the compact real form by multiplying $\Sigma_3$
and $\Sigma_1$ with $i$.  Setting $i\Sigma_1=\tilde{\Sigma}_1$,
$\Sigma_2=\tilde{\Sigma}_2$, $i\Sigma_3=\tilde{\Sigma}_3$ the
commutation relations become those of the special orthogonal group:

\beq
\label{eq:tildeSigmarels}
\begin{array}{c} 
[\tilde{\Sigma}_1,\tilde{\Sigma}_2]=-\frac{1}{\sqrt 2}\tilde{\Sigma}_3,\ \ \ \ \ \ \ 
[\tilde{\Sigma}_2,\tilde{\Sigma}_3]=-\frac{1}{\sqrt 2}\tilde{\Sigma}_1,\ \ \ \ \ \ \ 
[\tilde{\Sigma}_3,\tilde{\Sigma}_1]=-\frac{1}{\sqrt 2}\tilde{\Sigma}_2\end{array}
\eeq

The last commutation relation in eq.~(\ref{eq:Sigmarels}) has changed
sign whereas the others are unchanged.  $C^2_{31}$, $C^2_{13}$, and
consequently $g_{33}$ and $g_{11}$ change sign and we get the metric 
for ${\bf SO(3,R)}$:

\beq
\label{eq:SO3}
g_{ij}=\left( \begin{array}{ccc} -1 & 0 & 0 \\
                                 0 & -1 & 0 \\
                                 0 & 0 & -1 \end{array} \right)
\eeq 

This is the compact real form. The subspaces of the compact algebra
${\bf G}={\bf K}\oplus {\bf P}$ are ${\bf K}=\{\tilde{\Sigma}_2\}$ and
${\bf P}=\{\tilde{\Sigma}_3,\tilde{\Sigma}_1\}$.  Weyl's theorem
states that a simple Lie group $G$ is compact, if and only if the
Killing form on ${\bf G}$ is negative definite; otherwise it is
non--compact.  In the present example, we see this explicitly.

\subsection{The classification machinery}
\label{sec-realforms2}

To classify all the real forms of any complex Lie algebra, with
characters lying between the character of the normal real form and
the compact real form (the intermediate real forms obviously have an
indefinite metric), it suffices to enumerate all the involutive
automorphisms of its compact real form.  A detailed and almost
complete account of these procedures for the non--exceptional groups
can be found in \cite{Gilmore}, Chapter~9, paragraph~3. To summarize,
if ${\bf G}$ is the compact real form of a complex semisimple Lie
algebra ${\bf G^C}$, ${\bf G^*}$ runs through all its associated 
non--compact real
forms ${\bf G^*}$, ${\bf G'^*}$, ... with corresponding maximal
compact subgroups ${\bf K}$, ${\bf K'}$, ...  and complementary
subspaces $i{\bf P}$, $i{\bf P'}$, ...  as $\sigma $ runs through all
the involutive automorphisms of ${\bf G}$.

One such automorphism is complex conjugation $\sigma_1 =K$, which is
used to split the compact real algebra into subspaces ${\bf K}$ and
${\bf P}$ in eq.~(\ref{eq:compactrealform}). (To avoid confusion: the
generators can be complex even though the field of real numbers is
used to multiply the generators in a real form of an algebra. If the
generators are also real, we speak of a real representation. However,
whether we consider the field to be ${\bf R}$ and the generators to be
complex, or the opposite, also depends on our definition of basis;
cf. one of the footnotes in subsection \ref{sec-rootsp}).
The involutive automorphisms $\sigma $ satisfy $\sigma {\bf G}
\sigma^{-1} = {\bf G} $, $\sigma^2=1$, which implies that $\sigma $
either commutes or anticommutes with the elements of the compact
algebra ${\bf G}$: if $\sigma X \sigma^{-1} = X'$, then $\sigma X'
\sigma^{-1} = X$, and we get $X'=\pm X$ for $X,\ X'\in {\bf G}$ (see
the example below).  One can show \cite{Loos} (Ch.~VII), that it
suffices to consider the following three possibilities for $\sigma $:
$\sigma_1 =K$, $\sigma_2 = I_{p,q}$ and $\sigma_3 = J_{p,p}$ where

\beq
\label{eq:I_J_}
I_{p,q} = \left( \begin{array}{cc} I_p & 0 \\
                                   0  & -I_q \end{array} \right), 
\ \ \ \ \ \  J_{p,p} = \left( \begin{array}{cc} 0 & I_p \\
                                   -I_p  &  0 \end{array} \right) 
\eeq

and $I_p$ denotes the $p \times p$ unit matrix. By operating with one
(or two successive ones) of these automorphisms on the elements of
${\bf G}$, we can construct the subspaces ${\bf K}$ and ${\bf P}$, and
${\bf K}$ and $i{\bf P}$ of the corresponding non--compact real form
${\bf G^*}$.  A complex algebra and all its real forms (the compact
and the various non--compact ones) correspond to the same root lattice
and Dynkin diagram.

{\bf Example}: The normal real form of the complex algebra ${\bf
  G^C}={\bf SL(n,C)}$ is the non--compact algebra ${\bf G^*}={\bf
  SL(n,R)}$. As we saw in subsection \ref{sec-Inv}, this algebra can
be decomposed as ${\bf K}\oplus i{\bf P}$ where ${\bf K}$ is the
algebra consisting of real, skew--symmetric and traceless $n\times n$
matrices and $i{\bf P}$ is the algebra consisting of real, symmetric
and traceless $n\times n$ matrices.  Under the Weyl unitary trick we
constructed, in a previous example, this algebra from the compact real
form of ${\bf G^C}$, ${\bf SU(n,C)}={\bf G}={\bf K}\oplus {\bf P}$.

Starting with the compact real form ${\bf G}$, we can construct all
the various non--compact real forms ${\bf G^*}$, ${\bf G'^*}$,...
from it, by applying the involutive automorphisms $\sigma_1$,
$\sigma_2$, $\sigma_3$ to the elements of ${\bf G}$.  All the real
forms related to the root system $A_{n-1}$ are obtained by applying
the three involutions to ${\bf G}={\bf SU(n,C)}$:
 
$\sigma_1$) The involutive automorphism $\sigma_1=K$ (complex
conjugation) splits ${\bf G}={\bf SU(n,C)}$ into ${\bf K}\oplus {\bf P}$ 
(we recall this from the example in paragraph \ref{sec-Inv}).
The non--compact real form obtained this way, by the Weyl unitary
trick, is exactly the normal real form ${\bf G^*}={\bf K}\oplus 
i{\bf P}= {\bf SL(n,R)}$.

$\sigma_2$) A general matrix in the Lie algebra ${\bf SU(n,C)}$ can be
written in the form 

\beq
\label{eq:generalSU(n)}
X=\left( \begin{array}{cc} A & B \\
                          -B^\dagger & C \end{array}\right)
\eeq

where $A$, $C$ are complex $p\times p$ and $q\times q$ matrices
satisfying $A^\dagger = -A$, $C^\dagger = -C$, ${\rm tr}A+{\rm tr}C=0$
(since the determinant of the group elements must be $+1$), and $B$ is
an arbitrary complex $p\times q$ matrix ($p+q=n$). In
eq.~(\ref{eq:generalSU(n)}), the matrices $A$, $B$ and $C$ are all
linear combinations of submatrices in {\it both} subspaces ${\bf
  K}=\{\frac{1}{\sqrt{2}}(E_{\alpha_i}-E_{-\alpha_i})\}$ and ${\bf
  P}=\{iH_j,\frac{i}{\sqrt{2}}(E_{\alpha_i}+E_{-\alpha_i})\}$. The
action of the involution $\sigma_2=I_{p,q}$ on $X$ is

\beq
I_{p,q}XI_{p,q}^{-1}= \left( \begin{array}{cc} A & -B \\
                          B^\dagger & C \end{array}\right)
\eeq

Therefore, we see that the subspaces ${\bf K'}$ and ${\bf P'}$ are given by
the matrices

\beq
\left( \begin{array}{cc} A & 0 \\
                          0 & C \end{array}\right) \in {\bf K'},
\ \ \ \ \ \ \ 
\left( \begin{array}{cc} 0 & B \\
                          -B^\dagger & 0 \end{array}\right) \in {\bf P'}
\eeq

Indeed, we see that $I_{p,q}$ transforms the Lie algebra elements in 
${\bf K'}$ into themselves, and those in ${\bf P'}$ into minus themselves.
The transformation by $I_{p,q}$ mixes the subspaces ${\bf K}$ and ${\bf P}$,
and splits the algebra in a different way into ${\bf K'}\oplus {\bf P'}$. 
The matrices

\beq
\left( \begin{array}{cc} A & iB \\
                         -iB^\dagger & C \end{array}\right)
\in {\bf K'}\oplus i{\bf P'}
\eeq

define the non--compact real form ${\bf G'^*}$. 
This algebra is called ${\bf SU(p,q;C)}$ and its maximal 
compact subalgebra ${\bf K'}$ is ${\bf SU(p)\otimes SU(q)\otimes U(1)}$. 

$\sigma_3$) By the involutive automorphism $\sigma_3\sigma_1=J_{p,p}K$
one constructs in a similar way (for details see \cite{Gilmore}) a
third non--compact real form (for even $n=2p$) ${\bf G''^*}={\bf
  K''}\oplus i{\bf P''}$ associated to the algebra ${\bf G}={\bf
  SU(2p,C)}$.  ${\bf G''^*}$ is the algebra ${\bf SU^*(2p)}$ and its
maximal compact subalgebra is ${\bf USp(2p)}$.  \footnote{ The algebra
  ${\bf SU^*(2p)}$ is represented by complex $2p\times 2p$ matrices of
  the form \beq
  X=\left(\begin{array}{cc} A & B \\
      -B^* & -A^*\end{array}\right) \eeq where ${\rm tr}A+{\rm
    tr}A^*=0$.  ${\bf USp(2p)}$ denotes the complex $2p\times 2p$
  matrix algebra of the group with both unitary and symplectic
  symmetry (${\bf USp(2p,C)}$ can also be denoted ${\bf U(p,Q)}$ where
  $Q$ is the field of quaternions).  A matrix in the algebra ${\bf
    USp(2p,C)}$ can be written as \beq
  X=\left(\begin{array}{cc} A & B \\
      -B^{\dagger } & -A^R\end{array}\right) \eeq where
  $A^\dagger=-A$, $B^R=B$, and the superscript $^R$ denotes reflection
  in the minor diagonal. }

This procedure, summarized in the formula below, 
exhausts all the real forms of the simple algebras.  

\beq
{\bf G^C} \to {\bf G}={\bf K}\oplus {\bf P}  
\begin{array}{l}{\buildrel{\scriptstyle \sigma_1} \over \nearrow }\\
 {\buildrel{\scriptstyle \sigma_2} \over \to } \\
 {\buildrel{\scriptstyle \sigma_3} \over \searrow }
\end{array}
\begin{array}{l}
{\bf G^*}={\bf K}\oplus i{\bf P}\\ \\
{\bf G'^*}={\bf K'}\oplus i{\bf P'}\\ \\
{\bf G''^*}={\bf K''}\oplus i{\bf P''}
\end{array}
\eeq

{\bf Example:} Note that it may not always be possible to apply all
the above involutions $\sigma_1$, $\sigma_2$, $\sigma_3$ to the
algebra. For example, complex conjugation $\sigma_1$ does not do
anything to ${\bf SO(2n+1,R)}$, because it is represented by real
matrices, neither is $\sigma_3$ a symmetry of this algebra, since the
adjoint representation is odd--dimensional and $\sigma_3$ has to act
on a $2p\times 2p$ matrix. The only possibility that remains is
$\sigma_2=I_{p,q}$.  For a second, even more concrete example, let's
look at the algebra ${\bf SO(3,R)}$, belonging to the root lattice
$B_1$. This algebra is spanned by the generators $L_1$, $L_2$, $L_3$
given in subsection \ref{sec-cosets}. A general element of the algebra
is

\beq
X={\bf t\cdot L}=\frac{1}{2}\left(\begin{array}{ccc}     & t^3 & t^2\\
                   -t^3 &     & t^1\\
                   -t^2 & -t^1& \end{array}\right)=
\frac{1}{2}\left(\begin{array}{ccc}     & t^3 & \\
                   -t^3 &     & \\
                        &     & \end{array}\right) \oplus
\frac{1}{2}\left(\begin{array}{ccc}     &     & t^2\\
                        &     & t^1\\
                   -t^2 & -t^1&    \end{array}\right)
\eeq

This splitting of the algebra is caused by the involution $I_{2,1}$ acting
on the representation:

\beq
I_{2,1}XI_{2,1}^{-1}=\left(\begin{array}{ccc} 1 &    &    \\
                                           &  1 &    \\
                                           &    & -1 \end{array}\right)
\frac{1}{2}\left(\begin{array}{ccc}     & t^3 & t^2\\
                                   -t^3 &     & t^1\\
                                   -t^2 & -t^1& \end{array}\right)
\left(\begin{array}{ccc} 1 &    &    \\
                           &  1 &    \\
                           &    & -1 \end{array}\right)
=\frac{1}{2}\left(\begin{array}{ccc}     & t^3 & -t^2\\
                                    -t^3 &     & -t^1\\
                                     t^2 & t^1& \end{array}\right)
\eeq

and it splits it into 
${\bf SO(3)}={\bf K}\oplus {\bf P}={\bf SO(2)}\oplus {\bf SO(3)/SO(2)}$.
Exponentiating, as we saw in subsection \ref{sec-cosets}, the coset 
representative is a point on the 2--sphere

\beq
M=\left(\begin{array}{ccc}
.&.&t^2\frac{{\rm sin}\sqrt{(t^1)^2+(t^2)^2}}{\sqrt{(t^1)^2+(t^2)^2}}\\
.&.&t^1\frac{{\rm sin}\sqrt{(t^1)^2+(t^2)^2}}{\sqrt{(t^1)^2+(t^2)^2}}\\
.&.&{\rm cos}\sqrt{(t^1)^2+(t^2)^2}\end{array}\right)=
\left(\begin{array}{ccc} . & .  & x\\
                         . & .  & y\\
                         . &  . & z   \end{array}\right);\ \ \ \ \ \ \  
x^2+y^2+z^2=1
\eeq

By the Weyl unitary trick we now get the non--compact real form 
${\bf G^*}={\bf K}\oplus i{\bf P}$: 
${\bf SO(2,1)}={\bf SO(2)}\oplus {\bf SO(2,1)/SO(2)}$. This algebra is
represented by 

\beq
\label{eq:SO(2,1)}
\left(\begin{array}{ccc}     & t^3 & it^2\\
                   -t^3 &     & it^1\\
                   -it^2 & -it^1& \end{array}\right)=
\left(\begin{array}{ccc}     & t^3 & \\
                   -t^3 &     & \\
                        &     & \end{array}\right) \oplus
\left(\begin{array}{ccc}     &     & it^2\\
                        &     & it^1\\
                   -it^2 & -it^1&    \end{array}\right)
\eeq

and after exponentiation of the coset generators

\beq
M=\left(\begin{array}{ccc} . & . & it^2\frac{{\rm sinh}\sqrt{(t^1)^2+(t^2)^2}}
                             {\sqrt{(t^1)^2+(t^2)^2}} \\
                      . & . & it^1\frac{{\rm sinh}\sqrt{(t^1)^2+(t^2)^2}}
                             {\sqrt{(t^1)^2+(t^2)^2}}\\
                      . & . & {\rm cosh}\sqrt{(t^1)^2+(t^2)^2}  \end{array}\right)=
\left(\begin{array}{ccc} .  & .  & ix\\
                         .  & .  & iy\\
                         .  & .  & z   \end{array}\right);\ \ \ \ \ \ \  
(ix)^2+(iy)^2+z^2=1
\eeq

The surface in ${\bf R}^3$ consisting of points $(x,y,z)$ satisfying this
equation is the hyperboloid $H^2$. Similarly, we get the isomorphic space
$SO(1,2)/SO(2)$ by applying $I_{1,2}$: ${\bf SO(1,2)}={\bf \tilde{K}}\oplus 
i{\bf \tilde{P}}={\bf SO(2)}\oplus {\bf SO(1,2)/SO(2)}$ and in terms of the 
algebra

\beq
\tilde{X}=
\frac{1}{2}\left(\begin{array}{ccc}  &      &     \\
                                     &      & t^1 \\
                                     & -t^1 &     \end{array}\right) \oplus
\frac{1}{2}\left(\begin{array}{ccc}     & -it^3 & -it^2\\
                                  it^3  &       &      \\
                                  it^2  &       & \end{array}\right)
\eeq

\section{The classification of symmetric spaces}
\label{sec-claSS}
\setcounter{equation}{0}

In this section we introduce the curvature tensor and the sectional
curvature of a symmetric space, and we extend the family of symmetric
spaces to include also flat or Euclidean--type spaces. These are
identified with the subspace ${\bf P}$ of the Lie algebra itself, and
the group that acts on it is a semidirect product of the subgroup $K$
and the subspace ${\bf P}$. As we will learn, to each compact subgroup
$K$ corresponds a triplet of symmetric spaces with positive, zero and
negative curvature.  The classification of these symmetric spaces is
in exact correspondence with the new classification of random matrix
models to be discussed in Part~II. These spaces exhaust the
Cartan classification and have a definite metric. They are listed in
Table~\ref{tab1} together with some of their properties.

In paragraph \ref{sec-restricted} we introduce restricted root
systems.  In the same way as a Lie algebra corresponds to a given root
system, the ``algebra'' (subspace ${\bf P}$ or $i{\bf P}$) of each
symmetric space corresponds to a restricted root system.  These root
systems are of primary importance in the physical applications to be
discussed in Part~II.  The restricted root system can be of an
entirely different type from the root system inherited from the
complex extension algebra, and its rank may be different. We work out
a specific example of a restricted root system as an illustration. In
spite of their importance, we have not been able to find any explicit
reference in the literature that explains how to obtain the restricted
root systems. Instead, we found that they are often referred to in
tables and in mathematical texts without explicitly mentioning that
they are restricted, which could easily lead to confusion with the
inherited root systems. In reference
\cite{Gilmore} the root system that is associated to each symmetric
space is the one inherited from the complex extension algebra, whereas
for example in Table~B1 of reference~\cite{OlshPere} and in
\cite{Loos} the restricted root systems are listed.
 
There are also symmetric spaces with an indefinite metric, so called
pseudo--Riemannian spaces, corresponding to a maximal {\it
  non--compact} subgroup $H$. For completeness, we will briefly
discuss how these are obtained as real forms of symmetric spaces
corresponding to compact symmetric subgroups.  This does not require
any new tools than the ones we have already introduced, namely the
involutive automorphisms.

\subsection{The curvature tensor and triplicity}
\label{sec-curv}

Suppose that ${\bf K}$ is a maximal compact subalgebra of the
non--compact algebra ${\bf G^*}$ in the Cartan decomposition ${\bf
  G^*} = {\bf K} \oplus i{\bf P}$, where $i{\bf P}$ is a complementary
subspace. ${\bf K}$ and ${\bf P}$ (alternatively ${\bf K}$ and $i{\bf
  P}$) satisfy eq.~(\ref{eq:commrel}):

\beq
\label{eq:commrel2}
[{\bf K},{\bf K}]\subset {\bf K},\ \  [{\bf K},{\bf P}]\subset {\bf P},\ \  
[{\bf P},{\bf P}]\subset {\bf K} 
\eeq

${\bf K}$ is called a symmetric subalgebra and the coset spaces $\exp
({\bf P})\simeq G/K$ and $\exp (i{\bf P})\simeq G^*/K$ are globally
symmetric Riemannian spaces. Globally symmetric means that every point
on the manifold can be moved to any other point by a particular group
operation (we discussed this in paragraph \ref{sec-cosets}; for a
rigorous definition of globally symmetric spaces see Helgason
\cite{Helgason}, paragraph IV.3).  In the same way, the metric can be
defined in any point of the manifold by moving the metric at the
origin to this point, using a group operation (cf. eq.~(\ref{eq:g(M)})
in paragraph \ref{sec-metric}).  The Killing form restricted to the
tangent spaces ${\bf P}$ and $i{\bf P}$ at any point in the coset
manifold has a definite sign. The manifold is then called
``Riemannian''. The metric can be taken to be either plus or minus the
Killing form so that it is always positive definite (cf.
paragraph~\ref{sec-metric}).
 
A {\it curvature tensor} with components $R^i_{jkl}$ can be defined on
the manifold $G/K$ or $G^*/K$ in the usual way
\cite{Helgason,FosNigh}. It is a function of the metric tensor and its
derivatives. It was proved for instance in \cite{Helgason}, Ch. IV,
that the components of the curvature tensor at the origin of a
globally symmetric coset manifold is given by the expression

\beq
\label{eq:curv}
R^n_{ijk}X_n= [X_i,[X_j,X_k]] = C^n_{im}C^m_{jk}X_n
\eeq

where $\{X_i\}$ is a basis for the Lie algebra. The sectional
curvature at a point $p$ is equal to

\beq
\label{eq:K}
{\cal K}=g([[X,Y],X],Y)
\eeq

where $g$ is an arbitrary symmetric and nondegenerate metric (such a
metric is also called a pseudo--Riemannian structure, or simply a
Riemannian structure if it has a definite sign) on the tangent
space at $p$, invariant under the action of the group elements.  In
(\ref{eq:K}), $g(X_i,X_j)\equiv g_{ij}$ and $\{ X,Y\} $ is an
orthonormal basis for a two--dimensional subspace $S$ of the tangent
space at the point $p$ (assuming it has dimension $\geq 2$).  The
sectional curvature is equal to the gaussian curvature on a
2--dimensional manifold. If the manifold has dimension $\geq 2$,
(\ref{eq:K}) gives the sectional curvature along the section $S$.

Eqs.~(\ref{eq:curv}) and (\ref{eq:K}), together with
eq.~(\ref{eq:commrel2}) show that the curvature of the spaces $G/K$
and $G^*/K$ has a definite and opposite sign (\cite{Helgason},
par.~V.3). Thus, we see that if $G$ is a compact semisimple group, to
the same subgroup $K$ there corresponds a positive curvature space
$P\simeq G/K$ and a dual negative curvature space $P^*\simeq G^*/K$.
The reason for this is exactly the same as the reason why the sign changes 
for the components of the metric corresponding to the generators in $i{\bf P}$
as we go to the dual space ${\bf P}$. We remind the reader that the sign of the
metric can be chosen positive or negative for a compact space. The issue here
is that the sign changes in going from $G^*/K$ to $G/K$. 

{\bf Example:} We can use the example of $SU(2)$ in paragraph
\ref{sec-realforms1} to see that the sectional curvature is the
opposite for the two spaces $G/K$ and $G^*/K$. If we take $\{
X,Y\}=\{\Sigma_3, \Sigma_1\}$ as the basis in the space $i{\bf P}$ and
$\{\tilde{\Sigma}_3, \tilde{\Sigma}_1\}$ ($\tilde{\Sigma}_i\equiv
i\Sigma_i$) as the basis in the space ${\bf P}$, we see by comparing
the signs of the entries of the metrics we computed in eqs.~(\ref{eq:SO21})
and (\ref{eq:SO3}) that the sectional curvature ${\cal K}$ at the origin 
has the opposite sign for the two spaces $SO(2,1)/SO(2)$ and $SO(3)/SO(2)$.

Actually, there is also a zero--curvature symmetric space $X^0=G^0/K$
related to $X^+=G/K$ and $X^-=G^*/K$, so that we can speak of a {\it
  triplet} of symmetric spaces related to the {\it same} symmetric
subgroup $K$. The zero--curvature spaces were discussed in
\cite{OlshPere} and in Ch.~V of Helgason's book \cite{Helgason}, where
they are referred to as ``symmetric spaces of the euclidean type''.
That their curvature is zero was proved in Theorem~3.1 of
\cite{Helgason}, Ch.~V.

The flat symmetric space $X^0$ can be identified with
the subspace ${\bf P}$ of the algebra. The group $G^0$ is a semidirect
product of the subgroup $K$ and the invariant subspace ${\bf P}$ of
the algebra, and its elements $g=(k,a)$ act on the elements of $X^0$
in the following way:

\beq
g(x)=kx+a,\ \ \ \ k\in K,\ \ \ \  x,a \in X^0
\eeq

if the $x$'s are vectors, and 

\beq
\label{eq:matrices}
g(x)=kxk^{-1}+a,\ \ \ \ k\in K,\ \ \ \  x,a \in X^0
\eeq

if the $x$'s are matrices. We will see one example of each below.

The elements of the algebra ${\bf P}$ now define an {\it abelian
additive group}, and $X^0$ is a vector space with euclidean geometry.
In the above scenario, the subspace ${\bf P}$ contains only the
operators of the Cartan subalgebra and no others: ${\bf P}={\bf H_0}$,
so that ${\bf P}$ is a subalgebra of ${\bf G^0}$. The algebra ${\bf
G^0}={\bf K}\oplus {\bf P}$ belongs to a non--semisimple group $G^0$,
since it has an abelian ideal ${\bf P}$: $[{\bf K},{\bf K}]\subset
{\bf K}$, $[{\bf K},{\bf P}]\subset {\bf P}$, $[{\bf P},{\bf
P}]=0$. Note that ${\bf K}$ and ${\bf P}$ still satisfy the
commutation relations (\ref{eq:commrel2}).  In this case the coset
space $X^0$ is flat, since by (\ref{eq:commrel2}), $R^n_{ijk}=0$ for
all the elements $X\in {\bf P}$.  Eq.~(\ref{eq:curv}) is valid for any
space with a Riemannian structure.  Indeed, it is easy to see from
eqs. (\ref{eq:curv}), (\ref{eq:K}) that $R^n_{ijk}={\cal K}=0$ if the
generators are abelian.  Even though the Killing form on
non--semisimple algebras is degenerate, it is trivial to find a
non--degenerate metric on the symmetric space $X^0$ that can be used
in ~(\ref{eq:K}) to find that the sectional curvature at any point is
zero.  For example, as we pass from the sphere to the plane, the
metric becomes degenerate in the limit as $[L_1,L_2]\sim L_3\to
[P_1,P_2]=0$ (see the example below). Obviously, we do not inherit
this degenerate metric from the tangent space on ${\bf R^2}$ like in
the case of the sphere, but the usual metric for ${\bf R^2}$,
$g_{ij}=\delta_{ij}$ provides the Riemannian structure on the plane.

{\bf Examples:} An example of a flat symmetric space is $E_2/K$, where
$G^0=E_2$ is the euclidean group of motions of the plane ${\bf R^2}$:
$g(x)= kx+a$, $g=(k,a)\in G^0$ where $k\in K=SO(2)$ and $a\in {\bf
  R}^2$.  The generators of this group are translations $P_1$, $P_2
\in {\bf H_0}= {\bf P}$ and a rotation $J\in {\bf K}$ satisfying
$[P_1,P_2]=0$, $[J,P_i]=-\epsilon^{ij}P_j$, $[J,J]=0$, in agreement
with eq.~(\ref{eq:commrel2}) defining a symmetric subgroup. The
abelian algebra of translations $\sum_{i=1}^2t^iP_i$, $t^i\in {\bf
  R}$, is isomorphic to the plane ${\bf R^2}$, and can be identified
with it.

The commutation relations for $E_2$ are a kind of limiting case of the
commutation relations for ${\bf SO(3)} \sim {\bf SU(2)}$ and ${\bf
  SO(2,1)}$. If in the limit of infinite radius of the sphere $S^2$ we
identify $\tilde{\Sigma}_1$ with $P_1$, $\tilde{\Sigma}_2$ with $P_2$,
and $\tilde{\Sigma}_3$ with $J$, we see that the commutation relations
resemble the ones described in eq.~(\ref{eq:Sigmarels}) and
(\ref{eq:tildeSigmarels}) -- we only have to set
$[\tilde{\Sigma}_1,\tilde{\Sigma}_2]=0$, which amounts to setting
$C_{12}^3=-C_{21}^3\to 0$. From here we get the degenerate metric of
the non--semisimple algebra ${\bf E_2}$:

\beq
g_{ij}=\left(\begin{array}{ccc} -1 & & \\ & 0 & \\ & & 0 \end{array}\right)
\eeq

where the only nonzero element is $g_{33}$. This is to be confronted
with eqs.~(\ref{eq:SO21}) and (\ref{eq:SO3}) which are the metrics for
${\bf SO(2,1)}$ and ${\bf SO(3)}$. This is an example of contraction of 
an algebra.

An example of a triplet $\{X^+, X^0, X^-\}$ corresponding to the same 
subgroup $K=SO(n)$ is: 

1) $X^+=SU(n,C)/SO(n)$, the set of symmetric unitary matrices with
unit determinant; it is the space ${\rm exp}({\bf P})$ where ${\bf P}$
are real, symmetric and traceless $n \times n$ matrices.  (Cf. the
example in subsection \ref{sec-Inv}.)

2) $X^0$ is the set ${\bf P}$ of real, symmetric and traceless
$n\times n$ matrices and the non--semisimple group $G^0$ is the group
whose action is defined by $g(x)=kxk^{-1}+a$, $g=(k,a)\in G^0$ where
$k\in K=SO(n)$ and $x,a\in X^0$.  The involutive automorphism maps
$g=(k,a)\in G^0$ into $g'=(k,-a)$.

3) $X^-=SL(n,R)/SO(n)$ is the set of real, positive, symmetric
matrices with unit determinant; it is the space ${\rm exp}(i{\bf P})$
where ${\bf P}$ are real, symmetric and traceless $n \times n$
matrices.

We remark that the zero--curvature symmetric spaces correspond to the
integration manifolds of many known matrix models with physical
applications.

The pairs of dual symmetric spaces of positive and negative curvature
listed in each row of Table~\ref{tab1} originate in the same complex extension
algebra \cite{Gilmore} with a given root lattice. This ``inherited''
root lattice is listed in the first column of the table.  In our
example in paragraph \ref{sec-realforms2} this was the root lattice of
the complex algebra ${\bf G^C}= {\bf SL(n,C)}$. The same root lattice
$A_{n-1}$ characterizes the real forms of ${\bf SL(n,C)}$: as we saw
in the example these are the algebras ${\bf SU(n,C)}$, ${\bf
  SL(n,R)}$, ${\bf SU(p,q;C)}$ and ${\bf SU^*(2n)}$, and we have seen
how to construct them using involutive automorphisms.

However, also listed in Table~\ref{tab1} is the {\it restricted root system}
corresponding to each symmetric space. This root system may be
different from the one inherited from the complex extension algebra.
Below, we will define the restricted root system and see an explicit
example of one such system. While the original root lattice
characterizes the complex extension algebra and its real forms, the
restricted root lattice characterizes a particular symmetric space
originating from one of its real forms.  The root lattices of the
classical simple algebras are the infinite sequences $A_n$, $B_n$,
$C_n$, $D_n$, where the index $n$ denotes the rank of the
corresponding group.  The {\it root multiplicities} $m_o$, $m_l$,
$m_s$ listed in Table~\ref{tab1} (where the subscripts refer to ordinary, long
and short roots, respectively) are characteristic of the restricted
root lattices.  In general, in the root lattice of a simple algebra
(or in the graphical representation of any irreducible
representation), the roots (weights) may be degenerate and thus have a
multiplicity greater than 1. This happens if the same weight $\mu =
(\mu_1,...,\mu_r)$ corresponds to different states in the
representation. In that case one can arrive at that particular weight
using different sets of lowering operators $E_{-\alpha}$ on the
highest weight of the representation. Indeed, we saw in the example of
$SU(3,C)$ in subsection \ref{sec-rootsp}, that the roots can have a
multiplicity different from 1. The same is true for the restricted
roots.

The sets of simple roots of the classical root systems (briefly listed
in subsection \ref{sec-srs}) have been obtained for example in
\cite{Gilmore,Georgi}. In the canonical basis in ${\bf R^n}$, the
roots of type $\{ \pm e_i \pm e_j, i\neq j\}$ are ordinary while the
roots $\{ \pm 2e_i \}$ are long and the roots $\{ \pm e_i \}$ are
short.  Only a few sets of root multiplicities are compatible with the
strict properties characterizing root lattices in general.

\subsection{Restricted root systems}
\label{sec-restricted}

The restricted root systems play an important role in connection with
matrix models and integrable Calogero--Sutherland models (these models
will be introduced in section \ref{sec-intmod}).  We will discuss this
in detail in Part~II. In this subsection we will explain how
restricted root systems are obtained and how they are related to a
given symmetric space.\footnote{The authors are indebted to Prof. Simon
Salamon for explaining how the restricted root systems are obtained.}

As we have repeatedly seen in the examples using the compact algebra
${\bf SU(n,C)}$ (in particular in subsection \ref{sec-realforms2}),
the algebra ${\bf SU(p,q;C)}$ (${\bf p}+{\bf q}={\bf n}$) is a
non--compact real form of the former. This means they share the same
rank--$(n-1)$ root system $A_{n-1}$. However, to the {\it symmetric
space} $SU(p,q;C)/(SU(p)\otimes SU(q)\otimes U(1))$ one can associate
another rank--$r'$ root system, where $r'={\rm min}(p,q)$ is the rank
of the symmetric space. For some symmetric spaces, it is the same as
the root system inherited from the complex extension algebra (see
Table~\ref{tab1} for a list of the restricted root systems), but this
need not be the case.  For example, the restricted root system is, in
the case of $SU(p,q;C)/(SU(p)\otimes SU(q)\otimes U(1))$,
$BC_{r'}$. When it is the same and when it is different, as well as
why the rank can change, will be obvious from the example we will give
below.

In general the restricted root system will be different from the
original, inherited root system if the Cartan subalgebra is a subset
of ${\bf K}$. The procedure to find the restricted root system is then
to define an {\it alternative Cartan subalgebra} that lies partly (or
entirely) in ${\bf P}$ (or $i{\bf P}$).

To achieve this, we first look for a different representation of the
original Cartan subalgebra, that gives the same root lattice as the
original one (i.e., $A_{n-1}$ for the ${\bf SU(p,q;C)}$ algebra). In
general, this root lattice is an automorphism of the original root
lattice of the same kind, obtained by a permutation of the roots.
Unless we find this new representation, we will not be able to find a
new, alternative Cartan subalgebra that lies partly in the subspace
${\bf P}$.

Once this has been done, we take a maximal abelian subalgebra of ${\bf
  P}$ (the number of generators in it will be equal to the rank $r'$
of the symmetric space $G/K$ or $G^*/K$) and find the generators in
${\bf K}$ that commute with it. These generators will be among the
ones that are in the new representation of the original Cartan
subalgebra.  These commuting generators now form our new,
alternative Cartan subalgebra that lies partly in ${\bf P}$,
partly in ${\bf K}$.  Let's call it ${\bf A_0}$.

The new root system is defined with respect to the part of the maximal
abelian subalgebra that lies in ${\bf P}$. Therefore its rank is
normally smaller than the rank of the root system inherited from the
complex extension. We can define raising and lowering operators
$E'_\alpha$ in the {\it whole} algebra ${\bf G}$ that satisfy

\beq
[X'_i,E'_\alpha]=\alpha'_i E'_\alpha \ \ \ \ \ \ \ (X'_i\in {\bf A_0}\cap
{\bf P})
\eeq

The roots $\alpha'_i$ define the restricted root system.  

{\bf Example:} Let's now look at a specific example. We will start
with the by now familiar algebra ${\bf SU(3,C)}$. As before, we use
the convention of regarding the $T_i$'s as the generators, without the
factor of $i$ (recall that the algebra consists of elements of the form
$\sum_at^aX_a=i\sum_at^aT_a$; cf. the footnote in conjuction with
eq.~(\ref{eq:Gell-Mann})).  In subsection \ref{sec-rootsp} we
explicitly constructed its root lattice $A_2$. Let's write down the
generators again:

\beq
\label{eq:Gell-Mann'}
\begin{array}{l}
T_1=\frac{1}{2}\left(\begin{array}{ccc} 0 & 1 & 0 \\
                                   1 & 0 & 0 \\
                                   0 & 0 & 0 \end{array}\right), \ \ \ \
T_2=\frac{1}{2}\left(\begin{array}{ccc} 0 & -i & 0 \\
                                   i & 0 & 0 \\
                                   0 & 0 & 0 \end{array}\right), \ \ \ \
T_3=\frac{1}{2}\left(\begin{array}{ccc} 1 & 0 & 0 \\
                                   0 & -1 & 0 \\
                                   0 & 0 & 0 \end{array}\right), \nonumber \\
\\
T_4=\frac{1}{2}\left(\begin{array}{ccc} 0 & 0 & 1 \\
                                   0 & 0 & 0 \\
                                   1 & 0 & 0 \end{array}\right), \ \ \ \  
T_5=\frac{1}{2}\left(\begin{array}{ccc} 0 & 0 & -i \\
                                   0 & 0 & 0 \\
                                   i & 0 & 0 \end{array}\right), \ \ \ \
T_6=\frac{1}{2}\left(\begin{array}{ccc} 0 & 0 & 0 \\
                                   0 & 0 & 1 \\
                                   0 & 1 & 0 \end{array}\right), \nonumber \\ 
\\ 
T_7=\frac{1}{2}\left(\begin{array}{ccc} 0 & 0 & 0 \\
                                   0 & 0 & -i \\
                                   0 & i & 0 \end{array}\right), \ \ \ \
T_8=\frac{1}{2\sqrt{3}}\left(\begin{array}{ccc} 1 & 0 & 0 \\
                                                0 & 1 & 0 \\
                                                0 & 0 & -2 
                                   \end{array}\right) \end{array}
\\ \nonumber
\eeq

The splitting of the ${\bf SU(3,C)}$ algebra in terms of the subspaces
${\bf K}$ and ${\bf P}$ was given in eq.~(\ref{eq:SU3}):

\beq
{\bf K}=\{iT_2,iT_5,iT_7\} ,\ \ \ \ \ \ \
{\bf P}=\{iT_1,iT_3,iT_4,iT_6,iT_8\}
\eeq

The Cartan subalgebra is $\{ iT_3,iT_8\} $.
The raising and lowering operators were given in (\ref{eq:raislowSU3})
in terms of $T_i$:

\beq
\label{eq:raislowSU3'}
\begin{array}{l}
E_{\pm(1,0)}=\frac{1}{\sqrt{2}}(T_1\pm iT_2) \\ \\
E_{\pm(\frac{1}{2},\frac{\sqrt{3}}{2})}=\frac{1}{\sqrt{2}}(T_4\pm iT_5) \\ \\
E_{\pm(-\frac{1}{2},\frac{\sqrt{3}}{2})}=\frac{1}{\sqrt{2}}(T_6\pm iT_7) 
\end{array} 
\eeq

Now let us construct the Cartan decomposition of ${\bf G'^*}={\bf
  K'}\oplus i{\bf P'}= {\bf SU(2,1;C)}$. We know from paragraph
\ref{sec-realforms2} that ${\bf K'}$ and ${\bf P'}$ are given by
matrices of the form

\beq
\left( \begin{array}{cc} A & 0 \\
                          0 & C \end{array}\right) \in {\bf K'},
\ \ \ \ \ \ \ 
\left( \begin{array}{cc} 0 & B \\
                          -B^\dagger & 0 \end{array}\right) \in {\bf P'}
\eeq

where $A$ and $C$ are antihermitean and ${\rm tr}A+{\rm tr}C=0$.
Combining the generators to form this kind of block--structures (or
alternatively, using the involution $\sigma_2=I_{2,1}$) we need to
take linear combinations of the $X_i$'s, with real coefficients, and
we then see that the subspaces ${\bf K'}$ and $i{\bf P'}$ are spanned
by

\beq 
\begin{array}{l}
{\bf K'}=\left\{ 
\frac{i}{2}\left(\begin{array}{ccc}  0 & 1 &   \\
                                     1 & 0 &   \\
                                       &   & 0 \end{array}\right),
\frac{1}{2}\left(\begin{array}{ccc}  0 & 1 &   \\
                                    -1 & 0 &   \\
                                       &   & 0 \end{array}\right),
\frac{i}{2}\left(\begin{array}{ccc}  1 & 0  &   \\
                                     0 &-1 &   \\
                                       &   & 0 \end{array}\right),
\frac{i}{2\sqrt{3}}\left(\begin{array}{ccc} 1  & 0 &   \\
                                            0  & 1 &   \\
                                               &   & -2 \end{array}\right)
\right\} \\ \\
=\{ iT_1,iT_2,iT_3,iT_8\}\\  \\ 
i{\bf P'}=\left\{ 
\frac{1}{2}\left(\begin{array}{ccc}    &   & 1  \\
                                       &   & 0  \\
                                     1 & 0 &    \end{array}\right),
\frac{i}{2}\left(\begin{array}{ccc}    &   & -1  \\
                                       &   & 0  \\
                                    1  & 0 &    \end{array}\right),
\frac{1}{2}\left(\begin{array}{ccc}    &   & 0  \\
                                       &   & 1  \\
                                     0 & 1 &    \end{array}\right),
\frac{i}{2}\left(\begin{array}{ccc}    &   & 0  \\
                                       &   & -1  \\
                                    0  & 1 &    \end{array}\right)
\right\}\\  \\
=\{ T_4,T_5,T_6,T_7\}\end{array}
\eeq

where the block--structure is evidenced by leaving blank the remaining
zero entries. ${\bf K'}$ spans the algebra of the symmetric subgroup
$SU(2)\otimes U(1)$ and $i{\bf P'}$ spans the complementary subspace
corresponding to the symmetric space $SU(2,1)/(SU(2)\otimes U(1))$.
$i{\bf P'}$ is spanned by matrices of the form

\beq
\left(\begin{array}{cc} 0 & \tilde{B}\\ \tilde{B}^\dagger & 0\end{array}\right)
\eeq

We see that the Cartan subalgebra $i{\bf H_0}=\{iT_3,iT_8\}$ lies
entirely in ${\bf K'}$.  It is easy to see that by using the
alternative representation

\beq
T'_3=\frac{1}{2}\left(\begin{array}{ccc}  1 &   &   \\
                                            &0  &   \\
                                            &   & -1 \end{array}\right),
\ \ \ \ \ \ \ 
T'_8=\frac{1}{2\sqrt{3}}\left(\begin{array}{ccc}  1 &   &   \\
                                                    &-2 &   \\
                                                    &   & 1 \end{array}\right)
\eeq

of the Cartan subalgebra (note that this is a valid representation of
${\bf SU(3,C)}$ generators) while the other $T_i$'s are unchanged, we
still get the same root lattice $A_2$. The eigenvectors under the
adjoint representation, the $E_\alpha $'s, are still given by
eq.~(\ref{eq:raislowSU3'}). However, their eigenvalues (roots) are
permuted under the new adjoint representation of the 
Cartan subalgebra, so that they no longer correspond to the root
subscripts in (\ref{eq:raislowSU3'}).  This permutation is a Weyl reflection;
more specifically, it is the reflection in the hyperplane orthogonal 
to the root $(-\frac{1}{2},\frac{\sqrt{3}}{2})$.

Now we choose the alternative Cartan subalgebra to consist of
the generators $T_4$, $T'_8$: 

\beq
{\bf A_0}=\{T_4,T'_8\},\ \ \ [T_4,T'_8]=0, \ \ \ \ \ \ \ iT_4\in {\bf P'},
\ iT'_8\in {\bf K'}
\eeq

(Note that unless we first take a new representation of the original
Cartan subalgebra, we are not able to find the alternative Cartan
subalgebra that lies partly in ${\bf P'}$.) The restricted root system
is now about to be revealed.  We define raising and lowering operators
$E'_\alpha$ in the whole algebra according to

\beq
E'_{\pm 1}\sim (T_5\pm iT_3)\ \ \ \  
E'_{\pm \frac{1}{2}}\sim (T_6\pm iT_2)\ \ \ \  
\tilde{E}'_{\pm \frac{1}{2}}\sim (T_7\pm iT_1) 
\eeq

The $\pm\alpha $ subscripts are the eigenvalues of $T_4\in i{\bf
  P'}$ in the adjoint representation:

\bea
\label{eq:BC_1}
[T_4,E'_{\pm 1}]=\pm E'_{\pm 1},\ \ \ [T_4,E'_{\pm \frac{1}{2}}]=\pm
\frac{1}{2} E'_{\pm \frac{1}{2}},\ \ \ [T_4,\tilde{E}'_{\pm \frac{1}{2}}]=
\pm \frac{1}{2}\tilde{E}'_{\pm \frac{1}{2}}
\eea

These roots form a one--dimensional root system of type $BC_1$. We see
that the multiplicity of the long roots is $1$ and the multiplicity of
the short roots is $2=2(p-q)$. This result is general (cf. Table~\ref{tab1}).
If we had ordinary roots, their multiplicity would be $2$, but for
this low--dimensional group we can have only $3$ pairs of roots. Note
that we can rescale the lengths of all the roots together by rescaling
the operator $T_4$ in (\ref{eq:BC_1}), but their characters as long
and short roots can not change. The root system $BC_1$ is with respect
to the part of the Cartan subalgebra lying in $i{\bf P'}$ only, thus
it is called restricted.

According to eq.~(\ref{eq:KAK}), every element $p$ of $P\simeq G/K$ is
conjugated with some element $h={\rm e}^H$ ($H\in {\bf H_0}$) through
$p=khk^{-1}$, where $k\in K/M'$ and $H$ is defined up to the elements
in the factor group $M'/M$. Thus, the decomposition $p=khk^{-1}$ is
not unique.  The factor group $M'/M$ transforms a Cartan subalgebra
${\bf H_0}\subset {\bf P}$ into another Cartan subalgebra ${\bf
  H_0'}\subset {\bf P}$ conjugate with the original one.  This amounts
to a permutation of the roots of the restricted root lattice
corresponding to Weyl reflections. The factor group $M'/M$ then
coincides with the Weyl group of the restricted root system.  If we
fix the Weyl chamber of $H$, $H$ is unique and $k$ is defined up to
transformations by the subgroup $M$.
\newpage

\begin{table}[ht]
\caption{Irreducible symmetric spaces of positive and negative
curvature originating in simple Lie groups. In the fourth and fifth
columns the symmetric spaces $G/K$ and $G^*/K$ are listed for all the
entries except the simple Lie groups themselves, for which the
symmetric spaces $G$ and $G^C/G$ are listed.  Note that there are also
zero curvature spaces corresponding to non--semisimple groups and
isomorphic to the subspace ${\bf P}$ of the algebra, when ${\bf P}$ is
an abelian invariant subalgebra. These are not listed in the table,
but can be constructed as explained in subsection \ref{sec-curv}. The
root multiplicities listed pertain to the {\it restricted} root
systems of the pairs of dual symmetric spaces with positive and
negative curvature.\label{tab1}}
\vskip5mm

\hskip-.8cm
\begin{tabular}{|l|l|l|l|l|l|l|l|}
\hline
$\begin{array}{c}Root\\ space\end{array}$&
$\begin{array}{c}Restricted\\ root\ space\end{array}$ & $\begin{array}{c}Cartan\\ class \end{array}$ & $G/K\ (G)$ & $G^*/K\ (G^C/G)$ & $m_o$ & $m_l$ & $m_s$ \\
\hline

$A_{N-1}$ & $A_{N-1}$     & A    & $SU(N)$                 & $\frac{SL(N,C)}{SU(N)}$ & 2 & 0 & 0 \\   
          & $A_{N-1}$     & AI   & $\frac{SU(N)}{SO(N)}$   & $\frac{SL(N,R)}{SO(N)}$ & 1 & 0 & 0 \\  
          & $A_{N-1}$     & AII  & $\frac{SU(2N)}{USp(2N)}$ & $\frac{SU^*(2N)}{USp(2N)}$ & 4 & 0 & 0  \\
          &\hskip-2mm $\begin{array}{l} BC_q\ {\scriptstyle (p>q)} \\ C_q\  {\scriptstyle (p=q)} \end{array}$ 
                          & AIII & $\frac{SU(p+q)}{SU(p)\times SU(q)\times U(1)}$ & $\frac{SU(p,q)}{SU(p)\times SU(q)\times U(1)}$ & 2 & 1 & $2(p-q)$ \\
\hline

$B_N$ &$B_N$          & B    & $SO(2N+1) $               & $\frac{SO(2N+1,C)}{SO(2N+1)}$ & 2 & 0 & 2 \\

\hline

$C_N$  & $C_N$    & C    & $USp(2N)$                               &$\frac{Sp(2N,C)}{USp(2N)}$ & 2 & 2 & 0 \\
       & $C_N$    & CI   & $\frac{USp(2N)}{SU(N)\times U(1)}$      & $\frac{Sp(2N,R)}{SU(N)\times U(1)}$& 1 & 1 & 0  \\
       &\hskip-2mm $\begin{array}{l} BC_q\  {\scriptstyle (p>q)} \\  C_q\  {\scriptstyle (p=q)} \end{array}$ 
                  & CII & $\frac{USp(2p+2q)}{USp(2p)\times USp(2q)}$ & $\frac{USp(2p,2q)}{USp(2p)\times USp(2q)}$& 4 & 3 & $4(p-q)$ \\
\hline

$D_N$& $D_N$     & D          &   $SO(2N)$                            & $\frac{SO(2N,C)}{SO(2N)}$ & 2 & 0 & 0 \\
     &$C_N$     & DIII-even  & $\frac{SO(4N)}{SU(2N)\times U(1)}$     & $\frac{SO^*(4N)}{SU(2N)\times U(1)}$ & 4 & 1 & 0 \\
     &$BC_N$    & DIII-odd   & $\frac{SO(4N+2)}{SU(2N+1)\times U(1)}$ & $\frac{SO^*(4N+2)}{SU(2N+1)\times U(1)}$& 4 & 1 & 4 \\

\hline
\hskip-2mm $\begin{array}{l} B_N\ {\scriptstyle (p+q=2N+1)}\\ D_N\ {\scriptstyle (p+q=2N)}\end{array}$  
          &\hskip-2mm $\begin{array}{l} B_q\ {\scriptstyle (p>q)}\\ D_q\ {\scriptstyle (p=q)} \end{array}$
            & BDI & $\frac{SO(p+q)}{SO(p)\times SO(q)}$   & $\frac{SO(p,q)}{SO(p)\times SO(q)}$ & 1 & 0 & $p-q$ \\

\hline
\end{tabular}
\end{table}
  
 \newpage 
\subsection{Real forms of symmetric spaces}
\label{sec-realformsSS}

Involutive automorphisms were used to 
split the algebra ${\bf G}$ into orthogonal subspaces
to obtain the real forms ${\bf G}$, ${\bf G^*}$, ${\bf G'^*}$... 
of a complex extension algebra ${\bf G^C}$. By re--applying the same 
involutive automorphisms to the spaces ${\bf K}$,
${\bf P}$, and $i{\bf P}$, these spaces with a definite metric tensor can in 
turn be split into subspaces with eigenvalue $+1$ and $-1$ under this 
new involutive automorphism $\tau $. Thus,

\beq
\begin{array}{l}
\sigma : {\bf G}\to {\bf K}\oplus {\bf P},\\
\tau :{\bf K}\to {\bf K_1}\oplus {\bf K_2},\\
\tau :{\bf P}\to {\bf P_1}\oplus {\bf P_2},\\
\tau :i{\bf P}\to i{\bf P_1}\oplus i{\bf P_2},\end{array}\ \ 
\begin{array}{l}
{\bf G^*}={\bf K}\oplus i{\bf P}\\
{\bf H}={\bf K_1}\oplus i{\bf K_2}\\
{\bf M}={\bf P_1}\oplus i{\bf P_2}\\ 
i{\bf M}=i{\bf P_1}\oplus {\bf P_2} \end{array}
\eeq

As we already know, $K$ is a compact subgroup, and $\exp ({\bf P})$
and $\exp (i{\bf P})$ define symmetric spaces with a {\it definite}
metric (Riemannian spaces). In the same way, $H$ is a non--compact
subgroup, $\exp ({\bf M})$ and its dual space $\exp (i{\bf M})$ define
symmetric spaces with an {\it indefinite} metric. These are
pseudo--Riemannian symmetric coset spaces of a non--compact group by a
maximal non--compact subgroup\footnote{Note that not all the theorems
governing symmetric spaces corresponding to maximal compact subgroups
apply to the case at hand. A prime example is the decomposition
involving radial coordinates in subsection \ref{sec-radial}. We will
not discuss the symmetric spaces involving maximal non--compact
subgroups in any detail in this paper.}.  The original algebra ${\bf G}$
is thereby split into four components ${\bf K_1}$, ${\bf K_2}$, ${\bf
P_1}$, ${\bf P_2}$, depending on their eigenvalues ($++,+-,-+,--$)
under the two successive automorphisms $\sigma $, $\tau $. By applying
all the possible $\sigma $'s and all the possible $\tau $'s, or by
replacing either $\sigma $ or $\tau $ by the involutive automorphism
$\sigma \tau = \tau \sigma $, we obtain all the possible {\it real
forms of the symmetric spaces} associated with the compact algebra
${\bf G}$.

{\bf Example:} The complex algebra ${\bf SO(3,C)}$ has a root system
of type $B_n$. Its compact real form is ${\bf SO(3,R)}$, and its only
non--compact real form is ${\bf SO(p,q;R)}\simeq {\bf SO(q,p;R)}$
(${\bf p}+{\bf q}=3$), obtained by applying the involution $\sigma $=$I_{p,q}$
($I_{q,p}$) to ${\bf SO(3,R)}$. In paragraph \ref{sec-realforms2} we
constructed two Riemannian symmetric spaces associated with the
algebra ${\bf SO(3)}$, the sphere $SO(3)/SO(2)$ and the
double--sheeted hyperboloid $SO(2,1)/SO(2)$.  The Killing form has a
definite opposite sign for the two spaces.

The single--sheeted hyperboloid, described by the equation
$-x^2+y^2+z^2=1$ in ${\bf R}^3$, corresponds to the pseudo--Riemannian
symmetric space $SO(2,1)/SO(1,1)$ associated with the same algebra.
It is obtained by applying two consecutive involutive automorphisms
$\sigma $=$I_{2,1}$, $\tau $=$I_{1,2}$ to the algebra ${\bf G}={\bf
  SO(3,R)}$. Like in eq.~(\ref{eq:SO(2,1)}), $I_{2,1}$ and the Weyl
unitary trick transforms ${\bf G}$ into ${\bf G^*}$. Let's now apply
$I_{1,2}$ to ${\bf G^*}$:

\beq \begin{array}{l}
I_{1,2}XI_{1,2}^{-1}=\left(\begin{array}{ccc} 1 &    &    \\
                                           &  -1 &    \\
                                           &    & -1 \end{array}\right)
\frac{1}{2}\left(\begin{array}{ccc}     & t^3 & it^2\\
                   -t^3 &     & it^1\\
                   -it^2 & -it^1& \end{array}\right)
\left(\begin{array}{ccc} 1 &    &    \\
                                           &  -1 &    \\
                                           &    & -1 \end{array}\right) \\
\\
=\frac{1}{2}\left(\begin{array}{ccc}     & -t^3 & -it^2\\
                   t^3 &     & it^1\\
                   it^2 & -it^1& \end{array}\right)=
\frac{1}{2}\left(\begin{array}{ccc}     & -t^3 & \\
                   t^3 &     & \\
                    & & \end{array}\right)\oplus
\frac{1}{2}\left(\begin{array}{ccc}     &  & -it^2\\
                    &     & it^1\\
                   it^2 & -it^1& \end{array}\right)\\
\\
=({\bf K_1}\oplus {\bf K_2})\oplus i({\bf P_1}\oplus {\bf P_2})
\end{array}
\eeq

where in this example, ${\bf K_1}$ is empty. The spaces ${\bf K_1}$, 
${\bf K_2}$, ${\bf P_1}$, ${\bf P_2}$ consist of the generators in ${\bf G}$
with the following combinations  of eigenvalues under the two successive 
involutions, $\sigma \tau$:

\beq
{\bf K_1}: ++\ \ \ \ \ {\bf K_2}: +-\ \ \ \ \ {\bf P_1}: -+\ \ \ \ \ {\bf P_2}: --
\eeq

Thus we see that  ${\bf K_1}$ is empty and the others are spanned by 

\beq
{\bf K_2}=\left\{\frac{1}{2}\left(\begin{array}{ccc} & 1 & \\
                                                   -1 & & \\
                                          & & \end{array}\right)\right\},\ \ \ 
{\bf P_1}=\left\{\frac{1}{2}\left(\begin{array}{ccc} &  & \\
                                                    & &1 \\
                                        &-1 & \end{array}\right)\right\},\ \ \ 
{\bf P_2}=\left\{\frac{1}{2}\left(\begin{array}{ccc} &  & 1\\
                                                    & & \\
                                         -1 & & \end{array}\right)\right\}
\eeq

The new symmetric space is obtained by doing the Weyl unitary trick on 
the split spaces (${\bf K_1}\oplus {\bf K_2}$) and (${\bf P_1}\oplus 
{\bf P_2}$):

\beq \begin{array}{l}
{\bf H}={\bf K_1}\oplus i{\bf K_2}
=\frac{1}{2}\left(\begin{array}{ccc} & it^3 & \\
                               -it^3 & & \\
                                     & & \end{array}\right)\\ \\ 
{\bf M}={\bf P_1}\oplus i{\bf P_2}=
\frac{1}{2}\left(\begin{array}{ccc} &  & it^2\\
                                    & & t^1\\
                              -it^2 & -t^1 & \end{array}\right)
\end{array}
\eeq

The second involution $\tau $ (plus the Weyl trick) gives rise to a
non--compact subgroup $H$=$SO(1,1)$ and to the symmetric space $M\sim 
{\rm exp}{\bf M}$ and its dual $M^*\sim {\rm exp}(i{\bf M})$. The coset  
$M\sim SO(2,1)/SO(1,1)$ is represented by 

\beq
{\rm exp}{\bf M}=\left(\begin{array}{ccc} . & .  & ix\\
                         . & .  & y\\
                         -ix &  -y & z   \end{array}\right);\ \ \ \ 
(ix)^2+y^2+z^2=1
\eeq

The real forms of the simple Lie groups do not include all the
possible Riemannian symmetric coset spaces. For example, the compact Lie group 
$G$ is itself such a space, and so is its dual $G^C/G$ (here the algebra
${\bf G^C}={\bf G^*}\oplus i{\bf G^*}$ is the complex extension of
all the real forms ${\bf G^*}$). By starting with a compact algebra 
${\bf G}$ and applying to it all the combinations of the two involutive 
automorphisms $\sigma $, $\tau $, we construct, in the way just described,
all the remaining pseudo--Riemannian
symmetric spaces associated to the corresponding root system.
A complete list of these spaces can be found in Table~9.7 of 
reference~\cite{Gilmore}. 

Note that all the properties of the Lie algebra ${\bf G}$ (Killing form, 
rank, and so on) can be transferred to the vector 
subspaces ${\bf P}$, $i{\bf P}$ \cite{Gilmore}.
The only difference is that the subspaces are not closed under commutation.

In this section of the paper we have discussed symmetric 
spaces of positive, zero and
negative curvature that will be relevant for matrix models of the
circular, gaussian, and transfer matrix type, respectively. In
Part~II we will define and discuss various types of random matrix
ensembles and their applications to various physical problems, and we
will associate them to the corresponding symmetric spaces in Table~\ref{tab1}.

\section{Operators on symmetric spaces}
\label{sec-Operators}
\setcounter{equation}{0}

The differential operator uniquely determined by the simplest Casimir
operator on a symmetric space (and especially its radial part) plays
an important role both in mathematics and in the physical applications
of symmetric spaces.  Its eigenfunctions provide a complete basis for
the expansion of an arbitrary square--integrable function on the
symmetric space, and are therefore important in their own right.
Their importance in the applications to be discussed in Part~II is
evident when considering that the radial part of the Laplace--Beltrami
operator on an underlying symmetric space determines the dynamics of
the transfer matrix eigenvalues of the Dorokhov--Mello--Pereyra--Kumar
equation (DMPK equation for short) in the theoretical description of
quantum wires, and maps onto the Hamiltonians of integrable
Calogero--Sutherland models.  Here we will define some concepts
related to the Laplace--Beltrami operator and discuss its
eigenfunctions.

\subsection{Casimir operators}
\label{sec-Casi}

Let ${\bf G}$ be a semisimple rank--$r$ Lie algebra. A {\it Casimir operator} 
(invariant operator) $C_k$ ($k=1,...,r$)
associated with the algebra ${\bf G}$ is a homogeneous polynomial 
operator that satisfies

\beq
[C_k,X_i]=0
\eeq

for all $X_i\in {\bf G}$. The simplest (quadratic) Casimir
operator associated to the adjoint representation of the 
algebra ${\bf G}$ is given by 

\beq
\label{eq:C}
C=g^{ij}X_iX_j
\eeq

where $g^{ij}$ is the inverse of the metric tensor defined in (\ref{eq:metric})
and the generators $X_i$ are in the adjoint representation.
More generally, it can be defined for any representation  $\rho $ of ${\bf G}$ 
by

\beq
C_\rho =g_\rho^{ij}\rho (X_i)\rho (X_j)
\eeq

where $g_\rho^{ij}$ is the inverse of the metric (\ref{eq:g_rho})
for the representation $\rho$ (cf.
subsection {\ref{sec-metric}).  The Casimir operators lie in the
enveloping algebra obtained by embedding ${\bf G}$ in the associative 
algebra defined by the relations

\beq
\label{eq:asso}
X(YZ) = (XY)Z  \ \ \ \ \ \ \ [X,Y] = XY-YX 
\eeq

(note that in general, $XY$ makes no sense in the algebra ${\bf G}$).

The number of functionally independent Casimir operators is equal to
the rank $r$ of the group.  Other Casimir operators can be formed by
taking polynomials of the independent Casimir operators $C_k$
($k=1,...,r$).  Since the Casimir operators commute with all the
elements in ${\bf G}$, they make up the center of the associative
algebra (\ref{eq:asso}).

Note that Casimir operators are defined for {\it semisimple} algebras,
where the metric tensor has an inverse. This does not prevent one from
finding operators that commute with all the generators of
non--semisimple algebras.  For example, for the euclidean group $E_3$
of rotations $\{J_1,J_2,J_3\} $ and translations $\{P_1,P_2,P_3\} $,
${\bf P}^2=\sum P_iP_i$ and ${\bf P\cdot J}=\sum P_iJ_i$ commute with
all the generators (cf. the comments in the paragraph following
eq.~(\ref{eq:matrices})). Also the operators that
commute with all the generators of a non--semisimple algebra are
often referred to as Casimir operators.

All the independent Casimir operators of the algebra ${\bf G}$ can be
obtained as follows.  Suppose $\rho $ is an $n$--dimensional
representation of the rank--$r$ Lie algebra ${\bf G}$.  The {\it
  secular equation} for the algebra ${\bf G}$ is defined as the
eigenvalue equation

\beq
{\rm det}\left(\sum_{i=1}^{{\rm dim}{\bf G}} 
t^i\rho(X_i) -\lambda I_n\right)=\sum_{k=0}^n (-\lambda)^{n-k}\varphi_k(t^i)=0
\eeq

where the $\varphi_k(t^i)$ are functions of the real coordinates
$t^i$. In general, they will not all be functionally independent (for
example, $\varphi_0(t^i)$ is a constant).  There will be $r$
functionally independent coefficients $\varphi_k(t^i)$ multiplying the
powers of $-\lambda $ \cite{Gilmore}. When writing down the secular
equation, it is easiest to take a low--dimensional representation. By
making the substitution $t^i\to X_i$ in the functionally independent
coefficients, they become the functionally independent Casimir
operators of the algebra ${\bf G}$:

\beq
\label{eq:substitutionC}
\varphi_k(t^i)\begin{array}{c}{\scriptstyle t^i\to X_i}\\ 
\longrightarrow \\ {} \end{array} C_l(X_i)
\eeq

{\bf Example:} The generators $L_1$, $L_2$, $L_3$ of the ${\bf SO(3)}$
algebra were given explicitly in the adjoint representation in
equations (\ref{eq:L1L2}), (\ref{eq:L3}) in subsection
\ref{sec-cosets}. The secular equation for this algebra is then

\beq 
{\rm det}\left({\bf t\cdot L}-\lambda I_3\right)=
\left|\begin{array}{ccc} -\lambda & t^3/2 & t^2/2 \\
                                   -t^3/2 & -\lambda & t^1/2 \\
                                   -t^2/2 & -t^1/2 & -\lambda \end{array}\right|
=(-\lambda)^3+(-\lambda)\frac{1}{4}{\bf t}^2=0
\eeq

The equation has one functionally independent coefficient, which is
proportional to the trace of the matrix $({\bf t\cdot L})^2$. It equals
$\varphi_1({\bf t}) =\frac{1}{4}{\bf t}^2$. The rank of $SO(3)$ is $1$
and the only Casimir operator is

\beq
C_1\sim {\bf L}^2=L_1^2+L_2^2+L_3^2
\eeq

obtained by the substitution $t^i\to L_i$ in $\varphi_1({\bf t})$.
The Casimir operator can also be obtained from eq.~(\ref{eq:C}) by using
the metric $g_{ij}=-\frac{1}{2}\delta_{ij}$ for ${\bf SO(3)}$ given in
the example in subsection \ref{sec-metric}. 
We know from elementary quantum mechanics that 

\beq
\label{eq:L^2commrel}
[{\bf L}^2,L_1]=[{\bf L}^2,L_2]=[{\bf L}^2,L_3]=0
\eeq

is an immediate consequence of the commutation relations, so we see
that this operator indeed commutes with all the generators.  Even
though the commutation relations are not the same in polar
coordinates or after a general coordinate transformation, 
(\ref{eq:L^2commrel}) will nevertheless be true.

{\bf Example:} $SU(3)$ is a rank--2 group and therefore its characteristic
equation will have two independent coefficients. If we denote a general 
${\bf SU(3)}$ matrix $(a_{ij})$ we get the characteristic equation

\beq
\begin{array}{l}
{\rm det}\left( \begin{array}{ccc} a_{11}-\lambda & a_{12} & a_{13} \\
                                   a_{21} & a_{22}-\lambda & a_{23} \\
                                   a_{31} & a_{32} & a_{33}-\lambda 
\end{array}\right)
=(-\lambda )^3 + (-\lambda )^2 (a_{11}+a_{22}+a_{33})\\ \\ +
(-\lambda )(a_{11}a_{22}+a_{22}a_{33}+a_{33}a_{11}-a_{12}a_{21}-a_{23}a_{32}-
a_{31}a_{13})\\ \\ +(a_{11}(a_{22}a_{33}-a_{23}a_{32})+a_{12}(a_{23}a_{31}-
a_{21}a_{33})+a_{13}(a_{32}a_{21}-a_{31}a_{22}))=0
\end{array}
\eeq

The term proportional to $(-\lambda )^2$ vanishes, because the trace
of any matrix in the ${\bf SU(3)}$ algebra is zero. The two
independent coefficients are then $\varphi_2(a_{ij})$ and
$\varphi_3(a_{ij})$. Substituting the values in terms of the
coordinates $t^i$ of the algebra $\sum_i t^iT_i$ for the $a_{ij}$ (for
example, $a_{11}=t^3+\frac{1}{\sqrt{3}}t^8$, $a_{12}=t^1+it^2$, etc.),
we see that the expression for $\varphi_2(t^i)$ becomes

\beq
\varphi_2(t^i)=\sum_{i=1}^8 (t^i)^2
\eeq

and therefore the substitution (\ref{eq:substitutionC}) gives the 
first Casimir operator 

\beq
C_1=H_1^2+H_2^2+\sum_\alpha 
(E_\alpha E_{-\alpha }+E_{-\alpha }E_\alpha )=
\sum_{i=1}^8 T_i^2
\eeq

as expected. Making the same substitution in $\varphi_3(t^i)$ gives the second
Casimir operator for $SU(3)$, which has a more complicated form.

\subsection{Laplace operators}
\label{sec-Laplaceop}

The Casimir operators can be expressed as differential
operators in the local coordinates on the symmetric space. This is due
to the fact that each infinitesimal generator $X_\alpha \in {\bf G}$ is a
contravariant vector field on the group manifold. An element in the Lie algebra
can be written 

\beq
X=\sum_\alpha X^\alpha(x) \partial_\alpha \equiv \sum_\alpha X^\alpha(x) \frac{\partial }{\partial x^\alpha }
\eeq

where $x^\alpha $ are local coordinates \cite{Helgason,SattW} (for
example, $L_1=({\bf r \times P})_1= x^2\partial_3-x^3\partial_2$).
That the generators transform as lower index objects follows from the
commutation relations.

{\bf Example:} As an example we take the group $SO(3)$.  Under a
rotation $R=R(t^1,t^2,t^3)={\rm exp}(\sum t^kL_k)$, the vector ${\bf
  x}=x^i\hat{e}_i\ \in {\bf R}^3$ transforms as

\beq
{\bf x} {\buildrel{\scriptstyle R} \over \longrightarrow } {\bf x'}=x'^i\hat{e}'_i
\eeq

where the transformation laws for the components and the natural basis vectors
are 

\beq
x'^i=R^i_{\ j}x^j,\ \ \ \ \hat{e}'_i=\hat{e}_jR^j_{\ i}
\eeq

and $R^{-1}=R^T$.  The one--parameter subgroups of $SO(3)$ are
rotations

\beq
R(t^n)={\rm exp}(t^nL_n), \ \ \ \ \ \ \ (n=1,2,3) 
\eeq

(no summation) where $L_n$ are $SO(3)$ generators. It is easy to show
using the commutation relations for $L_n$ (given after
eq.~(\ref{eq:fixedNP})) that under infinitesimal rotations the $L_n$
transform like the lower index objects $\hat{e}_i$:

\beq
RL_iR^{-1}=L_jR^j_{\ i}
\eeq

Expressed in local coordinates as differential operators, the Casimirs
are called {\it Laplace operators}. In analogy with the Laplacian in
${\bf R^n}$,

\beq
{\bf P}^2=\Delta=\sum_{i=1}^n\frac{\partial^2}{\partial {x^i}^2}
\eeq

which is is invariant under the group $E_n$ of rigid motions
(isometries) of ${\bf R^n}$, the Laplace operators on
(pseudo--)Riemannian manifolds are invariant under the group of
isometries of the manifold.  The isometry group of the symmetric space
$P\simeq G/K$ is $G$, since $G$ acts transitively on this space and
preserves the metric, so the Laplace operators are invariant under the
group operations $g\in G$.
 
The number of independent Laplace operators on a Riemannian symmetric
coset space is equal to the rank of the space. As we defined in paragraph
\ref{sec-action}, the rank of a symmetric space is the maximal number
of mutually commuting generators $H_i$ in the subspace ${\bf P}$ (cf. also
subsection \ref{sec-restricted}).  If $X_\alpha$, $X_\beta,... \in
{\bf K}$ and $X_i$, $X_j,... \in {\bf P}$, it is also equal to the
number of functionally independent solutions to the equation

\beq
{\rm det}\left(\sum_{k=1}^{{\rm dim}{\bf P}} 
t^k\rho(X_k) -\lambda I_n\right)=\sum_{l=0}^n (-\lambda)^{n-l}\varphi_l(t^k)=0
\eeq

where $n$ is the dimension of the representation $\rho $ and 
where now in the determinant we sum over all $X_k\in {\bf P}$.  This
is equivalent to setting the coordinates $t^\gamma $ for all the
$X_\gamma\in {\bf K}$ equal to zero in the secular equation. In the
example in the preceding paragraph, the rank of the symmetric space
$SO(3)/SO(2)$ (the 2--sphere) is $1$, which in this case is also the
rank of the group $SO(3)$.

\vskip 0.5cm
The {\it Laplace--Beltrami operator} on a symmetric space is the
  special second order Laplace operator defined (when acting on a
  function (0--form) $f$) as

\beq
\label{eq:L-B-op}
\Delta_Bf=g^{ij}D_iD_jf=g^{ij}(\partial_i\partial_j-\Gamma^k_{ij}\partial_k)f
=\frac{1}{\sqrt{|g|}}\frac{\partial }{\partial x^i} g^{ij}
\sqrt{|g|}\frac{\partial }{\partial x^j}f,\ \ \ \ \ \ \ 
g\equiv {\rm det}g_{ij}
\eeq

Here $D_i$ denotes the covariant derivative on the symmetric space
and $g^{ij}$ are the components of the inverse of the metric tensor.
(The metric has an inverse because it is
non--degenerate on a semisimple algebra and can be mapped over the
entire symmetric space. For euclidean type spaces, we have the usual
metric $\delta_{ij}$.)  $D_i$
is defined in the usual way \cite{Helgason,FosNigh,3w}, for example
it acts on the components $x^j$ of a contravariant vector field in the
following way:
 
\beq
D_ix^j=\partial_ix^j+\Gamma^j_{ki}x^k
\eeq

where $\Gamma^j_{ki}$ are Christoffel symbols (connection coefficients).
The last term represents the change in $x^j$
due to the curvature of the space. We remind the reader that on a Riemannian 
manifold, the $\Gamma^j_{ki}$ are expressible in terms of the metric tensor,
hence the formula in eq.~(\ref{eq:L-B-op}).

{\bf Example:} Let's calculate the Laplace--Beltrami operator on the
symmetric space $SO(3)/SO(2)$ in polar coordinates using
(\ref{eq:L-B-op}) and the metric at the point $(\theta,\phi)$ given in
the second example of subsection \ref{sec-metric}:

\beq
g_{ij}=\left(\begin{array}{cc} 1 & 0 \\
                       0 & {\rm sin}^2\theta \end{array}\right),\ \ \ \ \ \ \  
g^{ij}=\left(\begin{array}{cc} 1 & 0 \\
                       0 & {\rm sin}^{-2}\theta \end{array}\right) 
\eeq

Substituting in the formula and computing derivatives we obtain the 
Laplace--Beltrami operator on the sphere of radius $1$:

\beq
\label{eq:Delta_on_sphere}
\Delta_B=\partial_\theta^2+{\rm cot}\theta\, \partial_\theta+
{\rm sin}^{-2}\theta \, \partial_\phi^2
\eeq

Of course this operator is exactly ${\bf L}^2$. We can check this by 
computing $L_x= y\partial_z-z\partial_y$, $L_y= z\partial_x-x\partial_z$,
and $L_z= x\partial_y-y\partial_x$ in spherical coordinates (setting $r=1$)
and then forming the operator $L_x^2+L_y^2+L_z^2$, remembering that all the
operators have to act also on anything coming after the expression for each 
$L_i^2$. We find that  ${\bf L}^2$ in spherical coordinates, expressed
as a differential operator, is exactly the Laplace--Beltrami operator. 

In general, a Laplace--Beltrami operator can be split into a radial
part $\Delta_B'$ and a transversal part. The radial part acts on
geodesics orthogonal to some submanifold $S$, typically a sphere
centered at the origin \cite{Helgason2}.

{\bf Example:} For the usual Laplace--Beltrami operator in ${\bf
R}^3$ expressed in spherical coordinates,

\beq
\label{eq:Ex1}
\Delta_B =\partial_r^2+2r^{-1}\partial_r
+r^{-2}\left(\partial_\theta^2+{\rm cot}\theta\, \partial_\theta+
{\rm sin}^{-2}\theta \, \partial_\phi^2 \right)
\eeq
  
the first two terms 

\beq
\label{eq:radR3}
\Delta_B'=\partial_r^2+2r^{-1}\partial_r
\eeq

constitute the radial part with respect to a sphere centered at the
origin and the expression in parenthesis multiplied by $r^{-2}$ is the
transversal part. The transversal part is equal to the projection of
$\Delta_B $ on the sphere of radius $r$ and equals the
Laplace--Beltrami operator on the sphere, given for $r=1$ in
eq.~(\ref{eq:Delta_on_sphere}).  This is a general result. For any
Riemannian manifold $V$ and an arbitrary submanifold $S$, the
projection on $S$ of the Laplace--Beltrami operator on $V$ is the
Laplace--Beltrami operator on $S$ (see Helgason \cite{Helgason2},
Ch.~II, paragraph 3).

The radial part of the Laplace--Beltrami operator on a symmetric space
has the general form

\beq
\label{eq:DeltaB'} 
\Delta_B'= \frac{1}{J^{(j)}}\sum_{\alpha =1}^{r'}\frac{\partial }
{\partial q^\alpha }J^{(j)}\frac{\partial }{\partial q^\alpha }\ \ \ \ \ \ \ 
(j=0,-,+)
\eeq

where $r'$ is the dimension of the maximal abelian subalgebra ${\bf
H_0'}$ in the tangent space ${\bf P}$ (the rank of the symmetric
space) and $J^{(j)}$ is the Jacobian, to be given in equation~(\ref{eq:J_j}),
of the transformation to radial coordinates.  The sum goes over the
labels of the independent radial coordinates defined in subsection
\ref{sec-radial}: $q={\rm log}h(x)=(q^1,...,q^{r'})$ where $h(x)$ is
the exponential map of an element in the Cartan subalgebra and
$q^\alpha $ are canonical coordinates on ${\bf H_0'}$ (in
\cite{OlshPere} they were denoted $(q,\alpha)\equiv {\bf q\cdot \alpha
}$; we will see in a moment that they are indeed given by ${\bf q\cdot
\alpha }$ where $\alpha $ is a restricted root).

The adjoint representation of a general element $H$ in the
maximal abelian subalgebra ${\bf H_0'}$ follows from a form similar to
eq.~(\ref{eq:adjrad}) (with or without a factor of $i$ depending on
whether we have a compact or non--compact space), but now the roots are
in the {\it restricted} root lattice.  For a non--compact space of type
$P^*$

\beq
\label{eq:q^alpha}
{\rm log}h=H={\bf q\cdot H}=\left(\begin{array}{cccccc}
0 & & & & & \\ & \ddots & & & & \\ & & 0 & & & \\ & & & {\bf q\cdot \alpha } & & \\
 & & & & \ddots & \\ & & & & & -{\bf q\cdot \eta }  \end{array}\right) \equiv
\left(\begin{array}{cccccc}
0 & & & & & \\ & \ddots & & & & \\ & & 0 & & & \\ & & & q^{\alpha } & & \\
 & & & & \ddots & \\ & & & & & q^{-\eta }  \end{array}\right)
\eeq

Hence $q^\alpha ={\bf q\cdot \alpha }$ and 

\beq
h={\rm e}^{H}=\left(\begin{array}{cccccc}
1 & & & & & \\ & \ddots & & & & \\ & & 1 & & & \\ & & & {\rm e}^{q^\alpha } & & \\
 & & & & \ddots & \\ & & & & & {\rm e}^{q^{-\eta }}  \end{array}\right)
\eeq

{\bf Example:} For the simple rank--1 algebra corresponding to the
compact group $SU(2)$, the above formulas take the form 
(cf. eq.~(\ref{eq:SU2adjoint}))

\beq
H=\theta H_1=\theta \left(\begin{array}{ccc}
0 & &  \\  & 1 & \\ & & -1 \end{array}\right),
\ \ \ \ \ \ \ 
h={\rm e}^{i\theta H_1}=\left(\begin{array}{ccc}
1 & &  \\  & {\rm e}^{i\theta } & \\ & & {\rm e}^{-i\theta }\end{array}\right)
\eeq

The radial coordinate is $q=(q^1)=\theta $.

There is a general theory for the radial parts of Laplace--Beltrami
operators \cite{Helgason2}. It is of interest to consider the radial
part of the Laplace--Beltrami operator on a manifold $V$ with respect
to a submanifold $W$ of $V$ that is transversal to the orbit of an
element $w\in W$ under the action of a subgroup of the isometry
group of $V$. Of special interest to us is the case in which the
manifold is a symmetric space $G/K$ and the Lie subgroup is $K$.

The Jacobian $J^{(j)}=\sqrt{|g|}$ (where $g$ is the metric tensor at
an arbitrary point of the symmetric space) of the transformation to
radial coordinates takes the form

\beq
\label{eq:J_j}
\begin{array}{l}
J^{(0)}(q)=\prod_{\alpha \in R^+} (q^\alpha )^{m_\alpha }\\
\\
J^{(-)}(q)=\prod_{\alpha \in R^+} ({\rm sinh}(q^\alpha ))^{m_\alpha }\\ 
\\
J^{(+)}(q)=\prod_{\alpha \in R^+} ({\rm sin}(q^\alpha ))^{m_\alpha }\end{array}
\eeq

for the various types of symmetric spaces with zero, negative and
positive curvature, respectively (see \cite{Helgason2}, Ch.~I,
par.~5).  In these equations the products denoted $\prod_{\alpha \in
R^+}$ are over all the positive roots of the restricted root lattice
and $m_\alpha $ is the multiplicity of the root $\alpha $. The
multiplicities $m_\alpha $ were listed in Table~\ref{tab1}.  

A remark on equation (\ref{eq:J_j}) is in order here. Strictly
speaking, in the euclidean case we have not defined any restricted
root lattice.  The formula for the Jacobian $J^{(0)}(q)$ for the
zero--curvature space is understood as the infinitesimal version of
the formula pertaining to the negative--curvature space. In the proof
\cite{Helgason2} of the formula for the zero--curvature case we
consider the Jacobian for a mapping from $K/M \times {\bf H_0}$ onto
${\bf P}$ (where $M$ is the centralizer in $K$ of the Cartan
subalgebra ${\bf H_0}$), whereas in the proof of the formula for the
negative--curvature symmetric space we consider the Jacobian for a
mapping from $K/M \times \e^{i{\bf H_0}}$ onto a subset of the
symmetric space $\e^{i{\bf P}}$.  $m_\alpha $ can in both cases be
interpreted as the dimension of the subspace of raising operators
corresponding to the root $\alpha $ in the same algebra (this is, in
the case of $X^-$, the root multiplicity of the restricted root
$\alpha $; see \cite{Helgason2}, Ch. ~I, paragraph 5). In the example
illustrating the construction of restricted root systems in subsection
\ref{sec-restricted} this space was spanned by the raising operators
$E'_{\frac{1}{2}}=T_6+iT_2$ and $\tilde{E}'_{\frac{1}{2}}=T_7+iT_1$
for the root $\alpha =\frac{1}{2}$, whereas for the root $\alpha =1$
it was spanned by only one raising operator $E'_1=T_5+iT_3$. The above
can perhaps also be understood in terms of the limiting procedure
discussed in paragraph \ref{sec-gauss} (cf. also equations
(\ref{eq:J_ja},\ref{eq:-to0+}) in section \ref{sec-zonal}). The 
$m_\alpha $'s will in the following be referred to as ``root multiplicities'' 
also for the zero curvature spaces, keeping the above in mind.

{\bf Example:} On the hyperboloid $H^2$ with metric

\beq
g_{ij}=\left(\begin{array}{cc} 1 & 0 \\
                       0 & {\rm sinh}^2\theta \end{array}\right),\ \ \ \ \ \ \  
g^{ij}=\left(\begin{array}{cc} 1 & 0 \\
                       0 & {\rm sinh}^{-2}\theta \end{array}\right) 
\eeq

equations (\ref{eq:DeltaB'},\ref{eq:J_j}) give the radial part of
$\Delta_B$ for $H^2$. The radial coordinate is $\theta $ so we get, in
agreement with (\ref{eq:J_j})

\beq
\label{eq:radpart}
\begin{array}{c}
J^{(-)}=\sqrt{|g|}={\rm sinh}\theta ,\ \ \ \ 
\Delta_B'={1\over {{\rm sinh}\theta }}\, \partial_\theta \, 
{\rm sinh}\theta  \, \partial_\theta 
=(\partial_\theta^2+{\rm coth}\theta \, \partial_\theta )
\end{array}
\eeq

In the same way we can also easily derive the
equation~(\ref{eq:radR3}) and equation~(\ref{eq:Ex2}) below.  In
particular, comparing with (\ref{eq:radpart}) we immediately get the
radial part of the Laplace--Beltrami operator acting on the
two--sphere $S^2$ transversally to a one--sphere $S^1$ centered on the
north pole:

\beq
\label{eq:Ex2}
\Delta_B'=(\partial_\theta^2+{\rm cot}\theta\, \partial_\theta )  
\eeq

which is exactly the radial part appearing in 
equation~(\ref{eq:Delta_on_sphere}).


The eigenfunctions of the radial part of the Laplace--Beltrami
operator on a symmetric space are called {\it zonal spherical
functions}.  In the applications of symmetric spaces to random matrix
theory (for example in quantum transport), the properties of these
zonal spherical functions are of central importance. For this reason
we shall devote the following subsection (and the appendix at the end
of this review) to a detailed discussion of their properties.

The so--called DMPK operator will be discussed in section
\ref{sec-RMT}. The differential equation involving this operator
describes the evolution of the distribution of the set of eigenvalues
of the transfer matrix of a quantum wire with an increasing length of
the wire. One of the most interesting applications of symmetric spaces
in the random matrix theory of quantum transport lies in the
identification of the DMPK operator with a simple transformation of
the Laplace--Beltrami operator on the symmetric space defining the
random matrix universality class.  We will discuss this in more detail
in Part~II of this review (see subsection \ref{sec-DMPK}).  

\subsection{Zonal spherical functions}
\label{sec-zonal}

The properties of the so called zonal spherical functions are
important for the research results to be discussed in Part~II.  Since
there is a natural mapping from the Hamiltonians of integrable
Calogero--Sutherland systems onto the Laplace--Beltrami operators of
the underlying symmetric spaces, these eigenfunctions play an
important role in the physics of integrable systems. But they are also
relevant in transport problems in connection with the DMPK equation
for a quantum wire. The known asymptotic expressions for these
eigenfunctions allows one to solve this equation in general or in the
asymptotic regime, because of the simple mapping from the DMPK
evolution operator to the radial part of the Laplace--Beltrami
operator.  (For an example of their use see \cite{MCDMPK,MCdis,Ol}.)

When $\rho $ is an irreducible representation of an
algebra, the associated Casimir operators $C_{k,\rho} $ are multiples of the
identity operator \cite{SattW,Hermann} (Schur's lemma). This means
that it has eigenvalues and eigenfunctions. Since the Casimir
operators (and consequently the Laplace operators) form a commutative
algebra, they have common eigenfunctions. There exists an extensive
theory regarding invariant differential operators and their
eigenfunctions \cite{Helgason2}.  Of particular interest are the
differential operators on a group $G$ or on a symmetric space $G/K$
that are left--invariant under the group $G$ and right--invariant
under a maximal compact subgroup $K$.  

Suppose the smooth complex--valued function $\phi_\lambda (x)$ is an
eigenfunction of such an invariant differential operator $D$ on the
symmetric space $G/K$:

\beq
D \phi_\lambda (x)=\gamma_D(\lambda ) \phi_\lambda (x)
\eeq

Here the eigenfunction is labelled by the parameter $\lambda $ and
$\gamma_D(\lambda )$ is the eigenvalue.  If in addition $\phi_\lambda
(kxk') =\phi_\lambda (x)$ ($x\in G/K$, $k\in K$) and $\phi_\lambda (e)=1$
($e=$identity element), the function $\phi_\lambda$ is called {\it
spherical}. A spherical function satisfies \cite{Helgason2}

\beq
\label{eq:integralformula}
\int_K \phi_\lambda (xky)\, dk =\phi_\lambda (x)\phi_\lambda (y)
\eeq

where $dk$ is the normalized Haar measure on the subgroup $K$.  We
will see examples of this formula below.

The common eigenfunctions of the Laplace operators on the symmetric
space $G/K$ are invariant under the subgroup $K$. They are termed {\it
zonal spherical functions}. Because of the bi--invariance under $K$,
these functions depend only on the radial coordinates $h$:

\beq
\phi_\lambda (x)=\phi_\lambda (h)
\eeq

{\bf Example:} Let's study for a moment the eigenfunctions of the
Laplace operator on $G/K=SO(3)/SO(2)$.  We know from quantum mechanics
that the eigenfunctions of ${\bf L}^2$ are the associated Legendre
polynomials $P_l({\rm cos}\theta )$, and $-l(l+1)$ is the eigenvalue
under ${\bf L}^2$ (our definition of ${\bf L}$ differs by a factor of $i$
from the definition common in quantum mechanics):

\beq
\label{eq:eigenPl}
{\bf L}^2P_l({\rm cos}\theta )=-l(l+1)P_l({\rm cos}\theta )
\eeq

where ${\rm cos}\theta $ is the $z$--coordinate of the point
$P=(x,y,z)$ on the sphere of radius $1$.  In spherical coordinates,
$P=({\rm sin}\theta \, {\rm cos}\phi , {\rm sin}\theta \, {\rm
  sin}\phi , {\rm cos}\theta )$.  As we can see, the eigenfunctions
are functions of the radial coordinate $\theta $ only. The subgroup
that keeps the north pole fixed is $K=SO(2)$ and its algebra contains
the operator $L_z=\partial_\phi $. Indeed, $P_l({\rm cos}\theta )$ is
unchanged if the point $P$ is rotated around the $z$--axis.

{\bf Example:} In terms of Euler angles, a general $SO(3)$--rotation
takes the form

\beq
R(\alpha, \beta ,\gamma )=g(\alpha )k(\beta )h(\gamma )=
\left(\begin{array}{ccc} {\rm cos}\alpha & 0 & -{\rm sin}\alpha \\
                               0        & 1 & 0 \\
                               {\rm sin}\alpha & 0 & {\rm cos}\alpha 
\end{array}\right)
\left(\begin{array}{ccc} {\rm cos}\beta & -{\rm sin}\beta & 0 \\
                         {\rm sin}\beta & {\rm cos}\beta &0 \\
                           0 & 0 & 1 
\end{array}\right)
\left(\begin{array}{ccc} {\rm cos}\gamma & 0 & -{\rm sin}\gamma \\
                               0        & 1 & 0 \\
                               {\rm sin}\gamma & 0 & {\rm cos}\gamma 
\end{array}\right)
\eeq

where $g$ and $h$ are rotations around the $y$ axis by the
angles $\alpha $ and $\gamma $ respectively, and $k$ is a rotation
around the $z$ axis by the angle $\beta $. Under such a rotation, the
north pole $(0,0,1)$ goes into $(-{\rm cos}\alpha \, {\rm cos}\beta \,
{\rm sin}\gamma - {\rm sin}\alpha \, {\rm cos}\gamma , -{\rm sin}\beta
\, {\rm sin}\gamma , -{\rm sin}\alpha \, {\rm cos}\beta \, {\rm
sin}\gamma + {\rm cos}\alpha \, {\rm cos}\gamma )$.  This means that
eq.~(\ref{eq:integralformula}) takes the form

\beq
\frac{1}{2\pi }\int_0^{2\pi }P_l(-{\rm sin}\alpha \, {\rm cos}\beta \, 
{\rm sin}\gamma + {\rm cos}\alpha \, {\rm cos}\gamma )\, d\beta = 
P_l({\rm cos}\alpha )P_l({\rm cos}\gamma )
\eeq

(To avoid confusion,
note that the eigenvalue $l$ in eq.~(\ref{eq:eigenPl}) is not equal to
$\lambda $. In fact, $\lambda = l+1/2$.) 

{\bf Example:} For the symmetric space $G/K=E_2/SO(2)$ the spherical functions
are the plane waves:

\beq
\psi (r)={\rm e}^{ikr}
\eeq
 
where $k$ is a complex number.  If $g$, $h$ denote translations in the
$x$--direction by a distance $b$, $a$ respectively, and $k$ is a
rotation around the origin of magnitude $\phi $, then the
transformation $g(b)k(\phi )h(a)$ moves the point $x\in {\bf R}^2$ by
a distance $\sqrt{a^2+b^2+2ab{\rm cos}\phi }$.  Therefore we obtain
from (\ref{eq:integralformula})

\beq
\frac{1}{2\pi }\int_0^{2\pi }\psi \left(\sqrt{a^2+b^2+2ab{\rm cos}\phi }
\right) \, d\phi =\psi (a) \psi (b)
\eeq

We introduce a parameter $a$ into the the Jacobians (\ref{eq:J_j}) 
as in reference \cite{OlshPere},

\beq
\label{eq:J_ja}
\begin{array}{l}
J^{(0)}(q)=\prod_{\alpha \in R^+} (q^\alpha )^{m_\alpha }\\
\\
J^{(-)}(q)=\prod_{\alpha \in R^+} (a^{-1}{\rm sinh}(aq^\alpha ))^{m_\alpha }\\ 
\\
J^{(+)}(q)=\prod_{\alpha \in R^+} (a^{-1}{\rm sin}(aq^\alpha ))^{m_\alpha }\end{array}
\eeq

The parameter $a$ corresponds to a radius. For example, for the sphere
$SO(3)/SO(2)$ it is the radius of the 2--sphere.

The various spherical functions corresponding to the spaces of
positive, negative and zero curvature are then related to each other
by the simple transformations \cite{OlshPere}

\beq
\label{eq:-to0+}
\begin{array}{l}
\phi_\lambda^{(0)} (q)={\rm lim}_{a\to 0} \phi_\lambda^{(-)}(q) \\   \\
\phi_\lambda^{(+)} (q)=\phi_\lambda^{(-)}(q)|_{a\to ia} \end{array}
\eeq   

There exist integral representations of spherical functions for the
various types of spaces $G/K$ \cite{OlshPere,Helgason2}. We will list
an integral representation only of $\phi_\lambda ^{(-)}(q)$ below,
recalling that formulas for the other types of spherical functions can
be obtained by (\ref{eq:-to0+}). If $\phi_\lambda^{(-)} (x) $ is
spherical and $h$ is the spherical radial part of $x$,

\beq
\label{eq:sphericalfunctions-}
\phi_\lambda^{(-)} (x)=\phi_\lambda^{(-)} (h)=
\int_K{\rm e}^{( i\lambda  - \rho )H(kx)}dk 
\eeq

Here $\lambda $ is a complex--valued linear function on the maximal
abelian subalgebra ${\bf H_0'}$ of $i{\bf P}$ and $\rho $ is the
function defined below in eq.~(\ref{eq:rho}).  In
eq.~(\ref{eq:sphericalfunctions-}) they act on the unique element
$H(kx)\in {\bf H_0'}$ such that $kx=n{\rm e}^{H(kx)}k'$ in the Iwasawa
decomposition introduced in subsection \ref{sec-radial}.  It was shown
by Harish--Chandra \cite{HC} that two functions $\phi_\lambda^{(-)}
(x)$ and $\phi_\nu^{(-)} (x)$ are identical if and only if $\lambda
=s\nu $, where $s$ denotes a Weyl reflection. The Weyl group is the
group of reflections in hyperplanes orthogonal to the roots and was
defined in subsection \ref{sec-rootsp}, eq.~(\ref{eq:Weyl}), and discussed 
further in subsection \ref{sec-chambers}.

Equation (\ref{eq:sphericalfunctions-}) may seem a bit cryptic, but it
becomes much more clear if one uses the explicit expression for the
Iwasawa decomposition.  It essentially becomes the integral over the
product of all the possible lower principal minors (raised to suitable
powers) of the matrix $ke^{ah}k^{-1}$ (where $a$ is the free parameter
introduced above). The explicit expression can be found
in~\cite{OlshPere}. This integral representation is of great
importance, because in some cases it can be exactly integrated leading
to explicit expressions for the zonal spherical functions. This point
will be discussed in more detail in the appendix at the end of this
review.  

The eigenvalues of the radial part of the Laplace--Beltrami operator
corresponding to the eigenfunctions on zero, negative and positive
curvature symmetric spaces are given by the following equations (see
\cite{OlshPere} and \cite{Helgason2}, Ch.~IV, par.~5):

\beq
\label{eq:eigen}
\begin{array}{l}
\Delta_B'\phi_\lambda^{(0)}= -\lambda^2 \phi_\lambda^{(0)}\\ \\
\Delta_B'\phi_\lambda^{(-)}= (-\frac{\lambda^2}{a^2}-\rho^2)\phi_\lambda^{(-)}\\ \\
\Delta_B'\phi_\lambda^{(+)}= (-\frac{\lambda^2}{a^2}+\rho^2)\phi_\lambda^{(+)}
\end{array}
\eeq

where $\rho $ is the function defined by

\beq
\label{eq:rho}
\rho=\frac{1}{2} \sum_{\alpha \in R^+}m_\alpha \alpha
\eeq

{\bf Example:} Take the symmetric space $SO(3)/SO(2)$. From Table~\ref{tab1} we
see that this space has $p-q=2-1=1$ short root of length 1. Then 

\beq
\rho^2 =\left(\frac{1}{2}\right)^2 \cdot 1^2 \cdot |\alpha |^2 =\frac{1}{4}
\eeq

and setting $a=1$, the eigenvalue is $-\lambda^2+1/4=-l(l+1)$.   


Remarkably enough, for a few classes of symmetric spaces explicit
expressions for the zonal spherical functions can be obtained. We will
make use of this result in section \ref{sec-beta2} when discussing the
exact solution of the DMPK equation in the $\beta=2$ case. These
solutions play an important role in several branches of physics. For
this reason we decided to discuss them in some detail in the appendix
of this review.  

\subsection{The analog of Fourier transforms on symmetric spaces}
\label{sec-Fourier}

Much of the material presented in this subsection is taken from the book
by Wu--Ki Tung \cite{Wu}.

A continuous smooth ($C^\infty$) spherical function $f$ is said to be
{\it elementary} if it is an eigenfunction of any differential
operator that is invariant under left translations by $G$ and right
translations by $K$. Thus the eigenfunctions of the Laplace operators
are elementary.  The elementary spherical functions are related to
irreducible representation functions for the group $G$. The
irreducible representation functions are the matrix elements of the
group elements $g$ in the matrix representation $\rho $. Let's clarify
this statement by an example well--known from quantum mechanics.

{\bf Example:} The angular momentum basis for $SO(3)$ is defined by 

\beq 
\begin{array}{l}
{\bf L}^2|lm>=l(l+1)|lm> \\ 
L_3|lm>=m|lm> \\ 
L_\pm |lm>=\sqrt{l(l+1)-m(m\pm 1)}|lm> \end{array}
\eeq

where $l$ labels the representation.  The irreducible representation
functions are, in the angular momentum basis $|lm>$, the matrix
elements $D^l(R)^{m'}_{\ m}$ such that

\beq
R|lm>=|lm'> D^l(R)^{m'}_{\ m}
\eeq

where $R={\rm exp}({\bf t\cdot L})$ is a general $SO(3)$ rotation.  It
is known that for $R\in SO(3)$, if $\alpha $, $\beta $, $\gamma $ are
the Euler angles of the rotation $R=R(\alpha ,\beta ,\gamma)$, these 
matrix elements take the form

\beq 
D^l(R)^{m'}_{\ m}= D^l(\alpha ,\beta ,\gamma )^{m'}_{\ m}={\rm
 e}^{-i\alpha m'}d^l(\beta )^{m'}_{\ m}{\rm e}^{-i\gamma m},\ \ \ \ \
 \ \ d^l(\beta )^{m'}_{\ m}\equiv <lm'|{\rm e}^{-i\beta L_2}|lm> 
\eeq

The associated Legendre functions $P_l^m({\rm cos}\theta)$ and the
special functions $Y_l^m(\theta, \phi )$ called spherical harmonics
are essentially this kind of matrix elements:

\beq \begin{array}{l}
P_l^m({\rm cos}\theta)=(-1)^m\sqrt{\frac{(l+m)!}{(l-m)!}}d^l(\theta )^m_{\ 0}\\
Y_l^m(\theta, \phi )=\sqrt{\frac{2l+1}{4\pi }}\left[ D^l(\phi ,\theta ,0)^m_{\ 0}\right]^*\end{array}
\eeq

The irreducible representation functions  
$D^l(R)^{m'}_{\ m}$ satisfy orthogonality and completeness relations. In fact,
they form a complete basis in the space of square integrable functions defined
on the group manifold. This is the Peter--Weyl theorem. From here the 
corresponding theorems follow for the special functions of mathematical
physics.

{\bf Example:} For $SO(3)$ the orthonormality 
condition reads

\beq
\label{eq:completeness}
(2l+1)\int d\tau D_l^\dagger(R)^m_{\ n}D^{l'}(R)^{n'}_{\ m'}=\delta_l^{l'} 
\delta_n^{n'} \delta_{m'}^m \ \ \ \ \ \ \  D_l^\dagger(R)^m_{\ n}\equiv
[D^l(R)^n_{\ m}]^*
\eeq

where $R=R(\alpha ,\beta ,\gamma)$ is an $SO(3)$ rotation expressed in
Euler angles and $d\tau$ is the invariant group integration measure
normalized to unity, $d\tau=d\alpha d({\rm cos}\beta )d\gamma
/8\pi^2$. That the irreducible representation functions form a
complete basis for the square--integrable functions on the $SO(3)$
group manifold can be expressed as

\beq
f(R)=\sum_{lmn}f_{lm}^nD^l(R)^m_n
\eeq

where $f(R)$ is square--integrable. Using (\ref{eq:completeness}) we obtain

\beq
f_{lm}^n=(2l+1)\int d\tau D_l^\dagger(R)^n_{\ m}f(R)
\eeq

If $R(\alpha ,\beta ,\gamma)=R(\phi, \theta ,0)$ we get the special
case of the spherical harmonics on the unit sphere (setting
$\sqrt{4\pi /(2l+1)}f_{lm}^0\equiv \tilde{f}_{lm}$):

\beq
f(\theta ,\phi )=\sum_{lm}\tilde{f}_{lm}Y_{lm}(\theta ,\phi )
\eeq

\beq
\tilde{f}_{lm}=\int  f(\theta ,\phi ) Y_{lm}^*(\theta ,\phi ) d({\rm cos}\theta )d\phi
\eeq

and further, for  $R(\alpha ,\beta ,\gamma)=R(0,\theta ,0)$ we get the 
completeness relation for the associated Legendre polynomials 
$P_l({\rm cos}\theta )=Y_{l0}\sqrt{4\pi /(2l+1)}$ 

\beq
\label{eq:sum}
f(\theta )=\sum_l f_l P_l({\rm cos}\theta ) 
\eeq

\beq
f_l={(2l+1)\over 2}\int f(\theta ) P_l^*({\rm cos }\theta ) d({\rm cos}\theta )
\eeq

where $f_l=f_{l0}^0$.
These are analogous to Fourier transforms. 
In the above example, we considered a symmetric space with positive curvature. 
For a space with zero or negative curvature, we have an integral instead of a
sum in (\ref{eq:sum}):

\beq
\label{eq:int}
f(q)=\int \tilde{f}(\lambda )\phi_\lambda^{(j)}(q)d\mu (\lambda )\propto
w^2\int \tilde{f}(\lambda )\frac{\phi_\lambda^{(j)}(q)}{|c(\lambda )|^{2}}
d\lambda \eeq

\beq
\tilde{f}(\lambda )=\int f(q)\left[ \phi_\lambda^{(j)}(q)\right]^*J^{(j)}(q)dq
\eeq

where $q$ are canonical radial coordinates.  The integration measure
$d\mu (\lambda )$ was determined by Harish--Chandra \cite{HC} to be
well--defined and proportional to $w^2 |c(\lambda )|^{-2}d\lambda $,
where $c(\lambda )$, to be discussed in detail below,
is a known function whose inverse is analytic
(in this context, see also \cite{OlshPere,Helgason2}) and $w$ is the order of
the Weyl group (the number of distinct Weyl reflections).
$J^{(j)}(q)dq$ is the invariant measure on the space of radial
coordinates.  In equation (\ref{eq:int}) the arbitrary
square--integrable function $f(q)$ is expressed in terms of the
complete set of basis functions $\phi_\lambda^{(j)} (q)$.

One can show \cite{Helgason2,HC} that the dimension of the space of
eigenfunctions of $\Delta_B'$ is less than or equal to $w$. It is a
remarkable fact that the eigenfunctions of the radial part of the
Laplace--Beltrami operator $\Delta_B'$ have the property of being
eigenfunctions of the radial part of {\it any} left--invariant
differential operator on the symmetric space as well
(\cite{Helgason2}, Ch.~IV).
  
In the asymptotic expansions that we will discuss in subsection
\ref{sec-use} of Part~II of this review in the context of quantum 
transport, a crucial role is played by the function $c(\lambda )$ in
eq.~(\ref{eq:int}). It encodes all the information relating the
transport problem to an underlying symmetric space. The explicit form
of $c(\lambda)$ is

\beq
c(\lambda )=\prod_{\alpha \in R^+} c_\alpha (\lambda )
\eeq

with

\beq 
c_\al(\lambda)=\frac{\Gamma (i\lambda^\alpha /2)}{\Gamma (m_\alpha /2
+i\lambda^\alpha /2)}
\eeq 

where $\Gamma $ denotes the Euler gamma function, $\alpha $ is a
generic root belonging to the restricted root lattice of the symmetric
space, $m_\alpha $ denotes its multiplicity and the product is
restricted to the sublattice $R^+$ of positive restricted roots
only. We shall see an explicit example of these functions in
equations~(\ref{ss2}) and (\ref{ss2b}) for the three symmetric spaces
which are relevant for the random matrix description of quantum wires.

An important feature of the zonal spherical functions is that they
satisfy, for large values of $|h|$, the following asymptotic
expression for spaces of zero and negative curvature \cite{OlshPere,HC}:

\beq
\phi_\lambda (h)\sim \sum_{s\in W} c(s\lambda ){\rm e}^{(is\lambda -\rho )(H)}
\label{asyexp}
\eeq

where $h={\rm e}^H$ is a spherical coordinate, $H$ is an element of
the maximal abelian subalgebra, $\lambda $ is a complex--valued linear
function on the maximal abelian subalgebra, and the function $\rho $
was defined in eq.~(\ref{eq:rho}).

This expansion will play a major role later in this review. We
will refer to it in the following as the ``Harish--Chandra asymptotic
expansion''. 

\section{Integrable models related to root systems}
\label{sec-intmod}
\setcounter{equation}{0}

As mentioned in the introduction, an important role in our analysis is
played by the class of integrable models known as Calogero--Sutherland
(CS) models, which turn out to be deeply related to the theory of
symmetric spaces.  These models describe $n$ particles in one
dimension, identified by their coordinates $q^1,...,q^n$ and
interacting (at least in the simplest version of the models) through a
pair potential $v(q^i-q^j)$. The Hamiltonian of such a system 
is given by

\bea
\label{eq:calH}
{\cal H}=\frac{1}{2}\sum_{i=1}^n p_i^2 + \sum_{\alpha \in R^+} 
g_\alpha^2\, v(q^\alpha)\nonumber \\
p_i=-i\frac{\partial }{\partial q^i},\ \ \ \ 
q^\alpha = q\cdot \alpha =\sum_{i=1}^n q^i\alpha_i\\ \nonumber 
\eea

where the coordinate $q$ is $q=(q^1,...,q^n)$,
$p_1,...,p_n$ are the particle momenta, and the particle mass is
set to unity.  In eq.~(\ref{eq:calH}) $R^+$ is the subsystem 
of positive roots of the root system $R=\{\alpha^1,...,\alpha^{\nu }\}$ 
related to a specific simple Lie algebra or symmetric space, and $n$ is
the dimension of the maximal abelian subalgebra ${\bf H_0}$ and of its 
dual space ${\bf H_0^*}$. The components of the positive root $\alpha =
\alpha^k\in R^+$ are $\alpha^k_1,...,\alpha^k_n$. The number of positive 
roots is $\nu/2$, where $\nu $ is the total number of roots.
  In general, the coupling constants $g_\alpha $ are
the same for equivalent roots, namely those that are connected with
each other by transformations of the Weyl group $W$ of the root system
(see subsections \ref{sec-rootsp}, \ref{sec-radial} and
\ref{sec-restricted}). 
    
Several realizations of the potential $v(q^\alpha)$
have been studied in the literature (for a review see
ref.~\cite{OlshPere}):

\bea
\label{eq:I-V}
v_I(\xi )&=&\xi^{-2} \nonumber \\
v_{II}(\xi )&=&{\rm sinh}^{-2}(\xi ) \nonumber \\
v_{III}(\xi )&=&{\rm sin}^{-2}(\xi ) \nonumber \\
v_{IV}(\xi )&=&{\cal P}(\xi ) \nonumber \\
v_{V}(\xi )&=&\xi^{-2}+\omega^2\xi^2 \nonumber \\
v_{VI}(\xi )&=&\e^\xi \\ \nonumber 
\eea

Here ${\cal P}(q)$ denotes the Weierstrass ${\cal P}$--function, to be
defined in eq.~(\ref{eq:Weier}). We will mainly be interested in the
first three realizations. The potential expressed in terms of the
Weierstrass ${\cal P}$--function will be discussed in subsection
\ref{sec-Weierstrass}, while we will not deal with the last two cases which
we reported here only for completeness. The reader is referred to
\cite{OlshPere} for a discussion of these two potentials.

The most relevant features of the Calogero--Sutherland models
are: 
\begin{itemize}

\item Under rather general conditions (see \cite{OlshPere} for a
detailed discussion) they are completely integrable, in the sense that
they possess $n$ commuting integrals of motion.

\item For particular values of the coupling constants (i.e. for those
related to the root multiplicities by eq.~(\ref{eq:rootvalues_gen})
which we will discuss below) the Calogero--Sutherland Hamiltonians can
be mapped onto the radial parts of the Laplace--Beltrami operators on
suitably chosen symmetric spaces. As we will see, these spaces have
negative curvature for the ${\rm sinh}$--type models and positive
curvature for the ${\rm sin}$--type models.

\end{itemize}

\subsection{The root lattice structure of the CS models}

In the original formulation of the CS model, the interaction among the
particles was simply pairwise~\cite{cs}. Only later it was realized
that this particular choice was the signature of an underlying
structure, namely the root lattices of Lie algebras of type $A_n$ (see
the example below). Also, the model could be extended to any root
lattice canonically associated to a simple Lie algebra or symmetric
space, keeping its relevant properties: complete integrability and
mapping to the radial part of a Laplace--Beltrami operator for special
values of the couplings \cite{OlshPere}. This corresponds exactly to
the choice of Hamiltonian introduced in eq.~(\ref{eq:calH}).
 
Let us look at two examples which may clarify this construction.  In
these examples we use the root lattices $A_n$ and $C_n$ and the
potential of type II.  Remember that the coupling constants $g_\alpha
$ are the same for equivalent roots, thus we expect a single coupling
constant in the first example and only two copuling constants in the
second example.  In order to fix the notation, let us denote with
$\{e_1,...,e_n\}$ a canonical basis in the space ${\bf R}^n$.

\begin{description} 

\item{$A_n$:} As mentioned in subsection \ref{sec-srs}, the $A_n$ root
system is contained in the hyperplane in ${\bf R}^{n+1}$ with equation
$x_1+x_2+...+x_{n+1}=1$. The root system $R$ is given by
$R=\{\alpha^1,...,\alpha^{\nu }\}= \{e_i-e_j,~i\neq j\}$. In this case
$W$, the Weyl group, is the permutation group of the set
$\{e_i\}$. The corresponding CS Hamiltonian is in this case

\begin{equation}
{\cal H}=-\frac{1}{2}\sum_{i=1}^{n}\frac{\partial^2}{\partial (q^i)^2}+
\sum_{i<j}\frac{g^2}{\sinh^2(q^i-q^j)}
\label{CS2}
\end{equation}

The arguments of the $\sinh $--function are $q^\alpha_k=q\cdot
\alpha^k=(q^1,...,q^n)\cdot (e_i-e_j)=q^i-q^j$ ($i<j$) where $\alpha^k
$ is a positive root of the root lattice $R$. As anticipated we have
only pairwise interaction and a single coupling constant appears in
the model.  This is the model originally considered in~\cite{cs}.

\item{$C_n$:} The $C_n$ root system is $R=\{\pm 2e_i,\pm e_i \pm e_j,
i\neq j\}$. The Weyl group is the product of the permutation group and
the group of transformations that changes the sign of the vectors
$\{e_i\}$.  The corresponding Hamiltonian is:

\beq
\label{CS3}
{\cal H}=-\frac{1}{2}\sum_{i=1}^{n}\frac{\partial^2}{\partial (q^i)^2}+
\sum_i\frac{g_l^2}{\sinh^2(2q^i)}
+\sum_{i<j}\left(\frac{g_o^2}{\sinh^2(q^i-q^j)}
+\frac{g_o^2}{\sinh^2(q^i+q^j)}\right)
\eeq

As anticipated, we have different coupling constants for the ordinary
roots (corresponding to pairwise interaction) and for the long
roots. In the following sections we will come back to this particular
choice of Hamiltonian which turns out to be of great importance in the
random matrix description of quantum wires.  \end{description}

\subsection{Mapping to symmetric spaces}

The most interesting property of these Hamiltonians, which is a direct
consequence of the underlying structure of the symmetric space, is
that for the potentials of type I, II and III there exists an
exact transformation of the Hamiltonian ${\cal H}$ into the radial
part of the Laplace--Beltrami operator on the symmetric space
corresponding to the root lattice $R$ of ${\cal H}$. In particular 
the target spaces have negative curvature for the ${\rm sinh}$--type models and
 positive curvature for the ${\rm sin}$--type models. Denote the radial part of the
Laplace--Beltrami operator by $\Delta_B'$; the transformation is then
given by (see appendix D of ref. \cite{OlshPere} for a proof)

\beq
\label{eq:H-Delta}
{\cal H}=\xi(q)\frac{1}{2}(\Delta_B'\pm\rho^2)\xi^{-1}(q)
\ \ \ \ \ \ \ (+\ \ {\rm for\ \ II},\ \ -\ \ {\rm for\ \ III},\ \ \rho =0\ \ 
{\rm for\ \ I})
\eeq
  
if and only if the coupling constants $g_\alpha $ in ${\cal H}$
take the following {\it root values} 

\beq
\label{eq:rootvalues_gen}
g_\alpha^2=\frac{m_\alpha (m_\alpha +2m_{2\alpha}-2)|\alpha |^2}{8}
\eeq

In equation (\ref{eq:rootvalues_gen}) $m_\alpha $ is the 
multiplicity of the root $\alpha $ and $|\alpha |$ its length,
and $\rho $ in eq.~(\ref{eq:H-Delta}) is the vector defined in (\ref{eq:rho})

\beq
\rho = \frac{1}{2} \sum_{\alpha \in R^+} m_\alpha \alpha
\eeq

For the ordinary roots eq. (\ref{eq:rootvalues_gen}) simplifies to
\beq
\label{eq:rootvalues}
g_\alpha^2=\frac{m_\alpha (m_\alpha -2)|\alpha |^2}{8}
\eeq

The function $\xi (q)$ in (\ref{eq:H-Delta}) is related to the
Jacobian of the transformation to radial coordinates
(cf. equation~(\ref{eq:J_j})). It is given by

\beq
\label{eq:xiJ}
\begin{array}{c}        \\
                \xi (q) \\
                        \\ \end{array}=
\begin{array}{ll} \prod_{\alpha \in R^+} [q^\alpha]^{m_\alpha /2} & {\rm I} \\
  \prod_{\alpha \in R^+} [{\rm sinh} (q^\alpha )]^{m_\alpha /2}  & {\rm II} \\
  \prod_{\alpha \in R^+} [{\rm sin} (q^\alpha )]^{m_\alpha /2}  & {\rm III} \\
\end{array}
\eeq

for quantum systems with potentials of type I, II, and III,
respectively.  The radial part of the Laplace--Beltrami operator,
$\Delta_B'$, on the symmetric space can be expressed in terms of $\xi
(q)$ as follows (see equation~(\ref{eq:DeltaB'}) in subsection
\ref{sec-Laplaceop}):

\beq
\Delta_B'=\frac{1}{\xi^2(q)}\sum_i \frac{\partial }{\partial q^i}\xi^2(q)
\frac{\partial }{\partial q^i}
\eeq
 
and the symmetric space has zero, negative, and positive curvature for the
cases I, II, and III, respectively.

At this point a number of results can be obtained for the
corresponding quantum systems merely by using the theory of symmetric
spaces. A detailed collection of results pertaining to spectra, wave
functions, and integral representations of wave functions can be found
in the original article \cite{OlshPere}, and we will not further
elaborate on them here.  The crucial point in this relation for our
analysis is another one. From eq.~(\ref{eq:rootvalues}) we see that
whenever the ordinary root multiplicity is $m_\alpha =2$, the
interaction between particles in the Hamiltonian vanishes. The degrees
of freedom decouple and the problem can be solved explicitly. Such a
decoupling could hardly have been inferred by looking at the original
form of the radial part of the Laplace--Beltrami operator, equation
(\ref{eq:DeltaB'}). This decoupling is also the reason why in these
cases explicit expressions for the zonal spherical functions exist
(see paragraph \ref{sec-zonal} and, for a more detailed discussion,
the appendix at the end of this review).

\newpage
\begin{center}
\Large{\bf Part II}
\end{center}
In the second part of this review we will use the background material
presented in Part~I to make evident the close relationships between
symmetric spaces and random matrix ensembles.  To this end, we discuss
in detail the identifications that have to be made between matrix
ensembles and symmetric space characteristics in section
\ref{sec-RMT}. We show that the integration manifolds of random matrix
theories with physical applications in the description of complex
nuclei, gauge field theories and mesoscopic systems, can be exactly
identified with the irreducible symmetric spaces based on
non--exceptional groups that were classified by Cartan, and their
euclidean counterparts. We identify the random matrix eigenvalues with
the spherical radial coordinates frequently used in the theory of Lie
groups, and show that the Jacobians determining the probability
distribution functions of random matrix eigenvalues are exactly the
Jacobians appearing in equation~(\ref{eq:J_ja}) determined by the
metric $g$ on the symmetric space: $J=\sqrt{|g|}$.  The structure of
the restricted root lattice associated to the symmetric space
determines this Jacobian completely through the multiplicities of its
roots.  In connection with the discussion of transfer matrix
ensembles, we define the Dorokhov--Mello--Pereyra--Kumar (DMPK)
equation, and in subsection \ref{sec-Coulomb} we include a general
discussion of Fokker--Planck equations associated to random matrix
ensembles. We also discuss the so--called Coulomb gas analogy used in
random matrix theory. We show that the DMPK operator connected to any
random matrix ensemble is simply related to the radial part of the
Laplace--Beltrami operator (defined in section \ref{sec-Operators}) on
the appropriate symmetric space.  We conclude by summarizing the
results in Table~\ref{tab2}.

We devote section \ref{sec-appl} to some consequences of the
identifications between random matrix theories and symmetric spaces.
In subsection \ref{sec-classification} we discuss the new
classification of disordered systems that the equivalence gives rise
to, as a natural consequence of the Cartan classification of symmetric
spaces.  In subsections \ref{sec-symmetriesRMT}--\ref{sec-OP} we will
study how the properties of the symmetric spaces are reflected in the
orthogonal polynomials associated to a random matrix partition
function, and see how the reflection or translation symmetries of the
Jacobians are direct consequences of the correponding properties of
the restricted root lattices.

An introduction to the relationship between the restricted root
lattices of symmetric spaces and integrable Calogero--Sutherland
models describing interacting many--particle systems in one dimension
was given in section \ref{sec-intmod}.  Olshanetsky and Perelomov
\cite{OlshPere} showed that the dynamics of these systems are related
to free diffusion on a symmetric space. This relationship is due to
the fact that the Hamiltonians of integrable Calogero--Sutherland
models map onto the radial parts of the Laplace--Beltrami operators of
the underlying symmetric spaces, as explained in section
\ref{sec-intmod} of Part~I. 

Since the Dorokhov--Mello--Pereyra--Kumar equation for a disordered
conductor can be mapped onto a Schr\"odinger--like equation in
imaginary time (\cite{BeeRejaei}, see also \cite{MCdis}) featuring a
Calogero--Sutherland type Hamiltonian with root values of the coupling
constants, information on the solutions of the DMPK equation
\cite{MCDMPK,MCdis} can be extracted from known properties of the
zonal spherical functions of the underlying symmetric space.  As we
have seen in section \ref{sec-Operators}, these are the eigenfunctions
of the radial part of the Laplace--Beltrami operator. Some results
relating to this will be described in detail in subsection 
\ref{sec-use}, as well as an application of the theory of symmetric spaces
in a quantum transport problem involving the magnetoconductance.

Finally in section \ref{sec-beyond} we discuss some possible
extensions of the present results and some non-standard applications
of the symmetric space formalism. We go beyond the Cartan
classification and discuss non--Cartan parametrization of symmetric
spaces in subsection \ref{sec-nonCartan}. In subsection
\ref{sec-clustered} we discuss generalizations of the DMPK equation
and the alternative clustered solutions to the DMPK equation that are
a consequence of the exact integrability of Calogero--Sutherland
models.  Finally, in the last subsection \ref{sec-Weierstrass} we
discuss the Weierstrass ${\cal P}$--function and show that it
describes three types of Calogero--Sutherland potential in its various
limits, corresponding to symmetric spaces of different curvature.
These limiting potentials correspond in the Calogero--Sutherland model
to particles interacting on a circle, hyperbola, or line,
respectively, reflecting the corresponding triplet of symmetric spaces
underlying the respective Calogero--Sutherland models.

\section{Random matrix theories and symmetric spaces}
\label{sec-RMT}
\setcounter{equation}{0}

\subsection{Introduction to the theory of random matrices}
\label{sec-introRMT}

\subsubsection{What is random matrix theory?}

Random matrix theory has evolved into a rich and versatile field with
applications in several branches of physics and mathematics.  In the
theory of random matrices one studies the statistical properties of
the eigenvalues of large matrices with randomly distributed elements.
Historically, large random matrices were first employed by Wigner
\cite{Wigner} and Dyson \cite{Dyson} to describe the energy levels in
complex nuclei, where they modelled the Hamiltonian of the
system. Disregarding the continuous part of the spectrum, one may
represent the Hamiltonian of such a system quantum mechanically by a
large hermitean matrix acting on a finite--dimensional Hilbert
space. In principle, the eigenvalues of the Hamiltonian then give us
the energy levels of the nucleus, and the eigenvectors give us the
eigenstates.

However, for a system involving hundreds of nucleons (or, in a small
metal sample, hundreds of electrons) we do not know the form of the
Hamiltonian, and even if we did, the huge number of degrees of freedom
involved would prevent us from solving the Schr\"odinger equation
exactly.  Choosing instead to represent the Hamiltonians of an
ensemble of such systems by an ensemble of large hermitean matrices,
whose elements are random variables with some given distribution, we
obtain a statistical theory for its eigenvalues
$\lambda_1,...,\lambda_N$, where $N$ is a large number equal to the
size of the random matrices \footnote{For symplectic ensembles the
eigenvalues are two--fold degenerate.}. If the random matrices in the
chosen ensemble have the same global symmetry properties as the actual
Hamiltonian, this statistical theory describes also the eigenvalues of
the Hamiltonian, to the extent that they depend only on symmetry,
i.e. they are {\it universal}. Even systems with just a few degrees of
freedom lend themselves to such a description. In general, a description in
terms of random matrices is possible whenever we are dealing with a
chaotic or disordered system (see \cite{GMW} for a discussion).

Such random matrices were studied by Dyson in a series of papers
\cite{Dyson}. He showed that they fall into one of three classes named
the orthogonal, unitary and symplectic ensemble, depending on whether
or not the Hamiltonian possesses time--reversal invariance and
rotational invariance. This classification was reviewed in Ch.~2 of
the book by Mehta \cite{Mehta}. The corresponding random matrices are
real symmetric, complex hermitean, or self--dual quaternion,
respectively (see below).

\subsubsection{Some of the applications of random matrix theory}

In typical applications, large matrices with given symmetry properties
and randomly distributed elements substitute a physical quantum
operator. This operator could be a Hamiltonian like in the example of
nuclear energy levels or in quantum chaotic scattering (in the latter
case an effective Hamiltonian appears in the scattering matrix), or it
could be for instance a scattering matrix, transfer matrix, or Dirac
operator. Then one studies the statistics of its eigenvalue spectrum
and extracts the universal behavior.

An early application of the Wigner--Dyson ensembles was in the theory
of scattering in chaotic quantum systems. With the help of random
matrices one is able to study the statistics of resonance poles and
scattering phase shifts in nuclei, atoms, and molecules, as well as
microwave cavities and ballistic systems (for a self--contained review
see \cite{FyoSom}).

A major area where random matrices are employed is in the description
of the infrared limit of gauge theories, notably QCD. In these
applications an integration over an appropriate random matrix ensemble
replaces the integration over gauge field configurations in the
partition function. This is achieved by substituting a suitable random
matrix for the Dirac operator appearing in the fermion
determinant. Because the magnitude of the quark condensate is
proportional to the spectral density of the Dirac operator at the
origin of the spectrum, random matrices are useful in this
context in obtaining insights concerning the spontaneous
breaking of chiral symmetry. This is one of the most fundamental
phenomena in QCD, since it determines the hadronic mass spectrum. As
it is a phenomenon that takes place at very low energy, it is
inaccessible to perturbation theory and therefore one of the most
difficult to study from a theoretical point of view.  The random
matrix approach brings considerable simplification and has indeed
contributed to the understanding of this phenomenon. The corresponding
random matrix ensembles are called chiral ensembles, because of the
chiral symmetry of the Dirac operator.  We will discuss chiral random
matrix theories in more detail in subsection \ref{sec-chiral}.

Random matrix theories are also largely and successfully applied in
the theoretical description of mesoscopic systems, where they may be
used to model the properties of scattering and transfer matrices.  A
mesoscopic conductor is a micrometer--sized metal grain or wire at low
temperature with randomly distributed impurities that act as
scattering centers for electrons originating from current leads
(alternatively, random scattering takes place at the edges and depends
on the shape of the conductor).  Examples include disordered wires,
quantum dots, and normal metal--superconductor heterostructures.  

The theory of quantum transport is concerned with the statistics of
the transmission eigenvalues in such systems. The transmission
eigenvalues are related to the transfer matrix in a way which will be
discussed in detail in paragraph \ref{sec-transfer}. These eigenvalues
directly determine the physical conductance.  We will discuss quantum
wires in some detail below. We will also briefly discuss applications
in normal metal--superconductor heterostructures as well as in
ballistic quantum dots. The ensembles used in these applications are
related to yet other symmetric spaces than those pertinent to the
ensembles used in the study of the usual mesoscopic systems. The
reason is, of course, that the symmetries of the physical operator
whose eigenvalues we study are different.

The same chiral ensembles used in the description of chiral
gauge theories are also realized in the Hamiltonians of random flux
and random hopping problems \cite{M1,M4} in the theory of quantum
transport.

Some important applications of hermitean matrix models which will not be
treated here, are in the description of random surfaces in the field of
quantum gravity (for an excellent introduction see \cite{AmQG}), where
a $N^{-2}$ expansion amounts to a genus expansion of the random
surface, and in string theory. 
 
The wide range of applications is one of the most fascinating aspects
of random matrix theory. Notably, the classical random matrix theories
have no adjustable parameters. The reason for their success in
accurately characterizing such a wide range of systems (nuclei,
disordered metals, chaotic systems) lies in the universal behavior of
the eigenvalue correlators in an appropriate double scaling limit.
This is the subject of the following paragraph.

\subsubsection{Why are random matrix models successful?}

The statistical properties of a sequence of apparently random rumbers
like nuclear energy levels may be either of the local or of the global
type.  The local and global spectral characteristics are completely
disconnected. Systems with identical global characteristics may have
different local characteristics, and vice versa.

A typical example of a global, or {\it macroscopic}, property is the
mean level (eigenvalue) density, or one--point correlation function,
in the limit of large $N$. It is often denoted $\rho (\lambda )$. For
the gaussian Wigner--Dyson ensembles the mean level density approaches a
semicircle in the large $N$ limit. In general its shape depends on the
random matrix potential determining the probability distribution of
individual matrix elements. A global characteristic spectral property
does not change appreciably on the scale of a few level spacings. An
example of a local spectral property is the spacing between two
successive levels. It fluctuates from level to level.

One amazing feature of random matrix models is that some local
spectral characteristics are independent of the distribution of
individual matrix elements in the limit of large $N$, {\it if we
simultaneously rescale the eigenvalues in an appropriate way}. This is
called a double scaling limit.  In this limit local spectral
characteristics are determined solely by the global symmetries of the
random matrices.  This so called {\it microscopic universality} is
manifest only in the appropriate double scaling limit, where the
average level spacing becomes much smaller than the scale on which it
varies. 

The $n$--point correlation function, or joint probability density for $n$
eigenvalues $\rho_n (\lambda_1,...,\lambda_n )$ ($1\leq n\leq N$) is
central in the theory of random matrices and is the most important
example of such a microscopically universal quantity. It is defined as
follows: $\rho_n (\lambda_1,...,\lambda_n )d\lambda_1...d\lambda_n$ is
the probability of finding one eigenvalue in each of the intervals
$[\lambda_i,\lambda_i+d\lambda_i]$ ($1\leq i\leq n$).  This means that
if one rescales the eigenvalues by the average level
spacing\footnote{In general one can consider also other scales.} and
simultaneously takes the limit $N\to \infty$, the $n$--point function
$\rho_n$ for each ensemble becomes a universal function characteristic
of the ensemble. There are also other spectral characteristics that
depend on symmetry only. The number variance (i.e., the variance
related to the number of eigenvalues in a certain interval), the
so--called spectral rigidity $\Delta_3$ (this quantity was defined for
example in \cite{Mehta,statistics}), the distribution of the first
eigenvalue, and the distribution of spacings between adjacent levels
are examples of such quantities. 

In a hermitean matrix model the eigenvalues are real. The eigenvalue
density (one--point function) of such a model is typically zero
outside some interval or intervals around the origin. As we learnt
above, its macroscopic shape depends on the probability distribution,
while locally its microscopic shape is universal.  Spectrum edge
universality, i.e. universality of scaled correlation functions near
the edge of the eigenvalue spectrum, is known to differ from
universality in the bulk of the spectrum. One normally distinguishes
between three microscopic scaling regimes: the bulk, the hard edge
(vicinity of the origin) and the soft edge (tail) of the spectrum. The
rescaling of the eigenvalues must be chosen appropriately in each
regime, and in practice amounts to a rescaling by an appropriate
integer or fractal power of $N$ \cite{Univers,multi,wealth,gernot1},
which is proportional to the inverse local density of eigenvalues and
thus proportional to the average local level spacing. This procedure
in turn amounts to magnifying a local region in such a way that the
behavior of the eigenvalue correlators is universal.

In case of a (multi)critical hermitean matrix model, the macroscopic
spectral density develops extra (multiple) zeros on the limit between
a one--cut (single--interval) support and a multi--band support. This
is achieved by fine--tuning couplings in the random matrix potential
determining the probability ditribution.  In this case the scaling is
more subtle and the calculations more involved due to subleading terms
in the large--$N$ expansion \cite{multi}. 
 
Several attempts at theoretically demonstrate universality can be
found in references \cite{Univers,multi,wealth,gernot1,for}.  It is
also supported by a wealth of numerical \cite{numerical} evidence.  In
\cite{Univers}, the universality of correlation functions near the
origin of the eigenvalue spectrum of chiral unitary and unitary
ensembles of random matrices in the microscopic limit was
investigated. Universality was shown to follow from the fact that the
recursion relation for the orthogonal polynomials determining the
kernel from which all correlators are derived, reduces to a universal
differential equation (in this case for Bessel functions) in the
appropriate microscopic limit.

\subsection{The basics of matrix models}
\label{sec-basics}

In this subsection we will give a few basic definitions and results
regarding the simplest kinds of hermitean matrix models. Similar
results apply for all the matrix models we will study in this paper,
and the purpose here is to introduce the reader not at all familiar
with matrix models into the subject. 

A general hermitean matrix model is defined by a partition function 

\beq
\label{eq:partition}
Z\sim \int dS\, P(S)
\eeq

where $S$ is a square $N\times N$ hermitean matrix with randomly
distributed elements $S_{ij}$ and $dS$ is a Haar measure. For example,
for the gaussian unitary ensemble consisting of complex hermitean
matrices (this ensemble describes a system with no time--reversal
invariance) the matrix elements are complex numbers,
$S_{ij}=S^r_{ij}+iS^i_{ij}$ (except the ones on the diagonal that must
be real). $dS$ is then given by

\beq
dS=\prod_{i\leq j} dS^r_{ij}\prod_{k<l} dS^i_{kl}
\eeq

and the probability $P(S)dS$ that a system described by the gaussian 
unitary ensemble will belong to the volume element $dS$ is invariant 
under the automorphism

\beq 
\label{eq:rot} 
S\to U^{-1}SU 
\eeq

of the ensemble to itself, where $U$ is any unitary $N\times N$
matrix\footnote{If S is time--reversal invariant $U$ will be either a
real orthogonal matrix or a symplectic matrix, depending on whether
rotational symmetry is also present. In this case we are dealing with
the orthogonal or symplectic ensemble.}. This amounts to a change of
basis. The invariance of $P(S)dS$ under such automorphisms restricts
$P(S)$ to depend on the traces of the first $N$ powers of $S$. For the
gaussian matrix models reviewed for example in the book by Mehta,
there was another requirement, namely that the matrix elements should
all be statistically independent. This excludes everything except the
traces of the first two powers, and these may occur only in an
exponential \cite{Mehta}. However, this requirement is not physically
motivated and was subsequently relaxed as other types of matrix models
were introduced. A probability distribution of the form

\beq
P(S)\propto {\rm e}^{-c\, {\rm tr} V(S)}
\eeq

is often used. Here $c$ is a constant and $V(S)$ is a matrix
potential, typically a polynomial with a finite number of terms. 

By doing an appropriate similarity transformation (\ref{eq:rot}) on
the ensemble of random matrices, the Haar measure $dS$ and potential
$V(S)$ can be expressed in terms of the eigenvectors and eigenvalues
of the matrix $S$. This amounts to (block)diagonalizing $S$. At the
same time the Haar measure is factorized into a part that depends only
on the eigenvectors (the "angular" part) and a part that is a function
of the eigenvalues $\{ \lambda_i\} $ of $S$ only. Assume for
simplicity that the random matrix potential is given by

\beq
cV(S)=S^2
\eeq

This kind of potential is called gaussian, for obvious reasons.  The
Jacobian of the similarity transformation depends on $\{ \lambda_i\} $
only and is given by \cite{Mehta}

\beq
J(\{ \lambda_i\} ) \sim \prod_{i<j} (\lambda_i-\lambda_j)^2
\eeq

for the gaussian unitary ensemble. For the other gaussian ensembles
one has a similar Jacobian (see eq.~(\ref{eq4.2}) below). The part of $P(S)dS$
that depends on the eigenvalues then takes the form

\eq
P(\{\lambda_i\})\, d\lambda_1...d\lambda_N \propto \prod_{i<j} (\lambda_i-\lambda_j)^2\, \e^{-\sum_{j=1}^{N}\lambda_j^2}\, d\lambda_1...d\lambda_N
\en

whereas the "angular" degrees of freedom do not appear in the
integrand at all and can be integrated out to give a constant in front
of the integral.  This model is easily solvable. This is the strength
of the random matrix description of disordered systems. The eigenvalue
correlators ($k$--point functions) are defined as 

\beq
\rho_k(\lambda_1,...,\lambda_k)=\frac{N!}{(N-k)!}
\int \prod_{j=k+1}^N d\lambda_jP(\{\lambda_1,...,\lambda_N\})
\eeq

and can be calculated exactly by rewriting the Jacobian as a product
of Vandermonde determinants of a set of polynomials orthogonal with
respect to the measure $\e^{-cV(\lambda )}$.  This procedure was
reviewed in \cite {Mehta}. For the gaussian unitary ensemble these
polynomials are Hermite polynomials $H_n(\lambda )$ and satisfy

\beq
\int_{-\infty}^{+\infty} H_m(\lambda )H_n(\lambda )\e^{-\lambda^2}d\lambda
=h_n\delta_{mn}
\eeq

where $h_n$ is a normalization.  In practice one defines a kernel
$K_N(\lambda_i,\lambda_j)$ that turns out to be universal and whose
determinant gives all the correlation functions \cite{Mehta}. Defining
so--called oscillator wave functions that for the gaussian unitary
ensemble take the form

\beq
\varphi_j(\lambda )=h_j^{-\frac{1}{2}}H_j(\lambda )\e^{-\frac{\lambda^2}{2}}
\eeq

this kernel is given by

\beq
K_N(\lambda_i,\lambda_j)=\sum_{k=0}^{N-1}\varphi_k(\lambda_i)
\varphi_k(\lambda_j)
\eeq

It can be evaluated explicitly using the Christoffel--Darboux formula
for the sum of products of orthogonal polynomials.  The $k$--point
function is then

\beq
\rho_k(\lambda_1,...,\lambda_k)=
{\rm det}[K_N(\lambda_i,\lambda_j)]_{1\leq i,j\leq k}
\eeq

Different authors use different conventions for the normalization constants
involved in these formulas.

So far we have not said anything about the integration manifold in
equation (\ref{eq:partition}). As we have seen, we start out with some
given class $G$ of matrices $S$, and the integral is a priori over
this class.  However, by doing the similarity transformation in
eq.~(\ref{eq:rot}), we can perform the integration over the subgroup
$K$ to which the matrix $U$ belongs without further ado. We then end
up with an integral over a smaller manifold $G/K$, the manifold of the
random matrix eigenvalues, and as we will see in the following
subsection, this manifold turns out to be a symmetric space for all
the commonly used hermitean random matrix ensembles.  This gives rise
to the deep and useful connection between random matrix theories and
symmetric spaces that is the subject of this review.

Let us now look more in detail at some explicit examples of random
matrix theories. In the following subsections we will identify in turn
their integration manifolds as symmetric spaces, their eigenvalues as
radial coordinates on the respective symmetric spaces, and their Dyson
and boundary indices (to be defined) as determined by the
multiplicities of the corresponding restricted root lattices. As we
will see, the Jacobian of the similarity transformation to the space
of random matrix eigenvalues and eigenvectors is determined explicitly
by these root multiplicities and the curvature of the symmetric space
manifold.

\subsection{Identification of the random matrix integration manifolds}
\label{sec-id-manifolds}

In this subsection we introduce and explicitly define some  
random matrix ensembles. We will see how a description in terms of
symmetric spaces emerges naturally from the corresponding integration
manifolds.  We discuss this connection in detail for the gaussian and
circular ensembles, which are the simplest and most frequently used
random matrix theories.

Historically, the gaussian ensembles were studied first, and only
subsequently the circular ensembles, in which the requirement of
statistical independence of the matrix elements is relaxed.  We will,
however, start by addressing the circular ensembles which turn out to
be simpler to discuss from a symmetric space point of view.  The
circular ensembles are related to symmetric {\it coset spaces} of
positive curvature, while the gaussian ensembles are related to
euclidean type symmetric spaces of vanishing curvature, which are
identified with a {\it subspace of a Lie algebra}.  Therefore the
identification of the symmetric space is slightly more tricky for the
gaussian ensembles. As we will see, the identification with euclidean
spaces is due to the translational invariance of the ensembles.  In
the following subsections we then discuss, in less detail but in the
same spirit, a few further examples of random matrix ensembles which
we have selected for their widespread applications in various branches
of physics.

\subsubsection{Circular ensembles}
\label{sec-circ}

{\bf [1] Physical applications of the circular ensembles}

The circular ensembles are used in the description of physical systems
which can be characterized by a unitary matrix $S$, typically a
scattering matrix, each of whose elements give the transition
probability from one state to another.  For instance, these ensembles
describe the statistics of scattering phase shifts $\phi_i$ in the
theory of mesoscopic systems (for a review and references see
\cite{Beenakker}). These systems are modeled by a phase--coherent
disordered scattering region connected to ideal leads.  Scattering of
electrons against randomly distributed impurities takes place in the
disordered region.

The scattering matrix $S$ relates the incoming and the outgoing wave
amplitudes of the electrons. If we have $N$ propagating modes
(channels) at the Fermi level, we can describe them by a vector of
length $2N$ of incident modes $I$, $I'$ and a similar vector of
outgoing modes $O$, $O'$ in each lead, where unprimed letters denote
the modes in the left lead and primed letters the modes in the right
lead. Then the scattering matrix is defined by

\beq
\label{eq:scat}
S\left( \begin{array}{c} I \\ I' \end{array} \right) =
\left( \begin{array}{c} O \\ O' \end{array} \right)
\eeq

Following a standard notation \cite{BILP,MPK} the scattering matrix
has the following block structure

\beq
\label{eq:Smatrix}
S=\left( \begin{array}{cc} r & t' \\
                          t & r' \end{array} \right)
\eeq

where $r$, $r'$, $t$, $t'$ are $N\times N$ reflection and transmission
matrices. The unitarity of $S$, warranted by flux conservation
($|I|^2+|I'|^2=|O|^2+|O'|^2$), implies that the four matrices
$tt^{\dagger }$, $t't'^{\dagger }$, $1-rr^{\dagger }$,
$1-r'r'^{\dagger }$ have the same set of real eigenvalues $T_1,...,T_N$
called transmission eigenvalues ($0 \leq T_i \leq 1$). These determine
the physical conductance of the scattering region and will be
discussed further in connection with transfer matrix ensembles.

{\bf [2] Definiton of the circular ensembles}

The circular ensembles were introduced by Dyson \cite{Dyson} in the
early sixties. They are the simplest possible matrix models, since the
probability distribution is a constant. By probability distribution we
mean, in general, a distribution $P(S)$ such that $P(S)dS$ is the
probability that a matrix from the ensemble takes a value between $S$
and $S+dS$ (for an introduction see \cite{Mehta}, Ch.~9). The
non--trivial features of the model only appear in the probability
distribution of the {\it eigenvalues}.

Mimicking the gaussian case (which we will discuss in the next
subsection) Dyson proposed three ensembles labelled by an index $\beta
$, which takes one of the values $\beta=$1, 2, and 4 and counts the
number of degrees of freedom characterizing each matrix element.  The
index $\beta$ is called the Dyson index of the matrix model. For the
gaussian matrix models, the three cases correspond exactly to systems
with time--reversal invariance and rotational invariance ($\beta=1$),
no time--reversal invariance ($\beta=2$), and time--reversal
invariance and no rotational invariance ($\beta=4$) \cite{Mehta}.

\begin{itemize} 

\item In the {\it circular orthogonal ensemble} labelled by $\beta=1$,
$S$ is a unitary symmetric $N\times N$ random matrix.

\item In the {\it circular unitary ensemble} ($\beta=2$) $S$ is a
unitary $N\times N$ random matrix.

\item Finally in the case of the {\it circular symplectic ensemble}
($\beta=4$) $S$ is a unitary self-dual random matrix with $N\times N$
{\it quaternionic} matrix elements. Generally speaking, a quaternion
can be expressed in terms of $2\times 2$ matrices as $Q=a_0\sigma_0
+i\vec{a}\cdot \vec{\sigma }$, where the $a_i$ are real or complex
numbers and $(\sigma_0,\vec{\sigma })=(1,\sigma_1, \sigma_2,
\sigma_3)$ are the $2\times 2$ unit and Pauli matrices. (For an
introduction to quaternion algebra, see \cite{Mehta} or \cite{Mehta1},
Ch.~8.)  By self--dual we mean $S^R=S$ where
$(S^R)_{ij}=\sigma_2(S^T)_{ij}\sigma_2$ (see \cite{Mehta,Mehta1} for
further details).  

The terminology for the circular ensembles may seem odd. It stems from
the gaussian ensembles, where the stability subgroups of invariance
are orthogonal, unitary, and symplectic, respectively, for the values
1, 2 and 4 of the Dyson index.

\end{itemize}

{\bf [3] Properties of the circular ensembles}

Below we recollect some general properties of the circular ensembles.  As we
will see, these properties have analogous realizations in all the
ensembles that we will study.

{\bf Invariance.} It is easy to see that $P(S)dS$ for the three
circular ensembles is invariant under the similarity transformations

\beq
\label{eq:Sinv}
S\to USU^{-1}\ \ \ \ (\beta=1,4)\ \ \ \ \ \ \ S\to VSW\ \ \ \ (\beta=2)
\eeq

where $U$ is unitary ($\beta =1$) or unitary with quaternion elements 
($\beta =4$) and $V$, $W$ are unitary. 

{\bf Diagonalization.} In each case the matrix $S$ can be diagonalized
by a similarity transformation like the ones in
eq.~(\ref{eq:Sinv}). The eigenvalues lie on a unit circle and take the
form ${\rm e}^{i\phi_i}$ (thereof the name "circular ensembles").
They are doubly degenerate in the symplectic case $\beta =4$.

{\bf Probability distribution of the eigenvalues.} A simple
calculation \cite{Mehta} shows that the Jacobian of the transformation
from the space of unitary matrices to eigenvalue space is given by:

\beq
J_\beta (\phi_i)\propto \prod_{i<j} |\e^{i\phi_i}-\e^{i\phi_j}|^{\beta}
\eeq

It gives rise to correlations between eigenvalues.  Since the original
distribution function $P(S)$ of the random matrices is constant, the
Jacobian coincides with the probability distribution function of the
eigenvalues:

\beq
\label{eq:circularP}
P_\beta (\phi_i)\sim J_\beta (\phi_i)
\eeq

{\bf [4] The symmetric spaces associated to the circular ensembles}

In this paragraph our main interest is identifying the integration 
manifold of the three circular ensembles. 

{\it Orthogonal ensemble.} Every symmetric unitary matrix $S$ can be written 
as 

\eq
S=U^TU
\label{dy1}
\en

where $U$ is a generic unitary matrix. However, this mapping is not
one--to--one. If we assume that $S=U^TU=V^TV$, then it is easy to see
that the matrix relating the two expressions, $R=VU^{-1}$, is unitary
and satisfies $R^TR=1$. Hence $R$ must be real and orthogonal.  Thus
we see that the manifold of the unitary symmetric matrices is actually
the coset $U(N)/O(N)$, due to the above mentioned degeneracy. From the
point of view of the physical properties of the ensemble nothing
changes if we perform the restriction to an irreducible symmetric
space.\footnote{In the partition function

\beq
Z\sim \int_{G/K} dS\, P_{\beta }(S)
\eeq

extracting such a $U(1)$ factor from the integration manifold just
amounts to redefining $Z$ by a constant.} Then the manifold becomes
$SU(N)/SO(N)$.  We will systematically perform the reduction to
irreducible symmetric spaces in the following.

{\it Unitary ensemble.} Finding the integration manifold is trivial in
the $\beta=2$ case, where we simply have unitary matrices without any
further constraint and the manifold is simply the group $U(N)$. Like
for $\beta =1$, if one is interested in irreducible symmetric spaces a
$U(1)$ factor must be taken out, and the manifold becomes $SU(N)$.

{\it Symplectic ensemble.} This case strictly resembles the $\beta=1$ case
discussed above. Any self-dual unitary quaternion matrix can be
written as

\eq
S=U^RU
\en 

where $U$ this time is a $2N\times 2N$ unitary matrix. By following
the same reasoning as in the case of the orthogonal ensemble, we see
that the same matrix $S$ can be obtained using a new unitary matrix
$V$ obtained from the previous one by the transformation $V=BU$ where
$B$ is constrained by $B^RB=BB^R=1$.  By definition of the duality
operator $^R$ this means that $B$ is a symplectic matrix. Thus,
extracting a $U(1)$ factor, the manifold coincides with
$SU(2N)/Sp(2N)$.

By comparing with Table~\ref{tab1} in subsection \ref{sec-restricted}
in Part~I of the present review, we see that the integration manifolds
of the three circular ensembles are exactly the first three coset
spaces (described in the Cartan notation as A, AI and AII) of positive
curvature in the list of possible irreducible symmetric spaces. This
is the main result of this paragraph.
   
\subsubsection{Gaussian ensembles}
\label{sec-gauss}

{\bf [1] Physical applications of the gaussian ensembles}

Some of the physical applications of the gaussian ensembles were
discussed already in the introduction to this section, in paragraph
\ref{sec-introRMT}. In general gaussian ensembles describe hamiltonian
ensembles. This is because they correspond to flat symmetric spaces,
as we will see below. The corresponding positive curvature ensembles
are the ensembles of the scattering matrices of the same systems,
since in general the scattering matrix is expressed as $S=\e^{i{\cal
H}}$.  As we have seen, applications include spectra resulting from
complicated many--body interactions like those taking place in neutron
resonances in atomic nuclei, the electronic energy levels inside tiny
metal grains at low temperature (that depend on the shape of the
grain), or generally the spectra of electrons moving in a random
potential with no further symmetries present.

The energy spectrum of classically chaotic systems is another example
of apparently gaussian random behavior. A realization of such a system
is a chaotic billiard. The motion of the billiard ball is determined
by the shape of the billiard table (which is chosen to be irregular in
some way) and it is drastically different for small variations in the
initial trajectory. Corresponding quantum billards may also be
considered, represented by a free quantum particle confined to a
finite part of space. Its discrete energy spectrum will be determined
by the Laplacian on the space, and may possess varying degrees of
randomness. In all these systems the statistical properties of the
energy spectrum can be described by a gaussian random matrix ensemble.
  
{\bf [2] Definition of the gaussian ensembles}

Historically the first type of ensembles to be studied, the gaussian
ensembles introduced by Wigner \cite{Wigner} and Dyson
\cite{DysonSS,Dyson} are defined by the probability distribution

\beq
P_\beta (H) \sim \e^{-\beta \, \tr V(H)}
\eeq

where $H$ is a hermitean $N\times N$ matrix and $V(H)$ is a
quadratic potential. We can define a partition function 

\beq
Z=\int dH \, P_\beta (H)
\eeq

where $dH$ is an invariant Haar measure.  It can be shown \cite{Mehta}
that the form of $P_\beta (H)$ is automatically restricted to the form

\beq
P_\beta (H)={\rm exp}(-a\, {\rm tr}H^2+b\, {\rm tr}H+c)
\eeq

($a>0$) if one postulates statistical independence of the matrix
elements $H_{ij}$.   Note that $P_\beta (H)$ can be cast in the form

\beq
\label{eq4.1}
P_\beta (H) \sim \e^{-a \, \tr H^2}
\eeq

by simply completing the square in the exponent.  

Depending on the nature of $H$ we can distinguish three cases labelled
by $\beta =$1, 2, and 4: 

\begin{itemize} 

\item In the {\it orthogonal ensemble} ($\beta =1$) $H$ is an $N\times
N$ hermitean symmetric random matrix.  It immediately follows that the
probability distribution $P_1(H)$ and the integration measure $dH$ are
invariant under all real orthogonal transformations of $H$. This
ensemble is used when the random matrix $H$ models the time--reversal
and rotation invariant Hamiltonian of a system with integral spin
(\cite{Dyson}, \cite{Mehta}, Ch.~2).

\item In the {\it unitary ensemble} ($\beta =2$) $H$ is an $N\times N$
hermitean random matrix and $P_2(H)dH$ is invariant under all unitary
transformations of $H$. A system without time--reversal invariance has
this type of Hamiltonian \cite{Dyson,Mehta}.

\item In case of the {\it symplectic ensemble} ($\beta =4$) $H$ is an
$N\times N$ hermitean self-dual random matrix with quaternionic
elements. It can be shown that the entries of such a matrix are all
real quaternions which can be expressed in terms of $2\times 2$
matrices as $Q=a_0\sigma_0 +i\vec{a}\cdot \vec{\sigma }$, where the
$a_i$ are real numbers and $(\sigma_0,\vec{\sigma })=(1,\sigma_1,
\sigma_2, \sigma_3)$ are the $2\times 2$ unit and Pauli
matrices. $P_4(H)dH$ is invariant under all symplectic transformations
of $H$.  The Hamiltonian of a system with time--reversal invariance,
no rotational invariance and half--odd integral spin is of this type
\cite{Dyson,Mehta}.

\end{itemize}

As we can see, the symmetry classes are distinguished by their
behavior under time reversal (TR) and spin rotation (SR). 

{\bf [3] Properties of the gaussian ensembles}

As we did for the circular ensembles, let us review some general
properties of the gaussian ensembles.

{\bf Statistical independence of the entries.} With the choice of
potential in eq.~(\ref{eq4.1}) the three ensembles have independently
distributed elements, since the potential ${\rm tr}H^2=
\sum_{ij} |H_{ij}|^2$.

{\bf Invariance.} As implied above, $P_\beta (H)$ and the integration
measure $dH$ are separately invariant under the transformation

\beq
\label{eq:UHU}
H \to UHU^{-1},
\eeq

where $U$ is an orthogonal, unitary or symplectic $N\times N$ matrix
depending on the value of $\beta $.  However, it is important to
notice that the symmetry group of $P_\beta (H)\,dH$ is larger and
consists of rotations by the matrix $U$ like in eq.~(\ref{eq:UHU}),
and addition by square hermitean matrices:

\beq
H\to UHU^{-1}+H' 
\label{UHUH}
\eeq

This will play an important role in the identification of these
ensembles with a symmetric space.

{\bf Diagonalization.} For each matrix $H$ there is a matrix $U$ that
maps it onto its eigenvalues:

\beq
\label{eq:Hinv}
H=U\Lambda U^{-1},\ \ \ \ \ \ \Lambda ={\rm diag}(\lambda_1,...,\lambda_N)
\eeq

where $\lambda_i$ ($i=1,...,N$) are real eigenvalues (if $\beta =4$,
they are twofold degenerate.) A simple calculation shows that the
Jacobian of the transformation from the space of hermitean matrices
$H$ to eigenvalue space is given by

\beq
\label{eq:Jbeta}
J_\beta (\{\lambda_i\})\propto \prod_{i<j} |\lambda_i-\lambda_j|^{\beta }
\eeq

where $\beta=$1, 2, and 4 in the orthogonal, unitary and symplectic case, 
respectively (see \cite{Mehta} and \cite{Hua}, Ch. 3).

{\bf Probability distribution of the eigenvalues.} Under the
transformation (\ref{eq:Hinv}), the integration measure $dH$
factorizes into an integral over the symmetry subgroup $K$ and an
integral over the eigenvalues with integration measure $J_\beta
(\{\lambda_i\})\, \prod_i d\lambda_i $. The integral over the subgroup
can be performed immediately and gives a constant equal to the volume
of the subgroup. With this in mind and the random matrix potential
given by $V(H)=\frac{1}{2}H^2$ \footnote{This form of the potential
gave rise to the name "gaussian ensembles". Of course, a gaussian
random matrix potential can be used in any context, as it often is,
because it is the simplest one to deal with.}, the joint probability
density for the eigenvalues becomes

\eq
\label{eq4.2}
P_\beta (\{\lambda_i\})\propto \prod_{i<j} |\lambda_i-\lambda_j|^{\beta }
\e^{-\frac{\beta }{2}\sum_{j=1}^{N}\lambda_j^2}
\en

The proportionality constant is not relevant if one normalizes the
integral to unity (however, if needed, exact normalization constants
for many integrals like the one in (\ref{eq4.2}) were given by Hua in
\cite{Hua}).

The spectral correlations are due to the Jacobian. Eq.~(\ref{eq4.2})
can be rewritten in the general form

\beq
\label{eq:PA}
P_\beta (\{\lambda_i\})\sim \e^{-\beta  \left( \sum_i 
V(\lambda_i) -\sum_{i<j} \ln |\lambda_i-\lambda_j|\right)}
\eeq

The logarithmic pair potential leads to repulsion between the
eigenvalues.  This is the so called {\it spectral rigidity} which is one of
the most important features of random matrix theory. It is not present
in uncorrelated Poisson distributions. The interpretation of
(\ref{eq:PA}) is that the probability of finding the $i$'th eigenvalue
in the interval between $\lambda_i$ and $\lambda_i+{\rm d}\lambda_i$
is proportional to $P_\beta(\{\lambda_i\}) \prod_i {\rm d}\lambda_i$.

{\bf [4] The symmetric spaces associated to the gaussian ensembles}

Let us now rephrase the above results in a group theoretical language.
This will allow us to make contact with the analysis in section
\ref{sec-claSS}.  Our goal here is to show that the integration
manifolds of the above ensembles are symmetric spaces
of zero curvature. This is essentially due to the translational invariance 
of eq.~(\ref{UHUH}).

Let us start with the $\beta =1$ case.  As a first step let us limit
ourselves to traceless matrices only. This is the equivalent in the
present context of the choice that we made in the previous subsection
when we eliminated the $U(1)$ factor from the coset. Also in the
present case the reason is that we want to obtain a description in
terms of {\it irreducible} symmetric spaces. The remaining degree of
freedom acts trivially and is usually neglected.  

We now try to describe the set of all real, symmetric, and traceless
matrices as a symmetric space $X=G/K$.  By comparing with the example
in subsection \ref{sec-Inv} in Part~I of the review, this can easily
be done by identifying it with the {\it algebra subspace} ${\bf
SL(N,R)/SO(N)}$. This simply amounts to taking a matrix from the
set of generic, traceless, real $N\times N$ matrices, write it as a
sum of a symmetric and an antisymmetric matrix, and then eliminate the
antisymmetric part.  The resulting space is exactly the set of all
real, symmetric and traceless matrices that we were looking for. By
taking only traceless matrices, we have eliminated an ${\bf R^+}$
factor from the algebra ${\bf GL(N,R)}$, corresponding to the absolute
value of the determinant, and we end up with the algebra ${\bf
SL(N,R)}$, of which we then choose only the symmetric generators.

Following the discussion of flat symmetric spaces in paragraph
\ref{sec-curv}, the set of all real, symmetric and traceless matrices
(let us call it ${\bf P}$) can be realized as a zero--curvature
symmetric space by the following two steps:

\begin{description}
  
\item{(1)} The group $G^0$ of affine transformations of ${\bf P}={\bf
    SL(N,R)/SO(N)}$ into itself is given by the semidirect product of
  the subgroup $K=SO(N)$ and the invariant algebra subspace ${\bf P}$.
  The action of $G^0$ on ${\bf P}$ is given by $g(p)=kpk^{-1}+a$,
  where $g=(k,a)\in G^0$, $k\in K$, $p,\ a\in X^0=G^0/K={\bf P}$.  But
  this affine transformation is exactly of the same type as the
  symmetry transformation of eq.~(\ref{UHUH}) that leaves the gaussian
  ensembles invariant (in fact, for the algebra subspace under
  discussion it is this transformation for $\beta=1$).
  
\item{(2)} The algebra subspace ${\bf P}$ now forms an abelian
  subalgebra (identical to the Cartan subalgebra) of the algebra ${\bf
    G^0}$, because it is additive: $[{\bf P},{\bf P}]=0$.  Therefore
    ${\bf G^0}={\bf K}\oplus {\bf P}$ has an abelian ideal, and is
    non--semisimple.  The curvature tensor of these spaces is zero by
    equations (\ref{eq:commrel2}) and (\ref{eq:curv}) of subsection
    \ref{sec-curv}. Also, the commutation relations (\ref{eq:commrel})
    defining a symmetric space are satisfied for the algebra elements
    in ${\bf K}$ and ${\bf P}$.  Hence $X^0=G^0/K={\bf P}$ is a flat
    symmetric space with a euclidean geometry.

\end{description}

By performing a similar analysis for $\beta=$2, 4 we obtain the
following general result: The gaussian ensembles labelled by $\beta
$=1, 2, and 4 consist of hermitean square matrices belonging to
algebra subspaces ${\bf SL(N,R)/SO(N)}$, ${\bf SL(N,C)/SU(N)}$, and
${\bf SU^*(2N)/USp(2N)}$, respectively. The algebra ${\bf SL(N,C)}$ is
the algebra of hermitean, complex and traceless $N\times N$ matrices,
and the subalgebra ${\bf SU(N)}$ defines the subgroup of invariance of
the unitary ensemble.  The algebras ${\bf SU^*(2N)}$ and ${\bf
USp(2N)}$ were formally defined in a footnote in subsection
\ref{sec-realforms2}.  Again, they are exactly the algebra of
hermitean, self--dual $N\times N$ quaternion matrices and the
subalgebra of unitary symplectic matrices.  From Table~\ref{tab1} we
see that these three symmetric spaces correspond to algebra subspaces
in Cartan classes A, AI and AII. The integration manifolds of the
circular ensembles are the positive curvature coset spaces
corresponding to the same Cartan classes.

{\bf [5] Remarks concerning flat symmetric spaces}

The flat symmetric spaces may be seen as limiting cases of their curved
counterparts. Below we make a few observations that illustrate this.

As we discussed in subsection \ref{sec-curv}, the symmetric spaces are
naturally organized in triplets of negative, zero and positive
curvature.  The simplest way to look at the zero curvature symmetric
spaces is to see them as limiting cases of one of the other two
symmetric spaces (with nonzero curvature) in the same triplet. This
limiting procedure is related to group contraction.  As the simplest
example of this, we can take the group $SO(3)$.  As we saw in
subsections \ref{sec-realforms1} and \ref{sec-curv}, the algebras
${\bf SO(3,R)}$, ${\bf SO(2,1;R)}$ and ${\bf E_2}$ are all related to
each other.  The coset $SO(3)/SO(2)$ can be identified with the unit
2--sphere $S^2$, a positive curvature space (cf. subsection
\ref{sec-cosets}).  By performing the Weyl unitary trick on the
algebra ${\bf SO(3)}$ we obtain the non--compact algebra ${\bf
SO(2,1;R)}$.  As we showed in subsection \ref{sec-realforms2}, the
coset space $SO(2,1;R)/SO(2)$ corresponds to the hyperboloid $H^2$, a
negative curvature symmetric space.  The commutation relations for the
non--semisimple euclidean group $E_2$ are a limiting case of
the commutation relations for ${\bf SO(3)}$ and ${\bf SO(2,1)}$ in the
limit of infinite radius (cf. subsection \ref{sec-curv}).  The
symmetric space $E_2/SO(2)$ is thus a zero curvature symmetric space,
isomorphic to the euclidean plane and a limiting case of the
corresponding positive and negative curvature spaces in the same
triplet.

A simple and important example of a zero curvature symmetric space,
which has several analogies with the case at hand, is the flat
euclidean space in four dimensions.  It is well known that this space
can be realized as the coset of the euclidean Poincare' group
$\tilde{P}$ with respect to $SO(4)$: ${\bf P}\sim
\tilde{P}/SO(4)$. The translations of the Poincare' group play the
role of ${\bf P}$, they are isomorphic to euclidean space and have all
the characteristics of a symmetric space of vanishing curvature.  The
above observation that the zero curvature spaces can be obtained as
limits of positive curvature spaces can be exemplified as follows.  We
can realize the euclidean Poincare' group as a suitable limit of the
$SO(5)$ group.  In this limit the coset $SO(5)/SO(4)$ (which is
nothing else than the four dimensional unit sphere) exactly becomes
the euclidean four--dimensional space.
  
We can rephrase the above result in still another (more geometric)
way.  The zero curvature symmetric spaces can be seen as tangent
spaces of their curved partners. This is clear both from the above
example and from the algebraic structure of the space.

\subsubsection{Chiral ensembles}
\label{sec-chiral}

A natural generalization of the previous results is the extension to
{\it rectangular} random matrices. The study of this class of
ensembles was initiated in the mid--1980's by Cicuta et al.
\cite{Cicuta:1987} and further discussed in reference \cite{ns93}
where they were named Laguerre ensembles (due to the fact that the
orthogonal polynomials associated to them are Laguerre polynomials).
An updated review on this subject can be found for instance in~\cite{revVerb}.

{\bf [1] Physical applications of the chiral ensembles}

One of the most relevant applications of the chiral ensembles is in the
study of the infrared limit of gauge theories. In particular, the
universal properties of Dirac spectra that depend only on the global
symmetry of the euclidean Dirac operator $i\!\!\not{\!\!\!D}$, can be
reproduced by substituting the integral over gauge fields in the
euclidean partition function by an integral over an appropriate random
matrix ensemble. The work by J.~Verbaarschot et al.  \cite{VerI}
related to this subject dates back to the 1990's. It is in this
framework (for reasons which will soon be clear) that these ensembles
came to be named ``chiral''.  In this paragraph we discuss some of
their characteristics with emphasis on their connection with the
theory of symmetric spaces.

Chiral ensembles are also relevant in connection with random flux and
random hopping problems. In this subsection we will discuss two
papers by Mudry, Brouwer, Simons and Altland \cite{M1,M4} where chiral
ensembles are realized. 

{\bf Chiral random matrix theory and QCD.}  The interest in the
low--lying spectrum of the Dirac operator is due to the Banks-Casher
formula \cite{BC},

\begin{equation}
\langle \bar{q}q \rangle = \frac{\pi \rho(0)}{V}
\end{equation}

relating the magnitude of the quark condensate in QCD in a finite volume
$V$ to the spectral density of the Dirac operator at the origin. The
part of the Dirac spectrum near the origin is therefore of interest in
studying the mechanism of spontaneous breaking of chiral symmetry,
since the quark condensate is an order parameter for the chiral transition.
Since we are interested in the connection with symmetric spaces in
this review, we will not further report the results of these important
applications. We only give a sketchy description below and refer to
the literature cited for more detailed information.

Due to the presence of chiral symmetry ($\{\gamma_5,i\!\not{\!\! D}\}=0$)
in four space--time dimensions, the Dirac operator has the block structure 

\beq
\label{eq:block}
i\!\not{\!\! D}=\left( \begin{array}{cc} 0 & W \\ W^\dagger & 0\end{array}
\right)
\eeq

where $W$ is rectangular, say $p\times q$ ($p>q$), thus $\nu \equiv
p-q$ corrsponds to the number of fermionic zero modes. In QCD, these
may describe the zero modes in the field of an instanton. Then $\nu
$ equals the winding number (topological charge). Zero modes may also
originate in the bulk of the spectrum due to repulsion between
eigenvalues. The total size $N=p+q$ of the matrix in (\ref{eq:block})
corresponds to the finite space--time volume $V$ in the gauge
theory. In the calculations to be sketched below, the rectangular
submatrices $W$ in the euclidean Dirac operator $i\!\not{\!\!D}$ will
be replaced by random matrices $T$ chosen from an appropriate
ensemble.  The symmetries of the {\it chiral ensembles} of random
matrices (i.e. the ones having a block--structure like the one in
eq.~(\ref{eq:block})) are chosen on the basis of the fermion
representation (fundamental, adjoint) and the number of quark
colors. The various possibilities lead to the presence (or absence) of
an anti--unitary symmetry

\beq
\label{eq:anti}
[i\! \not{\! \! D},Q]=0
\eeq

where $Q$ is an anti--unitary operator. The presence or absence of the
symmetry (\ref{eq:anti}) and the explicit form of $Q$ determines
whether the Dirac operator can be represented as a matrix with real
($\beta =1$), complex ($\beta =2$) or quaternion real ($\beta =4$)
elements \cite{VerZ4}. The corresponding random matrices are in Cartan
class BDI, AIII, and CII.  In the gauge theory partition function in
euclidean space

\begin{equation}
Z=\int DA \, {\rm e}^{-S[A]} \prod_f {\rm det}(\not{\!\! D} + m_f) 
\end{equation}

(where the integral is over the gauge field configurations, $S[A]$ is
a gauge field action, the product of fermion determinants is over the
flavor degree of freedom and $m_f$ is the fermion mass for flavor $f$)
the integral over gauge fields is then substituted with a gaussian
average over a random matrix $T$. Since the Dirac operator has the
block--structure in equation (\ref{eq:block}), we can substitute $W$
with $T$ to get a much simpler model

\begin{equation}
\label{eq:Z}
Z=\int DT \, {\rm e}^{-\frac{q\beta \Sigma^2}{2} {\rm tr}T^\dagger T} \prod_f
\det \left(\begin{array}{cc}m_f & iT\\ iT^\dagger & m_f \end{array}\right) 
\end{equation}

with the same symmetries as the original partition function ($DT$ is a
Haar measure and the size of $T$ is $p\times q$). From this random matrix
theory we obtain the eigenvalue density and correlators, in particular
their universal microscopic limit.

Both the random matrix theory and the field theory map onto the same
low-energy effective partition function in the mesoscopic regime describing
static Goldstone modes in a finite volume. This partition function is
expressed as an integral over a coset manifold $G'/K'$ related to
spontaneous symmetry breaking and the emergence of Goldstone modes.
Here $G'$ is the original symmetry group of the Lagrangian and $K'$ is
the unbroken subgroup. It is worth noting that these Goldstone
manifolds obtained in \cite{VerZ4,UM,VerZ3} for $\beta =1$, $2$, and
$4$ for QCD in both three (non--chiral ensembles) and four (chiral
ensembles) space--time dimensions are saddle--point manifolds that
appear in the large--$N$ limit and should not be confused with the
symmetric spaces $G/K$ identified with the original ensembles.  These
Goldstone manifolds are the fermionic parts $M_F$ of the symmetric
supermanifolds $M_B\times M_F$ discussed by Zirnbauer in \cite{Zirn}.

In addition to the finite volume partition function of static
Goldstone modes expressing the quark mass dependence, one can obtain
the flavor symmetry (or parity in odd space--time dimensions) breaking
pattern. One can also derive sum rules constraining the eigenvalues of
the Dirac operator in a finite volume \cite{VerZ4,UM,VerZ3}.

{\bf Chiral ensembles in the context of random flux and random hopping
problems.}  It is of interest in the theory of quantum transport to
study how the conductance of a system behaves analytically as one or
more of the dimensions of the system go to infinity. This approach is
called the scaling theory of localization, and deals with the
conditions under which a system is metallic, insulating, or undergoes
a transition between the two states. Such a transition can be
disorder--induced, i.e. it can depend on impurities in the
metal. Another important factor is dimensionality.  We postpone a
further discussion of transport problems to the subsection on transfer
matrix ensembles.  In this paragraph we will only mention some results
relating to chiral ensembles.

In the random magnetic flux problem one studies the localization
properties of a spinless electron moving in a plane perpendicular to a
static magnetic field of random amplitude and vanishing mean. In two
dimensions, it is difficult to reach a conclusion about the
localization properties of the states based on numerical data.
Therefore it is easier to study a quasi--one dimensional
configuration.  This was done by Mudry et al. in reference~\cite{M1},
where an $M\times N$ lattice ($M>>N>>1$) corresponding to a thick
quantum wire with weak disorder was studied. Although the study of
such wires is a vast field, we will limit ourselves here to reporting
only one of the results of this interesting paper, so as not to go
beyond our scope, which is to give examples of physical manifestations
of chiral random matrix symmetry classes.

In the case at hand, the system is governed by a Hamiltonian ${\cal
H}$ (whose detailed form the interested reader can look up in the
original paper) through a Schr\"odinger equation ${\cal
H}\psi=\epsilon \psi$.  Away from the band center $\epsilon =0$, the
localization properties of the particle are those of the standard
gaussian unitary ensemble. In other words, the Hamiltonian ${\cal H}$
of the system is in Cartan class A.  Exactly at the band center
$\epsilon =0$, an additional symmetry of the transfer matrix, called
chiral or particle--hole symmetry, changes the symmetry class of the
Hamiltonian into the chiral unitary symmetry class AIII. (Note that
the presence of the magnetic field breaks time--reversal symmetry, so
the chiral orthogonal and the chiral symplectic class can not be
realized in this system.)  Thus the random magnetic flux problem
provides a physical realization of the chiral unitary ensemble.  We
will refer to the above results again in connection with transfer
matrices, when we enumerate some of the physical manifestations of
symmetric spaces of negative curvature.

Related to the random flux problem is the random hopping problem, in
which a particle hops on a lattice with random hopping
amplitudes. Also in this problem one finds that the point at the
center of the band is special. A delocalization transiton in such a
system in one dimension goes back to work by Dyson (see the discussion
and references in \cite{M4}).  A quasi--one dimensional hopping model
with weak staggering was investigated in \cite{M4}. The lattice
consisted of $N$ coupled chains of length $L$, where $L>>N$.  In the
vicinity of the Fermi energy the lattice model was approximated by a
continuum model.  For $\beta =1,2,4$ respectively, the Hamiltonian has
the symmetries of a chiral ensemble and belongs to Cartan classes BDI,
AIII, and CII, respectively.  For our analysis the important point
here is that the Hamiltonian in each case is a realization of a chiral
symmetry class, just like the Dirac operator in four--dimensional QCD
and the Hamiltonian at the band center in the random flux problem. For
details and other results of the quasi--1d random hopping problem we
refer to the original paper.

{\bf [2] Some properties of the chiral ensembles}

{\bf Invariance and diagonalization.}  Just like in the circular and
gaussian ensembles, the rectangular random matrix $T$ above can be
diagonalized through a transformation with unitary matrices $U$ and
$V$:

\beq
T=U\Lambda V^{-1}
\eeq

The matrix $\Lambda $ is a diagonal real matrix with positive elements
corresponding to the non--zero eigenvalues of $T$.  The partition function 
in (\ref{eq:Z}) is invariant under this transformation. 

{\bf Probability distribution of the eigenvalues.}  It can be shown
that the Jacobian for the transformation from matrix to eigenvalue
space for the chiral ensembles is given by

\beq
\label{eq:chiralJ}
J_{\beta}^{\nu}(\{\lambda_i\})\propto \prod_{i<j} |\lambda_i^2-\lambda_j^2|^\beta \prod_{k}
\lambda_k^{\beta (\nu +1)-1}
\eeq

where $\{\lambda_i\}$ are the eigenvalues of the rectangular matrix
$T$.  In terms of the new variables $x_i\equiv \lambda_i^2$
(\ref{eq:chiralJ}) takes the well--known form

\beq
J_{\beta}^{\nu}(\{x_i\})\propto \prod_{i<j} |x_i-x_j|^\beta \prod_k x_i^
{\frac{\beta (\nu +1)-2}{2}} 
\eeq   

Here $\nu $ is the number of zero modes of the chiral matrices.

{\bf [3] The symmetric spaces associated to the chiral ensembles}

The chiral block--structure of the random matrices already gives a
hint that they must belong to the subspace ${\bf P}$ of some algebra
(see subsection \ref{sec-algstr}). Depending on the number of degrees
of freedom of the matrix elements (i.e. the value of $\beta $: 1, 2,
or 4), it is not hard to guess that this subspace is identified
respectively with ${\bf SO(p,q)/(SO(p)\otimes SO(q))}$, ${\bf
SU(p,q)/(SU(p)\otimes SU(q))}$, or ${\bf USp(p,q)/(USp(p)\otimes
USp(q))}$ (in this case ${\bf p}$, ${\bf q}$ have to be
even\footnote{Our notation is such that $USp(2p)\equiv USp(2p,C)\equiv
U(p,Q)$.}). Obviously, these algebra subspaces are symmetric spaces of
the euclidean (zero--curvature) type. They correspond to Cartan
classes BDI, AIII and CII in Table~\ref{tab1}. Above we have
enumerated several physical realizations of these symmetry classes.

\subsubsection{Transfer matrix ensembles}
\label{sec-transfer}

The transfer matrix ensembles appear in the theory of quantum
transport, in the random matrix theory description of so called
quantum wires.  In these pages we will only discuss the part of the
theory which is relevant for our purpose, the study of the mapping
between random matrix theory and symmetric spaces. For further
information on the experimental and theoretical issues we refer the
reader to the excellent introductory review on random matrix theory
and quantum transport by Beenakker \cite{Beenakker}.

{\bf [1] Physical context of the use of transfer matrix ensembles}

The natural theoretical framework for describing mesoscopic systems is
the Landauer theory~\cite{lan}. Within this approach Fisher and Lee
proposed the following expression for the conductance in a
two--probe geometry (a finite disordered section of wire to which
current is supplied by two semi-infinite ordered leads):

\eq
\label{eq:1a}
G=G_0~{\rm Tr}(tt^\dagger)\equiv G_0\sum_n T_n,~~~~~~~
G_0=\frac{2e^2}{h}
\en

where $t$ is the $N\times N$ transmission matrix of the conductor (see
eq.~(\ref{eq:Smatrix})), $N$ is the number of scattering channels at
the Fermi level and $T_1,T_2 \cdots T_N$ are the eigenvalues of the
matrix $tt^\dagger$. The $T_i$'s are usually referred to as transmission
eigenvalues.  The constants $e$ and $h$ denote the electronic charge
and Planck's constant, respectively. 

The transmission eigenvalues $T_i$ are related to the scattering
matrix and to the transfer matrix, which define equivalent
descriptions of the impurity scattering process. In the next paragraph 
we will define the relationship between the physical degrees of freedom
$T_i$ and the transfer matrix.

{\bf [2] Definition of the transfer matrix} 

The scattering matrix of a mesoscopic conductor was defined in
eq.~(\ref{eq:scat}) in the context of circular ensembles.  Like in
subsection \ref{sec-circ}, if we have $N$ propagating modes at the
Fermi level, we can describe them by a vector of length $2N$ of
incoming modes $I$, $I'$ and a similar vector of outgoing modes $O$,
$O'$ in each lead. Let again the unprimed letters denote the modes in
the left lead and the primed letters the modes in the right
lead. While the scattering matrix $S$ relates the incoming wave
amplitudes $I$, $I'$ to the outgoing wave amplitudes $O$, $O'$ (see
eq.~(\ref{eq:scat})), the transfer matrix $M$ relates the wave
amplitudes in the left lead to those in the right lead:

\beq
M\left( \begin{array}{c} I \\ O \end{array} \right) =
\left( \begin{array}{c} O' \\ I' \end{array} \right)
\eeq

Following the notation used in subsection \ref{sec-circ} and in
\cite{BILP,MPK}, the scattering matrix has the block structure

\beq
\label{eq:Smatrix'}
S=\left( \begin{array}{cc} r & t' \\
                          t & r' \end{array} \right)
\eeq

where $r$, $r'$, $t$, $t'$ are $N\times N$ reflection and transmission
matrices. As we saw in subsection \ref{sec-circ}, the unitarity of $S$
implies that the four matrices $tt^{\dagger }$, $t't'^{\dagger }$,
$1-rr^{\dagger }$, $1-r'r'^{\dagger }$ have the same set of
eigenvalues $T_1,...,T_N$ called transmission eigenvalues ($0 \leq T_i
\leq 1$).  The eigenvalues $\lambda_i$ of the matrix

\beq
\label{eq:Qlam}
Q=\frac{1}{4}(M^{\dagger}M +(M^{\dagger}M)^{-1}-2)
\eeq

are related to the transmission eigenvalues by

\beq
\label{eq:Tlambda}
\lambda_i=\frac{1-T_i}{T_i}
\eeq
 
The $\lambda_i$ are non--negative. In terms of these, $M$ can be parametrized 
as \cite{MPK}

\beq
\label{eq:Mpara}
M=\left( \begin{array}{cc} u & 0 \\
                           0 & u' \end{array} \right) 
\left( \begin{array}{cc} \sqrt{1+\Lambda} & \sqrt{\Lambda} \\
                         \sqrt{\Lambda} & \sqrt{1+\Lambda}\end{array} \right) 
\left( \begin{array}{cc} v & 0 \\
                         0 & v' \end{array} \right) \equiv U\Gamma V
\eeq

where $u$, $u'$, $v$, $v'$ are unitary $N \times N$ matrices (related
by complex conjugation: $u'=u^*$, $v'=v^*$ if $M\in Sp(2N,R)$ or $M\in
SO^*(4N)$, see below) and $\Lambda = {\rm
  diag}(\lambda_1,...,\lambda_N)$.  In case spin--rotation symmetry is
broken, the number of degrees of freedom in (\ref{eq:Mpara}) is
doubled and the matrix elements are real quaternions.

{\bf [3] The symmetric spaces associated to the transfer matrix ensembles}

Transfer matrices are strongly constrained by various physical
requirements. As a result they belong to one of three different
symmetric spaces. We discuss in some detail below how
these symmetric spaces are obtained.

The physical requirements of flux conservation, presence or absence of
time--reversal symmetry, and presence or absence of spin--rotation
symmetry lead to conditions on the transfer matrix. These conditions
determine the group $G$ to which $M$ belongs.  Flux conservation leads
to the following condition on $M$ \cite{MPK}

\beq
\label{eq:fluxcons}
M^{\dagger }\Sigma_z M =\Sigma_z, \ \ \ \ \ \ \Sigma_z=\left( \begin{array}{cc}
1 & 0 \\ 0 & -1 \end{array} \right)
\eeq

i.e. $M$ preserves the ($2N\times 2N$) metric $\Sigma_z$. This means
that $M$ belongs to the pseudo--unitary group $SU(N,N)$ ($M$ has to be
continuously connected to the unit matrix so we take the connected
component of $U(N,N)$). If the Hamiltonian is invariant under
time-reversal, the condition on the transfer matrix is \cite{MPK}

\beq
\label{eq:timerev}
M^*= \Sigma_x M \Sigma_x, \ \ \ \ \ \ \Sigma_x=\left( \begin{array}{cc}
0 & 1 \\ 1 & 0 \end{array} \right)
\eeq

It is easy to check that together with the condition of flux conservation, 
this implies

\beq
M^TJM=J, \ \ \ \ \ \ J=\Sigma_x \Sigma_z = \left( \begin{array}{cc}
0 & -1 \\ 1 & 0 \end{array} \right) 
\eeq

But $J$ is the skew--symmetric form invariant under the non--compact
symplectic group $Sp(2N,R)$ (see e.g. \cite{Huff,SattW,Hamermesh}), thus
$M$ belongs to this group in case time--reversal symmetry is present.

If the Hamiltonian contains a spin--orbit interaction (i.e. in case of
magnetic impurities in the conductor), spin--rotation symmetry is
broken. In this case the presence of an extra spin degree of freedom
doubles the number of components in the vectors $I$, $I'$, $O$, $O'$
of incoming/outgoing wave amplitudes (the components become spinors).
Formally, the same conditions (\ref{eq:fluxcons}), (\ref{eq:timerev})
on $M$ are valid for flux conservation and time--reversal symmetry,
but $M$ is now a matrix of $N \times N$ real quaternion elements
\cite{ChMac}.  If time--reversal symmetry is broken, we get $M \in
SU(2N,2N)$ like before.  If it is conserved, a condition analogous to
(\ref{eq:timerev}) is valid, with the only difference that the
matrices now act on an $N$--dimensional vector space of real
quaternions.  In this case $M$ belongs to the group $SO^*(4N)$. This
is the connected component of the group of linear transformations that
preserve a skew--symmetric bilinear form on a quaternionic vector
space (see. e.g. \cite{FulHar}, paragraph 7.2; cf. also H\"uffmann 
\cite{Huff}). 

{\bf Invariance of the transfer matrix ensembles.}  As we can check
using the parametrization eq.~(\ref{eq:Mpara}), rotating $M$ by a
matrix $W\in U(N)$, $SU(N) \times SU(N) \times U(1)$, or $U(2N)$ (if
$M\in Sp(2N,R)$, $SU(N,N)$, or $SO^*(4N)$, respectively), gives a new
transfer matrix $M'=WMW^{-1}=U'\Gamma V'$ with the same physical
degrees of freedom $\{\lambda_1,...,\lambda_N\}$, since the matrix
$\Gamma $ is unchanged.  This means that in each case $\Gamma $
belongs to a coset space $G/K$, where $M\in G$ and $W\in K$. These
three ensembles of transfer matrices, corresponding to different
physical symmetries, are usually named (in analogy with the
nomenclature of the standarad gaussian ensembles) the ``orthogonal''
($Sp(2N,R)/U(N)$), ``unitary'' ($SU(N,N)/SU(N) \times SU(N) \times
U(1)$) and ``symplectic'' ($SO^*(4N)/U(2N)$) ensemble. This
nomenclature is a bit unfortunate, since the stability subgroups do
not correspond to it. These coset spaces are irreducible and
non--compact. Thus they are symmetric spaces of negative curvature, as
is also evident from Table~\ref{tab1}.  They correspond to Cartan
classes CI, AIII, and CII, respectively.

{\bf New transfer matrix ensembles in the context of random
tight--binding models.} In connection with the chiral ensembles we
discussed the Hamiltonian at the band center in a random magnetic flux
problem. It is of the chiral symmetry class AIII \cite{M1}. This means
that the transfer matrix for the same physical system belongs to
Cartan class A, i.e. it is a physical realization of the coset space
$SL(N,C)/SU(N)$.  We also discussed the random hopping problem
\cite{M4} in the same context and saw that the Hamiltonian for $\beta
=1,2,4$ has chiral symmetry. This means the transfer matrices of the
systems belong to coset spaces $SL(N,R)/SO(N)$ (Cartan class AI),
$SL(N,C)/SU(N)$ (Cartan class A), and $SU^*(2N)/USp(2N)$ (Cartan class
AII). These are new realizations of transfer matrix ensembles not
found in the applications discussed above. In the same type of
problems, also other transfer matrix ensembles are manifest. We refer
to \cite{TitBrou,M1,M4,M,M5,M2} for more details. Of course, to each
hamiltonian ensemble corresponds a transfer matrix ensemble and a
scattering ensemble, so in a certain sense, all the boxes in
Table~\ref{tab2} should be filled with "physical" ensembles.  As we
have already noted, the correspondence between hamiltonian ensembles
and the transfer matrix ensembles for the same systems was given in
Table~I of reference \cite{TitBrou}.
  
By inspection of the parametrization of eq.~(\ref{eq:Mpara}), we
realize that the (generalized) eigenvalues $\lambda_i$ are not exactly
the radial coordinates in the three symmetric spaces.  By a change of
coordinates (see eq.~(\ref{defofx}) below), we will be able to
``disentangle'' the coordinates in such a way that this
parametrization, and the DMPK equation to be discussed in the next
paragraph, become much more tractable.

\subsubsection{The DMPK equation}
\label{sec-DMPK}

What is needed at this point to complete the transfer matrix
description of the quantum wire, is the explicit expression for the
probability distribution of the $\{\lambda_i\}$ as a function of an
external parameter, the length $L$ of the the quantum wire.  The
standard approach to this rather non--trivial problem has been to find
some dynamical principle so as to obtain an ``equation of motion'' for
the probability distribution of the $\{ \lambda_i \}$ as a function of
$L$, and then (hopefully) to solve the equation and obtain the probability
distribution.

The construction of the equation for the probability distribution was
completed during the eighties, at least in the case of
quasi--one--dimensional wires ($L>>W$ where $W$ is the thickness of
the wire), by Dorokhov~\cite{dor}, and independently by Mello,
Pereyra, and Kumar~\cite{MPK} (for $\beta=1$) by studying the
infinitesimal transfer matrix describing the addition of a thin slice
to the wire. The resulting evolution equation for the eigenvalue
distribution $P(\{\lambda_i\},s)$, where $s$ is the dimensionless
length of the wire, is usually known as the
Dorokhov--Mello--Pereyra--Kumar (DMPK) equation.  The only assumptions
which are needed to obtain this equation are: 1) that the conductor is
weakly disordered, so that the scattering in the thin slice can be
treated by using perturbation theory, and 2) that the flux incident in
one scattering channel is, on average, equally distributed among all
outgoing channels. It is exactly this second assumption which
restricts the DMPK equation to the quasi--1D regime, where the finite
time scale for transverse diffusion can be neglected. The results
of~\cite{MPK} were subsequently generalized to $\beta =2,4$ in
refs.~\cite{ms,ChMac}. The DMPK equation can be written as

\begin{equation}
\frac{\partial P}{\partial s}=DP
\label{DMPK}
\end{equation}

where $s$ is the wire length measured in units of the mean free
path $l$: $s\equiv L/l$, and $D$ can be written in terms of the
$\{\lambda_i\}$ as follows:

\begin{equation}
{}~D~=
\frac{2}{\gamma}\sum_{i=1}^{N}
\frac{\partial}{\partial\lambda_{i}}\lambda_{i}(1+\lambda_{i})
J(\lambda)\frac{\partial}{\partial\lambda_{i}}J(\lambda)^{-1},
\label{DMPK2}
\end{equation}

with $\gamma \equiv \beta N+2-\beta$.  $\beta $ is the symmetry index
of the ensemble of scattering matrices, in analogy with the
well--known Wigner--Dyson classification, and $J(\lambda )\equiv
J(\{\lambda_{n}\})$ is given by

\begin{equation}
J(\lambda )=\prod_{i<j}|\lambda_{j}-\lambda_{i}|^{\beta}
\label{jacobian}
\end{equation}

The solution of this equation will be discussed in the forthcoming
subsections.  For the moment, let us only point out that
$J(\lambda )$, which is the Jacobian of the transformation from
random matrices to eigenvalues and eigenangles in the gaussian
ensembles, {\it has nothing to do} with the Jacobian between the space
of transfer matrices and the space of radial coordinates in the
present case. Its appearance in the DMPK equation has historical
reasons, as the authors tried to mimic the well--known gaussian
ensembles.  This choice of coordinates makes the DMPK equation
asymmetric and ultimately hard to solve.

{\bf The relationship between the Jacobian and the radial part of the
Laplace--Beltrami operator.} The DMPK equation can be rewritten in a
more tractable form by making the change of coordinates

\eq
\label{defofx}
\lambda_i=\sinh^2x_i
\en

This introduces the ``right'' Jacobian for the transformation to
eigenvalue space in the transfer matrix ensemble. This Jacobian turns
out to be the function $\xi^2(\{ x_i\} )\equiv \xi^2(x)$ given by

\begin{equation}
\label{eq:Jxi}
J(\{x_i\})=\xi^2(x)=\prod_{i<j}|\sinh^{2}x_{j}-\sinh^{2}x_{i}|^\beta
\prod_{k}|\sinh 2x_{k}|
\label{BR2}
\end{equation}

Let us introduce a new operator $B$ defined as

\begin{equation}
B=\sum_{k=1}^{N}\frac{\partial}{\partial x_k}\xi^2(x)
\frac{\partial}{\partial x_k}\xi^{-2}(x)~~,
\label{defb}
\end{equation}
 
It is easy to see, by direct substitution, that  $B$ is 
related to the DMPK operator by:

\begin{equation}
D=\frac{1}{2\gamma}\, B
\label{DtoB}
\end{equation}

Thus we can rewrite the DMPK equation as

\begin{equation}
\frac{\partial P}{\partial s}=
\frac{1}{2\gamma}\, BP
\label{DMPKbis}
\end{equation}

By comparing eq.~(\ref{defb}) with eq.~(\ref{eq:DeltaB'}) of section
\ref{sec-Laplaceop} we see that the operator $B$ is related to the
radial part of the Laplace--Beltrami operator $\Delta_B'$ on the
negative curvature symmetric spaces associated to the transfer matrix
ensembles by $\Delta_B'=J^{-1}BJ$.  This is the first indication of a
general, important relation which we will discuss in more detail in
section \ref{sec-Coulomb} below.

Similar Fokker--Planck equations can be derived for all the symmetry
classes. In \cite{M4} the equation was derived in
conjunction with a tight--binding model for the cases in which the
transfer matrix is in the standard ensembles and the Hamiltonian in
the chiral ensembles. In \cite{M5} it was derived for systems with
Hamiltonians of the Bogoliubov--de Gennes (BdG) type (see the next
subsection). The latter apply to quasiparticle transport at the Fermi
level in a disordered superconducting wire. In this case charge is not
conserved, but one can study transport of heat and spin. 
 
In standard systems, localization occurs when the wave function of an
electron undergoes a transition from a Bloch wave extending throughout
the sample to a localized exponentially decaying form $\psi(r) \sim
\e^{-r/\xi }$, where $\xi$ is the localization length.  As a result,
in the localized regime the conductivity is dominated by the lowest
eigenvalue and decreases exponentially with the length of the wire. In
the context of disordered wires, the seven universality classes
pertaining to chiral and BdG Hamiltonians are more interesting than
the standard ones in that they can display a departure from
exponential localization \cite{M5}.

\subsubsection{BdG and $p$--wave ensembles}
\label{sec-BdG-p}

{\bf [1] Physical applications of the BdG ensembles}

The so called NS ensembles are examples of Bogoliubov--de Gennes, or
BdG, ensembles.  They get their name from the physical structures they
describe \cite{AltZ}.  The symmetry classes of these new ensembles are
realized in normal metal--superconductor (NS) heterostructures. These
are mesoscopic systems composed by a normal conductor in conjunction with a
superconductor.  Because of Andreev reflection at the NS interface,
this system is different from the conventional one.  Just like the
additional chiral symmetry of the Dirac operator and the symmetries of
the transfer matrix of a normal mesoscopic conductor give rise to new
ensembles, so the symmetries of the so called Bogoliubov--de Gennes
(BdG) Hamiltonian give rise to four new symmetry classes depending on
whether time--reversal (TR) and/or spin--rotation (SR) symmetry is
present.  The BdG Hamiltonian is a first--quantized version of the BCS
Hamiltonian in a mean--field approximation. It has an additional
particle--hole grading which is absent in the Hamiltonian for a normal
metal. This gives rise to a discrete particle--hole symmetry.

{\bf [2] The symmetric spaces associated to the BdG ensembles}

In \cite{AltZ} the authors show that the BdG Hamiltonian ${\cal H}$
(actually $i{\cal H}$) belongs to one of four symmetry classes
depending on which symmetries are present. Since the presentation in
\cite{AltZ} is excellent, we will just summarize the results here. The
space to which $i{\cal H}$ belongs is an algebra or an algebra
subspace and is either ${\bf SO(4N)}$ (no TR and no SR), ${\bf
USp(2N)}$ (only SR), ${\bf SO(4N)/U(2N)}$ (only TR), or ${\bf
USp(2N)/U(N)}$ (both TR and SR).  All these algebra subspaces can be
considered to be symmetric spaces of zero curvature, by the
construction we have repeatedly discussed.  By comparing with
Table~\ref{tab1} we see that they can be identified with Cartan
classes D, C, DIII and CI respectively.  The Jacobian for the
transformation by the stability subgroups to radial coordinates is
\cite{AltZ}

\beq
\label{eq:BdGJ}
J_{r,s}(\{q_i\})\propto \prod_{i<j}|q_i^2-q_j^2|^r
\prod_k |q_k|^s
\eeq

where the pair $(r,s)$ takes the values $(2,0)$, $(2,2)$, $(4,1)$, and
$(1,1)$ for ${\bf SO(4N)}$, ${\bf USp(2N)}$, ${\bf SO(4N)/U(2N)}$, and
${\bf USp(2N)/U(N)}$, respectively.  For a review of NS junctions and
the BdG equation, see \cite{Beenakker}.

{\bf Scattering matrices of the BdG type.}  The scattering matrix for
an NS--type heterostructure is obtained by exponentiation of the
Hamiltonian: $S={\rm e}^{i{\cal H}}$. Since $i{\cal H}$ is in the
algebras or tangent spaces listed above for the respective symmetry
classes, the scattering matrix is in the corresponding symmetric
spaces of positive curvature. Further information on these ensembles
can be found in \cite{Beenakker}.

{\bf [3] The symmetric spaces associated to the $p$--wave ensembles}

Recently, Ivanov \cite{Ivanov} found realizations of Cartan classes B
and DIII--odd in the algebra ${\bf SO(2N+1)}$ and the algebra subspace
${\bf SO(4N+2)/U(2N+1)}$ in $p$--wave superconductors.  The
corresponding ensembles are called $p$--wave ensembles and are
characterized by the presence of a zero mode. We recall that
$\nu=|p-q|$ zero modes are also present for the chiral ensembles for
$p\neq q$ (for notation see Table~\ref{tab1}).  These two classes were
obtained for disordered vortices in $p$--wave superconductors with or
without time--reversal symmetry, using the same Bogoliubov--de Gennes
Hamiltonian as in \cite{AltZ}.  As in the case of BdG ensembles, the
identification of the $p$--wave ensembles with algebra (sub)spaces
maps them onto flat symmetric spaces.  For more comments on these
ensembles see \cite{Ivanov}.


{\bf [4] Vicious walks and BdG ensembles}

Recently a new and completely independent realization of the gaussian
BdG ensembles has been proposed in the context of the two--matrix
model description of the vicious walk in presence of a wall
(see~\cite{ktnk03} and references therein).  In this case the peculiar
symmetries of the BdG ensemble are a direct consequence of the
boundary conditions imposed by the presence of the wall in the model.


\subsubsection{S--matrix ensembles} \label{sec-S-mat}

{\bf [1] Physical applications of the S--matrix ensembles}

The important class of matrix models referred to as S--matrix ensembles
was introduced a few years ago in~\cite{JalaPB+BarMello,JalP} to
describe the behavior of ballistic chaotic quantum dots. These are
microstructures in which impurity scattering can be neglected, and
only boundary scattering is considered. 

{\bf [2] Definiton of the S--matrix ensembles}

One assumes that the scattering matrix, given in
eq.~(\ref{eq:Smatrix'}), belongs to the circular ensemble.  Like in
subsection \ref{sec-transfer}, the transmission eigenvalues $T_n$ are
the eigenvalues of the submatrix combination $tt^\dagger$, and these
are not simply related to the scattering phase shifts. (Recall that in
the circular ensembles, the probability distribution
(\ref{eq:circularP}) in the original formulation by Dyson is expressed
in terms of scattering phase shifts.) By expressing $P_\beta $ in
terms of the variables $\{\lambda_i\}$ of eq.~(\ref{eq:Tlambda}),

\beq
\lambda_i=\frac{1-T_i}{T_i}
\eeq

one obtains a representation suitable for studying the transport
properties through a quantum dot.

The resulting distribution $P_\beta(\{\lambda_i\})$ takes the form of
a Gibbs disribution. A direct calculation \cite{JalaPB+BarMello} shows that

\bea
\label{jpb2}
P_\beta (\{\lambda_i\})&\sim &\e^{-\beta \left( \sum_i V_\beta 
(\lambda_i) -\sum_{i<j} \ln |\lambda_i-\lambda_j|\right)} \nonumber \\
V_\beta (\lambda_i)&=&\left(N+\frac{2-\beta }{2\beta }\right)\ln (1+\lambda_i)
\\ \nonumber
\eea

{\bf [3] The symmetric space associated to the S--matrix ensemble for $\beta = 2$} 

It appears that the S--matrix ensemble for $\beta = 2$ corresponds to
the compact symmetric space $SU(2N)/SU(N)\times SU(N)\times U(1)$ (for
$p=q$ in $SU(p+q)/SU(p)\times SU(q)\times U(1)$, which corresponds to
the multiplicity of the short roots $m_s=2(p-q)=0$, i.e. a root
lattice of type $BC_n$ that reduces to $C_n$ for $p=q$).
Interestingly, as we will see in subsection \ref{sec-jac}, the two
other ensembles (for $\beta = 1,\ 4$) do not find correspondence in
the Cartan classification \cite{MCclass}.  This issue will be
discussed further in section \ref{sec-beyond}.

\subsection{Identification of the random matrix eigenvalues and universality indices}
\label{sec-RMT-SS}

Just like the integration manifolds of random matrix theories have
been identified with symmetric spaces in the previous subsection, we
will now identify the random matrix eigenvalues with the spherical
radial coordinates, the Jacobian from matrix to eigenvalue space with
the Jacobian of the transformation to spherical coordinates, and the
random matrix symmetry indices with the multiplicities of the
restricted roots of the pertinent symmetric spaces.

{\bf Radial coordinates and the Weyl group.}  In subsection
\ref{sec-radial} of this paper we discussed the important role played
by the reference frame of spherical coordinates in the study of
symmetric spaces.  In particular the {\it radial spherical
coordinates} turned out to be given by the Cartan subalgebra of the
subspace ${\bf P}$. We saw that $P=\e^{\bf P}$ could be naturally
divided into conjugation classes by the action of the stability
subgroup $K$ of the symmetric space. More precisely, we showed that
any element $p\in P$ is conjugated with some element of the Cartan
subalgebra by means of the adjoint representation of $K$: $p=khk^{-1}$
where $k\in K$, $h=\e^H$ and $H$ is in the Cartan subalgebra. The
radial coordinates of $p$ coincide with the set of
eigenvalues of the matrix $H$.

In the case of a gaussian ensemble, from eq.~(\ref{eq4.1}) we see that
the probability density depends only on these radial coordinates. This is not a
coincidence.  It happens in all the random matrix models that we will
study, and can be considered one of their characteristic features.
This does not mean that random matrix models in which also the angular
coordinates play a role are not important. They appear in several
interesting contexts, but they are in general much more difficult to
study and require methods which are outside the scope of this review.

We observe that the Weyl group is the symmetry group of the root
system of the symmetric space and of the radial coordinates at the
same time. The Weyl group can easily be
found. It is simply given by all the possible permutations of the
eigenvalues. The Weyl chamber (defined in section \ref{sec-chambers}) is 
fixed once we choose a particular ordering for the eigenvalues. We
may for example order them according to increasing magnitude.

{\bf Jacobians and the root lattice.}  In the previous subsection we
have discussed a number of different random matrix ensembles. For each
of these ensembles we identified the integration manifold with a
suitable symmetric space of positive, negative or zero curvature.  For
each ensemble we now identify the Jacobian of the transformation to
eigenvalue--eigenvector space with the Jacobian (\ref{eq:J_ja})
(paragraph \ref{sec-zonal}) of the transformation to spherical
coordinates on the symmetric space identified with the integration
manifold. In random matrix theory this Jacobian gives rise to the
familiar geometric interactions between eigenvalues, and determines
the form of the joint probability distribution function of the latter.
As we will see, the detailed form of the Jacobian fits exactly into
the common and elegant framework provided by the classification of
symmetric spaces. The curvature of these spaces, together with the
restricted long, ordinary and short roots and their respective 
multiplicities, exactly determine its form.

{\bf A few important remarks.}  Before discussing the various
ensembles, let us make two observations which will hopefully help in
clarifying the pattern which is emerging.  In general, symmetric
spaces of positive curvature correspond to ensembles of circular type,
those of zero curvature to ensembles of gaussian type
(i.e. hamiltonian ensembles), and those of negative curvature to
transfer matrix ensembles.  Algebraically, the curvature of the spaces
is reflected in the explicit form of the Jacobian of the
transformation to spherical coordinates.  This can be understood by
comparing with eq.~(\ref{eq:J_ja}) which we reproduce here for
comparison:

\beq
\label{eq:J_ja'}
\begin{array}{l}
J^{(0)}(q)=\prod_{\alpha \in R^+} (q^\alpha )^{m_\alpha }\\
\\
J^{(-)}(q)=\prod_{\alpha \in R^+} (a^{-1}{\rm sinh}(aq^\alpha ))^{m_\alpha }\\ 
\\
J^{(+)}(q)=\prod_{\alpha \in R^+} (a^{-1}{\rm sin}(aq^\alpha ))^{m_\alpha }\end{array}
\eeq

Note that the root multiplicities are the same for the spaces of
positive and negative curvature in the triplet corresponding to the
same symmetric subgroup.  Even though we have not defined a restricted
root system for the zero--curvature spaces arising from non--semisimple
groups, the same multiplicities characterize also the Jacobian
pertaining to the zero--curvature member of the triplet.  This should
be understood as explained in the remark following
equation~(\ref{eq:J_j}) in subsection \ref{sec-Laplaceop}.

\subsubsection{Discussion of the Jacobians of various types of matrix 
ensembles}
\label{sec-jac}

Let us now come back to the classification scheme.  The root
multiplicities are listed in Table~\ref{tab1}, to which we
systematically refer in the following. This means that a few integers
and the sign of the curvature ($+,-,0$) are enough to completely
characterize a matrix ensemble.

1) {\bf The gaussian ensembles} 

As we saw in subsection \ref{sec-gauss}, the standard gaussian
ensembles are labelled by the Dyson index $\beta $. It is noteworthy
that this index is exactly equal to the multiplicity $m_o$ of ordinary
roots of the restricted root lattice characterizing a triplet of
symmetric spaces (see comments in the previous subsection concerning
the ``root multiplicities'' characterizing a non--semisimple algebra
${\bf G^0}$). The Jacobian in the gaussian case is given by
eq.~(\ref{eq:Jbeta}):

\beq
\label{eq:gaussJ}
J^{(0)} (\{\lambda_i\})\sim \prod_{i<j} |\lambda_i-\lambda_j|^{m_o} \nonumber
\eeq

This case corresponds to the $A_{n-1}$ root lattices with only
ordinary roots.  The Jacobian in (\ref{eq:gaussJ}) has exactly the
form given in eq.~(\ref{eq:J_ja'}) for zero curvature spaces with a
root lattice of type $A_{n-1}$ ($m_l=m_s=0$).  The positive roots in
this case are $\alpha = e_i-e_j$ ($i>j$) and the expressions entering
in the Jacobian are the radial coordinates $\lambda_\alpha\equiv {\bf
\lambda \cdot \alpha}=\lambda_i-\lambda_j$ (see paragraph \ref{sec-Laplaceop}).
The absolute value in eq.~(\ref{eq:gaussJ}) corresponds to a choice of
Weyl chamber (cf. paragraph \ref{sec-chambers}) 
$\lambda_1 \geq \lambda_2 \geq ... \geq \lambda_N$.

2) {\bf The circular ensembles}

The Jacobian of the circular ensembles of unitary scattering matrices
was given in (\ref{eq:circularP}):

\beq
J_\beta (\{ \phi_i \} )\propto \prod_{i<j} |\e^{i\phi_i}-\e^{i\phi_j}|^{\beta}
\eeq

It is not hard to see that this Jacobian can be rewritten in the form

\beq
J^{(+)}(\{\phi_i\})\sim \prod_{i<j} \left|\, 2\, {\rm sin}\left(
\frac{\phi_i-\phi_j}{2}\right)\right|^{m_0}
\eeq

which is exactly the form given by eq.~(\ref{eq:J_ja'}) for the
Jacobian pertaining to the transformation to spherical coordinates on a
symmetric space of type $A_{n-1}$ with only ordinary roots and of
positive curvature, if we choose the radius $a=1/2$. This allows us to
identify these ensembles with the positive curvature symmetric spaces
of Cartan classes A, AI and AII, in agreement with what we had
already concluded from the corresponding integration manifolds.

3) {\bf The chiral ensembles} 

The Jacobian given for the chiral ensembles in eq.~(\ref{eq:chiralJ}) 

\beq
J_{\beta }^{\nu }(\{\lambda_i\})\propto \prod_{i<j}
|\lambda_i^2-\lambda_j^2|^\beta \prod_{k}|\lambda_k|^{\beta (\nu +1)-1}
\eeq

corresponds in the most general case to a root lattice with all types
of roots.  The chiral ensembles are algebra subspaces, and like for
the gaussian ensembles we associate to it the restricted root lattice
of the curved symmetric spaces in the same triplet. The restricted
root lattices for the chiral ensembles are of type $B_q$ or $D_q$ for
$\beta =1$ and $BC_q$ or $C_q$ for $\beta =2,4$.  In case the root
lattice is not of type $A_{n-1}$, not only is $e_i-e_j$ a positive
root but also $e_i+e_j$.  Thus the positive roots are as follows: for
$B_q$ $\{ e_i, e_i\pm e_j\}$, for $D_q$ $\{ e_i\pm e_j\}$, for $BC_q$
$\{ e_i, e_i\pm e_j, 2e_i\}$, for $C_q$ $\{ e_i\pm e_j,2e_i\}$ ($i\neq
j$ always). Using the root multiplicities $m_o=\beta $, $m_l=\beta
-1$, $m_s=\beta |p-q|\equiv \beta \nu $ we see that the Jacobian is of
type (\ref{eq:J_ja'}).  It can be rewritten

\beq
\label{eq:J0}
J^{(0)}(\{\lambda_i\})\sim \prod_{i<j} |\lambda_i^2-\lambda_j^2|^{\beta }
\prod_k |\lambda_k|^\alpha,\ \ \ \ \ \ \ \beta \equiv m_o,\ \ \alpha \equiv m_s+m_l
\eeq

From the Jacobian it is evident that in addition to the usual
repulsion between different eigenvalues, $\lambda_i$ also repels its
mirror image $-\lambda_i$, and the eigenvalues are no longer
translationally invariant. This kind of ensembles are therefore called
boundary random matrix theories.  The boundary random matrix theories
include chiral and NS ensembles.

4) {\bf The transfer matrix ensembles} 

The ``orthogonal'' and ``symplectic'' transfer matrix ensembles
correspond to the same classes of symmetric spaces as the BdG
ensembles (see below), but with negative curvature. The ``unitary''
transfer matrix ensemble also corresponds to a symmetric space of
negative curvature, but it is of the chiral, not BdG type (see
Table~\ref{tab1}). The relevant root systems are all of type $C_n$ with long
and ordinary roots. The multiplicities of these can be read from
Table~\ref{tab1}. $m_l$ is always equal to 1 and $m_o=\beta $. Therefore 
the Jacobian given in eq.~(\ref{BR2}),

\beq
\label{eq:transferJ'}
J_\beta (\{\lambda_i\})\propto \prod_{i<j} |{\rm sinh}^2\lambda_i-
{\rm sinh}^2\lambda_j|^\beta \prod_{k}{\rm sinh}(2\lambda_k)
\eeq

is of the general form $J^{(-)}(\{\lambda_i\})$ if we note that the
positive roots are $\{ e_i\pm e_j, 2e_i\} $ and use the following
identity for hyperbolic functions:

\beq
\label{eq:sinh-id}
{\rm sinh}(\lambda_i-\lambda_j)\, {\rm sinh}(\lambda_i+\lambda_j)=
{\rm sinh}^2\lambda_i-{\rm sinh}^2\lambda_j
\eeq

5) {\bf The BdG ensembles}

In exactly the same way as above, using Table~\ref{tab1} and the trigonometric identity
similar to equation (\ref{eq:sinh-id}) 

\beq
{\rm sin}(\lambda_i-\lambda_j)\, {\rm sin}(\lambda_i+\lambda_j)=
{\rm sin}^2\lambda_i-{\rm sin}^2\lambda_j
\eeq

we check that the Jacobian for the BdG ensembles of scattering matrices

\beq
\label{eq:BdGJ'}
J_{r,s}(\{\lambda_i\})\propto \prod_{i<j}|{\rm sin}^2\lambda_i-{\rm sin}^2
\lambda_j|^r \prod_k {\rm sin}^s(2\lambda_k)
\eeq

is the one corresponding to positive curvature spaces
($J^{(+)}(\{\lambda_i\})$) and determined by the corresponding
restricted root systems of type $D_n$ (only ordinary roots) or $C_n$
(long and ordinary roots) according to equation (\ref{eq:J_ja'}).
This Jacobian corresponds to {\it scattering} matrices of the BdG
type, $S={\rm e}^{i{\cal H}}$.  For the corresponding {\it algebra}
subspaces to which $i{\cal H}$ belongs, the Jacobian was given in
eq.~(\ref{eq:BdGJ}). Obviously, its structure is the one given by the
same restricted root lattices according to eq.~(\ref{eq:J_ja'}), but
for zero curvature spaces ($J^{(0)}(\{\lambda_i\})$).

6) {\bf The S--matrix ensembles}

In terms of the transmission eigenvalues $T_n$ the probability
distribution of equation (\ref{jpb2}) (which is the same as the
Jacobian for circular ensembles) can be rewritten as:

\begin{equation}
\label{jpb3}
P(\{T_i\})\, dT_i\equiv J(\{T_i\})\, dT_i~\sim~\prod_{i<j}\left| T_i - T_j
\right|^\beta \prod_{k} |T_{k}|^{\frac{\beta-2}{2}} d{T_k}
\end{equation}

Changing variables and setting $T_i=\sin^2\theta_i$ we find

\begin{equation}
\label{jpb4}
J(\{\theta_i\})\, d\theta_i~\sim~\prod_{i<j} \left| \sin^2\theta_i - 
\sin^2\theta_j \right|^\beta \prod_k |\sin \theta_k|^{\beta -2}
|\sin 2\theta_k| 
\, d{\theta_k}
\end{equation}

As we have by now understood, this is the typical Jacobian for a
positive curvature space, described by a root lattices of the $BC_n$
type with root multiplicities $m_o=\beta ,~m_s=\beta -2,~m_l=1$, where
the Dyson index as usual can take the three values $\beta
=1,~2,~4$. However, in this case we see a rather unexpected feature:
it seems that only one of the three ensembles can be mapped onto a
symmetric space.  It is the unitary ensemble ($\beta =2$), which is
described by a $C_n$ root lattice, since for $\beta =2$, $m_s=0$. The
other two cases correspond to choices of the root multiplicities which
lie outside the classification of Table~\ref{tab1}.  We will come back to this
problem in subsection \ref{sec-nonCartan}, where we will see that the
ensembles characterized by Dyson index $\beta =1$ and $\beta =4$ can
be understood in terms of a {\it non-Cartan parametrization} of
standard symmetric spaces.

\subsection{Fokker--Planck equation and the Coulomb gas analogy}
\label{sec-Coulomb}

An important tool in the theory of random matrices is the so called
``Coulomb gas analogy'' due to Dyson (see \cite{Mehta}, Ch.~8 for a
review). The content of this analogy is the following. The probability
distribution 

\beq
\label{eq:CG_prob}
P(\lambda_1,...,\lambda_N)\propto \e^{-\beta W(\{\lambda_i\})}
\eeq

of random matrix eigenvalues is identical to the probability density of the
positions $\{\lambda_i\}$ of $N$ unit charges in one dimension in a
stationary potential that is given by

\beq
\label{eq:CG_pot}
W(\{\lambda_i\})=\frac{1}{2}\sum_{i=1}^N \lambda_i^2-
\sum_{i<j}{\rm ln}|\lambda_i-\lambda_j|
\eeq

if the random matrix potential is of the gaussian type.
The temperature of this system is $kT=\beta^{-1}$.  The force exerted
by the potential $W$ is of the Coulomb type. If in addition to this force
one considers the Brownian motion of the unit charges in the
presence of a time--dependent random fluctuating force and a
frictional force, one can derive a Fokker--Planck type equation
describing the evolution of the Coulomb gas with time $t$ \cite{Mehta}.

This analogy allows us to write the probability distribution of
eigenvalues in the gaussian and circular ensembles as the asymptotic
limit of a Brownian motion process. Similar equations can be written
also for all the other ensembles of positive and zero curvature listed
in the previous section.  The relevant feature of these equations for
the purpose of the present review is that the differential operator
which describes the Brownian motion in the case of the positive
curvature ensembles can be exactly related to the radial part of the
Laplace--Beltrami operator on the corresponding symmetric space.
Remarkably enough, if one performs a similar mapping for the negative
curvature symmetric spaces (i.e. the transfer matrix ensembles) one
exactly finds the DMPK equation. Below we will discuss all this in
more detail, but let us first summarize the results: \begin{itemize}
\item The DMPK equation is the exact analog for the negative curvature
symmetric spaces of the Coulomb gas equation introduced by Dyson for
the positive and zero curvature ensembles.

\item
 All these equations
describe the free diffusion -- this is the ultimate meaning of the
Laplace--Beltrami operator -- of the eigenvalues on the corresponding
symmetric spaces.

\item
For the positive curvature ensembles, in the large time asymptotic
limit the eigenvalues reach the distribution dictated by the
Jacobian. In fact, if we had the whole Laplace--Beltrami operator (not
just its radial part) we would expect a uniform distribution on the
symmetric space, which is exactly the probability distribution of the
random matrices in the circular ensembles. Since the Dyson operator is the
radial projection of the Laplace--Beltrami operator, the final result
is that the eigenvalues are distributed according to the Jacobian of
the transformation to radial coordinates. 

\item
 For the negative curvature
ensembles, in the asymptotic limit the eigenvalues depart from each
other exponentially. This behavior has a deep physical
meaning. Indeed, it is well known that for quasi--one--dimensional
wires, as the length of the wire (which corresponds to the time in the
Brownian motion analogy) increases, the wire eventually reaches the
insulating regime.  This perfectly agrees with the Brownian motion
picture.  As the time increases the eigenvalues depart arbitrarily far
from each other. As a result the conductivity is dominated by the
lowest eigenvalue and decreases exponentially, in agreement with the
fact that we have reached the insulating regime. 
\end{itemize}

 Let us now discuss
these results in detail.

\subsubsection{The Coulomb gas analogy}

Let us concentrate on the gaussian case for definiteness.  As we
mentioned above, the probability density $P_\beta (\{\lambda_i\})$ of
eq.~(\ref{eq4.2}) and (\ref{eq:CG_prob}) can also be interpreted as
the probability density for the positions of $N$ unit charges constrained to
move on a line and interacting through the potential (\ref{eq:CG_pot}).
The integral

\eq
Z(\beta)= \int\prod_i d\lambda_i e^{-\beta W(\{\lambda_i\})}
\en

for the probability can thus be interpreted as a standard
thermodynamic partition function with $\beta $ playing the role of
inverse temperature. A nice feature of this representation is that we
may obtain the distribution $P_\beta(\{\lambda_i\})$ as the asymptotic
limit of a Brownian motion type process. To this end the following
steps are needed:

First we must add a fictitious time dependence in $P_\beta $. Then we 
must assume that the out--of--equilibrium dynamics of the Coulomb gas is of the
Brownian motion type (free diffusion). As we implied above, this involves
introducing, in addition to the Coulomb force, a time--dependent rapidly
fluctuating force giving rise to the Brownian motion, and a frictional force.
As a consequence the time dependence of $P_\beta(\{\lambda_i\},t)$
must be described by the following Fokker--Planck equation \cite{Mehta}

\eq
f\frac{\partial P}{\partial t}=\sum_{j=1}^N
\left\{\frac1\beta\frac{\partial^2 P}{\partial \lambda_j^2}-
\frac{\partial}{\partial \lambda_j}\left[ E(\lambda_j)P\right]\right\}
\en

where $f$ is the friction coefficient setting the time scale for the
diffusion process, and

\eq
E(\lambda_j)\equiv -\frac{\partial W(\{\lambda_i\})}{\partial \lambda_j}=
-\lambda_j+\sum_{i\not= j}\frac{1}{\lambda_j-\lambda_i}
\en

is the electric Coulomb--type force experienced by the unit charge at
$\lambda_j$.  Standard manipulations show that the Fokker--Planck
equation can be equivalently rewritten as

\eq
\label{eq4.4}
\beta f\frac{\partial P}{\partial t}=\sum_{j=1}^{N}
\frac{\partial }{\partial \lambda_j}\tilde{J}
\frac{\partial}{\partial \lambda_j} \tilde{J}^{-1} P
\en

where $\tilde{J}$ exactly coincides with the joint probability density for the
eigenvalues of eq.~(\ref{eq4.2})

\eq
\tilde{J}(\{\lambda_i\})\equiv P_\beta(\{\lambda_i\})
=C \prod_{i<j} |\lambda_i-\lambda_j|^{\beta }
e^{-\beta\sum_{j=1}^{N}\lambda_j^2/2}
\en

with $C$ an undetermined normalization constant. We have

\eq
\lim_{t\to\infty} P_\beta(\{\lambda_i\},t) = P_\beta(\{\lambda_i\})
\en

One can write down similar Fokker--Planck equations for the all the
zero and positive curvature spaces. All these Fokker--Planck equations
describe the approach to equilibrium of the eigenvalue
distribution. As mentioned above, the equation for the negative
curvature spaces coincides with the DMPK equation.  

\subsubsection{Connection with the Laplace--Beltrami operator}
\label{sec-LB-connection}

Comparing the Fokker--Planck equation (\ref{eq4.4}) in the case of
positive curvature ensembles (in this case the potential is absent and
$\tilde{J}(\{\lambda_i\})= J^{(+)}(\{\lambda_i\})$, that is, the joint
probability distribution of the eigenvalues is identified with the
Jacobian of the transformation to radial coordinates) with equation
(\ref{eq:DeltaB'}) we see that the Fokker--Planck operator

\eq
F\equiv\sum_{j=1}^{N}
\frac{\partial }{\partial \lambda_j}\tilde{J} 
\frac{\partial}{\partial \lambda_j}\tilde{J}^{-1}
\en

is related to the radial part of the Laplace--Beltrami operator,

\beq 
\Delta_B'= \frac{1}{J^{(+)}}\sum_{\alpha =1}^{r'}\frac{\partial }
{\partial q^\alpha }J^{(+)}\frac{\partial }{\partial q^\alpha }
\eeq

where $q^\alpha $ are the radial coordinates identified with the
eigenvalues $\lambda_j$. More precisely, the two operators are related
by

\eq
\Delta_B'=J^{-1}FJ
\en

Similarly, in case of negative curvature spaces the operator $B$ of
equations (\ref{defb}, \ref{DtoB}), that is proportional to the DMPK
operator

\begin{equation}
B=\sum_{k=1}^{n}\frac{\partial}{\partial x_k}\xi^2(x)
\frac{\partial}{\partial x_k}\xi^{-2}(x)=2\gamma D\ \ \ \ \ 
(\gamma \equiv \beta N+2-\beta) 
\end{equation}

is mapped to $\Delta_B'$ by the following relation

\eq
\label{eq:mainresult}
\Delta_B'=\xi^{-2}(x)~B~\xi^2(x)
\en

after identifying $\xi^2(x)$ with the Jacobian
$J(\{x_i\})=\prod_{i<j}|\sinh^{2}x_{j}-\sinh^{2}x_{i}|^\beta \prod_{i}|\sinh
2x_{i}|$ of the transformation to radial coordinates
(eq.~(\ref{eq:Jxi})) and the radial coordinates themselves with the
eigenvalues $x_i$ related to $\lambda_i$ by (\ref{defofx}):
$\lambda_i=\sinh^2x_i$ .

Equation (\ref{eq:mainresult}), that relates the DMPK operator to the
free diffusion on a symmetric space, is one of the main results of the
mapping between random matrix ensembles and symmetric spaces. As we
shall see it also represents the starting point for the exact solution
of the DMPK equation for the transfer matrix ensembles.

\subsubsection{Random matrix theory description of parametric correlations}

The original motivation behind the Coulomb gas approach to random
matrix theory was to obtain the equilibrium probability distribution
of the eigenvalues in a different and more physical way. In this
respect the only interesting regime of the corresponding
Fokker--Planck equations was the asymptotic $t\to\infty $
limit. However, it was later realized that this equation contains
interesting physical information for all values of $t$. In particular
it could be used for describing the adiabatic response to an external
perturbation of the energy spectrum of a mesoscopic
system~\cite{brphysa}.  In this approach the role of the fictitious
time is played by the perturbation parameter.

An interesting application appears if the role of the external
perturbation is played by an external magnetic field.  In a random
matrix description of a disordered conductor, there is a smooth
transition between ensembles characterized by various values of $\beta
$ as a function of the magnetic field $B$ around $B\sim 0$ (for a
review see \cite{Beenakker}). When such a transition is completed, the
level {\it distribution} becomes independent of the magnetic
field. The random fluctuation of {\it individual} energy levels is
still present, however, as a function of $B$. Such random fluctuations
are described by Fokker--Planck type equations similar to the DMPK
equation, but containing the Laplace--Beltrami operator on symmetric
spaces of positive curvature \cite{Beenakker}.

\subsection{A dictionary between random matrix ensembles and symmetric spaces}
\label{sec-dictionary}

Probably the simplest way to summarize what we have said in this
section is to write down the equivalences between random matrix and
symmetric space concepts in a table. This is what we have done in
Table~\ref{tab2}.

\begin{table} 
\begin{center}
\setlength\tabcolsep{0.15cm}
\caption{
The correspondence between random matrix ensembles and symmetric 
spaces\label{tab2}} 
\begin{tabular}{|l|l|}
\hline
{\bf Random Matrix Theories (RMT)}           &  {\bf Symmetric Spaces (SS)} \\
\hline
\hline
circular or scattering ensembles           & positive curvature spaces      \\
\hline
gaussian or hamiltonian ensembles           & zero curvature spaces      \\
\hline
transfer matrix ensembles           & negative curvature spaces      \\
\hline
\hline
random matrix eigenvalues  & radial coordinates \\
\hline
probability distribution of eigenvalues  & Jacobian of transformation to radial
coordinates \\
\hline
Fokker--Planck equation        & radial Laplace--Beltrami equation         \\
\hline
Coulomb gas analogy              & Brownian motion on the symmetric space  \\
\hline
\hline
ensemble indices  &  root multiplicities \\
\hline
Dyson index $\beta $ &  multiplicity of ordinary roots ($\beta =m_o$)\\
\hline
boundary index $\alpha =\beta (\nu+1)-1$ & multiplicity of short and long roots
($\alpha=m_s+m_l$)\\
\hline
\hline
translationally invariant ensembles & SS with root lattice of type $A_n$ \\
\hline
boundary matrix ensembles  &  SS with root lattices of type $B_n$, $C_n$, $D_n$ or $BC_n$\\
\hline
\hline
pair interaction between eigenvalues & ordinary roots \\
\hline
\end{tabular}
\end{center}
\end{table}
\smallskip

\section{On the use of symmetric spaces in random matrix theory}
\label{sec-appl}
\setcounter{equation}{0}

In this section we discuss some of the applications of the mapping
between random matrix ensembles and irreducible symmetric spaces
outlined in section \ref{sec-RMT}.  The main application, which is a
natural consequence of the Cartan classification of symmetric spaces,
is a tentative classification of the random matrix
ensembles. We will discuss this important issue in subsection
\ref{sec-classification}, while in subsection \ref{sec-symmetriesRMT}
we discuss how the symmetries of the spaces are reflected in the
random matrix ensembles. A second natural application is related to
the orthogonal polynomial approach to the construction of eigenvalue
correlation functions of random matrix ensembles~\cite{Mehta}. As it
turns out, the orthogonal polynomials associated to random matrix
ensembles whose integration manifolds are symmetric spaces, are all of
the classical type and can be directly constructed from the knowledge
of the curvature and root multiplicities of the underlying
symmetric space. We will discuss this issue in paragraph \ref{sec-OP}.
Finally in paragraph \ref{sec-use} we discuss the applications of
symmetric spaces to the transfer matrix ensembles which appear in the
description of quantum transport. This is probably the field in which
the knowledge of the mathematical structure of the underlying
symmetric spaces is most helpful.

\subsection{Towards a classification of random matrix ensembles}
\label{sec-classification}

The most interesting consequence of the identification of
random matrix ensembles with irreducible symmetric spaces is that the
Cartan classification of the latter naturally induces a classification
of the matrix ensembles.  It is important to observe, however, that we
should look at this Cartan classification more as a useful framework
for organizing the various matrix ensembles than as a rigid
classification of all possible ensembles.  It is not hard to construct
ensembles that fall outside of the Cartan grid.  This can be done by
suitably constraining the random matrices. There is nothing wrong with
the ensembles that go beyond the Cartan classification. However, a
consequence of such constructions is that for these ensembles, the
beautiful and powerful properties of symmetric spaces are no longer
applicable, and it becomes much more difficult to extract meaningful
information or predictions for the random matrix theory\footnote{See
the discussion in section \ref{sec-nonCartan} below for a remarkable
exception to this fact.}.

At the same time it appears that most of the matrix ensembles which
have physically interesting realizations belong to the Cartan
framework. It is possible that this is merely a consequence of the
fact that, in trying to describe a physical problem, one prefers to
use ensembles that are simple to deal with, even though this may imply
stronger approximations.  Let's mention an example which illustrates
the issue at hand.  Recently various generalizations of the DMPK
transfer matrix ensembles discussed in this review have been
proposed~\cite{m2001} as an attempt to avoid the
``quasi--one--dimensional'' approximation involved in the standard
DMPK equation.  The resulting equations cannot be mapped to a
symmetric space, and thus cannot be solved exactly or even
asymptotically.  In spite of this, important information on the
expected behavior of the eigenvalues can all the same be obtained by
means of suitable perturbative expansions~\cite{m2001}.

In Table~\ref{tab3} we have included the random matrix ensembles
discussed in section \ref{sec-RMT} in the Cartan classification of
irreducible symmetric spaces. This classification was presented in
Table~\ref{tab1}.  The scattering matrix for an NS--type
heterostructure is obtained by exponentiation of the Hamiltonian
$S={\rm e}^{i{\cal H}}$. Since $i{\cal H}$ is in the algebras or
tangent spaces for the respective symmetry classes, the scattering
matrix is in the corresponding symmetric spaces of positive
curvature. The scattering matrix ensembles of NS systems have been
listed in Table~\ref{tab3} with the notation $B^+_{m_o,m_l,m_s}$ and
are of the same type as the circular and S--matrix
ensembles. We list
the ensembles using this kind of notation, which is more consistent
than some of the traditional names given to ensembles in the past with
abuse of language (for example, an ensemble was called ``unitary'' if
$\beta =2$, even though the stability subgroup was not unitary).

The notation used here is as follows. An ensemble is labelled by a
letter indicating the type of ensemble and alluding to the traditional
name. Let C stand for circular ensembles, G for gaussian ensembles, P
for the $p$--wave ensembles of Ivanov (these are of the BdG type, but
have a zero mode), B for Bogoliubov--de Gennes ensembles, T for
transfer matrix ensembles, S for S--matrix ensembles, and $\chi $ for
chiral ensembles), a superscript taking the values $+$, $0$, $-$
indicating the curvature of the space, and three subscripts
$m_o,m_l,m_s$. Since the pair $\{ \beta ,\alpha \}$ can be the same
for some pairs of distinct ensembles, it is better to keep all three
root multiplicities in the label.  For example, B$^+_{2,0,0}$
indicates the BdG (NS) ensemble corresponding to a symmetric space of
positive curvature with root multiplicities $\{ m_o,m_l,m_s\} =\{
2,0,0\}$.  In this way each ensemble is uniquely labelled.

Let us stress that the empty spaces of Table~\ref{tab3} do not mean
that a corresponding random matrix ensemble doesn't exist. Following
the previous discussion, it is easy to see that for each symmetric
space (of arbitrary curvature) one can construct a perfectly
consistent matrix ensemble.  The empty boxes simply mean that such an
ensemble still has not found a relevant physical application or
realization (or, possibly, that we are not aware of such a
realization).  It is likely that with time all the empty boxes in
Table~\ref{tab3} will be occupied with physically interesting
applications. Note also that the type of restricted root system
changes within the same Cartan class for the ensembles labelled by two
integers $\{p,q\}$, depending on whether $p>q$ or $p=q$. The S--matrix
and transfer matrix ensembles have $p=q$ ($\nu =0$) and have been
written on the corresponding line, while the chiral ensembles may have
$p>q$ or $p=q$.

It is important to note that the symmetric space associated to a
random matrix theory can be given either for an ensemble of random
{\it Hamiltonians} ${\cal H}$, for an ensemble of random {\it transfer
matrices} ${\cal M}$, or for an ensemble of random {\it scattering
matrices} $S$.  Thus, the symmetric space associated to the transfer
matrix group of a given system (for example a quantum wire) is {\it
different} from the symmetric space associated to the Hamiltonian
${\cal H}$ or the scattering matrix $S$ of the {\it same} physical
system.  A table of correspondences between the ${\cal M}$ and ${\cal
H}$ descriptions was given in \cite{TitBrou}.

\begin{table}
\caption{Irreducible symmetric spaces and some of their random
matrix theory realizations. The random matrix ensembles with known
physical applications are listed in the columns labelled $X^+$, $X^0$
and $X^-$ and correspond to symmetric spaces of positive, zero and
negative curvature, respectively.  Extending the notation used in the
applications of chiral random matrices in QCD, where $\nu $ is the
winding number, we set $\nu \equiv p-q$.  The notation is C for
circular, G for gaussian, $\chi $ for chiral, B for Bogoliubov--de
Gennes, P for $p$--wave, T for transfer matrix and S for S--matrix
ensembles.  The upper indices indicate the curvature, while the lower
indices correspond to the multiplicities of the restricted roots
characterizing the spaces with non--zero curvature. To the euclidean
type spaces $X^0\sim G^0/K$, where the non--semisimple group $G^0$ is
the semidirect product $K\otimes {\bf P}$, we associate the root
multiplicities of the algebra ${\bf G}={\bf K}\oplus {\bf P}$.
\label{tab3}}
\vskip5mm

\hskip-1cm
\begin{tabular}{|l|l|l|l|l|l|l|l|l|l|}
\hline
$\begin{array}{c}Restricted\\ root\ space\end{array}$ & $\begin{array}{c}Cartan\\ class \end{array}$ & $G/K\ (G)$ & $G^*/K\ (G^C/G)$ & $m_o$ & $m_l$ & $m_s$ & $X^+$ & $X^0$ & $X^-$\\
\hline

$A_{N-1}$     & A    & $SU(N)$                 & $\frac{SL(N,C)}{SU(N)}$ & 2 & 0 & 0 & C$^+_{2,0,0}$ & G$^0_{2,0,0}$ & ${\rm T}^-_{2,0,0}$ \\ 
$A_{N-1}$     & AI   & $\frac{SU(N)}{SO(N)}$   & $\frac{SL(N,R)}{SO(N)}$ & 1 & 0 & 0 & C$^+_{1,0,0}$ & G$^0_{1,0,0}$ & ${\rm T}^-_{1,0,0}$\\ 
$A_{N-1}$     & AII  & $\frac{SU(2N)}{USp(2N)}$ & $\frac{SU^*(2N)}{USp(2N)}$ & 4 & 0 & 0 & C$^+_{4,0,0}$&G$^0_{4,0,0}$& ${\rm T}^-_{4,0,0}$\\
\hskip-2mm $\begin{array}{l} BC_q\ {\scriptstyle (p>q)} \\ C_q\  {\scriptstyle (p=q)} \end{array}$ 
                          & AIII & $\frac{SU(p+q)}{SU(p)\times SU(q)\times U(1)}$ & $\frac{SU(p,q)}{SU(p)\times SU(q)\times U(1)}$ & 2 & 1 & $2\nu $  & \hskip-2mm $\begin{array}{l}  \\ {\rm S}^+_{2,1,0}\end{array}$  & 
 $\chi^0_{2,1,2\nu }$ & 
 \hskip-2mm $\begin{array}{l}  \\ {\rm T}^-_{2,1,0} \end{array}$  \\
\hline

$B_N$          & B    & $SO(2N+1) $               & $\frac{SO(2N+1,C)}{SO(2N+1)}$ & 2 & 0 & 2  &   & P$^0_{2,0,2}$ &  \\ 

\hline

$C_N$ & C    & $USp(2N)$                               &$\frac{Sp(2N,C)}{USp(2N)}$ & 2 & 2 & 0 & B$^+_{2,2,0}$& B$^0_{2,2,0}$ & ${\rm T}^-_{2,2,0}$\\
$C_N$ & CI   & $\frac{USp(2N)}{SU(N)\times U(1)}$      & $\frac{Sp(2N,R)}{SU(N)\times U(1)}$& 1 & 1 & 0  & B$^+_{1,1,0}$ & B$^0_{1,1,0}$  & T$^-_{1,1,0}$  \\
\hskip-2mm $\begin{array}{l} BC_q\  {\scriptstyle (p>q)} \\  C_q\  {\scriptstyle (p=q)} \end{array}$ 
                  & CII & $\frac{USp(2p+2q)}{USp(2p)\times USp(2q)}$ & $\frac{USp(2p,2q)}{USp(2p)\times USp(2q)}$& 4 & 3 & $4\nu $ &   & $\chi^0_{4,3,4\nu }$ & \hskip-2mm $\begin{array}{l} \\ {\rm T}^-_{4,3,0}\end{array}$   \\
\hline

$D_N$    & D          &   $SO(2N)$                            & $\frac{SO(2N,C)}{SO(2N)}$ & 2 & 0 & 0 & B$^+_{2,0,0}$  &  B$^0_{2,0,0}$  & ${\rm T}^-_{2,0,0}$\\
$C_N$    & DIII  & $\frac{SO(4N)}{SU(2N)\times U(1)}$     & $\frac{SO^*(4N)}{SU(2N)\times U(1)}$ & 4 & 1 & 0 & B$^+_{4,1,0}$ & B$^0_{4,1,0}$ & T$^-_{4,1,0}$  \\
$BC_N$   & DIII  & $\frac{SO(4N+2)}{SU(2N+1)\times U(1)}$ & $\frac{SO^*(4N+2)}{SU(2N+1)\times U(1)}$& 4 & 1 & 4 &  & P$^0_{4,1,4}$ &   \\
\hline

\hskip-2mm $\begin{array}{l} B_q\ {\scriptstyle (p>q)}\\ D_q\ {\scriptstyle (p=q)} \end{array}$
            & BDI & $\frac{SO(p+q)}{SO(p)\times SO(q)}$   & $\frac{SO(p,q)}{SO(p)\times SO(q)}$ & 1 & 0 & $\nu $ &   & $\chi^0_{1,0,\nu }$ & $\begin{array}{l} \\ {\rm T}^-_{1,0,0} \end{array}$\\

\hline
\end{tabular}
\end{table}

\subsection{Symmetries of random matrix ensembles}
\label{sec-symmetriesRMT}

Some known symmetries of the random matrix ensembles can be understood
in terms of the symmetries of the associated restricted root
lattice. In particular, ensembles of $A_n$ type are characterized by
translational invariance of the eigenvalues. This translational
symmetry is seen to originate in the root lattice: all the restricted
roots of the $A_n$ lattice are of the form $(e_i-e_j)$.  The
Wigner--Dyson (circular and gaussian) ensembles are of the translation
invariant type.

For all the other types of restricted root lattices ($B_n$, $C_n$,
$D_n$ and $BC_n$) this invariance is broken and substituted by a new
$Z_2$ symmetry giving rise to the reflection symmetry of the
eigenvalues that we discussed in the context of the Jacobians.  Since
these ensembles are characterized by the presence of a boundary (not
always immediately evident) with respect to which they are reflection
invariant, they are called boundary random matrix theories (BRMT in
the following). They include all the remaining ensembles that are not
of the circular or gaussian type.

\subsection{Orthogonal polynomials}
\label{sec-OP}

An important role in the study of matrix ensembles is played by the
set of polynomials orthogonal with respect to the random matrix theory
integration measure.  These polynomials come into the picture when
rewriting the Jacobian for the transformation to eigenvalue space in
terms of a product of Vandermonde determinants. By adding linear
combinations of the rows, the determinant can be written as a
determinant of monic polynomials (a polynomial $P_n(x)$ is called
monic if $P_n(x)=x^n+{\cal O}(x^{n-1})$), for example

\beq
\label{eq:vander}
\prod_{i<j}^N (x_i-x_j) \sim \det_{1\leq i,j\leq N}x_j^{i-1}=
\det_{1\leq i,j\leq N} P_{i-1}(x_j)
\eeq

If these polynomials are then chosen orthogonal with respect to the
measure\footnote{The factor $N$ in the exponent is common in QCD
related applications. If the weight function is odd, so called
pseudo--orthogonal polynomials may be used. We will not discuss them
here, since our goal is just to remind the reader of the general
mechanisms that make orthogonal polynomials useful.}  $w(x)dx= {\rm
e}^{-NV(x)}dx$, where $V(x)$ is the random matrix potential

\beq
\int_I {\rm e}^{-NV(x)}P_m(x)P_n(x)dx=h_n\delta_{mn}
\eeq

(here $I$ is some interval on the real axis and $h_n$ is a
normalization factor), the eigenvalue correlation functions can be
expressed in terms of a Christoffel--Darboux kernel \cite{Mehta}.
An arbitrary $k$--point correlation function is defined as

\beq
\rho (x_1,...,x_k)=\int_{-\infty }^\infty \prod_{j=k+1}^N dx_j\, 
P(x_1,...,x_N)
\eeq

where $P(x_1,...,x_N)$ is the joint eigenvalue distribution of the
random matrix model. The quantity $\rho (x_1,...,x_k)dx_1...dx_k$ equals the
probability of finding one eigenvalue in each of the intervals between
$x_j$ and $x_j+dx_j$ ($j=1,...,k$). 

The general formula for the $k$--point function turns out to be 

\beq
\label{eq:rhoK}
\rho (x_1,...,x_k)=\frac{N^k(N-k)!}{N!} \det_{1\leq i,j\leq N}K_N(x_i,x_j)
\eeq

In particular, the spectral density is simply

\beq
\label{eq:rhoK'}
\rho (x)=K_N(x,x)
\eeq

In eqs.~(\ref{eq:rhoK}) and (\ref{eq:rhoK'}), $K_N(x_i,x_j)$ is the 
Christoffel--Darboux kernel defined in terms of orthogonal polynomials

\beq
\label{eq:kernel}
K_N(x_i,x_j)=N^{-1} \e^{\frac{N}{2}(V(x_i^2)+V(x_j^2))} \sum_{k=0}^{N-1}
h_k^{-1}P_k(x_i)P_k(x_j)
\eeq

It turns out that for all the matrix ensembles related to symmetric
spaces, the associated orthogonal polynomials belong to the set of so
called classical orthogonal polynomials. 

With the term {\it classical} one usually denotes three families of
 orthogonal polynomials: the Jacobi, Laguerre and Hermite polynomials,
 whose unifying feature is the so called Rodriguez
 formula\footnote{Actually this formula is a generalization due to
 Tricomi of the original formula obtained by Rodriguez for the case of
 Legendre polynomials.} which allows to construct the polynomials once
 the weight function $p(x)$ and the function $X(x)$ (which specifies
 the domain of support of the polynomials) are given:

\beq
P_n(x)=\frac{1}{A_n}\frac{1}{p(x)}\frac{d^n}{dx^n}\left\{p(x) X^n(x)\right\}
\eeq

Here $A_n$ is a normalization constant which can be obtained explicitly, 
but is irrelevant for our purposes.

The domain--specifying function $X(x)$ is a polynomial in $x$ of degree
$\leq 2$. The three possibilities are:

\begin{itemize}
\item
if $X(x)$ is simply a constant, then the polynomials are defined on the whole
real line, the weight function must be $p(x)=e^{-x^2}$, and we find the Hermite
polynomials.

\item
if $X(x)$ is a polynomial of first degree, one can always shift the
origin so as to obtain $X(x)=x$. In this case the orthogonal
polynomials are defined on the positive real axis, the weight function
must be $p(x)=x^{\lambda}e^{-x}$, and we find the Laguerre
polynomials.

\item
if $X(x)$ is a polynomial of second degree, then one can always
normalize it so as to obtain $X(x)=(1-x^2)$. In this case the
orthogonal polynomials are defined on the interval $[-1,1]$, the
weight function must be $p(x)=(1-x)^{\sigma}(1+x)^{\rho}$, and we find
the Jacobi polynomials. The well known Gegenbauer, Chebyshev and
Legendre polynomials are only special cases of Jacobi polynomials.
\end{itemize}

Each one of these polynomial families is in one--to--one
correspondence with a particular random matrix ensemble of
table~\ref{tab3}. The mapping is complete, i.e. all the ensembles of
table~\ref{tab3} are covered.
 
In particular the Hermite polynomials are related to the gaussian
Wigner--Dyson ensembles (i.e., those corresponding to symmetric spaces
defined by an $A_n$ root lattice).  Simple changes of variables allow
one to show that the Laguerre polynomials are related to the gaussian
BRMT's (these are the chiral, BdG and $p$-wave ensembles of zero
curvature spaces) and the Jacobi polynomials to the circular BRMT's
(i.e., ensembles related to symmetric spaces of positive curvature not
of the $A_n$ type, like for instance the BdG scattering ensembles). A
thorough discussion of the correlation functions for these ensembles
using the orthogonal polynomials listed above can be found for
instance in~\cite{for} (see also~\cite{duenez01}).

These results are well--known (for instance, in the early papers on
chiral matrix ensembles, these ensembles were named Laguerre
ensembles) and do not require any particular reference to the
symmetric space description of random matrix theories.  What is
interesting in our framework is that the parameters which define the
polynomials can be explicitly related to the multiplicities of short
and long roots of the underlying symmetric space and thus, by the
identification in Table~\ref{tab2}, with the boundary universality
indices of the BRMT.  The relation is the following:

{\bf Laguerre polynomials:}

\bea
L^{(\lambda)}(x)=\frac{x^{-\lambda}e^x}{n!}\frac{d^n}{dx^n}
(x^{n+\lambda} e^{-x})\ \ \ \ \ (x\geq 0) \nonumber \\
\lambda \equiv \frac{m_s+m_l-1}{2} \\ \nonumber
\eea

{\bf Jacobi polynomials:}

\bea
P^{(\rho,\sigma )}(x)=\frac{(-1)^n}{2^nn!}
\frac{(1-x)^{-\sigma }}{(1+x)^\rho }\frac{d^n}{dx^n}
\left( \frac{(1+x)^{n+\rho }}{(1-x)^{-n-\sigma }} \right) \ \ \ \ \ 
(-1\leq x\leq 1)
\nonumber \\
\rho \equiv \frac{m_s+m_l-1}{2},\ \ \sigma \equiv \frac{m_l-1}{2}\\ \nonumber
\eea

We see that $\lambda $ and $\rho $ have the same expression in terms
of $m_s$ and $m_l$. Thus the BRMT's corresponding to Laguerre and
Jacobi polynomials with the same $\lambda =\rho $ indices belong to
the same triplet in the classification of Table~\ref{tab3}. They are
respectively the zero curvature (Laguerre) and positive curvature
(Jacobi) elements of the triplet. This explains the so called ``weak
universality'' which was observed a few years ago for the boundary
critical indices of these ensembles~\cite{for}, i.e. the fact that the
form (near the boundary) of scaled $k$--level correlators is the same
for Laguerre and Jacobi ensembles if the symmetry parameter $\beta$ is
the same (see~\cite{for} for a discussion of this point).  The weak
universality turns out to be simply a consequence of the organization
in triplets of the symmetric spaces!

\subsection{Use of symmetric spaces in quantum transport}
\label{sec-use}

One of the most interesting applications of symmetric spaces is in the
transformation of the Fokker--Planck equations in random matrix theory
into Schr\"odinger equations in imaginary time (where, as before, the
time coordinate in transfer matrix ensembles is identified with the
dimensionless length $s$ of a quantum wire).  As a consequence of
this transformation, in the case $\beta =2$ the degrees of freedom in
the Schr\"odinger equation decouple and it can be solved exactly.

This result traces back to the original work by Dyson~\cite{Dyx} and
was later extended to boundary random matrix theories by various
authors. We mentioned this important result already at the end of
section \ref{sec-intmod}.  In paragraph \ref{sec-beta2} we will study
this mapping in detail in a case which is particularly relevant from a
physical point of view, namely for the DMPK equation of transfer
matrix ensembles.

Another important consequence of the mapping is that in the
``interacting cases'' (for $\beta =1$ and $4$) one can use the results
discussed in subsections \ref{sec-zonal} and \ref{sec-Fourier} on the
zonal spherical functions to obtain important information on the
asymptotic behavior of the solutions to the DMPK equation.  We will
discuss this issue in paragraph \ref{sec-beta14}. In
subsection \ref{sec-9.4.3} we will see an example of a change of
symmetry class (hence also of the underlying symmetric space
description) induced in a quantum wire by switching on an external
magnetic field. 
Finally in
subsection \ref{sec-9.4.4} we shall discuss a general
scheme (based on the DMPK equation)
for constructing the scaling equations for the density of
states of a  quantum wire that covers all
the Cartan symmetry classes.
\subsubsection{Exact solvability of the DMPK equation in the $\beta =2$ case}
\label{sec-beta2}

The exact solution of the DMPK equation in the $\beta=2$ case was
first obtained in a remarkable paper \cite{BeeRejaei} by Beenakker and
Rejaei.  Here we review their derivation in a slightly different
language, trying to stress the symmetric space origin of their result.

The starting point is the mapping discussed in paragraph
\ref{sec-DMPK} which we briefly recall here. By setting 
$\lambda_n \equiv {\rm sinh}^2x_n$ (cf. eq.~(\ref{defofx})) the
DMPK equation can be rewritten as (see equations (\ref{BR2}) and
(\ref{eq:mainresult}))

\begin{equation}
\frac{\partial P}{\partial s}=
\frac{1}{2\gamma}\, [\xi(x)]^{2}~\Delta_B'~[\xi(x)]^{-2}~~P
\label{DMPKbis2}
\end{equation}

where

\begin{equation}
\xi(\{x_i\})=\prod_{i<j}|\sinh^{2}x_i-\sinh^{2}x_j|^{\frac{\beta}{2}}
\prod_{k}|\sinh 2x_{k}|^{\frac{1}{2}}
\label{nn1}
\end{equation}

and $\Delta_B'$ is the radial part of the Laplace--Beltrami operator on the
underlying symmetric space.

At this point one can follow two equivalent ways. In the first one
(which was the one followed in~\cite{BeeRejaei}) one makes use of the
results discussed in section~\ref{sec-intmod}, in particular
eq.~(\ref{eq:H-Delta}), to map the DMPK equation into a Schr\"odinger
equation. By comparing with eq.~(\ref{eq:H-Delta}) we see that this
simply requires the substitution

\beq
\label{eq:subst}
P(\{x_n\},s) = \xi(\{x_n\}) \Psi(\{x_n\},s)
\eeq

A straightforward calculation shows that the DMPK equation then takes
the form of a Schr\"odinger equation in imaginary time:

\beq
\label{eq:DMPKSchr}
-\frac{\partial \Psi}{\partial s} = ({\cal H}-U)\Psi
\eeq

where $U$ is a constant and ${\cal H}$ is a Hamiltonian of the form

\beq
{\cal H} = -\frac{1}{2\gamma}\sum_i \left( \frac{\partial^2}{\partial x_i^2}
+ {\rm sinh}^{-2}(2x_i) \right) + \frac{\beta (\beta -2)}{2\gamma}\sum_{i<j}
\frac{{\rm sinh}^2(2x_i) + {\rm sinh}^2(2x_j)}{({\rm cosh}(2x_i) -
{\rm cosh}(2x_j))^2}
\eeq

At this point the main goal has already been reached: it is easy to see
that if $\beta =2$ the equation decouples and an exact solution can be obtained
\cite{BeeRejaei}. Before going into the details of this solution, let's
remark that the above equation can be recast in a slightly different
form, thus completing the chain of identifications DMPK equation ---
radial part of the Laplace--Beltrami operator --- Calogero--Sutherland
model. By using simple identities for hyperbolic functions, this
Hamiltonian becomes~\cite{MCDMPK}

\beq
\label{eq:HC_n}
\gamma{\cal H} = \sum_i \left( -\frac{1}{2}\frac{\partial^2}{\partial x_i^2}
+ \frac{g_l^2}{{\rm sinh}^2(2x_i)} \right) + \sum_{i<j} \left( 
\frac{g_o^2}
{{\rm sinh}^2(x_i-x_j)} + \frac{g_o^2}{{\rm sinh}^2(x_i+x_j)}\right) +c
\eeq

where $g_l^2\equiv -1/2$, $g_o^2\equiv\beta (\beta -2)/4$ and $c$ is
an irrelevant constant.  This Hamiltonian, apart from an overall
factor $1/\gamma $ and the constant $c$, exactly coincides with the
Calogero--Sutherland Hamiltonian (\ref{CS3}) discussed in section
\ref{sec-intmod}, corresponding to a root lattice $R=\{\pm 2x_i,\pm
x_i \pm x_j,i\neq j\}$ of type $C_n$ with root multiplicities
$m_o=\beta $, $m_l=1$.  The values of the coupling constants $g_o$,
$g_l$ are exactly the root values given in eq.~(\ref{eq:rootvalues})

\beq
\label{eq:rootvalues'}
g_\alpha^2=\frac{m_\alpha (m_\alpha -2)|\alpha |^2}{8}
\eeq

of section \ref{sec-intmod}, for which the transformation from ${\cal
H}$ into $\Delta_B'$ is possible. 

Let us now come back to the exact solution of the DMPK equation
following Beenakker and Rejaei.

As we have seen, for $\beta=2$ ${\cal H}$ is reduced to a sum of
single-particle Hamiltonians ${\cal H}_{0}$,

\begin{equation}
{\cal H}_{0}=-\frac{1}{2\gamma}\frac{\partial^{2}}
{\partial x^{2}}-\frac{1}{2\gamma\sinh^{2}2x}.
\label{hh1}
\end{equation}

At this point, to solve the DMPK equation one simply has to construct
the Green's function $G_{0}$ of the single-particle Hamiltonian ${\cal
H}_{0}$. This requires solving the eigenvalue equation

\begin{equation}
{\cal H}_{0}\psi(x)=
\varepsilon\psi(x),\label{hh2}
\end{equation}

A standard analysis shows that the spectrum of ${\cal H}_{0}$ is
continuous, with positive eigenvalues $\varepsilon=\frac{1}{4}k^{2}/N$
and that the eigenfunctions $\psi_{k}(x)$ are real functions given by

\begin{equation}
\psi_{k}(x)=\left[\pi k\tanh\left(\frac{\pi k}{2}\right)\sinh 2x\right]^{1/2}\,
{\rm P}_{\frac{1}{2}({\rm i}k-1)}(\cosh 2x).\label{hh3}
\end{equation}

where ${\rm P}_{\nu}(z)$ denotes the Legendre functions of the first
kind.  From this we obtain the spectral representation of the
single-particle Green's function $G_{0}$

\begin{eqnarray}
G_{0}(x,s\,|\, y)&=&
(2\pi)^{-1}\int_{0}^{\infty}\!\!dk\,\exp\left( -\frac{k^2s}{4N}\right) \,
\psi_{k}(x)\psi_{k}(y)\nonumber\\
&=&\frac{1}{2}(\sinh 2x\sinh 2y)^{1/2}\int_{0}^{\infty}\!\!dk\,
\exp\left( -\frac{k^2s}{4N}\right) k\tanh\left(\frac{\pi k}{2}\right)
\nonumber\\
&&\hspace{3cm}\mbox{}\times{\rm P}_{\frac{1}{2}({\rm i}k-1)}(\cosh 2x)
{\rm P}_{\frac{1}{2}({\rm i}k-1)}(\cosh 2y)
\label{eq:hh4}
\end{eqnarray}

The $N$--particle Green's function $G$ is related to the
single--particle Green's function $G_0$ through a Slater determinant,
and the probability distribution of eigenvalues is related to $G$
through a similarity transformation by the antisymmetrized eigenstate
$\Psi_0(x)=\xi (x)$ ($\beta =2$) of the $N$--fermion Hamiltonian (for
details see \cite{BeeRejaei}).  From the expression (\ref{eq:hh4}),
imposing so called ballistic initial conditions (which essentially
amount to requiring that all the eigenvalues are concentrated at the
origin for $s=0$), one finally obtains the probability distribution
$P(\{x_{n}\},s)$ for the eigenvalues

\begin{eqnarray}
P(\{x_{n}\},s)&=&C(s)\prod_{i<j}(\sinh^2x_j-\sinh^2x_i)
\prod_{k}(\sinh 2x_k)\nonumber\\
&&\mbox{}\times{\rm Det}\left[\int_{0}^{\infty}\!\!dk\,
\exp\left( -\frac{k^2s}{4N}\right) \tanh\left(\frac{\pi k}{2}\right) k^{2m-1}\,
{\rm P}_{\frac{1}{2}({\rm i}k-1)}(\cosh 2x_{n})\right]\nonumber \\
\label{final}
\end{eqnarray}

This is the {\it exact} solution of the DMPK equation for $\beta =2$
\footnote{We remark that in \cite{M4}, using the same technique as in
\cite{BeeRejaei}, the Fokker--Planck equation for the probability
distribution of eigenvalues in systems with a chiral Hamiltonian was
solved exactly in the case $\beta=2$, and in \cite{M5}, the equation
corresponding to a system with BdG Hamiltonian was solved in the
presence of time--reversal symmetry (for two of the four BdG symmetry
classes).}.

The second approach relies more heavily on the underlying symmetric
space structure. The starting point is again the identification made
in eq.~(\ref{DMPKbis2}) between the DMPK operator and the radial part
of the Laplace--Beltrami operator on the underlying symmetric
space. As a consequence of this identification, if $\Phi_k(x)$
($x=\{x_1,\cdots,x_N\}$, $k=\{k_1,\cdots,k_N\}$) is an eigenfunction
of $\Delta_B'$ with eigenvalue $k^2$, then $\xi(x)^2\Phi_k(x)$ will be
an eigenfunction of the DMPK operator with eigenvalue $k^2/(2\gamma)$.
The eigenfunctions of the $\Delta_B'$ operator (which are known in the
literature as zonal spherical functions) have been widely discussed in
subsections \ref{sec-zonal} and \ref{sec-Fourier}.  As we have seen,
by means of the zonal spherical functions one can define the analog of
the Fourier transform on symmetric spaces:

\begin{equation}
f(x)=\int \bar f(k) \Phi_k(x) \frac{dk}{|c(k)|^2}
\label{ss1}
\end{equation}

(here we have neglected an irrelevant multiplicative constant).  In
particular, for the three symmetric spaces which are of interest for
us one finds:

\begin{equation}
|c(k)|^2=|\Delta(k)|^2
\prod_{j}\left\vert\frac{\Gamma\left(
i\frac{k_j}{2}\right)}{\Gamma\left(\frac{1}{2}+
i\frac{k_j}{2}\right)}\right\vert^2
\label{ss2}
\end{equation}

with

\begin{equation}
|\Delta(k)|^2=\prod_{m<j}\left \vert\frac{\Gamma\left(
i\frac{k_m-k_j}{2}\right)\Gamma\left(
i\frac{k_m+k_j}{2}\right)}{\Gamma\left(\frac{\beta}{2}+
i\frac{k_m-k_j}{2}\right)\Gamma\left(\frac{\beta}{2}+
i\frac{k_m+k_j}{2}\right)}\right\vert^2
\label{ss2b}
\end{equation}

where $\Gamma $ denotes the Euler gamma function.  This is a completely
general result which we will use also in the next section.  The
problem is that in general the explicit form of the zonal spherical
functions involved in eq.~(\ref{ss1}) is not known. A remarkable
exception to this situation is represented exactly by the $\beta=2$
case for which we have~\cite{OlshPere,bk58} 
(see eq. (\ref{eq:app4} in the appendix):

\begin{equation}
\Phi_k(x)=\frac{{\rm det}\left[ Q_m^l\right]}
{\prod_{i<j}[(k_i^2-k_j^2)(\sinh^2 x_i
-\sinh^2 x_j)]}
\label{ss4}
\end{equation}

where the matrix elements of $Q$ are:

\begin{equation}
Q_m^l={\rm F}\left(\frac{1}{2}(1+ik_m),
\frac{1}{2}(1-ik_m),1;-\sinh^2x_l\right)
\label{ss4b}
\end{equation}

and $F(a,b,c;z)$ is the hypergeometric function.

Equations~(\ref{DMPKbis2},\ref{ss1}--\ref{ss2b}) allow us to write the
$s$-evolution of $P(\{x_n\},s)$ from given initial conditions
(described by the function $\bar f_{0}(k)$) as follows:

\begin{equation}
P(\{x_n\},s)=[\xi(x)]^2 \int \bar f_0(k)e^{-\frac{k^2}{2\gamma}s}
\Phi_k(x) \frac{dk}{|c(k)|^2}
\label{ss5}
\end{equation}

By inserting the explicit expression of $|c(k)|^2$ and by using the
identity:

\begin{equation}
\left\vert\frac{\Gamma\left(\frac{1}{2}+
i\frac{k}{2}\right)}
{\Gamma\left(
i\frac{k}{2}\right)}\right\vert^2=\frac{k}{2}\tanh\frac{\pi k}{2}
\end{equation}

we end up with the following general expression for $P(\{x_n\},s)$
with ballistic initial conditions (which, due to the normalization of
$\Phi_k(x)$, simply amount to choosing $ \bar{f}_0(k)$ equal to a
constant):

\eqa
P(\{x_n\},s)=[\xi(x)]^2 \int dk\,
{\rm e}^{-\frac{k^2}{2\gamma}s}
\frac{ \Phi_k(x)}{|\Delta(k)|^2} \prod_{j}k_j\tanh(\frac{\pi k_j}{2})
\label{ss6}
\ena

Inserting the explicit expression for $\Phi_k(x)$ from
equations~(\ref{ss4},\ref{ss4b}) into (\ref{ss6}) and using the
identity

\begin{equation}
{\rm P}_{\nu}(z)=F(-\nu,\nu+1,1;(1-z)/2)
\end{equation}

we exactly obtain, as expected, the solution (\ref{final}) found by
Beenakker and Rejaei. This is a remarkable and non trivial consistency
check of the correctness of this solution.

\subsubsection{Asymptotic solutions in the $\beta =1,4$ cases}
\label{sec-beta14}

The power of the description in terms of symmetric spaces becomes
evident in the $\beta =1$ and $\beta =4$ cases, in which the
interaction between the eigenvalues does not vanish and the first
approach discussed in the previous subsection does not apply. On the
contrary, the description in terms of zonal spherical functions (i.e.,
eq.~(\ref{ss5})) also holds in these two cases. Even though for
$\beta\not=2$ one does not know the explcit form of the zonal
spherical functions, one can use the powerful asymptotic expansion
(\ref{asyexp}) discussed at the end of subsection \ref{sec-Fourier} to
get asymptotic solutions. In our context this expansion reads:

\begin{equation}
\Phi_k(x)\sim \frac{1}{\xi(x)} \left(\sum_{r\in W} c(r k) e^{i(r k,x)}
\right)
\label{eq:ss3}
\end{equation}

where $rk$ is the vector obtained acting with $r\in W$ on $k$ ($W$
denotes the Weyl group of the symmetric space). The important feature
of eq.~(\ref{eq:ss3}) is that it is valid for {\it all values of k}, thus
it can be used both in the {\it metallic $(k\gg 1)$} and in the {\it
insulating $(k\ll 1)$} regimes. This leads to expressions for the
probability distribution of the eigenvalues and for the conductance,
which can then be compared to other theoretical results using
numerical simulations or experiments (see~\cite{MCDMPK} for a detailed
discussion).

\subsubsection{Magnetic dependence of the conductance}
\label{sec-9.4.3}

In \cite{MCnewuni}, the theory of symmetric spaces was applied to give
a possible explanation for the discrepancy between the random matrix
theory and non--linear sigma model analysis of the magnetoconductance
in the weakly insulating, localized regime close to the Anderson
transition of a disordered wire.  The Anderson transition is a
disorder--induced transition from the conducting to the insulating
regime.  For a review of the sigma model approach see \cite{useSusy}.

Magnetoconductance is the change in the conductance of the wire due to
the presence of a magnetic field (for a review see \cite{Beenakker}).
More precisely, it is the suppression of weak localization (a quantum
effect due to time--reversal symmetry) which appears when a magnetic
field destroys time--reversal invariance. The disagreement between the
random matrix theory approach and the non--linear sigma model approach
was evident in the prediction for the magnetoconductance in the
presence of strong spin--orbit scattering: the sigma model approach
gave a zero magnetoconductance, while the random matrix approach gave
a strong negative value due to suppression of the localization
length. A negative magnetoconductance has been observed in
experiments. However, as we will review below, also within the random
matrix approach a zero magnetoconductance can be expected, depending
on whether Kramers degeneracy of the eigenvalues is conserved or not.

The conductance in the insulating regime is related to the
localization length $\xi $ by $G=G_0{\rm exp}(-2L/\xi )$, where $L$ is
the length of the sample. In the standard random matrix approach, $\xi
$ is proportional to $\beta $. Thus the transition (due to the switching on 
of a magnetic field) between the ensemble characterized by $\beta =4$ and 
the ensemble characterized by $\beta =2$ means a negative contribution 
to $G$.

The key observation made in \cite{MCnewuni} was that the matrix
ensemble to which the transfer matrix belongs corresponds, for
$\beta =4$, to a symmetric space $SO^*(4N)/U(2N)$ with a root lattice
of type $C_N$.  This root lattice is characterized by two types of
roots, long and ordinary, and therefore by {\it two} indices $\beta
\equiv m_o = 4$ and $\eta \equiv m_l=1$. In the ensemble labelled by
$\beta =4$ the eigenvalues are twofold degenerate. Therefore, if there
are $N$ degenerate scattering channels for $\beta =4$ (so called
Kramers degeneracy), for $\beta =2$ there are $2N$ channels.

Using the mapping of the DMPK equation onto a Calogero--Sutherland
model (see section \ref{sec-intmod}), a generalized DMPK equation was
derived to take into account the new index $\eta $.  As a consequence,
it was shown that the localization length was not affected if one
performs the simultaneous change $\beta =4\ \to \ \beta =2$ and $N \to
2N$, while keeping $\eta $ fixed. This corresponds to the transition
between ensembles $SO^*(4N)/U(2N)\ \to \ SU(2N,2N)/(SU(2N)\times SU(2N)
\times U(1))$ (the latter ensemble is characterized by $\beta =2$,
$\eta =1$) and gives a zero magnetoconductance. Effectively, the {\it
  level statistics} depends on $\beta $ and $N$, but the localization
length depends more generally on $\beta $, $\eta $, and $N$.

Assuming instead that Kramers degeneracy is {\it conserved} ($N\ \to \
N$) we have a transition $SO^*(4N)/U(2N)\ \to \
USp(2N,2N)/(USp(2N)\times USp(2N))$ (the latter ensemble is
characterized by $\beta =4$, $\eta =3$) with the result that $\xi $ is
smaller in the presence of a magnetic field. In this case we have a
negative magnetoconductance. Thus we see that both possibilities, zero
and negative magnetoconductance, are feasible also within the
framework of random matrix theory.

\subsubsection{Density of states in disordered quantum wires.}
\label{sec-9.4.4}
Another  interesting application of the DMPK equation is the construction of the
scaling equation for the density of states of a disordered quantum wire which
was recently discussed by Titov, Brouwer, Furusaki and Mudry in
\cite{TitBrou,M2}. Computing the density of states $\rho (\epsilon )$
in the thermodynamic limit $N\to \infty $ in a quantum wire of
infinite length at energy $\epsilon $ turns out to be intimately related to the
solution of the DMPK equation in presence of absorption \cite{TitBrou,M2}. 
Absorption is described by the addition of a spatially uniform imaginary
potential $i\omega$, $\omega>0$ to the Hamiltonian.

A DMPK-like equation can be obtained also in this case. It is very similar to
the standard one (see eq.s (\ref{defb}) and (\ref{DMPKbis})). The only change is
an additional term proportional to the absorption constant.

\begin{equation}
\frac{\partial P_{i\omega}}{\partial s}=
\left(\frac{1}{2\gamma}\,
B +\frac{l\omega}{v_F}\sum_{j=1}^{N}\frac{\partial}{\partial
x_j}\sinh2x_j\right)P_{i\omega}
\label{DMPKabs}
\end{equation}
  where $l$ is the mean free path and $v_F$ the Fermi velocity.

It is not difficult to see that this equation has the following
 stationary (i.e. $L$ independent) solution:
 \eq
 P_{i\omega}(\{x_i\})=\frac{|J|}{Z(a)}\prod_{j=1}^{N}e^{-a\cosh2x_j}
 \en 
 where $a=\gamma l \omega/v_F$ is the adimensional absorption constant and
 $Z(a)$ is a normalization constant which is needed to ensure that $P_{i\omega}$
 is normalized to one. 
 The key (and non-trivial) point is that, once $Z(a)$ is known
 it is possible to
 show (by making an analytical continuation from $i\omega$ to $\epsilon$) that
 the  density of states $\rho(\epsilon)$
 is obtained from $Z(a)$ as follows:
\eq
\rho(\epsilon)=-\frac{d}{\pi v_F} {\rm Re}\lim_{a\to-i\gamma l \epsilon/v_F}
 \frac{\partial}{\partial a}
\left[a \frac{\partial}{\partial a} \log Z(a)\right]
\en
(see \cite{TitBrou} for a detailed derivation).
It is interesting to observe that this procedure works 
for all the Cartan classes.

 In the non--standard symmetry classes of
the Hamiltonian, the density of states is singular at the band center
(for chiral Hamiltonians) or Fermi energy (for BdG Hamiltonians). This
point corresponds to extra symmetries of the Hamiltonian, an issue
that was discussed in the context of chiral ensembles. The precise
form of this singularity depends on the symmetry class and, for chiral
Hamiltonians, on the parity of $N$ (the number of channels).  
 Apparently, there is a connection between the anomalous
behavior of the density of states and the divergence of the
localization length (signaling criticality) at the singular point
\cite{M2}.

\section{Beyond symmetric spaces}
\label{sec-beyond}
\setcounter{equation}{0}

In the previous section we have discussed some of the possible
applications of the theory of symmetric spaces in random matrix
models.  In this last section we discuss three issues which in one way
or another go beyond the theory of symmetric spaces developed up to
now. We will see that the power of the group theoretical methods
developed in the first part of this review allows us to obtain
interesting results also in some cases in which symmetric spaces
appear not to be useful.  At the same time the topics to be discussed
represent new open directions of research which we hope will lead to
future interesting results.

In subsection \ref{sec-nonCartan} we will discuss the non--Cartan
parametrization of symmetric spaces. We will see how it is possible to
map exotic random matrix ensembles to non--standard (in a sense that
will be clear) symmetric spaces. In subsection \ref{sec-clustered} we
will discuss how a wide set of non--isotropic solutions to the DMPK
equation can be constructed by simply resorting to the exact
integrability of the associated Calogero--Sutherland models. This new
tool might be useful in overcoming the quasi--1D constraint of the
DMPK description of quantum wires.  Finally, in subsection
\ref{sec-Weierstrass} we discuss the triplicity of (in a certain sense) 
the most general potential in a Calogero--Sutherland model, the
Weierstrass potential. We will see how this potential in various limits
reproduces the components of an arbitrary triplet of symmetric spaces.

\subsection{Non--Cartan parametrization of symmetric spaces and S--matrix ensembles}
\label{sec-nonCartan}

As we have seen in paragraphs \ref{sec-S-mat} and \ref{sec-jac},
apparently only the S-matrix ensembles labelled by $\beta =2$ (these
were introduced in references \cite{JalaPB+BarMello,JalP}) can be
associated to a symmetric space.  Let us briefly recall the
problem. The S--matrix ensembles correspond to root lattices of $BC_n$
type with the following root multiplicities: $m_o=\beta ,~m_s=\beta
-2,~m_l=1$, where $\beta $ takes values in the set $\{ 1,2,4\} $.  Only
the $\beta =2$ case corresponds to a set of root multiplicities
associated to a symmetric space.  In this case the restricted root
lattice degenerates into $C_n$, since the multiplicity of short roots
$m_s=0$.  It is instructive to re--obtain this symmetric space
description from the integration manifold of the random matrix
ensemble.  In the unitary case the scattering matrix is parametrized by
(cf. eq.~(\ref{eq:Smatrix}) in paragraph \ref{sec-transfer})

\begin{equation}
S=\left(\begin{array}{cc}r & t' \\t & r' \end{array}\right)
\end{equation}

where $r,t,r',t'$ are $N\times N$ matrices. One is interested in the
eigenvalues $\{ T_i\} $ of the product $tt^\dagger $, since these
determine a variety of transport properties.  This implies that there
is a hidden symmetry:

\begin{equation}
S\to S\left(\begin{array}{cc}v_3 & 0 \\0 & v_4 \end{array}\right)
\label{eq:2}
\end{equation}

with $v_3, v_4\in U(N)$. Indeed, under the transformation
(\ref{eq:2}), $t\to tv_3$ and the product $tt^\dagger $ is invariant
for unitary $v_3$. This additional symmetry defines the space in which
the matrices $S$ live. It is not $SU(2N)$ but the coset
$SU(2N)/S(U(N)\times U(N))$, which is exactly the
symmetric space described by the $C_N$ root lattice with
multiplicities $m_o=2$ and $m_l=1$ mentioned above.

We can write the coordinates $\{ T_i\} $ explicitly by
using the fact that any $2N\times 2N$ unitary matrix can be decomposed
as \cite{JalaPB+BarMello,JalP} (cf. eq.~(\ref{eq:Sinv}))

\bea
S=\left(\begin{array}{cc}v_1 & 0 \\0 & v_2 \end{array}\right)
\left(\begin{array}{cc}-\sqrt{1-\tau} & \sqrt{\tau}\\ 
\sqrt{\tau} & \sqrt{1-\tau}\end{array}\right)
\left(\begin{array}{cc}v_3 & 0 \\0 & v_4 \end{array}\right) \nonumber \\
\nonumber \\
=\left(\begin{array}{cc}-v_1\sqrt{1-\tau} & v_1\sqrt{\tau}\\ 
v_2\sqrt{\tau} & v_2\sqrt{1-\tau}\end{array}\right)
\left(\begin{array}{cc}v_3 & 0 \\0 & v_4 \end{array}\right)
\equiv S' \left(\begin{array}{cc}v_3 & 0 \\0 & v_4 \end{array}\right)
\\ \nonumber
\label{eq:3}
\eea

where $v_1, v_2, v_3, v_4\in U(N)$ (in the presence of time--reversal
symmetry $v_1, v_3$ and $v_2, v_4$ are related to each other by
transposition) and $\tau $ is a $N\times N$ diagonal matrix which
collects the coordinates $\{ T_i\} $. As was obvious to begin with,
they are not the radial coordinates corresponding to a Cartan
subalgebra in the symmetric space of the scattering matrix
(cf. paragraph \ref{sec-transfer}).

The same reasoning can be followed in the $\beta =1$ and $4$ cases.
In these cases the cosets which one obtains by imposing the gauge
symmetry discussed above turn out not to be symmetric spaces (paragraph
\ref{sec-jac}).  

We can look at these random matrix ensembles as plain
circular ensembles in which, for well--defined physical reasons, we
have chosen a set of parameters (the eigenvalues of the matrix
$tt^\dagger $) different from the eigenvalues of the matrix $S$ (i.e.,
the radial coordinates of the symmetric spaces associated to the Dyson
circular ensembles). 

Since the standard radial coordinates are related to the Cartan
generators, we can consider this non--standard choice of ``radial''
coordinates as a {\it non--Cartan parametrization} of the symmetric
space \footnote{We thank Martin Zirnbauer for suggesting this
possibility to us.} of the scattering matrix. As we will see in a
moment, the important point is that the construction of this
non--Cartan parametrization follows the same lines as the standard
parametrization (identification of an involutive automorphism,
separation of the generators in even and odd, and so on).  As a
consequence of this, several properties of the standard Cartan
parametrization are conserved, and allow us to treat these ensembles
in a way which is essentially the same as for those which correspond
to standard Cartan parametrization of symmetric spaces (see the
results of~\cite{JalaPB+BarMello,JalP}).

Below we will see in detail how this non-Cartan parametrization is
constructed.  To help the reader we will first briefly review the
general method (a detailed discussion can be found in Appendix A
of~\cite{hz95}) and then look at an example. As we will see, the
procedure has many similarities to the procedure we used in
constructing the restricted root lattice in paragraph
\ref{sec-restricted}, and like in subsection
\ref{sec-realformsSS}, we will use two successive involutions.

\subsubsection{Non-Cartan parametrization of $SU(N)/SO(N)$}

Let us begin with a compact\footnote{The fact that we choose a compact
space for definiteness is not important and we can equally well choose
a non--compact one.}  symmetric space $G/K$ and denote the
corresponding algebra subspace ${\bf G/K}$. Suppose $\sigma $ is the
involutive automorphism that splits the algebra ${\bf G}$ into ${\bf
K}\oplus {\bf P}$. Let's now operate on the subspaces ${\bf K}$ and
${\bf P}={\bf G/K}$ with a second involution $\tau $ exactly like in
paragraph \ref{sec-realformsSS}. Then the subspaces ${\bf K}$ and
${\bf P}$ are split into even and odd parts that were named ${\bf
K_1}$, ${\bf K_2}$, ${\bf P_1}$, ${\bf P_2}$ in subsection
\ref{sec-realformsSS}. (In reference \cite{hz95} these are the
subspaces ${\cal K}_e$, ${\cal K}_o$, ${\cal P}_e$, ${\cal P}_o$.)
Let ${\bf A}$ denote a maximal abelian subalgebra contained in ${\bf
P_2}$. That means its elements are odd under the transformations
$\sigma $ and $\tau $: $\sigma A \sigma =-A$, $\tau A \tau =-A$ for
$A\in {\bf A}$ (i.e. the elements of ${\bf A}$ anticommute with the
involutions).

Let ${\bf M}$ be the subspace of elements of ${\bf K_1}$ that commute
with all the $A\in {\bf A}$.  The subspace of ${\bf G}$ that is
invariant under $\tau $ is ${\bf K_1}\oplus {\bf P_1}\equiv {\bf
G_1}$.  The central ingredient in obtaining the non--Cartan
parametrization is the bijective mapping $\phi $ of the manifold
$G_1/M \times A^+$ into the symmetric space $G/K$ such that $\phi :
(gM,a) \to gaK$, where elements in the coset space are denoted $gK$
(cf. paragraph \ref{sec-cosets}). The non--Cartan ``radial
coordinates'' are encoded in the matrix $a$ belonging to a connected
open subset $A^+$ of $A=\e^{\bf A}$. In the tangent space this
corresponds to a mapping $T_a$ between algebra subspaces \cite{hz95}

\bea
\label{eq:T_a}
T_a : {\bf G_1/M} \times {\bf A} \to {\bf G/K} \nonumber \\
T_a(Z,H)=H+aZa^{-1}|_{\bf P} \\ \nonumber
\eea

where the notation $|_{\bf P}$ means the restriction to the subspace
${\bf P}$.  From the linear mapping $T_a$ we will obtain the Jacobian
as a determinant of the matrix expressing the differential of $\phi $
with respect to the new basis.  Keeping in mind the commutation
relations for symmetric subalgebras (eq.~(\ref{eq:commrel}) of
subsection \ref{sec-Inv}) and the fact that ${\bf A}$ anticommutes with
the two involutions $\sigma $ and $\tau $, we easily see that

\beq
\label{eq:Acommrel}
[{\bf A},{\bf P_1}]\subset {\bf K_2},\ \ \ 
[{\bf A},{\bf K_1/M}]\subset {\bf P_2/A},\ \ \ 
[{\bf A},{\bf K_2}]\subset {\bf P_1},\ \ \ 
[{\bf A},{\bf P_2/A}]\subset {\bf K_1/M}
\eeq

This means that the mapping ${\rm ad}(\ln a)$ from the algebra ${\bf G}$ 
to itself maps the four eigensubspaces of $\sigma \tau $ into each other
as follows 

\bea
\label{eq:Aeigensub}
{\rm ad}(\ln a) : {\bf K_1} \to {\bf P_2}  
\nonumber \\
{\rm ad}(\ln a) : {\bf P_2} \to {\bf K_1} 
\nonumber \\
{\rm ad}(\ln a) : {\bf K_2} \to {\bf P_1} 
\nonumber \\
{\rm ad}(\ln a) : {\bf P_1} \to {\bf K_2} 
\\ \nonumber
\eea

Denoting the adjoint action $aZa^{-1}$ in (\ref{eq:T_a}) with ${\rm
Ad}(a)Z$ and using ${\rm Ad}(a)={\rm exp}\, {\rm ad}(\ln a)= {\rm
cosh}\,{\rm ad}(\ln a)+{\rm sinh}\,{\rm ad}(\ln a)$, we see from
equation~(\ref{eq:Acommrel}) that if $Z=X+Y$ is the decomposition of
$Z\in {\bf G_1/M}$ into parts belonging to the subspaces ${\bf P_1}$
and ${\bf K_1/M}$ respectively,

\beq
T_a(Z,H)=H+{\rm cosh}\,{\rm ad}(\ln a)X-{\rm sinh}\,{\rm ad}(\ln a)Y
\eeq

(this follows from equations (\ref{eq:Aeigensub}) because ${\rm
cosh}x$ is an even and ${\rm sinh}x$ an odd function of $x$, and 
keeping in mind that in $T_a$ we take the projection on ${\bf P}$).

The Jacobian corresponding to the change of coordinates is the
determinant \cite{hz95}

\beq
J_{NC}(a)=
{\rm det} \left( {\rm cosh}\,{\rm ad}(\ln a)|_{{\bf P_1}\to {\bf P_1}}\right) 
{\rm det} \left( {\rm sinh}\,{\rm ad}(\ln a)|_{{\bf K_1/M}\to {\bf P_2/A}}
\right) 
\eeq

and it is obtained as the product of the eigenvalues. The eigenvalues of the
automorphism ${\rm ad}(\ln a)$ are nothing but the restricted roots
with respect to the abelian algebra ${\bf A}$. We therefore obtain
the general formula

\beq
\label{eq:generalJ}
J^{(j)}_{NC}(a) = \prod_{\alpha\in R_1^+} {\rm sinh}^{m_\alpha}\alpha(\ln a)
\prod_{\beta\in R_2^+} {\rm cosh}^{m_\beta}\beta(\ln a) 
\eeq

where the subscript $NC$ stands for ``non--Cartan'' and the index $+$
reminds us of the fact that the sum is over the positive roots only.
$\alpha(\ln a)$ is nothing else than the projection $q^\alpha $
introduced in equation (\ref{eq:q^alpha}) in subsection
\ref{sec-Laplaceop}. The positive roots have been divided into two
subsets $R_1^+$ and $R_2^+$ in an obvious notation. In case ${\bf
G}/{\bf K}$ is a compact space, the roots $\alpha $ and $\beta $ in
(\ref{eq:generalJ}) are purely imaginary. If we then set $\alpha
=i\alpha'$ and $\beta =i\beta'$ we obtain

\beq
\label{eq:general'J}
J^{(+)}_{NC}(a)=\prod_{\alpha'\in R_1^+} {\rm sin}^{m_{\alpha'}}\alpha'(\ln a)
\prod_{\beta'\in R_2^+} {\rm cos}^{m_{\beta'}}\beta'(\ln a) 
\eeq

whereas for real $\alpha $, $\beta $ (non--compact space) we get
hyperbolic functions in eq.~(\ref{eq:generalJ}) for $J^{(-)}_{NC}(a)$
(cf. the similar situation in eq.~(\ref{eq:-to0+})).
For comparison, recall that the Jacobian corresponding to the standard
Cartan parametrization of the $G/K$ space is given by equation
(\ref{eq:J_j}),

\beq
J^{(+)}(a) = \prod_{\alpha\in R^+} \sin^{m_\alpha}\alpha(\ln a)
\eeq

where the algebra corresponding to ${\bf A}$ in this case is a maximal
abelian subgroup of the whole subspace ${\bf P}$.

{\bf Example:} Let us now take as an example the $\beta =1$ S--matrix
ensemble, whose parametrization in subsection \ref{sec-jac} falls
outside of the Cartan classification of symmetric spaces. We will
identify it as a non--Cartan parametrization of a standard symmetric
space.

The starting point in this case is the circular
orthogonal ensemble of Dyson, i.e.  $G/K=SU(N)/SO(N)$. Let us assume
for completeness that we have a different number of left and right
scattering channels $p=N_L$ and $q=N_R$. In this case the involution
$\tau $ is given by $I_{p,q}=I_{N_L,N_R}={\rm diag}(1_{N_L},-1_{N_R})$
(cf. equation (\ref{eq:I_J_})) with fixed point set $G_1=SU(N_L)\times
SU(N_R)$ and the rank of a maximal abelian subgroup $A$ is $r = {\rm
min}(N_L,N_R)$.

We can choose $A$ to be generated by the matrices

       \[
       A_k=i\left( E_{k,N_L+k}+E_{N_L+k,k} \right)\ \ \ \ \
       (k=1,...,r)
       \]

where $E_{ij}$ is the matrix having a $1$ in the $i$th row and $j$th
column, and zeros elsewhere.  After finding the corresponding roots
like in the example in subsection \ref{sec-restricted}, the radial
coordinates $\alpha'\cdot x$ (leaving out the $i$) and the
corresponding root multiplicities turn out to be

       \begin{eqnarray}
       R_1^+ : \ 
       &&(x_k \pm x_l) \quad (m = 1) \quad (k > l)
       \nonumber \\
       &&x_k \quad (m = |N_L - N_R|) \quad (k=1,...,r)
       \nonumber \\
       R_2^+ : \ 
       &&(x_k \pm x_l) \quad (m = 1) \quad (k > l)
       \nonumber \\
       &&2x_k \quad (m = 1) \quad (k=1,...,r)
       \nonumber \\
       &&x_k \quad (m = |N_L - N_R|) \quad (k=1,...,r)
       \nonumber
       \end{eqnarray}

(note that the {\it total} root multiplicities
$m_o=2,~m_l=1,~m_s=2(N_L-N_R)$ are exactly the ones for a $BC_N$ type
restricted root lattice; cf. Table~\ref{tab1}).  These are the radial
coordinates in the Jacobian for the non--Cartan parametrization,
eq.~(\ref{eq:general'J})

\beq
J^{(+)}_{NC}(a)=\prod_{i>j}{\rm sin}(x_i-x_j){\rm sin}(x_i+x_j)
\prod_k {\rm sin}^{\nu }x_k \prod_{l>m}{\rm cos}(x_l-x_m){\rm cos}(x_l+x_m)
\prod_n {\rm cos}(2x_n) \prod_q {\rm cos}^{\nu }x_q 
\eeq

where $\nu \equiv |N_L-N_R|$.  By making now the variable substitution
$T_k=\sin^2 2x_k$ in this Jacobian, we obtain the radial measure in
the form

\beq 
J^{(+)}_{NC}(a) da = \prod_{i<j}|T_i-T_j| \prod_k
|T_k|^{(|N_L-N_R|-1)/2} dT_k ,
\eeq 

which agrees with equation (\ref{jpb3}) for the special case
$N_L=N_R$ and $\beta =1$. As anticipated in paragraph
\ref{sec-jac}, we conclude that equations (\ref{jpb3},\ref{jpb4}) 
represent the Jacobian for a non--Cartan parametrization (albeit
expressed in different variables) of a standard symmetric
space, which for $\beta =1$ is the space $SU(N)/SO(N)$.

\subsection{Clustered solutions of the DMPK equation}
\label{sec-clustered}

As we mentioned previously, there are two major drawbacks in the DMPK
approach to quantum wires. The first is that the DMPK description only
holds in the quasi--one dimensional limit. The second is that the
solution discussed in paragraphs \ref{sec-beta2} and \ref{sec-beta14}
does not allow studying the intermediate cross--over region between
the metallic and the insulating regimes, where no simplifying
approximation is allowed.  This is true even in the simplest case, for
$\beta =2$.

In the last few years various generalizations of the DMPK equation
have been suggested \cite{m2001} to avoid the quasi--one dimensional
limit. However, as we discussed at the beginning of subsection
\ref{sec-classification}, in all these generalized equations most of
the attractive properties of the DMPK equations are lost, mainly due
to the fact that the description in terms of symmetric spaces is no
longer valid.  Only a few pieces of information on the expected joint
probability density of the transmission eigenvalues can be obtained.

Recently a different strategy has been proposed in \cite{cc03} where
the DMPK equation is kept unchanged, but one looks for a set of
special solutions (with non trivial initial conditions) which break
the isotropy ansatz.  To this end one uses the exact integrability of
the Calogero--Sutherland models: recall that the DMPK equation can be
mapped into the evolution operator of a suitably chosen
Calogero--Sutherland model (for a review see \cite{OlshPere}). This is
a highly non--trivial property which is more general than the
underlying symmetric--space structure; in fact it holds also for
generic integer values of the root multiplicities \cite{OlshPere}.  It
is possible to show that as a consequence of their exact
integrability, in these models -- besides the well known symmetric
solution -- a wide class of non--trivial (but exact) solutions exists,
in which the particles are grouped into clusters.  Once the mapping to
the DMPK equation is performed, the clusters of particles become
clusters of eigenvalues. The exact integrability of the
Calogero--Sutherland model ensures that this asymmetric distribution
of eigenvalues survives in the asymptotic limit, and the remarkable
properties of the underlying symmetric space allow us to explicitly
write down such asymptotic expansions.

Let us discuss these solutions in more detail.  We assume the cluster
to be composed of the first $N'<N$ eigenvalues.  This means:

\bea  |x_i-x_{j}|< \infty, &i,j=1,\ldots , N'~~~~(i\not=j).
\\\nonumber  |x_i-x_{j}|\rw \infty,& i=1,\ldots ,
N;~~~j=N'+1,...,N~~ (i\not=j) \label{condcluster}
\eea

In the symmetric space framework we can identify the cluster by
selecting a subsystem of roots associated to the space. Let
$\Pi $ be the system of simple roots associated to the symmetric space
$X$, and $\Pi'$ a subsystem of simple roots which satisfies the
inequality

\beq 
\Pi'=\{\al \in \Pi | \lim_{|x|\rw \infty} x^\al <\infty\}
\eeq

where $x^\al=(x,\al)$.  At this point there are two possibilities.
Since $\Pi $ is a $C_N$ type lattice, $\Pi'$ can be either of type
$C_N$ (in this case it must also contain the long root, and the
ordinary roots must be chosen so as to preserve the $Z_2$ symmetry of
the lattice), or it can be of type $A_N$. In both cases one can
construct, from the ordinary roots of $\Pi'$, the differences
$x^\al=x_i-x_{i+1}$ which correspond to the nearest neighbor distances
between eigenvalues. It follows from the definition of $\Pi'$ that
these distances must remain finite in the asymptotic limit, so that
$\Pi'$ defines a cluster if it is connected, or a set of clusters
otherwise. If the cluster is of type $A_N$, there is no other
constraint and the cluster can in principle flow to an infinite
distance from the origin (while the eigenvalues inside the cluster are
kept at a finite distance from each other). On the contrary, if the
cluster is of type $C_N$, the eigenvalues are bounded by the lattice
structure of $\Pi'$ itself and consequently remain within a finite
distance from the origin.  In the following we denote the radial
coordinates outside the cluster by $\ti x$ and the ones inside the
cluster by $x'$.

The asymptotic expansion of the zonal spherical functions in the 
presence of such a cluster was obtained a few years ago by Olshanetsky in 
\cite{olshanetsky2}. It turns out to be a rather natural generalization of 
the 
Harish--Chandra asymptotic expansion,
eq.~(\ref{eq:ss3}):

\beq
\Psi_k(x)\sim \sum_{r\in W/W'}
c_z(\widetilde{rk})e^{i(\widetilde{rk},\ti x)}\Psi_{(rk)'}(x')
\label{eq:cluster}
\eeq

where $W'$ which appears in the coset $W/W'$ is the Weyl group
associated to the cluster, and $\Psi_k(x)\equiv \xi(x)\Phi_k(x)$
(cf. equation (\ref{eq:ss3})).  The function $c_z(k)$ is defined 
by

\beq 
c_z(k)=\prod_{\al \in R^+/R'^+} c_\al(k)
\eeq

where $R^{'+}$ is the set of positive roots associated to the cluster.
In (\ref{eq:cluster}) $(rk)'$ denotes the
projection of the vector $rk$ on the sublattice $\Pi'$ and
$\widetilde{rk}$ is its complement. The function $\xi (x)$ is given by
eq.~(\ref{eq:xiJ}):

\beq
\xi (x)=\prod_{i<j}|\sinh^2 x_j-\sinh^2
x_i|^{\frac{\beta}{2}}\prod_i|\sinh 2x_i|^{\frac{1}{2}}
\eeq

In spite of its apparent simplicity, the expression (\ref{eq:cluster})
is highly non--trivial.  Notice for instance that the symmetrization
with respect to the Weyl coset $W/W'$ acts not only on the part
containing the coordinates $\ti x$ but also on the momenta of the
zonal spherical function describing the cluster coordinates
$\Phi_{(rk)'}(x')$.  This means that the particles inside the cluster
do not move independently in a section of the whole space but they
``feel'' the presence of the other particles and are subject to the
symmetry group of the remaining space.  We refer the reader to the recent
paper \cite{cc03} for some explicit examples of this type of solutions and
for indications of how they could be used to address the two problems
mentioned at the beginning of this subsection. 

We only mention here that the clustered solutions described in this
section may find another natural application if one tries to model
systems in which the number of open channels is reduced by the
structure of the wire itself (cf. the wide--narrow--wide geometry of
\cite{beenakker3}).  In this case, one could consider a configuration
formed by a $C_N$ type cluster (bounded to the origin) made of $N'$
eigenvalues and let the remaining $N-N'$ eigenvalues flow to infinity.
One can choose $N'$ and $s$ (the dimensionless length of the wire) so as to
keep the cluster in the metallic regime, while the other eigenvalues
are in the insulating regime and do not contribute to the wire
conductance.  

\subsection{Triplicity of the Weierstrass potential}
\label{sec-Weierstrass}

The Calogero--Sutherland potentials of type IV in eq.~(\ref{eq:I-V}),

\beq
v_{IV}(\xi )={\cal P}(\xi ) 
\eeq

where ${\cal P}(\xi )$ is the Weierstrass ${\cal P}$--function, is the 
most general type of potential. The Weierstrass ${\cal P}$--function is defined
as

\beq
\label{eq:Weier}
{\cal P}(z;\omega_1,\omega_2)=\frac{1}{z^2}+\sum_{m,n}'\left( \frac{1}
{(z-2\omega_1m-2\omega_2n)^2}-  \frac{1}{(2\omega_1m+2\omega_2n)^2}\right)
\eeq

where the prime on the sum means we are summing over all pairs $(m,n)\in
{\bf Z^2}$ except $(m,n)=(0,0)$. The ${\cal P}$--function is doubly periodic 
with periods $2\omega_1$, $2\omega_2$, which can be seen by rearranging the 
sum.
The Weierstrass ${\cal P}$--function is an elliptic function. An elliptic 
function is defined to be a single--valued, doubly--periodic analytic function,
whose only singularities in the finite part of the complex plane are poles.   
We will show that as both, or just one, of the two
periods goes to infinity, we recover the potentials of type I, II and III, 
respectively. This fact was mentioned, but not proved, in \cite{OlshPere},
and a proof for the III type potential was briefly indicated in Appendix A 
of \cite{Langmann}. Set

\beq
\label{eq:Vsin}
V(z)=\sum_{m=-\infty}^\infty \frac{1}{4\, {\rm sin}^2\left(
\frac{z+i\beta m}{2}\right)}
\eeq

where $z$ is a complex variable. It is immediately obvious that $V(z)$ is
doubly periodic with periods $2\pi $ and $i\beta $, has a double pole
in each period--parallelogram at $z=2\pi k+i\beta m$ ($k,m\in {\bf Z}$)
and is analytic elsewhere.
The function $V(z)-1/z^2$ is analytic in $z=0$ and in a neighborhood around
this point. Expanding it around $z=0$ we find

\beq
\label{eq:k}
\left[ V(z)-\frac{1}{z^2} \right]_{z=0} = \frac{1}{12}-\sum_{m=1}^\infty 
\frac{1}{2\, {\rm sinh}^2\left( \frac{\beta m}{2}\right)}
\eeq

Let $w_{km}=2\pi k+i\beta m$ denote the position of the double poles of
$V(z)$. Since $V(z)$ is periodic, analytic except in the poles, and has 
double poles in $w_{km}$ it must be of the form

\beq
\label{eq:Vexp}
V(z)=\sum_{k,m}\left[ \frac{c_{-2}}{(z-w_{km})^2}+\frac{c_{-1}}{(z-w_{km})}+c_0
+\sum_{n=1}^\infty c_n(z-w_{km})^n \right]
\eeq

From the condition that $f(z)\equiv V(z)-1/z^2$ is analytic and equal to 
$\sum_{n=0}^\infty f^{(n)}(0)z^n/n!$ in and near $z=0$ we then get $c_{-2}=1$,
$c_{-1}=0$ by matching terms. 

We now see that the two elliptic functions $V(z)$ and 
${\cal P}(z;\pi,i\beta /2)$ have the same periods, poles, and principal 
parts at each pole. They then differ by a constant $k$ (see \cite{EMOT}, 
paragraph~13.11), so that $c_n=0$ for $n>0$ in (\ref{eq:Vexp}). 
So we have proved
that $V(z)-1/z^2+k={\cal P}(z;\pi,i\beta /2)-1/z^2$ where $k$ is defined by 
eq.~(\ref{eq:k}), since both sides have to have the same value at $z=0$:

\beq
\left[ V(z)-\frac{1}{z^2}\right] -\left[ V(z)-\frac{1}{z^2} \right]_{z=0}=
\sum_{k,m}'\left( \frac{1}
{(z-2\pi k-i\beta m)^2}-  \frac{1}{(2\pi k+i\beta m)^2}\right)
\eeq

or in other words

\beq
{\cal P}(z;\pi,i\beta /2) =
\sum_{m=-\infty}^\infty \frac{1}{4\, {\rm sin}^2\left(
\frac{z+i\beta m}{2}\right)} -
\frac{1}{12}+\sum_{m=1}^\infty 
\frac{1}{2\, {\rm sinh}^2\left( \frac{\beta m}{2}\right)}
\eeq

It is now a simple matter to show that in the limit $\beta \to \infty $,

\beq
V(z) \to \frac{1}{4\, {\rm sin}^2\left( \frac{z}{2}\right)}
\eeq

that is, apart from a constant $\frac{1}{12}$, 
${\cal P}(z;\pi,i\beta /2)$ in the limit $\beta \to \infty $ becomes a 
potential of type III in eq.~(\ref{eq:I-V}).

In a completely analogous way one shows that the hyperbolic potential 

\beq
\tilde V(z)=\sum_{m=-\infty}^\infty \frac{1}{4\, {\rm sinh}^2\left(
\frac{z+\alpha m}{2}\right)}
\eeq

is related to the Weierstrass ${\cal P}$--function by 

\beq
{\cal P}(z;\alpha /2,i\pi )=\sum_{m=-\infty}^\infty \frac{1}{4\, 
{\rm sinh}^2\left( \frac{z+\alpha m}{2}\right)} +
\frac{1}{12}-\sum_{m=1}^\infty 
\frac{1}{2\, {\rm sinh}^2\left( \frac{\alpha m}{2}\right)}
\eeq

As $\alpha \to \infty $, the potential $\tilde V(z)$ approaches

\beq
\tilde V(z) \to \frac{1}{4\, {\rm sinh}^2\left( \frac{z}{2}\right)}
\eeq

so that ${\cal P}(z;\alpha /2,i\pi )$ becomes a 
potential of type II in eq.~(\ref{eq:I-V}).
  
Also, if both periods go to infinity, we obtain a potential of type I:

\beq
{\cal P}(z;\alpha /2,i\beta /2 )\to \frac{1}{z^2}\ \ \ {\rm as}\ \ \ 
\alpha ,\beta \to \infty
\eeq

The important consequence of this analysis is that we can see the
triplicity of the symmetric spaces as a limiting procedure on the
general Weierstrass potential. This gives us a framework for
interpolating between spaces of different curvature (at fixed root
lattice), and may have relevant applications in the context of random
matrices. Indeed, in the last few years a substantial amount of
non--trivial mathematical results has been accumulated concerning
these ${\cal P}(z)$ type models~\cite{weierstrass}.  The hope is that
some of them will be useful for constructing non--trivial
generalizations of the known random matrix theories, while preserving
some of the attractive properties due to the underlying symmetric
space structure.

\section{Summary and conclusion}
\label{sec-concl}

In this review we have discussed the usefulness of viewing random
matrix ensembles as symmetric spaces. Random matrix theory, that has
evolved into an important branch of mathematical physics, is used in
the description of physical systems with chaotic behavior, disorder,
or a large number of degrees of freedom. The versatility of random
matrix theories allows for a parameter--free description of an
assortment of systems, ranging in size from nuclei to mesoscopic
conductors.  The unifying feature of these systems is chaotic behavior
effectively resulting in randomness, and the available information is
the universal statistical behavior of spectra.  On the other hand,
symmetric spaces are well--understood mathematical objects that can be
represented as coset spaces $G/K$ of a Lie group $G$ with respect to a
symmetric subgroup $K$ (or as Lie algebra subspaces). After a general
introduction to Lie algebras and root spaces, we have seen how to
construct symmetric spaces from semisimple Lie algebras using
involutions, and how to identify them with the integration manifolds
used in random matrix theory.  In the process we gave concrete
examples of all the mathematical concepts that were introduced. We
discussed coordinate systems on symmetric spaces and identified the
spherical radial coordinates with the physical degrees of freedom in
most matrix models (an exception is the transfer matrix ensembles
where the physically interesting degrees of freedom, in the Landauer
theory, are the transmission eigenvalues, which are not the same as
the eigenvalues of the transfer matrix).  We explained how the Dyson
and boundary indices of the random matrix ensemble are related to the
multiplicities of the restricted roots associated to this symmetric
space.

We have seen that the metric on a Lie algebra, defined in terms of the
adjoint representation, leads to the concept of curvature tensor on
the symmetric space, and we have shown that the symmetric spaces
classfied by Cartan appear in triplets of positive, zero, and negative
curvature corresponding to a given (restricted) root lattice (and
therefore to a given set of multiplicities of long, ordinary and short
roots).  Further we have discussed Casimir operators in a general
context and their representations in local coordinates (the Laplace
operators) and given a general formula for obtaining the radial part
of the Laplace--Beltrami operator in terms of the Jacobian resulting
from a transformation from random matrix to eigenvalue space. We
devoted a subsection and the appendix to their eigenfunctions, the
zonal spherical functions, that were later used in some of the
applications.

In section \ref{sec-RMT} we gave a general introduction to some
commonly used random matrix theories, explaining how universality of
the correlation functions leads to general universal predictions
concerning the statistics of the eigenvalues. The spectral properties
that can be described in this way are exactly those that are
independent of the detailed dynamics of the system. The only input is
the global symmetries.  The fact that many of the physical systems
described by random matrices fall into the universality classes of
Cartan's classification is a consequence of this fact. The eigenvalue
density is determined by the matrix potential and by the Jacobian
obtained in diagonalizing the random matrices on the symmetric
manifold. The Jacobian is the origin of spectral correlations in the
matrix models discussed here. It is completely determined by the
restricted root system of the underlying symmetric space and by its
curvature.

In this review we also discussed the mapping of Calogero--Sutherland
models onto (restricted) root systems of Lie algebras and symmetric
spaces. The mapping is based on  
the fact that the Hamiltonian of these models for certain
values of the coupling constants (determined by the root multiplicities) 
can be exactly transformed into
the radial Laplace--Beltrami operator. This mapping allows to obtain 
several exact results for the zonal
spherical functions of the corresponding symmmetric spaces (see the discussion
in the appendix).

The Dorokhov--Mello--Pereyra--Kumar equation is the differential
equation determining the joint probability distribution of the
transfer matrix as a function of an external parameter.  The operator
appearing in this equation is mapped (by a transformation involving
the Jacobian) onto the radial part of the Laplace--Beltrami operator
representing free diffusion on the symmetric space underlying the
random matrix ensemble. This can be used to solve the Fokker--Planck
equation exactly for Dyson index $\beta=2$, and to obtain approximate
solutions for the probability distribution and the conductance in
other cases using asymptotic expansions of the zonal spherical
functions. As further examples of applications we reviewed how the
connection from random matrices to symmetric spaces can explain a
discrepancy between random matrix theory and the non--linear sigma
model regarding the magnetoconductance, and further how the orthogonal
polynomials and the symmetries of the matrix models can be traced
directly to the root lattice. In the last section we discussed some
applications going beyond the Cartan classification, and we showed that the
Weierstrass potential of Calogero--Sutherland models in its various limits
reflects the possible signs of the curvature of the symmetric space.

An interesting project that might be worth exploring in the same
spirit as above concerns non--hermitean matrix models, i.e. random
matrices having complex eigenvalues. Recently there has been a boost
of activity in this field. Since non--hermitean random matrices are
useful in a number of contexts, but not much is known about them (see
however \cite{Gin,poly,Akem}), a thorough study of them would be
extremely important. A classification according to their symmetries
was attempted in \cite{bl2001}, where 43 spaces were
enumerated. However, a study of the resulting spaces in the spirit of
this work has not yet been performed. Some non--hermitean ensembles
have only recently been applied to physical systems. 

Exploring what is known about the corresponding manifolds, 
and how this knowledge could
be applied in the above mentioned contexts, would appear to be a
worthwhile effort.  We are not aware of the extent to which the
properties of such manifolds are known in mathematics, and how useful
this knowledge might be in the problems involving non--hermitean
random matrix theories. Let us mention a few physical problems in
which non--hermitean random matrices are present:

a) Non--hermitean matrices are important in schematic random
matrix models of the QCD vacuum \cite{schem} at non--zero temperature
and/or large baryon density (i.e. finite chemical potential). The
spontaneous breaking of chiral symmetry is one of the most important
dynamical properties of QCD, as it shapes the hadronic spectrum. The
study of the chirally symmetric phase of QCD is one of the primary
objectives of heavy ion colliders. One of the main problems with
including a chemical potential in the QCD action is that the fermion
determinant becomes complex, which makes lattice simulations extremely
difficult. By using schematic random matrix models, one can study
qualitative features of QCD at finite density and/or temperature.

b) In classical statistical mechanics, non--equilibrium processes can be
studied as the time--evolution of non--hermitean Hamiltonians.  As an
example we mention non--hermitean spin chains related to a
Kardar--Parisi--Zhang equation. Such an equation can describe for
example interface growth, problems in fluid dynamics, a driven lattice
gas, or directed polymers or quantum particles in a random environment
\cite{FEM}.

c) Non--hermitean effective Hamiltonians appear in the S--matrix
description of open systems connected to reservoirs. The S--matrix
characterizes scattering in an open chaotic system like for example a
ballistic microstructure pierced by a magnetic flux \cite{chaos}.

d) Neural networks are described by continuous local variables related
through a non--linear gain function to a local field that could
represent the membrane potential of a nerve cell. The dynamics of the
network is described by a large number of coupled first order
differential equations featuring (random) non--hermitean matrices.
These contain the parameters coupling the output of the $j$'th neuron
to the input of the $i$'th neuron.  In the context of neurobiology,
the study of chaos in neural networks is relevant to the understanding
of several features of neural assemblies \cite{SCS}.
 
e) Anderson localization in a conductor denotes the phenomenon when
the wave function of an electron becomes localized and
exponentially decaying. As a result the conductor becomes an
insulator.  In conventional conductors localization is known to happen
in dimensions $d>2$ ($d=2$ is the critical dimension).  New
interesting phenomena may occur in systems with non--hermitean quantum
mechanics. In \cite{HatNel}, the authors considered particles
described by a random Schr\"odinger equation with an imaginary vector
potential. The model was motivated by flux--line pinning in
superconductors. They found that delocalization transitions arise in
both one and two dimensions. Such models were discussed further in
\cite{201,249,Ef}.

Even if the suggested direction of research concerning non--hermitean matrix
models turns out not to be feasible, there are likely to be many more
applications for hermitean matrix models that can be pursued within
the framework of symmetric spaces.

\vskip 1cm
{\bf Acknowledgments}
This work was partially supported by the 
European Commission TMR program HPRN-CT-2002-00325 (EUCLID).
U.~M. wishes to thank Prof. S. Salamon for discussions.
\newpage
\appendix{}
\section{Appendix: Zonal spherical functions}
\label{app1}
\renewcommand{\theequation}{A.\arabic{equation}}
\setcounter{equation}{0}

A remarkable feature of the theory of zonal spherical functions (ZSF)
discussed in subsection \ref{sec-zonal} is that in some special cases
exact formulae exist for these functions. This happens in particular
for all the symmetric spaces with multiplicity of ordinary roots $m_o=2$
in the classification of Table~\ref{tab1} or Table~\ref{tab3}. The
ultimate reason for this becomes clear by inspection of the mapping
discussed in section \ref{sec-intmod}. The radial part of the
Laplace--Beltrami operator can be transformed into a suitable
Hamiltonian of $n$ interacting particles which decouple exactly for
$m_o=2$.  Thus we may expect to be able to write the ZSF's in these
particular cases as a suitable determinant of single-particle
eigenfunctions.

In the simplest case in which the multiplicity of both the short and long
roots is zero one can write

\eq
\phi_\lambda \propto [\xi(q)]^{-1}\sum_{s\in W}\det s \e^{i(s\lambda,q)}
\label{eq:app1}
\en

where $W$ denotes the Weyl group (which in this case is simply the permutation group, while
$\det s$ is simply the sign
of the permutation $s$), $\lambda$ labels the ZSF and is
related to its eigenvalue (see eq. (\ref{eq:eigen})), $\xi(q)$ is
given by eq. (\ref{eq:xiJ}) and $(\lambda,q)$ denotes the vector
product.   
In the above expression we recognize, as anticipated, the determinant 
$\det e^{\lambda_iq_j}$ written as an
alternating sum over the permutation group $W\equiv S_n$.
 Notice, as a side remark, that if the curvature of the symmetric space is
positive, then $\lambda$ is ``quantized" and must belong to the dual
weight lattice.

This expression fixes the $q$--dependence of the ZSF. However, it does
not completely fix the $\lambda$--dependence, which is hidden in the
proportionality constant. Fixing this constant turns out to be a
highly non--trivial task. In the case of the $A_n$ type spaces it can
be obtained by direct integration of eq. (\ref{eq:sphericalfunctions-})
\cite{OlshPere}. The final result (in which we have introduced
explicitly the parameter $a$ as in paragraph \ref{sec-zonal} so as to
describe the three possible curvatures in one single equation) is

\eq
\phi^{(-)}_\lambda(aq)=\frac{1!\,2!\,...\,(n-1)!\sum_{s\in W}
\det s \,\e^{i(s\lambda,q)}}
{\prod_{j<k}\frac{\lambda_j-\lambda_k}{a} ~\prod_{j<k} \sinh [a(q_j-q_k)]}
\label{eq:app2}
\en
 
The extension to other root lattices is slightly more involved.  Due
to the presence of the long and/or short roots or to the additional
reflection symmetry of the lattice, the corresponding ``single
particle'' Schr\"odinger equations are more complicated and the
solutions are no longer simple plane waves. The problem was solved in
1958 by Berezin and Karpelevick~\cite{bk58}. The solution they
obtained is valid for a generic $BC_n$ lattice and, remarkably enough,
holds for {\it any} value of $m_s$ and $m_l$, not only for those
related to the symmetric spaces (while obviously $m_o=2$ is
mandatory).
 
In this case there is no compact way to deal with the three possible curvatures
in a single formula, so let us look at the three cases separately:
 
{\bf Positive curvature spaces}  

\eq
\phi_\lambda^{(+)}(q)=C_+\frac{ \det\left[\frac{\eta_l!\,\Gamma(\alpha+1)}
{\Gamma(\eta_l+\alpha+ 1)}~P^{\alpha,\beta}_{\eta_l}(\cos2q_j)\right]}
{\prod_{j<k}(\lambda_j^2-\lambda_k^2)~
\prod_{j<k}(\sin^2q_j^2-\sin^2q_k^2)}
\label{eq:app3}
\en

where the $P^{\alpha,\beta}_{\eta_l}$ are Jacobi polynomials with
$\alpha=(m_s+ m_l + 1)/2$, $\beta=(m_l-1)/2$ and
$\eta_l=(2\lambda_l-m_s-2m_l)/4$ (recall that in this case $\lambda$
is quantized, thus $\eta_l$ turns out to be an integer number). The
normalization constant $C$ does not contain any further dependence on
$\lambda$ and $q$. Its value is $C=2^{n(n-1)}
\prod_{l=1}^{n-1}l!\,(m_s+ l)^{n-l}$.

{\bf Negative curvature spaces}

\eq
\phi_\lambda^{(-)}(q)=C_-\frac{ \det\left[F(a_l,b_l,c;-\sinh^2q_j)\right]}
{\prod_{j<k}(\lambda_j^2-\lambda_k^2)~
\prod_{j<k}(\sinh^2q_j^2-\sinh^2q_k^2)}
\label{eq:app4}
\en

where this time $F(a_l,b_l,c;-\sinh^2q_j)$ denotes the hypergeometric
function and the three parameters are :
$a_l=(m_s+2m_l+2i\lambda_l)/4$, $b_l=(m_s+2m_l-2i\lambda_l)/4$ and
$c=(m_s+m_l+1)/2$.  The normalization constant is the same as in the
positive curvature case. This is the formula that we used in paragraph
\ref{sec-beta2} to solve the DMPK equation in the $\beta=2$ case.

{\bf Zero curvature spaces}  

In this case the expression is slightly simpler:

\eq
\phi_\lambda^{(0)}(q)=C_0\frac{ \det\left[(\lambda_lq_j)^{-\gamma}
J_\gamma(\lambda_lq_j)\right]}
{\prod_{j<k}(\lambda_j^2-\lambda_k^2)~
\prod_{j<k}(q_j^2-q_k^2)}
\label{eq:app5}
\en

where $\gamma={(m_s+ m_l-1)/2}$ and

\eq
C_0=\frac{(-1)^{n(n-1)/2}\,2^{n(n+\hat{m}
-3/2)}\,\Gamma(\hat{m}+1/2)\,...\,\Gamma(\hat{m}+n-1/2)}{1!\,2!\,3!\,...\,
(n-1)!}
\en
 
with $\hat{m}=(m_s+m_l)/2$.

\subsection{The Itzykson--Zuber--Harish--Chandra integral}

A remarkable feature of the zero curvature case discussed above is
that not only the zonal spherical functions themselves are simpler,
but also the integral representation from which they are obtained
drastically simplifies. In fact, in this case (recall that $m_o=2$)
the symmetric spaces coincide with the classical Lie algebras and
eq. (\ref{eq:sphericalfunctions-}) can be written as a simple integral
over the group manifold. For instance in the unitary case, i.e. for
systems of type $A_n$, the integral representation becomes
 
\eq
\label{eq:app6}
\phi_\lambda^{(0)}(q)=\int D U e^{{\rm Tr} \left( \Lambda U Q U^{\dagger}
\right)}
\en

where $U\in SU(n)$, $\Lambda$ and $Q$ are diagonal matrices:
$\Lambda={\rm diag}(\lambda_1,...,\lambda_n)$ and $Q={\rm diag}(q_1,...,q_n)$.

On the other hand, if we inspect the $a\to0$ limit of
eq. (\ref{eq:app2}) we find:

\eq
\phi^{(0)}_\lambda(q)=\frac{1!\,2!\,...\,(n-1)!\sum_{s\in W}
\det s \,\e^{i(s\lambda,q)}}
{\prod_{j<k}(\lambda_j-\lambda_k) \prod_{j<k} (q_j-q_k)}
\label{eq:app7}
\en
 
In this last expression we recognize the $\det e^{\lambda_iq_j}$ written as an
alternating sum over the symmetric group $W\equiv S_n$, while the two products
in the denominator can be interpreted as Vandermonde determinants

\eq
\Delta(\lambda) = \prod_{i<j} (\lambda_i-\lambda_j)
\label{vandermond}
\en

Moreover, we easily see that eq. (\ref{eq:app6}) holds unchanged if
$\Lambda$ and $Q$ are two generic hermitean matrices with eigenvalues
$\{\lambda_i\}$ and $\{q_i\}$, respectively. At this point, equating
these two expressions for $\phi^{(0)}$ we find the remarkable
relation:

\eq
\label{eq:app8}
\int D U e^{{\rm tr} \left( \Lambda U Q U^{\dagger} \right)}
=\frac{1!\,2!\,...\,(n-1)!\det \e^{\lambda_iq_j}} {\Delta(\lambda)\Delta(q)}
\en

which is the well--known Itzykson--Zuber--Harish--Chandra (IZHC in the
following) integral.  This result was originally obtained by
Harish--Chandra in~\cite{hc57}. It was later rediscovered in the
context of random matrix models by Itzykson and Zuber~\cite{iz80} and
fully exploited by Mehta in~\cite{m81}. It plays a major role in
several physical applications of random matrix ensembles, ranging from
the exact solution of two--dimensional QCD to the study of parametric
correlations in random matrix theories.
 
It is not difficult, using equation (\ref{eq:app5}), to generalize
eq.~(\ref{eq:app8}) to other Lie groups, i.e. to other root
lattices. This general result was already present in the original
paper by Harish--Chandra~\cite{hc57}. A recent review with detailed
expression for $SO(n)$ (both for even and odd $n$) and $Sp(n)$ can be
found for instance in appendix A of~\cite{bhl99}.

\subsection{The Duistermaat--Heckman theorem}

The remarkable elegance and simplicity of the IZHC integral suggests
that there should be some deep mathematical principle underlying this
result. Indeed, in the last few years it has been realized that the
IZHC integral is a particular example of a wide class of integrals
which may be solved {\it exactly} with the saddle point method
provided one sums over all the critical points (and not only over the
maxima). In the IZHC case this means that one has to sum over all the
elements of the Weyl group (i.e over all the permutations in the
unitary case) and not only over the ones with positive signature (as
one would do in a standard saddle point approximation). 

The rationale behind this remarkable result is the
Duistermaat--Heckman theorem \cite{dh82}. It states that the saddle
point is exact (provided one sums over all the critical points) if the
integration is performed over an orbit with a symplectic structure.
As a matter of fact, it was later recognized that this is only an
instance of a wider class of results which are known as {\it 
localization theorems}. Their common feature is that they can be used
to reduce integrals over suitably chosen manifolds to sums over sets
of critical points. A more detailed dicussion of this beautiful branch
of modern mathematics goes beyond the scopes of the present review. To
the interested reader we suggest reference \cite{bt95}, where a
comprehensive and readable review on the localization formulae and
some their physical applications can be found.

\end{document}